%% file: thesis.tex
\documentclass[12pt,a4paper,oneside]{book}

\usepackage[T1]{fontenc} 

\usepackage[portuguese, english]{babel}

\usepackage{hyphenat}
\usepackage{natbib}
\setcitestyle{round}
\usepackage{graphicx}
\usepackage[table,xcdraw]{xcolor}
\usepackage{amsmath}	
\usepackage{tablefootnote}
\usepackage{amssymb}
\usepackage{titlesec}
\usepackage{placeins}
\usepackage{caption}
\usepackage{multirow}
\usepackage{float}
\usepackage{arydshln}
\usepackage{animate}
\usepackage[colorlinks = true, linkcolor = blue, citecolor = cyan]{hyperref}
\usepackage{lscape}
\addto\captionsportuguese{}
\addto\captionsportuguese{}
\addto\captionsportuguese{}
\addto\captionsportuguese{}
\usepackage{pdfpages}

\defcitealias{Somerville2015}{SD15}
\defcitealias{Naab2017}{NO17}
\defcitealias{Cimatti2020}{CFN20}
\defcitealias{Vogelsberger2020}{V20}

\begin{document}


\newcommand\aap{A\&A}                
\let\astap=\aap                          
\newcommand\aapr{A\&ARv}             
\newcommand\aaps{A\&AS}              
\newcommand\actaa{Acta Astron.}      
\newcommand\afz{Afz}                 
\newcommand\aj{AJ}                   
\newcommand\ao{Appl. Opt.}           
\let\applopt=\ao                         
\newcommand\aplett{Astrophys.~Lett.} 
\newcommand\apj{ApJ}                 
\newcommand\apjl{ApJ}                
\let\apjlett=\apjl                       
\newcommand\apjs{ApJS}               
\let\apjsupp=\apjs                       
\newcommand\apss{Ap\&SS}             
\newcommand\araa{ARA\&A}             
\newcommand\arep{Astron. Rep.}       
\newcommand\aspc{ASP Conf. Ser.}     
\newcommand\azh{Azh}                 
\newcommand\baas{BAAS}               
\newcommand\bac{Bull. Astron. Inst. Czechoslovakia} 
\newcommand\bain{Bull. Astron. Inst. Netherlands} 
\newcommand\caa{Chinese Astron. Astrophys.} 
\newcommand\cjaa{Chinese J.~Astron. Astrophys.} 
\newcommand\fcp{Fundamentals Cosmic Phys.}  
\newcommand\gca{Geochimica Cosmochimica Acta}   
\newcommand\grl{Geophys. Res. Lett.} 
\newcommand\iaucirc{IAU~Circ.}       
\newcommand\icarus{Icarus}           
\newcommand\japa{J.~Astrophys. Astron.} 
\newcommand\jcap{J.~Cosmology Astropart. Phys.} 
\newcommand\jcp{J.~Chem.~Phys.}      
\newcommand\jgr{J.~Geophys.~Res.}    
\newcommand\jqsrt{J.~Quant. Spectrosc. Radiative Transfer} 
\newcommand\jrasc{J.~R.~Astron. Soc. Canada} 
\newcommand\memras{Mem.~RAS}         
\newcommand\memsai{Mem. Soc. Astron. Italiana} 
\newcommand\mnassa{MNASSA}           
\newcommand\mnras{MNRAS}             
\newcommand\na{New~Astron.}          
\newcommand\nar{New~Astron.~Rev.}    
\newcommand\nat{Nature}              
\newcommand\nphysa{Nuclear Phys.~A}  
\newcommand\pra{Phys. Rev.~A}        
\newcommand\prb{Phys. Rev.~B}        
\newcommand\prc{Phys. Rev.~C}        
\newcommand\prd{Phys. Rev.~D}        
\newcommand\pre{Phys. Rev.~E}        
\newcommand\prl{Phys. Rev.~Lett.}    
\newcommand\pasa{Publ. Astron. Soc. Australia}  
\newcommand\pasp{PASP}               
\newcommand\pasj{PASJ}               
\newcommand\physrep{Phys.~Rep.}      
\newcommand\physscr{Phys.~Scr.}      
\newcommand\planss{Planet. Space~Sci.} 
\newcommand\procspie{Proc.~SPIE}     
\newcommand\rmxaa{Rev. Mex. Astron. Astrofis.} 
\newcommand\qjras{QJRAS}             
\newcommand\sci{Science}             
\newcommand\skytel{Sky \& Telesc.}   
\newcommand\solphys{Sol.~Phys.}      
\newcommand\sovast{Soviet~Ast.}      
\newcommand\ssr{Space Sci. Rev.}     
\newcommand\zap{Z.~Astrophys.}       

\newcommand{\sd}{\citepalias{Somerville2015}}
\newcommand{\naab}{\citepalias{Naab2017}}
\newcommand{\cfn}{\citepalias{Cimatti2020}}
\newcommand{\vogel}{\citepalias{Vogelsberger2020}}

\newcommand{\sdt}{\citetalias{Somerville2015}}
\newcommand{\naabt}{\citetalias{Naab2017}}
\newcommand{\cfnt}{\citetalias{Cimatti2020}}
\newcommand{\vogelt}{\citetalias{Vogelsberger2020}}

\newenvironment{simbolos}{%
\refstepcounter{section}
	\pretexto%
	\pdfbookmark[0]{\listsimbname}{listsimbname}
	\begin{singlespace}%
		{\centering\normalfont\normalsize\bfseries\MakeUppercase\listsimbname\\*[\baselsdefault]}%
		\vspace{1cm}
		\begin{supertabular}{lll}
}{
		\end{supertabular}
	\end{singlespace}%
	\pretexto%
}

\newenvironment{abreviaturasesiglas}{%
\refstepcounter{section}
	\pretexto%
	\pdfbookmark[0]{\nomeabreviaturasesiglas}{nomeabreviaturasesiglas}
	\begin{singlespace}%
		{\centering\normalfont\normalsize\bfseries\MakeUppercase\space\nomeabreviaturasesiglas\\*[\baselsdefault]}%
		\vspace{1cm}
		\begin{supertabular}{lll}
}{
		\end{supertabular}
	\end{singlespace}%
	\pretexto%
}



\input{capa}


\input{contra-capa}


\pagenumbering{roman}
\setcounter{page}{2}


\input{folha-de-rosto}


\includepdf[pages=-,offset=25mm 0mm, pagecommand={}]{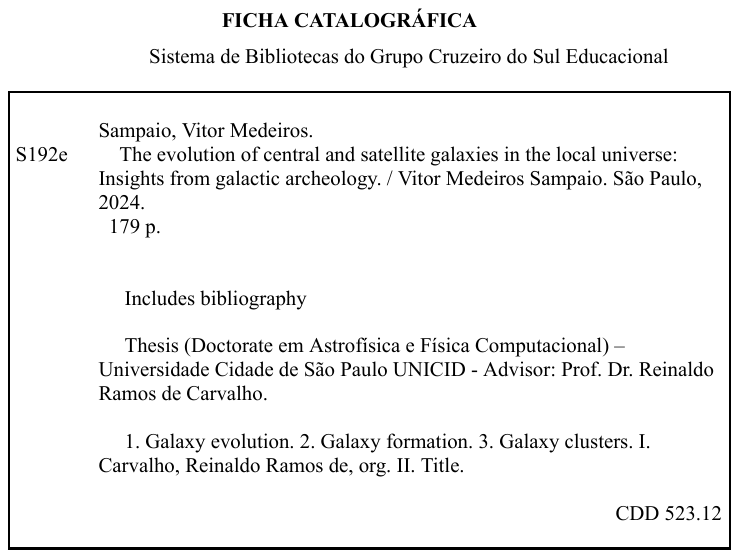} 

\newpage


\includepdf[pages=-,offset=25mm 0mm, pagecommand={}]{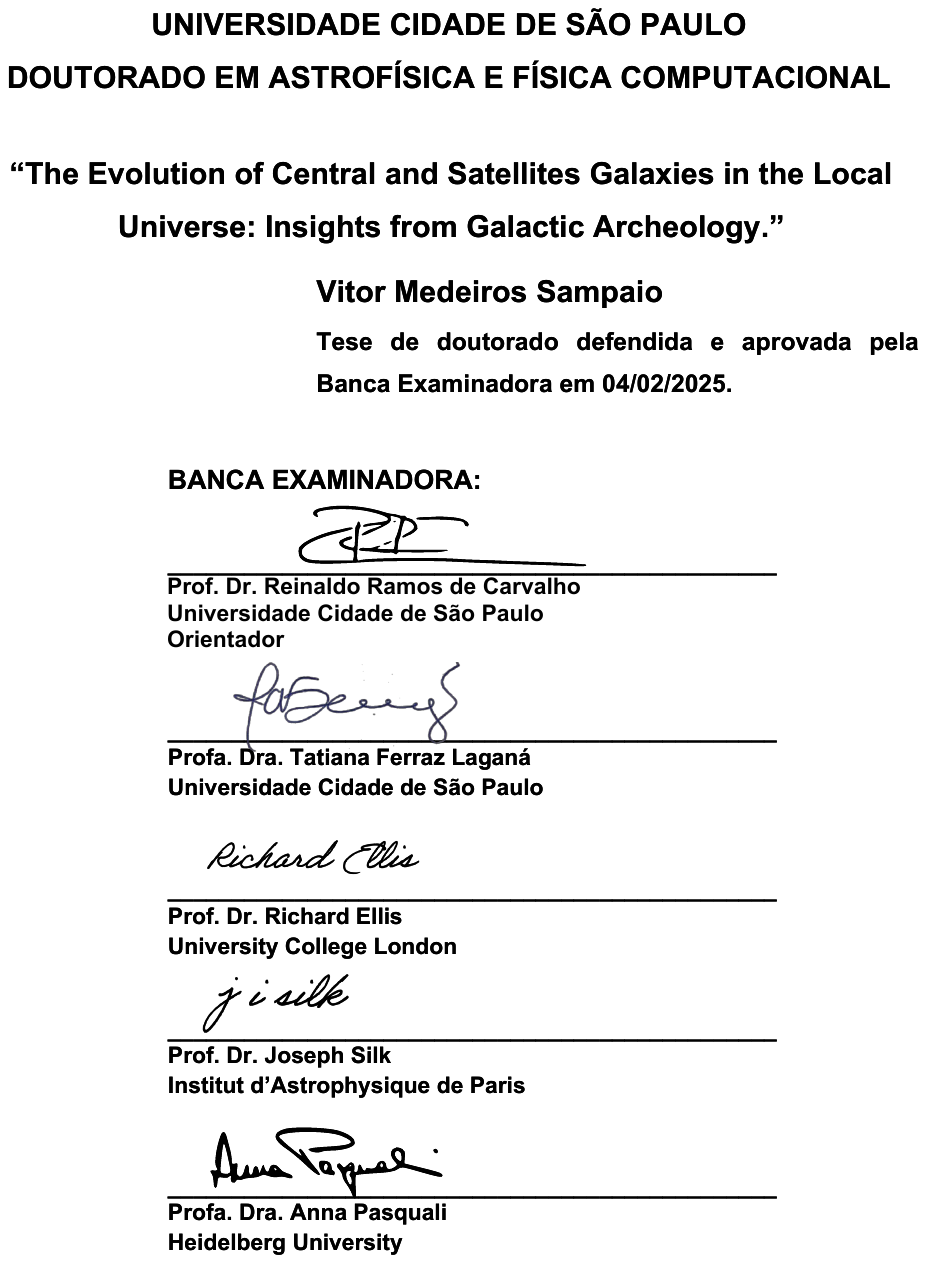}


\input{epigrafe}


\input{dedicatoria}

\input{agradecimentos}


\listoffigures

\thispagestyle{myheadings}


\listoftables

\thispagestyle{myheadings}%


\input{abreviacoes}


\input{simbolos}


\input{constantes}


\tableofcontents

\thispagestyle{myheadings}


\input{resumo_pt}


\input{resumo_en}


\pagebreak
\pagenumbering{arabic}


\input{Chapters/Chapter1}
\input{Chapters/Chapter2}

\input{Chapters/Chapter3}

\input{Chapters/Chapter4}

\input{Chapters/Chapter5}

\input{Chapters/Chapter6}

\input{Chapters/Chapter7}

\input{Chapters/Chapter8}

\input{Chapters/Chapter9}


\input{referencias}

\appendix

\input{Chapters/Appendix}

\end{document}

%% file: capa.tex
\begin{titlepage}
  \begin{center}
    \vspace{\fill}
    \LARGE{Vitor Medeiros Sampaio}
    \vspace{3.0cm}
    \par
    \LARGE{\bf The Evolution of central and satellite galaxies in the local universe: Insights from galactic archeology}
    \par\vfill
    \Large{São Paulo, Dezembro de 2024}
  \end{center}
\end{titlepage}

%% file: contra-capa.tex
\begin{titlepage}
  \begin{center}
    \large{\textsc{Universidade Cidade de São Paulo } \\
           \textsc{Programa de Pós-Graduação} \\ 
           \textsc{em Astrofísica e Física Computacional} \\
          }
    \par\vfill
    \LARGE{Vitor Medeiros Sampaio}
    \par\vfill
    \LARGE{\bf The Evolution of central and satellite galaxies in the local universe: Insights from galactic archeology}
    \par\vfill
    \Large{São Paulo, Dezembro de 2024}
  \end{center}
\end{titlepage}

%% file: folha-de-rosto.tex
\begin{center}
\Large{\textsc{\bf Vitor Medeiros Sampaio}}

\vfill

{\bf The Evolution of central and satellite galaxies in the local universe: Insights from galactic archeology}

\vspace{3.0cm}

\begin{flushright}
\begin{minipage}{0.60\textwidth}

{Tese apresentada ao Programa de Pós-graduação em Astrofísica e Física Computacional da Universidade Cidade de São Paulo, como requisito parcial para obtenção do título de Doutor, sob a orientação do Dr. Reinaldo Ramos de Carvalho.}

\end{minipage}
\end{flushright}

\vspace{3.0cm}

\vfill

São Paulo, Dezembro de 2024.

\end{center}

%% file: epigrafe.tex
\begin{flushright}
\begin{minipage}{1.0\textwidth}

\vspace{7.0cm} 

\itshape ``Bring forward what is true, write it so that it is clear, defend it to your last breath!''

\bigskip
\bigskip

\begin{flushright}
\textit{Ludwig Edward Boltzmann}
\end{flushright}

\end{minipage}
\end{flushright}

%% file: dedicatoria.tex
\begin{flushright}
\begin{minipage}{0.6\textwidth}

\vspace{20.0cm} 

{\it Este trabalho é dedicado ao meu irmão Eduardo.}

\end{minipage}
\end{flushright}

%% file: agradecimentos.tex
\chapter*{Acknowledgements}

\thispagestyle{myheadings}

\noindent First and foremost, I would like to thank my mother (master) Valéria, my father Mário, and my brother Eduardo for never doubting me and always encouraging me to pursue my dream of becoming a renowned astrophysicist. You were the ones who stood by me during my darkest moments, and for that, I am deeply grateful. I also want to express my heartfelt gratitude to my grandmother, Neide, for raising me with so much love and care. You are no longer with us, but I'm sure you would be proud.

I would like to thank my supervisor, Reinaldo, who has believed in my potential as a researcher since my undergraduate years. The path of a researcher is challenging, and along the way, I have learned many valuable lessons. Today, I can confidently say that I have improved significantly. I hope that one day I can become even a fraction of the remarkable researcher you are. I would also like to extend my gratitude to Prof. Sandro Rembold, for all the discussions and critical thinking.

I thank all the collaborators I've worked with so far, Dr. I. Ferreras, Dr. A. Aragón-Salamanca, Dr. M. R. Merrifield, $\rm Dr^{a}.$ L. C. Parker, $\rm Dr^{a}.$ T. F. Laganá, and Dr. A. L. B. Ribeiro. In particular, I appreciate the patience and interest that Prof. Dr. I. Ferreras shows in my work and potential, he was a constant figure in our discussions. A special thanks also to Prof. Dr. A. Aragón-Salamanca and Prof. Dr. M. R. Merrifield, that accepted me as a guest in their institute and supervised me during my doctoral stay, from which I've learned a lot.

I also thank FAPESP for my grant, which enabled me to focus solely on this project. Last, but not least, I need to say a huge thanks to all my friends that supported me throughout this journey. A special thanks to Cris, Luis and Felipe, my best friends for more than 20 years now.

%% file: abreviacoes.tex
\chapter*{Acronyms}

\thispagestyle{myheadings}


BAO   \dotfill   Baryonic Acoustic Oscilation \\
CMB   \dotfill   Cosmic Microwave Background \\
CDM   \dotfill   Cold Dark Matter \\
HDM   \dotfill   Hot Dark Matter \\
CBE   \dotfill   Collisionless Boltzmann Equation \\
IMF   \dotfill   Initial Mass Function \\
SMBH   \dotfill   Super Massive Black Hole \\
IGM   \dotfill   Intergalactic Medium \\
UV   \dotfill   Ultraviolet \\
IR   \dotfill   Infrared \\
ISM   \dotfill   Interstellar Medium \\
AGN   \dotfill   Active Galactic Nucleus \\
ICM   \dotfill   Intracluster Medium \\
RPS   \dotfill   Ram Pressure Stripping \\
CCD   \dotfill   Charge-Coupled Device \\
C   \dotfill   Concentration \\
A   \dotfill   Asymmetry \\
S   \dotfill   Smoothness \\
CNN   \dotfill   Convolutional Neural Network \\
SED   \dotfill   Spectrum Energy Distribution \\
SSP   \dotfill   Single Stellar Population \\
SFR   \dotfill   Star Formation Rate \\
AGB   \dotfill   Asymptotic Giant Branch \\
BPT   \dotfill   Baldwin, Phillips \& Terlevich \\
LINER   \dotfill   Low Ionization Nuclear Emission Regions \\
pPXF   \dotfill   Penalized PiXel-Fitting \\
SFG   \dotfill   Star Forming Galaxies \\
BC   \dotfill   Blue Cloud \\
PG   \dotfill   Passive Galaxies \\
RS   \dotfill   Red Sequence \\
SFMS   \dotfill   Star Formation Main Sequence \\
GV   \dotfill   Green Valley \\
LTG   \dotfill   Late-Type Galaxy \\
ETG   \dotfill   Early-Type Galaxy \\
S0   \dotfill   Lenticular Galaxy \\
NFW   \dotfill   Navarro-Frenk-White \\
YZiCS   \dotfill   Yonsei Zoom-in Cluster Simulation \\
PPS   \dotfill   Projected Phase Space \\
SDSS   \dotfill   Sloan Digital Sky Survey \\
G   \dotfill   Gaussian \\
NG   \dotfill   Non-Gaussian \\
BOSS   \dotfill   Baryon Oscillation Spectroscopic Survey \\
MPA-JHU   \dotfill   Max Planck Institut f\"ur Astrophysik -- John Hopkins University \\
MILES   \dotfill   Medium resolution INT Library of Empirical Spectras \\
Ell   \dotfill   Elliptical \\
Sab   \dotfill   a-b spiral \\
Scd   \dotfill   c-d spiral \\
KIAS-VAGC   \dotfill   Korea Institute for Advanced Study Value-Added Galaxy Catalog \\
NYU-VAGC   \dotfill   New York University - Value Added Catalog \\
BCG   \dotfill   Brightest Cluster Galaxy \\
Mpc   \dotfill   Megaparsec \\
RA   \dotfill   Right Ascension \\
DEC   \dotfill   Declination \\
LOS   \dotfill   Line-of-Sight \\
HD   \dotfill   Hellinger Distance \\
PDF   \dotfill   Probability Density Function \\
FWHM   \dotfill   Full Width at Half-Maximum \\
kpc   \dotfill   kiloparsec \\
B   \dotfill   Bright \\
F   \dotfill   Faint \\
MNRAS   \dotfill   Monthly Notices of the Royal Astronomical Society \\
Perm   \dotfill   Permutation \\
AD   \dotfill   Anderson-Darling \\
BW   \dotfill   Bandwidth \\
IQR   \dotfill   Interquartile Range \\
PNZ   \dotfill   Pasquali New Zones \\
sSFR   \dotfill   specific Star Formation Rate \\
Gyr   \dotfill   Gigayear \\
P+R   \dotfill   Passive+Retired \\

%% file: simbolos.tex
\chapter*{List of Symbols}

\thispagestyle{myheadings}

$L_{\rm CP}$   \dotfill   Characteristic length for the cosmological principle be valid \\
$t$   \dotfill   Time \\
$z$   \dotfill   Redshift \\
$\lambda$ \dotfill Wavelength \\
$\lambda_{\rm observed}$   \dotfill   Wavelength reaching the observer \\
$\lambda_{\rm emitted}$   \dotfill   Wavelength originally emitted by the source \\
$\Lambda$   \dotfill   Cosmological constant \\
$\Delta t$   \dotfill   Time interval \\
$T$   \dotfill  Temperature \\
$H$   \dotfill   Hubble parameter \\
$\rho$   \dotfill   Density \\
$r$   \dotfill   Radial coordinate \\
$\beta$   \dotfill   Anisotropy parameter \\
$\Gamma$   \dotfill   Cooling function \\
$n_{\rm i}$   \dotfill   Ion number density \\
$Z$   \dotfill   Atomic number \\
$n_{\rm e}$   \dotfill   Electron number density \\
$M_{\rm star}$   \dotfill   Star mass\\
$T_{\rm vir}$   \dotfill   Virial temperature\\
$M_{\rm halo}$   \dotfill   Dark matter halo mass \\
$n$   \dotfill   Number density of particles\\
$\phi$   \dotfill   Number density of galaxies with a given stellar mass\\
$M_{\star}$   \dotfill   Stellar mass \\
$M^*$   \dotfill   Characteristic stellar mass for the $\phi$ function\\
$\phi^*$   \dotfill   Normalization constant for the $\phi$ function\\
$\alpha$   \dotfill   Faint-end slope for the $\phi$ function\\
$\Psi$   \dotfill   Cosmic star formation density \\
$\rho_{\rm ICM}$   \dotfill   Intracluster medium density \\
$R_{\rm ICM}$   \dotfill   ICM radial distance from the cluster center \\
$\Sigma_{\rm gas}$   \dotfill   Surface density of atomic and molecular gas \\
$g_{i}$   \dotfill  Maximum restoring gravitational acceleration of the i-th component \\
$g_{DM}$   \dotfill  Maximum restoring gravitational acceleration of the Dark Matter component \\
$g_{\rm d,\star}$   \dotfill  Maximum restoring gravitational acceleration of the stellar disk component \\
$g_{\rm b}$   \dotfill  Maximum restoring gravitational acceleration of the bulge component \\
$g_{\rm HI}$   \dotfill  Maximum restoring gravitational acceleration of the atomic gas component \\
$g_{\rm H2}$   \dotfill  Maximum restoring gravitational acceleration of the molecular gas component \\
$m_{\rm x}$   \dotfill  Apparent magnitude in the broad-band x\\
$F_{\rm x}$   \dotfill  Observed flux in the broad-band x\\
$F_{0}$   \dotfill  Reference flux used to standardize measurements\\
$M_{\rm x}$   \dotfill  Absolute magnitude in the broad-band x\\
$d$   \dotfill  Object distance \\
$I_{\rm Sersic}$   \dotfill  Intensity as a function of radial distance in the case of a Sersic profile\\
$r_{\rm e}$   \dotfill  Effective radius\\
$I_{\rm e}$   \dotfill  Intensity at the effective radius\\
$n_{\rm Sersic}$   \dotfill  Sersic index\\
$r_{\rm e}$   \dotfill  Effective radius\\
$b_{n}$   \dotfill  Constant depending on the value of Sersic index \\
$C$   \dotfill  Concentration \\
$r_{20\%}$   \dotfill  Radius encompassing 20\% of total light\\
$r_{80\%}$   \dotfill  Radius encompassing 80\% of total light\\
$A$   \dotfill  Asymmetry \\
$I_{0}$   \dotfill  Original image \\
$I_{180}$   \dotfill  Image rotated by 180 degrees \\
$B_{0}$   \dotfill  Original background image \\
$B_{180}$   \dotfill  Background image rotated by 180 degrees \\
$I_{\rm S}$   \dotfill  Smoothed image \\
$B_{\rm S}$   \dotfill  Smoothed background image \\
$L({\rm H\alpha})$   \dotfill  $H\alpha$ luminosity \\
$A_{\rm ext}$   \dotfill  Extinction as a function of wavelength \\
$k$   \dotfill  Extinction curve as a function of wavelength \\
$F_{\rm H\alpha}$   \dotfill  Flux of $\rm H\alpha$ \\
$F_{\rm H\beta}$   \dotfill  Flux of $\rm H\beta$ \\
$(F_{\rm H\alpha}/F_{\rm H\beta})_{obs}$  \dotfill Observed ratio between the flux in $\rm H\alpha$ and $\rm H\beta$
$(F_{\rm H\alpha}/F_{\rm H\beta})_{obs}$  \dotfill Intrinsic ratio between the flux in $\rm H\alpha$ and $\rm H\beta$
$F_{\lambda}$   \dotfill  Flux at a specific wavelength $lambda$ \\
$F_{\rm cont}$   \dotfill  Continuum flux at a specific wavelength $lambda$ \\
$F_{\rm obs}$   \dotfill  Observed flux as a function of wavelength \\
$F_{i}^{\rm SSP}$   \dotfill  Flux as a function of wavelength for each SSP \\
$\omega_{i}$   \dotfill  Weight attributed to each single stellar population \\
$\sigma_{\rm galaxy}$  \dotfill  Central velocity dispersion of galaxies \\
$\rho_{\rm NFW}$  \dotfill  Dark matter density for a NFW profile \\
$\rho_{0}$  \dotfill  Characteristic density for a NFW profile \\
$r_{\rm s}$  \dotfill  Scale radius for a NFW profile \\
$\rho_{\rm obs}^{DM}$  \dotfill  Observed Dark matter profile \\
$f_{\rm MB}$  \dotfill  Maxwell-Boltzmann velocity distribution  \\
$D4000$  \dotfill  Break at 4000 \AA  \\
$R$  \dotfill  Resolution  \\
$\Delta \lambda$  \dotfill  Uncertainty in wavelength  \\
$m_{\rm r}$  \dotfill  Apparent magnitude in the SDSS r-band  \\
$EW({\rm H\alpha})$  \dotfill  Equivalent width in $\rm H\alpha$  \\
$Q_{16\%}$  \dotfill  16\% percentile  \\
$Q_{50\%}$  \dotfill  Median -- 50\% percentile  \\
$Q_{84\%}$  \dotfill  84\% percentile  \\
$\mu_{\delta}$  \dotfill  Mean of the uncertainty distribution  \\
$\sigma_{\delta}$  \dotfill  Standard deviation of the uncertainty distribution  \\
$[Z/H]$  \dotfill  Stellar metallicity  \\
$P({\rm Ell})$  \dotfill  Probability of a galaxy being an elliptical  \\
$P({\rm S0})$  \dotfill  Probability of a galaxy being a lenticular  \\
$P({\rm Sab})$  \dotfill  Probability of a galaxy being an ab-spiral  \\
$P({\rm Scd})$  \dotfill  Probability of a galaxy being an cd-spiral  \\
$\nabla(g-i)$ \dotfill (g-i) color gradient \\
$R_{\rm P}$  \dotfill  Petrosian radius  \\
$N_{\rm members}$  \dotfill  Number of galaxies in a given dark matter halo  \\
$M_{\rm r}$  \dotfill  Absolute magnitude in the SDSS r-band  \\
$^{(0,0)}(g-r)$ \dotfill (g-r) colour evolution and k-corrected to z = 0 \\
$^{(0,0)M_{\rm r}}$ \dotfill  Absolute magnitude in the SDSS r-band evolution and k-corrected to z = 0 \\
$z_{\rm c}$  \dotfill  Cluster median/central redshift  \\
$z_{\rm group}$  \dotfill  Group median/central redshift  \\
$\rho_{\rm c}$  \dotfill  Critical density  \\
$R_{200}$  \dotfill  Radius at which the cluster density is 200 times the critical density  \\
$R_{500}$  \dotfill  Radius at which the cluster density is 00 times the critical density  \\
$M_{200}$  \dotfill  Mass encompassed by an sphere of radius $R_{200}$  \\
$N_{200}$  \dotfill  Number of galaxies within an sphere of radius $R_{200}$  \\
$R_{\rm vir}$  \dotfill  Virial radius  \\
$\sigma_{\rm LOS}$  \dotfill  Velocity dispersion along the line-of-sight  \\
$M^{\rm Thresh}_{200}$  \dotfill  Minimum halo mass threshold  \\
$P_{1,2}$  \dotfill  A given discrete distribution  \\
$p_{1,2}$  \dotfill  Probability density function for the discrete distributions  \\
$\pi_{1,2}$  \dotfill  Amplitude for the (1) or (2) Gaussian distribution   \\
$\mu_{1,2}$  \dotfill  Mean for the (1) or (2) Gaussian distribution   \\
$\sigma_{1,2}$  \dotfill  Standard deviation for the (1) or (2) Gaussian distribution   \\
$\delta_{\rm modes}$  \dotfill  Distance between two Gaussian distributions (modes)   \\
$N_{\rm points}$  \dotfill  Number of points in a given distribution  \\
$N_{\rm G}$  \dotfill  Number of cluster with a Gaussian members velocity distribution\\
$N_{\rm NG}$  \dotfill  Number of cluster with a Non-Gaussian members velocity distribution\\
$N_{\rm G + NG}$  \dotfill  Total number of cluster with a reliable gaussianity measurements \\
$R_{\rm proj}$  \dotfill  Sky-projected distance to the cluster center \\
$V_{\rm LOS}$  \dotfill   Velocity along the line-of-sight \\
$t_{\rm inf}$  \dotfill   Infall time/Time since infall \\
$\Delta t_{\rm inf}$  \dotfill   Variation in Time since infall \\
$N_{\rm slice}$  \dotfill   Number of objects in a given slice \\
$N_{\rm bootstrap}$  \dotfill   Number of bootstrap samplings \\
$\Delta SFMS$  \dotfill   Distance from the star formation main sequence\\
$Q_{\sigma}$  \dotfill   Standard deviation estimated using quartiles\\
$\Delta \text{T--Type}$  \dotfill   Variation in T--Type\\
$\sigma_{\rm scatter}$  \dotfill   Scatter of a given distribution\\
$t_{\rm delay}$  \dotfill   Delay time in the slow-then-rapid quenching model\\
$F_{\rm STRM}$  \dotfill   Function parametrizing the slow-then-rapid quenching model\\
$a_{1}$  \dotfill   Slope of the galaxy properties variation prior to $t_{\rm delay}$\\
$b$  \dotfill   Intercept of the galaxy properties variation prior to $t_{\rm delay}$\\
$a_{2}$  \dotfill   Slope of the galaxy properties variation after $t_{\rm delay}$\\
$\Delta t_{\rm inf}^{\rm GV}$  \dotfill   Time-scale that galaxies spent in the green valley\\
$F^{\rm GV}$  \dotfill  Average green valey fraction \\
$a_{2}^{\rm BC}$  \dotfill   Slope of the blue cloud galaxies properties variation after $t_{\rm delay}$\\
$a_{2}^{\rm RS}$  \dotfill   Slope of the red sequence galaxies properties variation after $t_{\rm delay}$\\
$N_{\rm S0}$  \dotfill   Number of lenticular galaxies\\
$N_{\rm S}$  \dotfill   Number of spiral galaxies\\
$F_{\rm S0,S}$  \dotfill   Relative fraction between lenticulars and spirals\\
$\tau_{\rm 50\%}$  \dotfill   Time scale for the quenching or morphological transition of 50\% of the galaxies\\
$R_{20}$  \dotfill   Projected radius comprising 20\% of cluster stellar mass\\
$R_{80}$  \dotfill   Projected radius comprising 20\% of cluster stellar mass\\
$\sigma_{G-NG}$  \dotfill   Definition to define significance level differences between G and NG clusters\\
$\sigma_{i,{\rm G}}^{\chi}$  \dotfill   Errors associated with the coefficient $i$ of the parameters for G clusters\\
$\sigma_{i,{\rm NG}}^{\chi}$  \dotfill   Errors associated with the coefficient $i$ of the parameters for NG clusters\\
$\left < {\rm IR} \right>$  \dotfill Average infall rate  \\
$\left < {\rm IR} \right>_{G}$  \dotfill Average infall rate in Gaussian clusters  \\
$\left < {\rm IR} \right>_{NG}$  \dotfill Average infall rate in Non-Gaussian clusters  \\

%% file: constantes.tex
\chapter*{Physical Constants}

\thispagestyle{myheadings}

${\rm pc} = 3.086 \times 10^{16} $ m \\
$H_{0} = 72 \, {\rm km \, s^{-1} \, Mpc^{-1}}$ -- value adopted in this project. \\
$c = 2.998 \times 10^{8} \, {\rm m \, s^{-1}}$ \\
$G = 6.674 \times 10^{-11} \, {\rm m^{3} \, kg^{-1} \, s^{-2}}$ \\
$h = 6.626 \times 10^{-34} \, {\rm m^{2} \, kg \, s^{-1}}$ \\
$\pi = 3.1415$ \\
$k_{\rm B} = 1.381 \times 10^{-23} {\rm J \, K^{-1}}$ \\
$m_{\rm p} = 1.673 \times 10^{-27} \, {\rm kg}$ \\
$\sigma_{\rm T} = 6.65 \times 10^{-29} \, {\rm m^{2}}$ \\
${\rm M}_{\odot} = 1.989 \times  10^{30} \, {\rm kg}$ \\
\AA$=10^{-10} \, {\rm m}$

%% file: resumo_pt.tex
\chapter*{Resumo}

\thispagestyle{myheadings}

\noindent Esta tese investiga a evolução de galáxias em diferentes ambientes, utilizando dados do Sloan Digital Sky Survey (SDSS). A amostra cobre $0.03 \leq z \leq 0.1$ e magnitudes aparentes mais brilhantes que 17,78, garantindo completude espectroscópica e estimativas confiáveis de populações estelares. As galáxias são classificadas em sistemas de campo ou aglomerados/grupos, subdivididos entre satélites e centrais. Ao analisar o plano de taxa de formação estelar (SFR)–massa estelar, este trabalho identifica diferenças sistemáticas entre o blue cloud (BC), green valley (GV) e red sequence (RS) nos diferentes ambientes. Análises de morfologia e populações estelares mostram que transições de T--Type, metalicidade e idades estelares destacam o papel do esgotamento ambiental. Um novo diagrama de T--Type \emph{vs.} SFR específica indica que a transformação morfológica antecede o esgotamento total da formação estelar. A correlação entre propriedades das galáxias e o tempo desde a incorporação, via espaço de fase projetado, confirma o modelo "atrasado-então-rápido" para galáxias de baixa e média massa, estendendo-o à morfologia. Escalas de tempo para o esgotamento da formação estelar e transições morfológicas também são derivadas em função da massa estelar. Sobre AGNs, este trabalho sugere dois mecanismos para alimentar galáxias centrais: acreção de gás do halo em sistemas isolados de alta massa e interações entre centrais e satélites em pequenos grupos. Um excesso de galáxias Seyfert na GV reforça o vínculo entre AGNs e a supressão da formação estelar, com evidências de que AGNs impulsionam transformações morfológicas. Finalmente, ao explorar como a acreção de galáxias e grupos afeta a dinâmica de aglomerados, este trabalho compara sistemas Gaussianos e não-Gaussianos. Aglomerados não-Gaussianos mostram taxas maiores de infall e mais galáxias tênues recentemente incorporadas, refletindo históricos de montagem mais dinâmicos e oferecendo estimativas inéditas de infall nesses sistemas.

Palavras-chave: Evolução de galáxias; Formação de galáxias; Aglomerados de galáxias.

%% file: resumo_en.tex
\chapter*{Abstract}

\thispagestyle{myheadings}

\noindent This thesis investigates the evolution of galaxies in diverse environments, utilizing Sloan Digital Sky Survey (SDSS) data to explore the impact of environmental richness on central and satellite galaxies across stellar mass ranges, compared to isolated systems. The sample is limited to $0.03 \leq z \leq 0.1$ and apparent magnitudes brighter than 17.78, ensuring spectroscopic completeness and reliable stellar population estimates. Galaxies are categorized by environment as field or cluster/group systems, with further separation into satellites and centrals. By analyzing the star formation rate (SFR)-stellar mass plane, this work identifies systematic differences in the blue cloud (BC), green valley (GV), and red sequence (RS) across environments. Morphological and stellar population analyses reveal that T-type, metallicity, and stellar age transitions highlight the role of environmental quenching. A newly introduced T-Type \emph{vs.} specific SFR diagram provides evidence that morphological transformation precedes full quenching. Correlating galaxy properties with time since infall through projected phase space confirms the delayed-then-rapid quenching model for low- and intermediate-mass galaxies, extending it to morphology. Time-scales for quenching and morphological transitions are also derived as a function of stellar mass. Focusing on AGN activity, this thesis suggests two fueling mechanisms for central galaxies: gas accretion from the halo for isolated high-mass systems, and central-satellite interactions in smaller groups. An excess of Seyfert galaxies in the GV strengthens the connection between AGN activity and star formation suppression, with evidence linking AGN activity to morphological transformation. Finally, this work explores how galaxy and group accretion affect cluster dynamics, comparing Gaussian and non-Gaussian clusters. Non-Gaussian systems exhibit higher infall rates and a larger fraction of faint, recently infalling galaxies, reflecting a more dynamic assembly history and providing the first observationally motivated infall rate estimates for these clusters.

Keywords: Galaxy evolution; Galaxy formation; Galaxy clusters.

%% file: Chapters/Chapter1.tex
\chapter{Introduction}
\label{chap:introduction}

\thispagestyle{empty}

\noindent

Galaxy formation and evolution are based on a series of events happening at different epochs during the universe's cosmic time. The investigation of the evolution of the universe on large scales is the main objective of cosmology, which sets the theoretical framework to understand how galaxies form and evolve. A fundamental hypothesis in cosmology is the cosmological principle \citep[e.g.][]{1980lssu.book.....P}, which posits that the universe on large scales ($L_{\rm CP} \geq 1 {\rm Gpc}$) follows two main characteristics: 1) homogeneity -- the universe has the same energy density, composition, and structure everywhere; and 2) isotropy -- there are no preferred directions in space. The cosmological principle provides a basis for constructing and understanding cosmological models. Furthermore, cosmologists, up to date, identified that our universe is composed by 4 different components \citep[e.g.][]{2008cosm.book.....W}, which are: 1) radiation -- comprising relativistic\footnote{The classification of relativistic particles is made by comparing the rest energy, $mc^{2}$, with the squared linear momentum, $p^{2}$. If $p^{2} \gg mc^{2}$, the particle is said to be relativistic.} particles, mainly photons and neutrinos; 2) baryonic matter -- consists of protons, neutrons, electrons, and the elements they form, and it interacts through gravitational, electromagnetic, weak, and strong nuclear forces; 3) dark matter -- a form of matter that does not emit, absorb, or reflect photons, and only interacts through gravity; and 4) dark energy -- a form of energy that permeates all of space homogeneously and with a constant energy density.  

\section{Conditions for galaxy formation}
The radiation-dominated era set the initial conditions for the formation of large-scale structures by influencing the behavior of dark matter and baryonic matter. Dark matter, interacting only via gravity, began forming potential wells shortly after the Big Bang. In contrast, baryonic matter, coupled with radiation in a photon-baryon plasma, was subject to high photon pressure, preventing gravitational collapse and creating sound waves known as baryon acoustic oscillations (BAO). These oscillations left imprints on the matter distribution. As the universe expanded and cooled, recombination occurred around $z \sim 1100$, forming neutral hydrogen and allowing photons to decouple and propagate freely, observed today as the Cosmic Microwave Background (CMB). The CMB provides a snapshot of density fluctuations that seeded galaxy formation.

Following photon decoupling, density perturbations grew under gravity, with their evolution differing based on the properties of dark matter. Cold Dark Matter (CDM) supported a bottom-up structure formation model, where smaller systems formed first and merged to create larger structures like galaxies and clusters. This contrasts with Hot Dark Matter (HDM), which smoothed small-scale fluctuations and favored a top-down scenario, forming large structures first. Observations of high-redshift small galaxies and clusters strongly support the CDM model.

As density perturbations grew, their evolution was initially governed by linear theory, with perturbations collapsing when their size exceeded the Jeans length. During this phase, density contrasts increased proportionally to the scale factor. Once perturbations became non-linear, their dynamics were described by the collisionless Boltzmann Equation (CBE), evolving into stable structures such as galaxies. This process culminated in a dynamical equilibrium where gravitational forces balanced kinetic motions, forming galaxies and other cosmic structures.

\section{Including baryonic physics}
\label{sec:baryonic_physics}
Baryonic physics plays a vital role in galaxy formation, supplementing gravity with processes such as gas cooling, angular momentum conservation, star formation, and the growth of central supermassive black holes (SMBHs).

\textbf{Cooling Processes and Angular Momentum:} Cooling mechanisms allow gas clouds to lose thermal energy and condense, facilitating structure formation. Radiative cooling through processes like Bremsstrahlung emission, collisional excitation, and molecular transitions (e.g., CO and H2) dominates different temperature regimes. Additionally, angular momentum influences galaxy morphology; high specific angular momentum favors disk formation, while low angular momentum leads to compact spheroidal galaxies.

\textbf{Star Formation:} Collapsing gas clouds fragment into clumps that form stars, governed by the Initial Mass Function (IMF). Nuclear fusion ignites when the core reaches temperatures of $\sim 10^7$ K, producing energy and heavier elements through the proton-proton chain. Massive stars end their lives in supernovae, enriching the interstellar medium with heavy elements and forming a sequence of stellar populations: Population III (pristine gas, no metals), Population II (low metallicity, old), and Population I (metal-rich, young). This sequential enrichment establishes the stellar mass-metallicity relation in galaxies.

\textbf{Central SMBH Formation:} SMBHs, found in galaxy centers, likely form from Population III star remnants or direct collapse of massive gas clouds. These seeds grow via gas accretion and mergers. Accretion disks emit X-rays and UV radiation as gas spirals inward, regulated by the Eddington limit. SMBHs evolve through merging events in hierarchical structure formation, with dense environments accelerating their growth via enhanced gas accretion and dynamical perturbations.

The interplay between cooling, star formation, angular momentum, and SMBH growth shapes galaxy properties and evolution, leaving observable imprints such as the stellar mass-metallicity relation, star formation activity, and central black hole dynamics.

\section{Galaxy evolution across cosmic time}

\label{sec:galaxy_evolution}

Star formation events rely on the availability of gas and its ability to cool to a point where nuclear fusion is triggered. Gas from the intergalactic medium (IGM) flows into galaxies through several channels, including cold mode accretion, where gas enters directly without being significantly heated, and hot mode accretion, where gas is initially heated to the virial temperature \citep{2019igfe.book.....C}, 
\begin{equation}
\label{eq:virial_temperature}
    T_{\rm vir} \sim 36 \left(\frac{M_{\rm halo}}{10^{8} {\rm M}_{\odot}}\right)^{2/3} \left (\frac{1+z}{10}\right) {\rm K},
\end{equation}
of the galaxy's halo mass ($M_{\rm halo}$) and then cools over time. The time-scale needed for cooling can be expressed as the ratio of the thermal energy content of the gas to the cooling rate,
\begin{equation}
    t_{\rm cool} = \frac{\frac{3}{2} n k_{\rm B} T}{n_{\rm e} n_{\rm i} \Gamma(T)},
\end{equation}
where $n_{\rm e}$ and $n_{\rm i}$ are the number density of electrons and ion, respectively, and $\Gamma(T)$ is the cooling function, that quantifies the rate at which gas loses thermal energy per unit volume, and is a function of the gas temperature and metallicity.

The cooling time-scale has a direct impact in the structures observed in the local universe. Namely, due to the relation between virial temperature and halo mass, galaxies have an upper mass threshold, such that the gas can cool in a time-scale shorter than the Hubble time ($1/H_{0}$). The number density of galaxies as a function of stellar mass ($M_{\star}$) is commonly well parametrized by the Schechter function \citep{1976ApJ...203..297S}, written as
\begin{equation}
    \phi(M_{\star}) dM_{\star} = \phi^{*} \left ( \frac{M_{\star}}{M^{*}}  \right)^{\alpha} \exp{\left ( -\frac{M_{\star}}{M^{*}}\right )}\frac{dM_{\star}}{M^{*}}
\end{equation},
where $\phi(M_{\star})$ is the number density of galaxies with stellar mass ($M_{\star}$), $M^{*}$ is the characteristic stellar mass where the function transitions from a power law to an exponential cutoff, $\phi^{*}$ is a normalization constant describing the number density of galaxies at the characteristic mass, and $\alpha$ is the faint-end slope, describing the distribution of low-mass galaxies. 
\begin{figure}[ht]
    \centering
    \includegraphics[width=0.7\columnwidth]{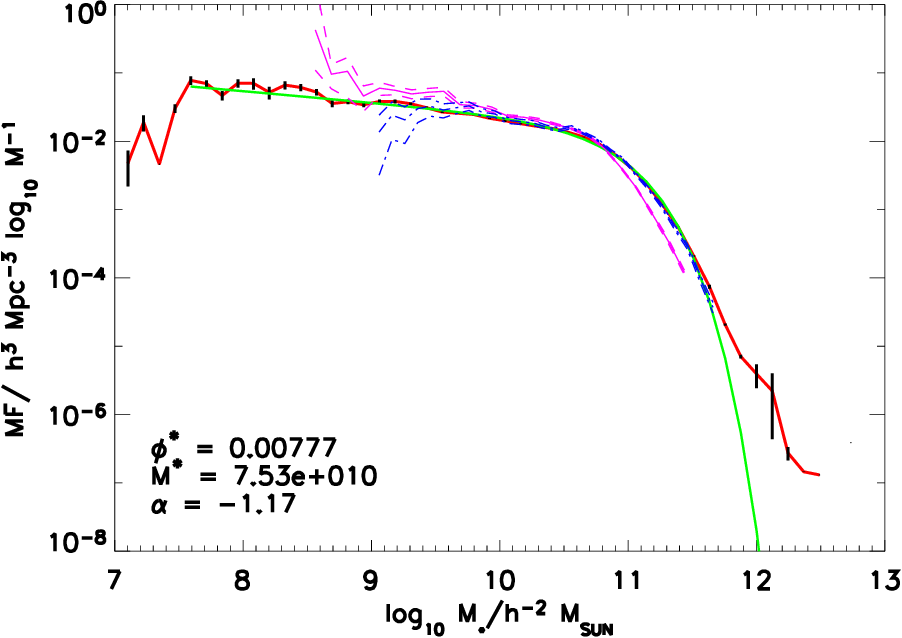}
    \caption{Mass function for the first release of the Sloan Digital Sky Survey (see Chapter \ref{chap:data}), with a Schechter Function fit shown as solid green line. Adapted from \citet{2003ApJS..149..289B}.}
    \label{fig:sdss_schechter}
\end{figure}
Typical values for the local universe ($z \sim 0$) are $\phi^{*} \sim 0.007 \, {\rm Mpc}^{-3}$, $M^{*} \sim 10^{10.7} {\rm M}_{\odot}$, and $\alpha \sim -1.2$. Fig.~\ref{fig:sdss_schechter} shows the number density of galaxies as a function of $M_{\star}$ for the first data release from the Sloan Digital Sky Survey \citep{2003AJ....126.2081A}. The sharp cutoff in the number density of galaxies with stellar mass greater than $\sim 10^{11}$ indicates the characteristic mass scale beyond which the formation of more massive galaxies is less efficient, likely due to the decreased efficiency of gas cooling in more massive halos, and possibly also due to feedback processes (as will be discussed in Chapter~\ref{Chap2_GalArch}).

Additionally, the cooling process depends on the radiative properties of the gas, and, in particular, on its metallicity. Metal-rich gas cools more efficiently due to the presence of metals that facilitate radiative cooling via emission lines. Therefore, due to the evolution of gas properties as a function of time, it follows naturally that the rate at which star formation occurs changes across cosmic time. 

\subsection{The Madau-diagram}

\label{subsec:madau_diagram}

Tracing star formation events requires observations in multiple wavelengths, particularly in the ultraviolet (UV) and the infrared (IR). UV light is directly emitted by young, massive stars, making it a tracer of recent star formation. However, UV light is easily absorbed by dust in the interstellar medium, leading to significant extinction and underestimation of the true star formation rate if only UV observations are used \citep{2000ApJ...533..682C}. Thus, infrared observations complement UV observations by capturing the light re-emitted by dust that has absorbed UV photons. When dust absorbs UV light, it heats up and radiates energy in the infrared. 

\begin{figure}[ht]
    \centering
    \includegraphics[width=\columnwidth]{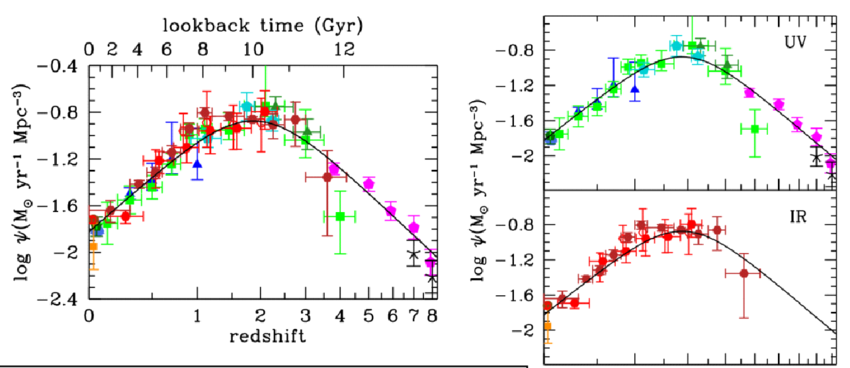}
    \caption{The Madau-diagram, tracing the cosmic star formation density ($\Psi$) as function of redshift (or lookback time, shown on top axis). Right panels show the data points estimated from UV (top panel) and IR (bottom panel) observations. Adapted from \citet{2014ARA&A..52..415M}.}
    \label{fig:madau_diagram}
\end{figure}

The Madau-diagram traces the cosmic star formation density ($\Psi$) as a function of redshift. This diagram is shown in Fig.~\ref{fig:madau_diagram}, and reveals a distinct pattern: the $\Psi$ rises from the early universe to a peak around redshift $z \sim 2-3$, known as ``cosmic noon'', and then declines towards the present day. This peak indicates that the universe was most active in forming stars about 10 billion years ago, a period when galaxies were rapidly converting gas into stars at rates much higher than what we observe today.

The differential cosmic star formation density as a function of redshift has a significant impact on observations in the present-day universe. Following the so-called ``Downsizing scenario'' \citep{1996AJ....112..839C}, most massive galaxies formed their stars earlier and more rapidly than less massive galaxies. During the peak epoch of star formation, massive galaxies experienced intense bursts of star formation, quickly building up their stellar populations. In contrast, smaller galaxies continued to form stars at lower rates over longer periods. The downsizing scenario also implies that different mechanisms regulate star formation in galaxies of different masses. Once a galaxy is formed and star formation is triggered, a series of physical mechanisms can directly affect its evolution, in particular by suppressing star formation events. These mechanisms can be roughly separated into two different categories \citep[e.g.][]{2010ApJ...721..193P}: nature -- that relies mainly on intrinsic properties of the galaxy; and nurture -- that is strongly dependent on the galaxy environment.

\subsection{Internal mechanisms driving galaxy evolution}

\label{subsec:internal_mechanisms}

A major regulator of star formation events are the ``feedback processes'' \citep{2012RAA....12..917S}. In the context of galaxy evolution ``feedback processes'' stands for mechanisms through which energy, momentum, and enriched material are transferred to the interstellar medium (ISM). If the effect of this transfer is an enhanced star formation rate, then the feedback is said to be ``positive'' \citep{2013ARA&A..51..511K}. If it suppresses star formation events, it is named ``negative'' \citep{1986ApJ...303...39D}. Feedback processes are classified according to their origin.

During the life-cycle of stars, different stages include an energetic output to the ISM, from which stellar winds, radiation pressure, and supernova explosions are noteworthy. All these mechanisms together are named ``stellar feedback''. Massive stars, in particular, play a crucial role due to their intense radiation and energetic winds, which can heat and ionize the surrounding gas, creating bubbles and cavities within the ISM. When these stars end their lives in supernova explosions, they release vast amounts of energy, generating shock waves that propagate through the galaxy. These shock waves can compress nearby gas clouds, triggering new episodes of star formation (positive feedback), or conversely, disperse gas, inhibiting star formation (negative feedback). Stellar feedback is thought to play a crucial role in suppressing the formation and growth of structures below a certain mass threshold in the universe. In low-mass galaxies, the gravitational potential is weaker, making it easier for the energy from supernovae to overcome the gravitational binding energy of the gas. As a result, supernova-driven winds can efficiently expel gas from these galaxies, reducing the amount of material available for star formation and effectively quenching star formation in these systems.

An additional form of feedback originates from the central SMBH. The friction and gravitational forces within the accretion disk cause the material to heat up and radiate energy to the ISM. Furthermore, if the SMBH is accreting material, it is said to be ``active'', and it is treated as an ``Active Galactic Nucleus'' (AGN). The distinction comes from the fact that AGNs can produce powerful jets and outflows of charged particles that travel at relativistic speeds. These jets can extend far beyond the host galaxy. Observational evidence support the idea that AGN feedback regulates galaxy growth, preventing the formation of excessively massive galaxies.

Besides feedback processes, other galaxy properties can prevent star formation events. For instance, morphological quenching refers to the suppression of star formation in galaxies due to their structural properties, particularly the presence of a significant central bulge or an elliptical morphology. In disk galaxies, star formation predominantly occurs in the cold, dense gas within the galactic disk. However, when a galaxy develops a substantial bulge or transitions into an elliptical shape, the dynamics and distribution of gas within the galaxy are altered. The increased gravitational potential and velocity dispersion in these morphologically transformed galaxies are dominated by random motions, instead of circular velocity, which stabilizes the gas against collapse and prevents the formation of dense molecular clouds. This stabilization effect reduces the overall star formation efficiency, leading to a quiescent galaxy despite the presence of sufficient gas. Finally, from equation \ref{eq:virial_temperature} it is noticeable the increase in the virial temperature, of which the accreting gas is heated to, and the halo mass. Thus, in massive galaxies, the combination of gravitational heating and AGN feedback effectively quenches star formation, leading to the formation of "red and dead" elliptical galaxies with old stellar populations and little ongoing star formation.

\subsection{Environmental star formation quenching}

\label{subsec:environmental_quenching}

In addition to the internal mechanisms described in the previous subsection, interaction between the galaxy and its host environment can also significantly alter their evolution. The environment of a galaxy can be classified according to its location along the large-scale structure in the universe \citep{1980lssu.book.....P}. In the field, galaxies are located in relatively isolated regions of space, far from the influence of other galaxies. Within the field environment, further distinction can be made regarding voids and filaments. The former is a large, nearly empty regions of space with very few galaxies, that experience minimal gravitational interaction. On the other hand, filaments are networks of galaxies connected by long, thread-like structures of dark matter and gas. Galaxies in filaments are more densely packed than in voids and can experience mild gravitational interactions. When located at the nodes of the dark matter web, the environment of galaxies can be roughly described by their host dark matter halo mass. Galaxies inhabiting halo masses $M_{\rm halo} \leq 10^{13.5} \, {\rm M}_{\odot}$ and dividing the same halo with at most tens other galaxies are said to be member of a ``group''. Above this threshold, galaxies usually share their dark matter halos with tens to thousands other systems, which then defines a galaxy ``galaxy cluster''.

In this regard, galaxy clusters are excellent laboratories for studying how galaxies interact with their environment, as they show a gradient between central and outer regions regions with respect to galactic density and velocity dispersion. Besides galaxies, the in-between member galaxies, the intracluster medium (ICM), is filled with hot gas\footnote{Since these are the most massive structures in the universe, the gas being accreted in the dark matter halo is heated up to extremely high temperatures, following equation \ref{eq:virial_temperature}.}. 

\begin{figure}[ht]
    \centering
    \includegraphics[width=0.7\linewidth]{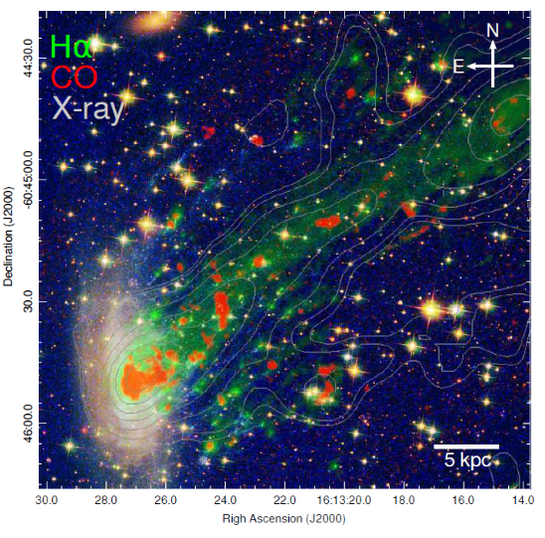}
    \caption{Example of a galaxy suffering ram pressure stripping. The galaxy is infalling towards the center of the Virgo cluster. Green maps the ionized hydrogen, red the star forming regions (through the CO emission) and in white the X-ray emission. Adapted from \citet{2014ApJ...792...11J}.}
    \label{fig:rps_example}
\end{figure}

A primary mechanism of environmental quenching is the so-called ram pressure stripping \citep[RPS,][]{1972ApJ...176....1G}. As a galaxy moves through the hot, dense ICM of a galaxy cluster, the pressure exerted by the ICM can strip away the galaxy's interstellar gas. This process is particularly effective in the outer regions of the galaxy, where the gravitational binding is weaker. For example, Fig.~\ref{fig:rps_example} shows an ionized hydrogen tail being stripped from the system, and star-forming regions (mapped by CO emission) shown in red. Mathematically, the condition for RPS to strip galaxies components can be written as \citep{2019ApJ...873...42R}
\begin{equation}
\rho_{\text{ICM}}(R_{\rm ICM})v^{2} > [g_{\text{DM}}(r) + g_{{\rm d},\star}(r) + g_{\rm b}(r) + g_{\text{HI}}(r) + g_{\text{H2}}(r)]\Sigma_{\text{gas}}(r);
\label{RPE}
\end{equation}
where $\rho_{\rm ICM}$ is the density of the intracluster medium as a function of distance from the cluster center, $R_{\rm ICM}$; $\Sigma_{\rm gas}(r) $ is the surface density of both atomic and molecular gas components as a function of distance from the galaxy center, $r$; and $g_{\rm i}$ is the maximum restoring gravitational acceleration -- i.e. the maximum gravitational acceleration that acts to retain the gas within the galaxy against external forces -- for all system components: Dark Matter (i = DM), Stellar Disk (i = ${\rm d}, \star$), Bulge (i = b), atomic gas (i = HI), and molecular gas (i = H2). By making hypothesis about the density profile of each component, it is possible to define a threshold at which the RPS becomes extremely efficient in removing galaxies gas and suppressing star formation events.

\begin{figure}[ht]
    \centering
    \includegraphics[width=0.7\linewidth]{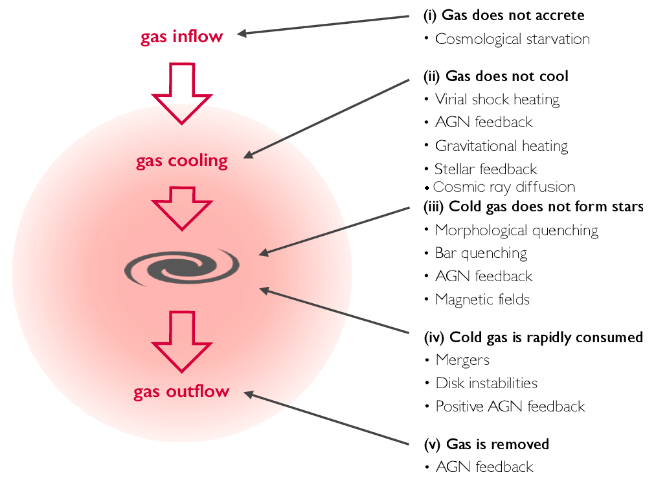}
    \caption{Schematic view of the different mechanisms that can suppress star formation. Adapted from \citet{2018NatAs...2..695M}.}
    \label{fig:sf_suppression}
\end{figure}

Despite RPS being pointed out as the primary mechanism driving star formation suppression in clusters, several other mechanisms can be relevant to galaxy evolution, from which it is noteworthy:
\begin{itemize}
    \item Strangulation \citep{1980ApJ...237..692L} -- the high temperature and pressure of the ICM can also prevent the cooling and condensation of infalling gas, which would otherwise replenish a galaxy's gas reservoir and fuel new star formation. This ``strangulation'' process gradually cuts off the galaxy's supply of cold gas, leading to a suppression of star formation;

    \item Tidal interactions \citep{1999ApJ...514..109G} --  when galaxies within a cluster experience strong gravitational forces from the cluster's overall gravitational potential, these tidal forces can distort a galaxy's shape and create tidal tails. While tidal interactions can trigger bursts of star formation by compressing the gas in a galaxy, they can also remove gas from the galaxy.

    \item Galaxy harassment \citep{1999ApJ...514..109G} -- frequent, high-speed encounters between galaxies in a cluster. These interactions can increase turbulence and heat in the gas, making it more difficult for the gas to cool and collapse into new stars.
\end{itemize}
We summarize all the quenching mechanisms in Fig.~\ref{fig:sf_suppression}. It is important to highlight that different mechanisms act at different stages of the star formation process (stop accretion, prevent cooling, expel gas).

\begin{figure}[ht]
    \centering
    \includegraphics[width=\linewidth]{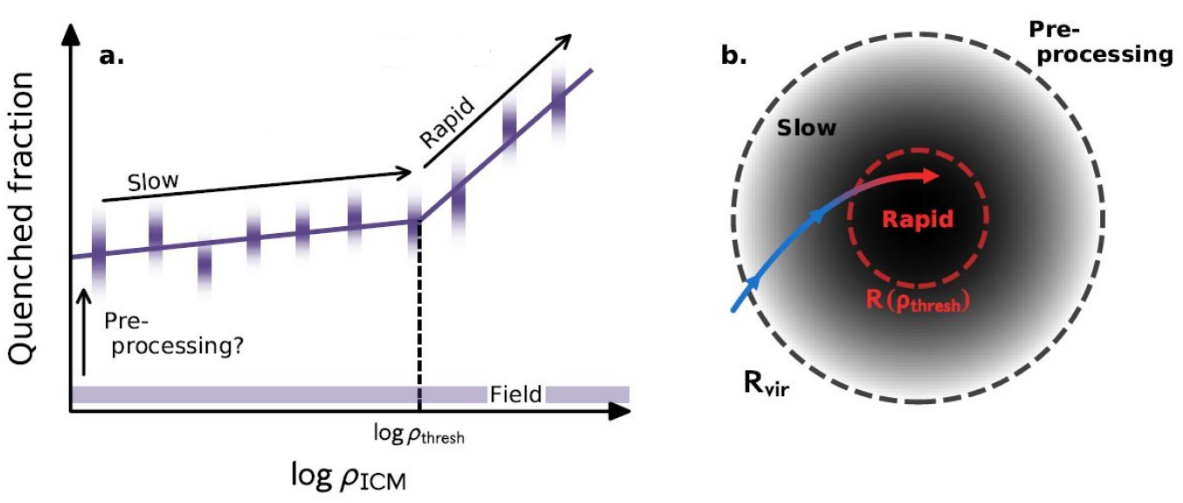}
    \caption{Schematic view of the delayed-then-rapid quenching model. Galaxies experience a delay time prior to star formation supperssion due to environmental effects. Adapted from \citet{2019ApJ...873...42R}.}
    \label{fig:slow_then_rapid}
\end{figure}

The combined effect of all these processes makes the modeling of galaxy evolution in clusters a challenging task, as the main quenching mechanism depends on several properties, including galaxy and host halo mass. The fraction of systems with negligible star formation rate (quenched galaxies) increases with stellar and host halo mass, and decreases with clustercentric distance. Different models tried to simplify the evolution of galaxies within clusters. One of the most successful models is the ``delayed-then-rapid quenching'' model \citep{2013MNRAS.432..336W}, in which a galaxy infalling in a cluster is at first unaffected by the high density environment and is mostly quenched due to starvation. After a delay time, environmental effects, especially RPS, rapidly halt galaxy star formation. This model is illustrated by Fig.~\ref{fig:slow_then_rapid}.

\section{Chapter summary}

\label{sec:chap_1_summary}

In this chapter, we presented the theoretical framework related to galaxy formation and evolution. In $\Lambda {\rm CDM}$ the universe comprises radiation, baryonic matter, dark matter, and dark energy. The Hubble law, which states that galaxy recessional velocity is proportional to distance, supports the expanding universe concept. This expansion suggests a denser, hotter past universe, leading back to the Big Bang singularity. The dynamics of the expanding universe are described by the Friedmann equations, which take into account the energy density and pressure of the universe. Initially dominated by radiation, the universe transitioned through a matter-dominated era before reaching the current dark-energy-dominated epoch.

Understanding the early universe, especially the radiation-dominated era, is crucial to comprehending large-scale structures observed today. Dark matter potential wells formed early, while the photon-baryon plasma within these wells experienced baryon acoustic oscillations, leaving imprints on the matter distribution. As the universe cooled, photons decoupled from matter, leading to the formation of the Cosmic Microwave Background. Galaxy formation theories include the bottom-up scenario, driven by Cold Dark Matter, and the top-down scenario, driven by Hot Dark Matter. Observations favor the cold dark matter model, where small structures form first and merge into larger ones. The linear growth of density perturbations eventually becomes non-linear, leading to the formation of stable structures like galaxies. The evolution of galaxies is influenced by factors such as angular momentum, cooling processes, and star formation, with feedback mechanisms from stars and supermassive black holes further regulating their growth. Environmental effects, particularly in galaxy clusters, also play a significant role in galaxy evolution, with mechanisms like ram pressure stripping and starvation impacting star formation.

%% file: Chapters/Chapter2.tex
\chapter{Observation in astrophysics: the basis for galactic archaeology}\label{Chap2_GalArch}

\thispagestyle{empty}

\noindent

Astrophysics stands out as a distinct branch of science. Unlike other scientific disciplines, astrophysicists cannot conduct controlled experiments. Instead, they rely on real-world observations and statistical inference to unveil the physical processes that shape our universe. 

\section{Insights from photometric observations}

The primordial way of observing celestial objects are through photometry, which is defined as the measurement of the flux or brightness of celestial bodies across different broad band filters covering part of the electromagnetic spectrum (see Fig.~\ref{fig:ugriz_bands} for an example). From the instrumental point of view, photometric measurements are typically made using Charge-Coupled Devices \citep[CCDs,][]{boyle1970charge} -- a highly sensitive semiconductor device used in digital imaging to convert light into electrical signals. The data collected through photometric observations can be used to determine, for a given object, its luminosity, color, size, and distance.

The luminosity of celestial objects is commonly expressed in apparent magnitude \citep{binney2021galactic}, given by
\begin{equation}
    m_{\rm x} = -2.5 \log \left( \frac{F_{\rm x}}{F_{0}} \right),
\end{equation}
where $m_{x}$ is the apparent in the $x$ band, $F_{x}$ is the object flux in the observing band, and $F_{0}$ is a reference flux, in order to standardize measurements. However, since the light flux of a given object depend on distance, it is also defined the ``absolute magnitude'' -- representing the magnitude the object would have if it were placed at a standard distance of 10 parsecs. The relation between apparent and absolute magnitude can be expressed as
\begin{equation}
    M_{\rm x} = m_{\rm x} - 5\log(d) + 5,
\end{equation}
where $M_{\rm x}$ is the absolute magnitude in the $x$ band, and $d$ is the distance to the object in parsecs. This allows for a comparison of the true luminosities of objects, regardless of their distance from the Earth.

\subsection{The galaxy morphology}
\label{subsec:galaxy_mophology}
Galaxy morphology provides critical insights into the processes of galaxy formation and evolution. The Hubble sequence establishes a guideline to classify galaxies into four main morphological types: spirals, lenticulars, ellipticals and irregulars. As presented in subsection~\ref{sec:baryonic_physics}, different morphology may result from different initial conditions of the collapsing gas cloud in the early universe \citep{1969ApJ...155..393P}. Still, the large variability observed in the local universe requires additional evolutionary processes shaping galaxy morphology. For instance, the morphology of a galaxy is closely linked to its formation history and environmental context.  

\begin{figure}[ht]
    \centering
    \includegraphics[width=0.5\linewidth]{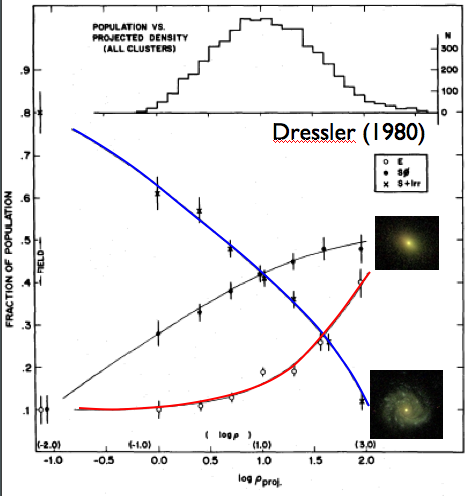}
    \caption{Fraction of galaxies with a given morphology as a function of environmental galactic density. While elliptical and lenticular galaxies are mainly found in high density environments, spiral systems dominate low density fields. Adapted from \cite{Dressler}.}
    \label{fig:md_relation}
\end{figure}

One of the most fundamental observational evidences of the environmental impact in galaxies morphology is the morphology-density relation \citep{Dressler}, which states that galaxies with different morphologies are not uniformly distributed in space. Fig.~\ref{fig:md_relation} shows the fraction of galaxies, divided according to morphological type, as a function of environmental galactic density. Namely, while elliptical and lenticular galaxies are mainly found in high density environments, spiral galaxies dominate low density fields.

However, classifying the morphology of galaxies is not straightforward as it may depend on several factors, from which are noteworthy:
\begin{itemize}
    \item Resolution -- the ability of a telescope to distinguish fine details in an image. Higher resolution allows astronomers to resolve smaller structures within galaxies, such as spiral arms, bars, and nuclear regions. Low resolution can blend these features together, leading to misclassification. For instance, a distant galaxy that might be classified as an elliptical due to unresolved spiral arms could actually be a spiral galaxy;

    \item Flux dimming -- the apparent brightness of an object decreases with the square of the distance from the observer ($\propto (1+z)^{-4}$). As galaxies are observed at higher redshifts (greater distances), their light is significantly dimmed, making it harder to detect and study faint features. This dimming can result in underestimating the star formation rates and missing low-luminosity components, such as faint spiral arms or diffuse outer regions;

    \item Decreasing angular scale with redshift -- This cosmological effect means that more distant galaxies are harder to resolve and study in detail. The decreased angular size of these galaxies makes it challenging to distinguish between different morphological features;

    \item Filter observed -- the choice of filter significantly affects the observation and classification of galaxies. Different filters can reveal various aspects of a galaxy. For example, UV filters highlight regions of active star formation, while IR filters penetrate dust and reveal the older stellar population. The use of specific filters can therefore bias the observed morphology. UV observations might show a galaxy as having more clumpy, irregular structures due to star-forming regions, whereas IR observations might show a smoother, more elliptical appearance.
\end{itemize}

Therefore, in the last decades, different methods have been used to classify galaxies morphology. A first, natural approach is to classify galaxies using visual inspection \citep[e.g.][]{1926ApJ....64..321H, tasca2009zcosmos, 2010ApJS..186..427N}. This method relies on the eye ability to recognize different structures composing galaxies. However, it is a time-consuming method, which makes it non-optimal for large datasets. Projects like the Galaxy Zoo \citep{2011MNRAS.410..166L, 2013MNRAS.435.2835W} have adopted the power of citizen science to overcome this time limitation, allowing the public to participate in the visual classification of galaxies from large surveys. Still, visual classification  can be subjective and labor-intensive, making it difficult to apply consistently across very large datasets.

A different approach relies on modeling the light distribution of a given galaxy using pre-defined function. One example is the Sérsic profile \citep{1963BAAA....6...41S}, which describes the light intensity as a function of radial distance from the galactic center. The profile is expressed as
\begin{equation}
I(r) = I_{\rm e} \exp \left\{ -b_n \left[ \left( \frac{r}{r_{\rm e}} \right)^{1/n} - 1 \right] \right\},
\end{equation}
where $I(r)$ is the intensity at a radius $r$ from the center, $r_{\rm e}$ is the effective radius -- the radius within which half of the total light of the galaxy is emitted, $I_{\rm e}$ is the intensity at the effective radius $r_{\rm e}$, $n$ is the Sersic index, which describes the concentration of the galaxy's light profile, and $b_{n}$ is a constant that depends on the Sérsic index $n$ and is approximately expressed as $b_{n} \sim 2n - 1/3$, for $n > 1$. In particular, the Sérsic profile can model a wide range of galaxy types with different light distributions, with $n\sim1$ for disk galaxies and $n > 4$ for elliptical galaxies. The parametric nature of the Sérsic profile allows for consistent and systematic comparisons between different galaxies. Still, the parametric method have intrinsic limitations due to assuming a pre-defined profile for galaxy light distribution. For instance, the Sérsic profile assumes a smooth and symmetric light distribution. This assumption can break down in the presence of irregularities, such as bars, spiral arms, or tidal features, which are common in real galaxies. As a result, the Sérsic profile might not capture these features accurately \citep{2001MNRAS.326..869T}, leading to oversimplified or incorrect classifications 

An alternative approach to avoid assumptions about the galaxy light profile is the use of non-parametric morphological indices \citep[e.g.][]{2014ARA&A..52..291C}. In the literature, the most common non-parametric system to characterize galaxy morphology is the CAS system \citep{2003ApJS..147....1C}, composed by:
\begin{itemize}
    \item Concentration (C) -- Measures the concentration of light towards the center of a galaxy, and is given by
    \begin{equation}    
        C = 5 \log_{10}\left(\frac{r_{80}}{r_{20}}\right),
    \end{equation}
    where $r_{x}$ is the radius containing x\% of the galaxy's total light;

    \item  Asymmtery (A) -- Asymmetry quantifies the degree to which the light distribution of a galaxy deviates from perfect symmetry. It is calculated by
    \begin{equation}
        A = \frac{\sum |I_0 - I_{180}|}{\sum |I_0|} - \frac{\sum |B_0 - B_{180}|}{\sum |I_0|},
    \end{equation}
where $I_{0}$ is the original image, $I_{180}$ is the image rotated by 180 degrees, and $B_{0}$ and $B_{180}$ are the original and rotated background images, respectively;

    \item Smoothness (S) -- Smoothness, or clumpiness, measures the fraction of light in a galaxy contained in clumpy distributions. It is given by
    \begin{equation}
        S = 10 \left(\frac{\sum |I_0 - I_{\rm S}|}{\sum |I_0|} - \frac{\sum |B_0 - B_{\rm S}|}{\sum |I_0|}\right),
    \end{equation}
    where $I_{\rm S}$ is the smoothed image, and $B_{0}$ and $B_{\rm S}$ are the original and smoothed background images, respectively.
\end{itemize}
Nevertheless, the CAS system have a limited performance when separating elliptical and spiral galaxies (see \ref{sec:JWST_morphology} for a critical view of the CAS system). Furthermore, the interpretation of the indices can be somewhat ambiguous. For instance, high asymmetry might indicate an ongoing merger, but it could also result from other processes such as strong star formation regions or noise in the data. 

In recent times, machine/deep learning algorithms \citep{2018MNRAS.476.3661D, 2020A&C....3000334B, kolesnikov2023unveiling}, particularly convolutional neural networks (CNNs), have become increasingly popular for classifying galaxy morphologies. These algorithms are trained on large labeled datasets to automatically classify galaxies based on their images. Although being highly efficient when dealing with large datasets, these algorithms are ultimately connect to the provided training samples, which can introduce biases and missclassifications if not reliably defined. 

Despite linked to galaxy formation evolution, a reliable classification of galaxy morphology is challenging, which comes from intrinsic evolutionary processes combined with observational limitations. All the methods presented show advantages and disavantages, which need to be balanced to guarantee a robust classification and avoid biases in morphological analysis.

\subsection{Color-Color diagrams}

By combining observations in different broad band filters, it is possible to define ``colors'' of galaxies as the difference between apparent magnitude in two different bands. It follows from stellar theory that different types of stars emit light at different wavelengths, which means their colors vary \citep[e.g.][]{2008ApJ...673..864J}. Young, massive stars emit more blue and ultraviolet light, giving them bluer colors. In contrast, older, cooler stars emit more red light, giving them redder colors. By plotting the colors of a galaxy in a color-color diagram \citep[e.g.][]{2001AJ....122.1861S, 2004ApJ...600..681B}, it is possible to define the dominant stellar population in a given galaxy.

\begin{figure}[ht]
    \centering
    \includegraphics[width=0.5\linewidth]{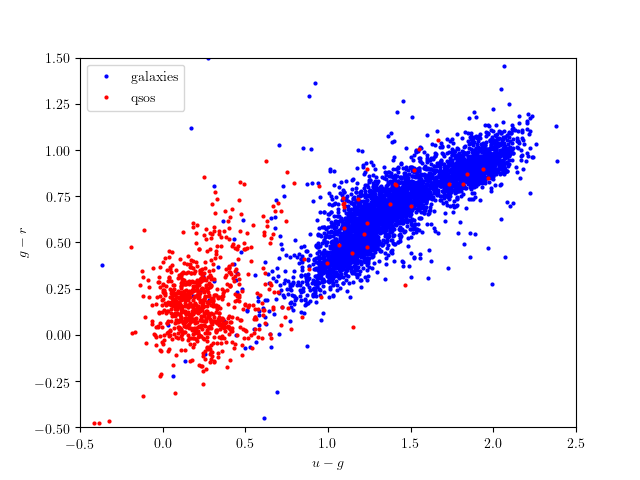}
    \caption{$(u-g)$ vs. $(g-r)$ diagram for a sample of 50,000 galaxies. Blue points denote galaxies, whereas red points denote quasars. Made by the author, following script from the \texttt{astroML} python-package \citep{astroMLText}.}
    \label{fig:color_diagram}
\end{figure}

Color-color diagrams can also be used to understand the effects of dust and extinction in galaxies. Dust in the interstellar medium can absorb and scatter light, causing the observed colors of stars and galaxies to reddening \citep{1994ApJ...429..582C}. This is particularly important for regions of intense star formation, where dust can obscure the light from young stars. Fig.~\ref{fig:color_diagram} shows the $(u-g)$ vs. $(g-r)$ diagram for a sample of 50,000 galaxies, blue colored. Galaxies that fall of the main sequence can be explained by the effects of dust. 

Color-color diagrams are fundamental in identifying quasars and active galactic nuclei (AGNs), which are distinguished by their unique spectral energy distributions. Quasars, a highly luminous and distant subset of AGNs, exhibit a significant UV excess that stellar emissions alone cannot explain \citep[e.g.][]{2004AJ....128.2603Y}. This UV excess positions quasars in a distinct region of color-color diagrams, making them easily identifiable. Specifically, the $(u-g)$ vs. $(g-r)$ diagram, illustrated in Fig.~\ref{fig:color_diagram}, effectively separates quasars from stars and galaxies. In this diagram, quasars stand out due to their prominent presence in the UV filter and a sharp drop-off in the redder filter, effectively highlighting their distinct spectral characteristics.

\subsection{Spectral energy distribution fitting}

Color-color diagrams provide a valuable means of analyzing the properties of stars and galaxies by comparing their colors in different wavelength bands. However, more information can be assessed when considering not only two pairs, but the magnitudes in all the available bands. In a first approximation, broad band filters can cover the emission of galaxies in a wide wavelength range. Since in the case of galaxies their luminosity is commonly dominated by the continuum emission of their stellar population, the combination of a large number of broad band observations can be used to infer, in a first approximation, stellar properties \citep[e.g.][]{2011Ap&SS.331....1W, 2013ARA&A..51..393C}.

\begin{figure}[ht]
    \centering
    \includegraphics[width=0.7\linewidth]{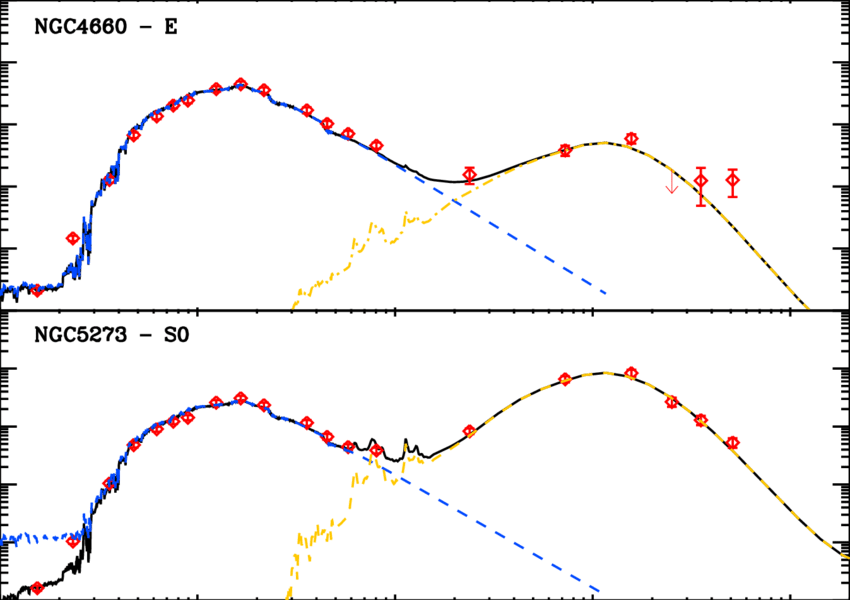}
    \caption{Example of SED fit done using the CIGALEMC code. For each galaxy, the data are observations in 10 different bands. The panels show the observations (red diamond), the best fit (black solid line), the stellar component not absorbed by dust (blue dashed line), and the dust emission model (orange dotted dashed line). Adapted from \citet{2014ApJ...783..135A}.}
    \label{fig:SED_fitting}
\end{figure}

This process, called spectral energy distribution (SED) fitting, is illustrated \citep{2014ApJ...783..135A} in Fig.~\ref{fig:SED_fitting} for two galaxies (NGC4660 - elliptical, and NGC5273 - lenticular/S0). However, the connection between magnitudes in different bands with stellar components heavily rely on modeling stellar evolution, dust absorption and even AGN activity in galaxies. 

In particular for the stellar component, this is done by considering ``single stellar populations'' (SSPs), defined as a set of stellar templates with known properties (age, metallicity, temperature, mass, gravity, etc.) and enable the comparison of observations with the expected when a star with given properties is present within the galaxy. An extended description of stellar population models is out of scope for this text. In a few words, these models \citep[e.g.][]{1999ApJS..123....3L, 2003MNRAS.344.1000B, 2005MNRAS.362..799M, 2006MNRAS.371..703S} incorporate libraries of stellar spectra, evolutionary tracks, and IMFs to predict the fluxes at various wavelengths and as a function of time. The models account for factors like star formation history, chemical enrichment, and dust attenuation.

SED fitting offers a more detailed insights into the stellar populations of galaxies in comparison to color-color diagrams, allowing the identification of young, hot stars that dominate the UV part of the spectrum, as well as older, cooler stars that contribute more to the infrared. Dust within galaxies can be inferred from characteristic absorption features and thermal emission in the infrared. By detailing the stellar population, SED fitting enables estimates of the current star formation rate, stellar mass, average stellar age, and metallicity. Additionally, this fitting process accounts for the displacement of spectral features due to redshift \citep{2000A&A...363..476B}, making it an efficient -- despite also having a large associated uncertainty -- method to infer the redshift of galaxies.

\section{Information encoded in a galaxy's spectrum}

Spectroscopy offers a complementary approach to photometric observations, although more time-consuming. By dispersing light from celestial objects into its constituent wavelengths, spectroscopy allows for detailed analysis of light intensity across different wavelengths. This is crucial for understanding the composition of celestial objects because it enables the identification of specific transitions as predicted by quantum mechanics. Analyzing absorption and emission lines in spectra allows for the determination of the elements present in stars and galaxies, their ionization states, and their kinematic properties. Moreover, spectroscopy provides one of the most reliable methods for measuring redshifts in galaxies and quasars, which is essential for determining their distances.

\subsection{Imprints of stellar population on spectrum features}

Stellar populations leave distinct imprints in the spectra of galaxies, which can be observed, for instance, as absorption and emission lines \citep[e.g.][]{2002ApJS..142...35K, 2008ApJ...681.1183K}. Emission lines are produced when gas is ionized by high-energy photons and then re-emits light as the ions recombine or transition to lower energy states. In particular in the optical wavelength range, common emission lines include those of hydrogen (${\rm H}{\alpha}$, ${\rm H}{\beta}$), oxygen ([O II], [O III]), nitrogen ([N II]), and sulfur ([S II]). 

These lines are typically associated with regions of active star formation, such as HII regions around young, massive stars. Among different approaches \citep{1998ARA&A..36..189K, 2004MNRAS.351.1151B}, measurement of the ${\rm H}{\alpha}$ flux stands out as a common way to estimate the current star formation rate (SFR) of a galaxy. The specific $H\alpha$ ($\lambda = 6563$\AA) line follows from the ionization of neutral hydrogen by energetic photons emitted by young massive blue stars. To convert the observed ${\rm H}{\alpha}$ flux to an SFR, empirical calibrations rely on assumptions about the IMF and the relationship between the ${\rm H}{\alpha}$ luminosity and the total ionizing photon production rate. A commonly used calibration is the Kennicutt law \citep{1959ApJ...129..243S, 1998ApJ...498..541K}, given by:
\begin{equation}
    \text{SFR} ({\rm M_{\odot} \, yr^{-1}}) = 7.9 \times 10^{-49} \times L({\rm H}\alpha)({\rm erg \, s^{-1}}), 
\end{equation}
where $L({\rm H\alpha})$ is the $\rm H\alpha$ luminosity in $\rm erg \, s^{-1}$. This method, however, also suffer from limitations. ${\rm H}{\alpha}$ flux can be significantly affected by dust extinction, which can lead to underestimations of the SFR if not properly corrected. 

A way to correct the SFR estimate rely on the ratio between $\rm H\alpha$ and $\rm H\beta$ ($\lambda = 4861$\AA) fluxes \citep[e.g.][]{1989agna.book.....O}. Under ideal, dust-free conditions, the ratio of $\rm H\alpha$ to $\rm H\beta$, known as the intrinsic Balmer decrement, is well-determined by atomic physics and is typically around $2.86$ for case B recombination at a temperature of $10,000$ K and electron density of $100 \, {\rm cm}^{-3}$. Dust in the interstellar medium preferentially absorbs and scatters shorter wavelength (bluer) light more than longer wavelength (redder) light. This means that $\rm H\beta$ is more affected by dust extinction than $\rm H\alpha$. Thus, any deviation from 2.86 indicates the presence of dust. The amount of dust extinction can be quantified using the difference between the observed and intrinsic Balmer decrements. The extinction at $\rm H\beta$, $A({\rm H}{\beta})$, can be calculated using the following relation:
\begin{equation}
A(\text{H}\beta) = \frac{2.5}{k(\text{H}\beta) - k(\text{H}\alpha)} \log \left( \frac{(F_{\text{H}\alpha}/F_{\text{H}\beta})_{\text{obs}}}{(F_{\text{H}\alpha}/F_{\text{H}\beta})_{\text{int}}} \right),
\end{equation}
where $k(\lambda)$ is the value from the extinction curve at a given wavelength; $(F_{\text{H}\alpha}/F_{\text{H}\beta})_{\text{obs}}$ is the observed flux ratio of H$\alpha$ to H$\beta$; $(F_{\text{H}\alpha}/F_{\text{H}\beta})_{\text{int}}$ is the intrinsic flux ratio of H$\alpha$ to H$\beta$ (2.86).

On the other hand, absorption lines are formed when photons emitted by a hot, underlying continuum source, such as a star or an active galactic nucleus, are absorbed by cooler gas along the line of sight. In the optical spectrum, the most prominent absorption lines are the following \citep{2000AJ....120..165T, 2003MNRAS.346.1055K}:
\begin{itemize}
    \item Balmer Series (e.g., H$\beta$ at 4861 \AA, H$\gamma$ at 4340 \AA, H$\delta$ at 4102 \AA) are produced by transitions of electrons from higher energy levels to the n=2 level. The Balmer lines are prominent in the spectra of young, hot stars and thus serve as indicators of recent star formation. In addition, the strength and profile of these lines can provide information on the age distribution of stellar populations;

    \item The Calcium H and K Lines (3968 \AA\ and 3934 \AA) arise from singly ionized calcium (Ca II) and are among the strongest absorption features in the optical spectra of many galaxies, following from old stellar populations. The Ca H and K lines are useful for probing the metallicity and age of the stellar population, as they are prominent in the spectra of F, G, and K-type stars, which are older and more metal-rich.

    \item Mgb Triplet (5167 \AA, 5173 \AA, 5184 \AA) lines are strong in the spectra of K and G giants and are used to study the metallicity and kinematics of the stellar populations;

    \item The Fe I Lines (e.g., 5270 \AA, 5335 \AA) are indicators of the iron abundance in a galaxy's stellar population. They are prominent in the spectra of cooler stars (G, K, and M types) and provide detailed information about the chemical composition and enrichment history of the galaxy.

\end{itemize}

In galaxies with minimal star formation, the atmospheres of cool stars (F, G, K, and M types) produce strong calcium and iron absorption lines around 4000 \AA \citep{1994ApJS...94..687W}. Additionally, young stars emit most of the blue energetic photons, so their absence results in reduced photon flux in the blue part of the spectrum. Together, these effects create the "D4000 break", a noticeable decrease in continuum flux around 4000 \AA. The strength of the D4000 break is thus a reliable indicator of the average age of the stellar population, reflecting the presence of older stars and the lack of young ones. Notably, galaxies can also show a break at 3646 \AA, the Balmer break, which arises from the termination of the hydrogen Balmer series and is prominent in young stellar populations with A-type stars, making it a key feature in distinguishing post-starburst galaxies from continuously star-forming systems.

\subsection{Diagnostic diagrams: combining permitted and forbidden emission lines}
\label{subsec:diagnostic_diagrams}
As presented in Section~\ref{sec:baryonic_physics}, the interstellar medium can be ionized due to photons originated by permitted or forbidden transitions. Permitted lines are mainly due to high-energy photons generated by young stars, thus serving as a proxy for star formation. The forbidden lines, mostly from ionised metals, originate mainly in collisionally excited states, which are usually related to the presence of AGN activity, traveling shocks, or the effect of evolved stars during a short, yet very hot and energetic phase known as post asymptotic giant branch \citep[post-AGB, e.g.][]{2019igfe.book.....C}. By combining permitted and forbidden emission lines, it is possible to differentiate between various ionization mechanisms within galaxies. 

\begin{figure}[ht]
    \centering
    \includegraphics[width=\linewidth]{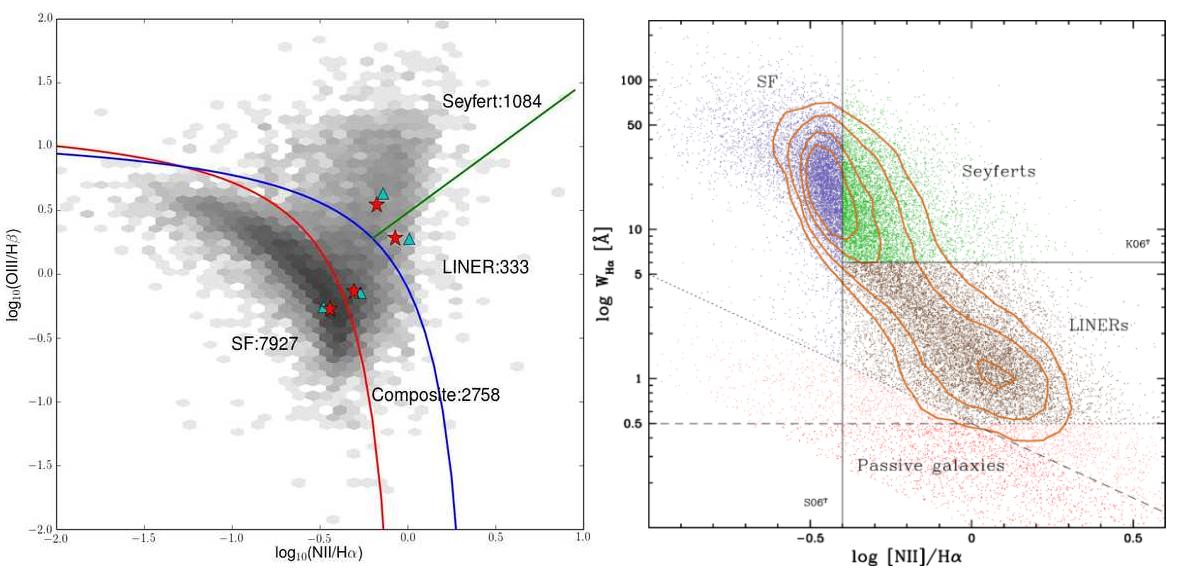}
    \caption{Example of diagnostic diagrams. Left: the BPT diagram. Right the WHAN diagram. Adapted from \citet{2018MNRAS.474.1873W} and \citet{2011MNRAS.413.1687C}, respectively.}
    \label{fig:diagnostic_diagrams}
\end{figure}

One of the most widely used diagnostic tools is the Baldwin, Phillips \& Terlevich (BPT) diagram, which typically utilizes the ratios of [OIII]$\lambda 5007$/H$\beta$ and [NII]$\lambda 6583$/H$\alpha$ \citep{1981PASP...93....5B}. Where both denominators are permitted lines, whereas both numerators are forbidden lines. By plotting these ratios, as shown in the left panel of Fig.~\ref{fig:diagnostic_diagrams}, the BPT diagram can effectively separate galaxies into distinct regions. The line demarcating the region where the dominant ionisation mechanism is star formation can be  derived theoretically as the upper limit for the line ratios that can arise in gas ionised by an starburst \citep{2001ApJ...556..121K}. A second line is defined observationally by \cite{2003MNRAS.346.1055K} separating systems whose interstellar gas is mainly collisionally ionized -- the intermediate region enclosed by these two lines contains galaxies with `composite' spectra, such that their interstellar gas is ionized by a combination of both radiative and collisional mechanisms. Finally, galaxies above and to the right of the \cite{2001ApJ...556..121K} line are further divided into two different classes,  Seyferts and Low Ionization Nuclear Emission Regions (LINERs), using another line defined by \cite{2001ApJ...556..121K}.

Recent studies have shown that, while the emission lines of Seyfert galaxies are well explained by the presence of a strong AGN, there is considerable uncertainty -- and some controversy -- on the origin of the emission lines observed in LINERs, as its ionization may follow from high-energy photons from post-AGB stars. The WHAN diagram \citep{2006MNRAS.371..972S, 2010MNRAS.403.1036C, 2011MNRAS.413.1687C}, shown in right panel of Fig.~\ref{fig:diagnostic_diagrams}, is built to address this complexity. This diagram is built using the same x-axis of the BPT diagram and the equivalent width of $\rm H\alpha$\footnote{The equivalent width (W) is a measure of the strength of an absorption or emission line in a spectrum and is given by \citep{2019igfe.book.....C} $\text{W} = \int \left( 1 - \frac{F_{\lambda}}{F_{\text{cont}}} \right) d\lambda
$, where $F_{\lambda}$ is the flux at a specific wavelength $\lambda$ within the line and $F_{\rm cont}$ is the flux of the continuum at the same wavelength.} ($W_{\rm H\alpha}$) in the y-axis. The separation between LINERs and passive/retired galaxies -- defined as galaxies with no AGN activity and negligible star formation rate -- is built considering the maximum ionization in $H\alpha$ that can be explained by photons from old, hot stellar population.

\subsection{Addressing the star formation history of a galaxy}

Despite the effectiveness of SED fitting in estimating star formation rates and stellar mass using galaxy magnitudes across various broad bands, this method relies solely on the continuum emission information from the stellar population of galaxies. To gain deeper insights into both the gas and stellar components, absorption and emission lines in galaxy spectra must be analyzed. Assessing a galaxy's past history using present-day observations requires identifying the imprints of different gas and stellar components within the host system spectrum. A more refined approach to estimate average stellar age and metallicity is the spectrum fitting technique \citep{2005MNRAS.358..363C, 2012ascl.soft10002C, 2017MNRAS.472.4297W}. This technique is based on the principle that the photospheres of stars at various evolutionary stages, ages, metallicities, and temperatures absorb photons at specific wavelengths. Consequently, different stellar populations leave distinct absorption line intensities in the observed spectrum of a galaxy. The spectrum fitting technique assumes that a galaxy's spectrum can be decomposed into a linear combination of SSPs, each representing stars of a particular age and metallicity. 
\begin{figure}[ht]
    \centering
    \includegraphics[width = \textwidth]{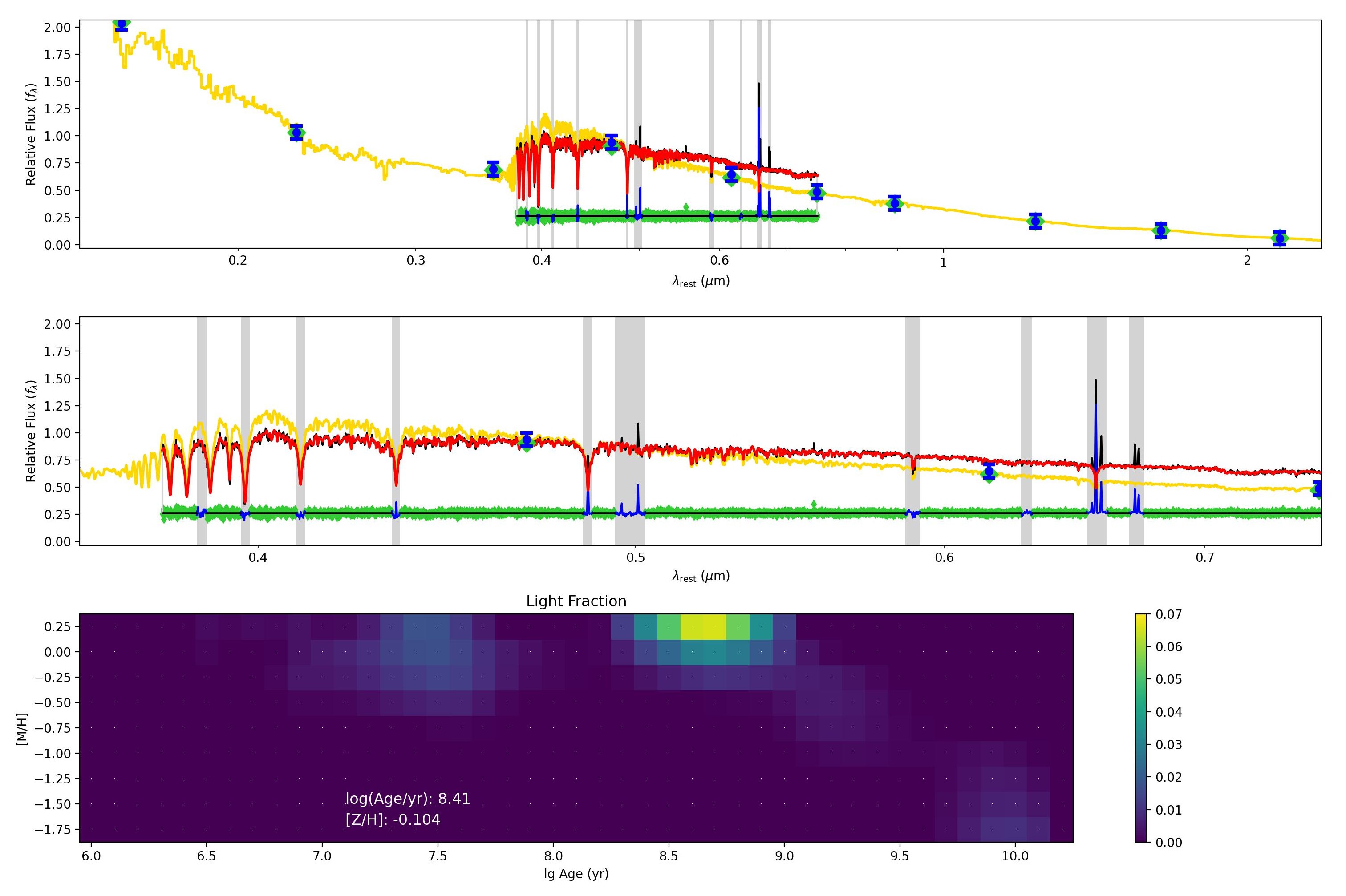}
    \caption{Example of the output of pPXF spectrum fitting code for a galaxy. Top row shows the spectrum (black line), broadband flux (blue points), fitting result considering only spectrum (red line) and adding photometric constraints (yellow line). Second row shows a zoom-in version of the region with spectrum information. Last row: distribution of light weights in the age vs metallicity diagram for the best fitting solution. Made by the author.
 }
    \label{fig:ppxf_example}
\end{figure}
In a mathematical form,
\begin{equation}
    F_{\text{obs}}(\lambda) = \sum_i w_i F^{\text{SSP}}_{i}(\lambda),
\end{equation}
where $F_{\text{obs}}(\lambda)$ is the observed galaxy flux as a function of wavelength $\lambda$, $w_i$ are the weights attributed to each SSP, and $F^{\text{SSP}}_{i}(\lambda)$ the flux as a function of $\lambda$ of each SSP. Therefore, by adopting a sufficiently complete set of SSPs, it is possible to estimate how many stars were formed as a function of lookback-time (mass assembly history) of galaxies by considering the weights given to each SSP. 

In a more refined version of the spectrum fitting technique, it is also possible to include photometric constraints from broad-band observations. An example of fitting procedure combining both spectrum and broad-band photometry using the penalized PiXel-Fitting (pPXF) code \citep{2023MNRAS.526.3273C} is shown in Fig.~\ref{fig:ppxf_example}.

Comparison between the two upper panels reveals that, although broadband photometry offers a significantly wider wavelength coverage than spectra, it primarily constrains the continuum emission, or the overall "shape" of the spectral energy distribution. In contrast, spectrum observations, despite having smaller coverage, contain crucial information about the properties of stellar components through imprints in absorption lines at specific wavelengths. This enables the tracing of stars with specific ages and metallicities, thereby constructing an age vs. metallicity grid, as illustrated in the bottom panel of Fig.~\ref{fig:ppxf_example}. In this grid, each pixel is colored according to the fraction of a given Single Stellar Population (SSP) used in the fit. By averaging these weights, it is possible to derive more robust light and/or mass\footnote{Since observations are made regarding flux, the translation from light to mass requires an additional step of assuming a light-to-mass ratio. This can further increase the complexity of the problem as this is not a well set parameter in astrophysics and may vary case by case.}-weighted stellar ages and metallicities.

Furthermore, since the result of the spectrum fitting technique contains the fraction of stars with different ages. It is possible to derive a first estimate of the mass-assembly history -- defined as the percentage of mass formed as a function of look-back time, and the star formation rate evolution -- defined as the time derivative of the mass assembly history. However, this procedure has several caveats, from which it is important to highlight: 1) when connecting the fraction of stars with a given age to look-back time, there is implicit assumption that the IMF does not vary with redshift, making every epoch comparable; 2) spectrum is limited in time resolution, such that imprints of recent star formation episodes can dominate observations and obscure previous events, which end up as missed information; and 3) the results from spectrum fitting can vary significantly when selecting different sets of SSPs -- in particular there are just a few SSPs of very old stars ($Age \geq 10$ Gyr).

\subsection{The persistent bimodality}

\begin{figure}[ht]
    \centering
    \includegraphics[width=\linewidth]{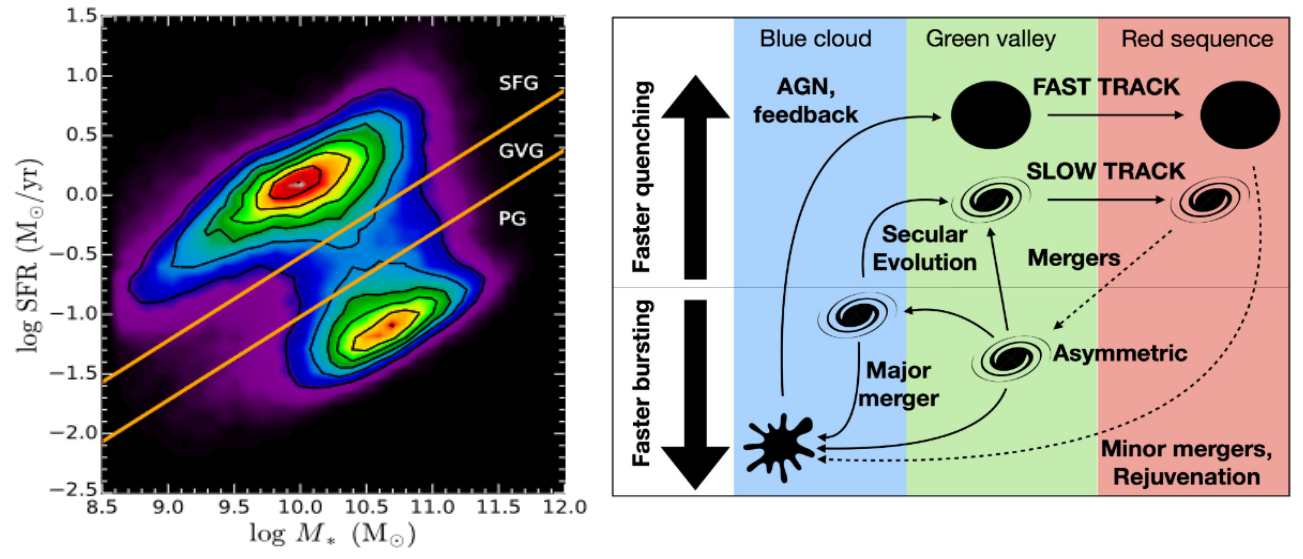}
    \caption{Left: density map of galaxy distribution in the star formation main sequence diagram. The diagonal lines divide the BC, GV and RS regions. Right: schematic view on how different mechanisms can affect the path of galaxy from the BC to the RS. Adapted from \citet{2020MNRAS.491.5406T} and \citet{2022MNRAS.509.3889D}, respectively.}
    \label{fig:sfms_diagram}
\end{figure}

When applied to galaxies in the local universe, all the aforementioned methods reveals a persistent bimodality in galaxies properties \citep[e.g.][]{Strateva1, 2020MNRAS.491.5406T}. The morphology-density relation, presented in subsection \ref{subsec:galaxy_mophology}, is a first indication that galaxies can be broadly divided into two major groups. The relation between star formation rate and stellar mass, shown in the left panel of Fig.~\ref{fig:sfms_diagram}, highlights that galaxies are mainly divided into two distinct regions: 1) one dominated by star forming galaxies (SFG), also called Blue Cloud (BC), comprising systems with ongoing star formation events; and 2) a set of passive galaxies (PG) with negligible star formation rate, defining the Red Sequence region (RS). 

Within the Blue Cloud, it is noticeable an tight correlation between star formation rate and stellar mass, with star formation rate increasing with stellar mass \citep{2007ApJ...660L..43N, 2014ApJS..214...15S,2014ApJ...795..104W}. This relation can be approximated by a linear relation, defining the star formation main sequence (SFMS). This sequence reveals that galaxies grow steadily by forming stars at rates proportional to their mass. The SFMS shows little scatter, suggesting that star formation in most galaxies is a relatively stable and self-regulated process. 

Furthermore, the two regions (BC and RS) are connected by an intermediate region, filled with galaxies that show reduced star formation rates, but haven't reached the passive stage yet. The connection between the two regions itself is an important evidence that galaxies evolve across cosmic time. In other words, the connection suggests that galaxies are progressively evolving from the BC to the RS. This intermediate -- sometimes hard to define \citep{2007ApJS..173..293W, 2015MNRAS.450..435S, 2019MNRAS.488L..99A}-- region is named ``Green Valley'' (GV).

The dichotomy between BC and RS in the star formation rate vs stellar mass diagram -- referred to as SFMS diagram -- is also extended to other properties, creating a well defined separation between systems experiencing different evolutionary stages. The BC is dominated by late-type galaxies (LTGs -- spirals), with young stellar population and low metallicity. On the other hand, the RS comprises mainly early-type galaxies (ETGs -- ellipticals + S0s), with old and high metallicity stellar population. The GV is a mixture of different morphologies, with some excess of S0 galaxies, characterized by intermediate age and metallicity.

The transition from the BC to the RS is characterized by the suppression of star formation rate. The different mechanisms presented in subsections \ref{subsec:internal_mechanisms} and \ref{subsec:environmental_quenching} can be responsible for driving a galaxy to a quiescent state. However, they do not act separately and what is observed is a mixture of different mechanisms driving the galaxy evolution. The right panel of Fig.~\ref{fig:sfms_diagram} show an schematic view of different mechanisms transitioning galaxies from the BC to the RS. Disentangling the different mechanisms is challenging, as most of them are related to dissipative process -- which means that energy (and therefore information) is lost. 

An example of success \citep{1998AJ....115.2285M, 2000ApJ...539L...9F, 2000ApJ...539L..13G, 2002ApJ...574..740T, 2009ApJ...698..198G} is the relation between central SMBH mass ($M_{\rm BH}$) and galaxies velocity dispersion ($\sigma_{\rm galaxy}$). This relation shows an increasing SMBH mass with velocity dispersion, thus suggesting a co-evolution of SMBHs and their host galaxies. The relation follows from a connection between velocity dispersion of the bulge stars and the galaxy's gravitational potential, which is also related to the mass of the SMBH \citep{2016MNRAS.463.3948D, 2022PhR...973....1D}. This relation also supports the idea that AGN feedback plays a critical role in shaping the evolution of massive galaxies.

\section{Combining simulations and observations}

The evolution of computers has profoundly transformed astrophysics through the use of simulations \citep[e.g.][]{2015ARA&A..53...51S, 2020NatRP...2...42V}. Early computational models were limited by the processing power and memory capacity of computers. However, the development of supercomputers and advancement of parallel processing techniques have enabled astrophysicists to create accurate simulations of cosmic phenomena. Cosmological simulations can model the behavior of vast numbers of particles and the interactions of physical forces over cosmic timescales, providing insights into the formation and evolution of galaxies, star clusters, and the large-scale structure of the universe. For example, projects like the Eagle Simulation \citep{2015MNRAS.450.1937C, 2015MNRAS.446..521S}, Millennium Simulation \citep{2005Natur.435..629S, 2006astro.ph..8019L} and the Illustris Project \citep{2014MNRAS.444.1518V, 2015A&C....13...12N} allow studies on the formation of galaxies and dark matter distribution. Thus, simulations provide additional insights when interpreting observational data.

\subsection{N-body simulations to trace dark matter profiles}

One of the key insights gained from simulations is the development of density profiles that describe the distribution of dark matter within halos. Assuming that dark matter particles interact only through gravity, N-body simulation are a sufficient tool to describe dark matter profiles within halos. The Navarro-Frenk-White (NFW) profile results from large-scale simulations \citep{1996ApJ...462..563N, 1997ApJ...490..493N, 2001ApJ...554..903K}, and describes the dark matter density as a function of radius $r$ from the center of the halo:
\begin{equation}
\rho_{\rm NFW}(r) = \frac{\rho_0}{\frac{r}{r_s}\left(1 + \frac{r}{r_s}\right)^2},
\end{equation}
where $\rho_{0}$ is a characteristic density and $r_{\rm s}$ is a scale radius. The NFW profile suggests that dark matter density increases steeply towards the center of the halo and falls off as $\rho_{\rm NFW} \propto r^{-3}$ at large radii, indicating a ``cuspy'' inner region and a gradually declining outer region.

Simulations have demonstrated that the NFW profile is not only applicable to individual galaxy halos but also to the larger cosmic structures. However, the ``cuspy'' -- $\rho_{NFW} \propto r^{-1}$ -- nature of the NFW profile in inner regions contrasts with ``cored'' profiles -- $\rho_{obs} \propto {\rm cte.}$ -- observed in dwarf galaxies. This discrepancy suggests that baryonic processes, such as feedback from star formation and supernovae, can alter the distribution of dark matter.

\subsection{Insights on the dynamical evolution of clusters: the projected phase space}

From a theoretical perspective, the dynamical evolution of a physical system can be described by phase space, a 6-D space constructed using three positional and three velocity coordinates. In some cases, such as for an ideal gas, the dynamical equilibrium is represented by a well-defined velocity distribution of particles, known as the Maxwell-Boltzmann distribution \citep{1860HMPS....1..148M, boltzmann1970weitere},
\begin{equation}
f(v) = \left(\frac{m}{2 \pi k_B T}\right)^{3/2} \exp \left(-\frac{m v^2}{2 k_B T}\right),
\end{equation}
where $f(v)$ is the distribution function of velocities, $m$ is the mass of a particle, $k_{\rm B}$ is the Boltzmann constant, $T$ is the temperature of the gas, and $v$ is the module of the velocity of a particle.

The extension of this analysis to large scale structures such as galaxy clusters is not straightforward. This follows from a set of interactions that may highly increase the complexity of the description of a cluster's dynamical stage. Utilising a series of assumptions, it is possible to show that, for systems bound by gravity, the equilibrium stage is also described by a Maxwell-Boltzmann distribution \citep{1957SvA.....1..787O, 1957SvA.....1..748O, 1967MNRAS.136..101L}. However, the assumptions made are unrealistic with real world observations, making the description of the dynamical evolution of galaxy clusters an open problem.

In that regard, cosmological simulations provide a suitable tool to gain a first insight on the dynamical evolution of clusters. One useful case is the Yonsei Zoom-in Cluster Simulation (YZiCS) simulation \citep{2018ApJ...866...78H}, which is designed to study the formation and evolution of galaxy clusters. The simulation is classified as a ``zoom-in'' simulation, in which it starts with a N-body simulation of a large cosmological volume and, then, overdensity regions are selected to perform a zoom-in in and include baryonic physics, such as gas cooling, star formation, supernova and AGN feedback, and chemical enrichment.

\begin{figure}[ht]
    \centering
    \includegraphics[width=\linewidth]{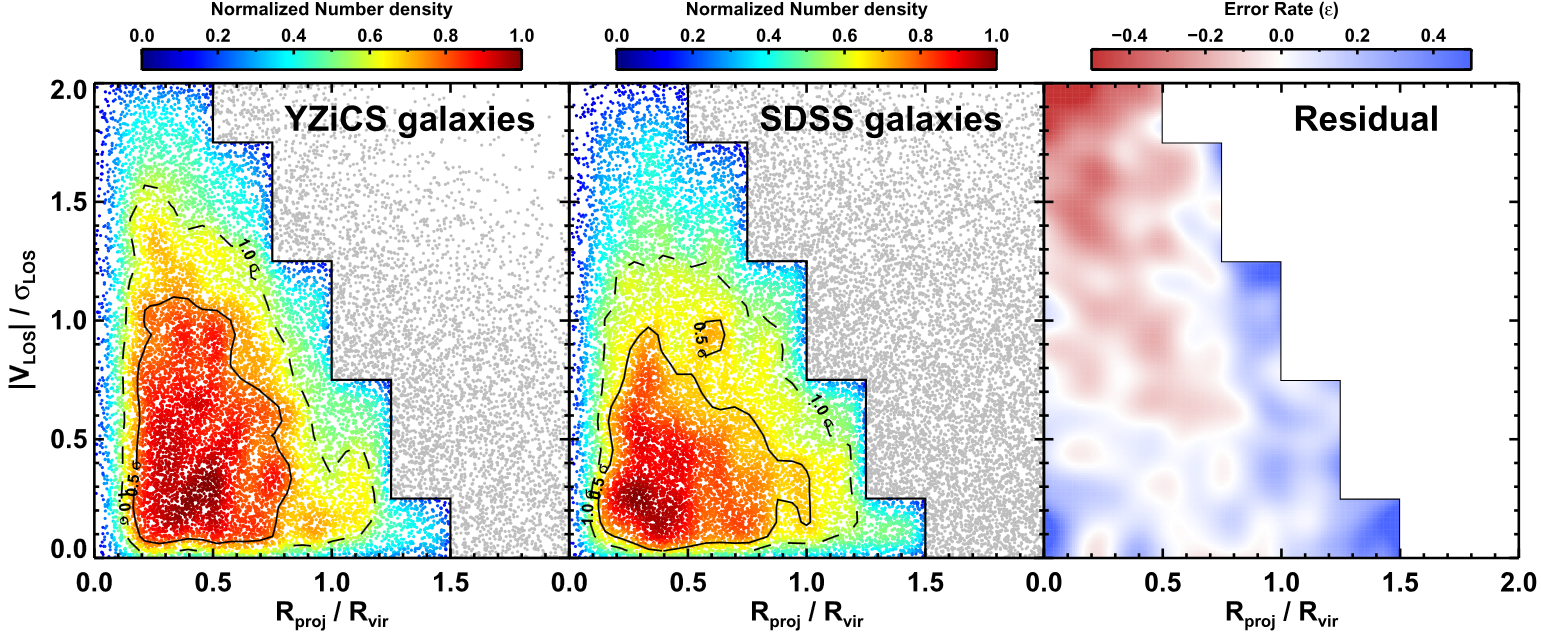}
    \caption{Comparison between the distribution of galaxies in the PPS for the YZiCS simulation (left) and real galaxies (center). The right panel shows the difference between the two distributions. Adapted from \citet{2020ApJS..247...45R}.}
    \label{fig:pps_comparison}
\end{figure}

When it comes to observationally defining the dynamical stage of clusters, the limitation to a single line-of-sight impose the use of a projected along the line-of-sight phase space, the projected phase space (PPS) -- built with the projected distance towards the cluster center in the x-axis and the velocity along the line-of-sight in the y-axis. Therefore, comparisons with simulations should be done in this space to ensure connection with real world observations. Fig.~\ref{fig:pps_comparison} shows the comparison between PPS resulting from the YZiCS simulation (left) and the one with real data (center) from the Sloan Digital Sky Survey (SDSS, see Chapter \ref{chap:data}).  The right panel show the residual between the two PPSs.

The dynamical evolution of clusters is intimately related to the amount of matter being accreted as a function of time. Therefore, a particularly useful approach is the use of simulations to define different regions in the PPS constraining member galaxies to a narrow range of time since infall/infall time -- defined as the time since the galaxy first crossed the virial radius of the cluster.  

\begin{figure}[ht]
    \centering
    \includegraphics[width=\linewidth]{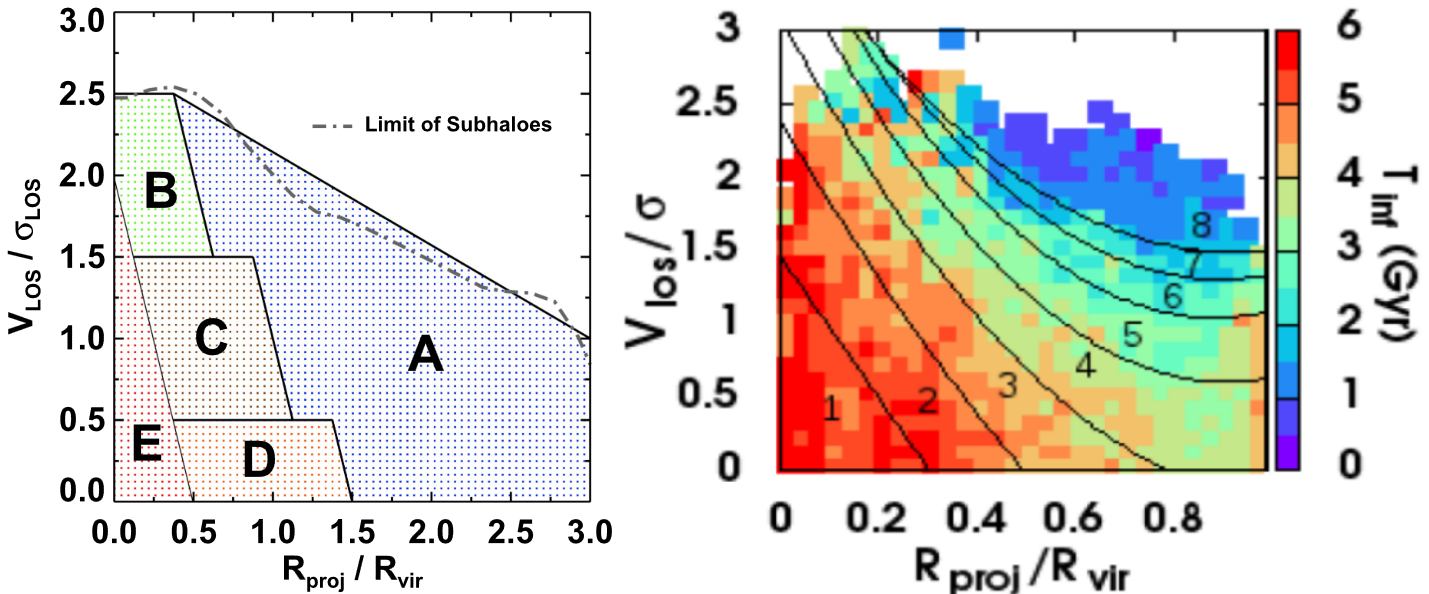}
    \caption{Different ways to slice the PPS according to infall time. Left: adopt a probabilistic approach, in which galaxies within each region have a probability of having a given infall time. Right: adopt a direct approach by characterizing each region by the median infall time value. Adapted from \citet{2017ApJ...843..128R} and \citet{2019MNRAS.484.1702P}, respectively.}
    \label{fig:infall_regions}
\end{figure}

Fig.~\ref{fig:infall_regions} shows different ways of dividing the PPS according to members infall time. Left panel adopt a probabilistic approach, in which galaxies within each region have a probability of having a given infall time \citep{2017ApJ...843..128R}, whereas the regions in the right panel a characterized by a median value \citep{2019MNRAS.484.1702P}. The division shown in the right panel is particularly important for this work, as it enables an investigation not only of the dynamical stage of clusters, but also galaxy properties evolution as a function of infall time.  

\section{This project: addressing nature vs. nurture}

Understanding the balance between intrinsic (nature) and extrinsic (nurture) factors driving galaxy evolution is a significant challenge in astrophysics. Intrinsic factors, such as the galaxy's mass, morphology, and internal dynamics, play a crucial role in determining its evolutionary path. On the other hand, extrinsic factors, including interactions with other galaxies, the influence of the intergalactic medium, and the local environment (e.g., galaxy clusters), can significantly alter a galaxy's structure and star formation activity. The main objective of this project is to provide a comprehensive analysis of the environmental effects on galaxy evolution and the related time-scales, focusing on star formation quenching, morphological transitions, and AGN activity. 

\subsection{Galaxy evolution in clusters: comparing star formation suppression and morphological transition}

The complex relation between infall time, galaxy properties and quenching mechanisms translates to a variety of paths for galaxy's transition from the BC to RS. Consequently, it is not straightforward to reliably define the BC, GV and RS regions, for which different works adopt different parameter spaces and methodologies. For instance, \cite{Strateva1} use the color-color diagram, \cite{2020MNRAS.491.5406T} adopt Stellar Mass versus Star Formation Rate and \cite{2019MNRAS.488L..99A} define quenching according to the spectroscopic measurement of the D4000 break. Although it is known the that morphology correlates with galaxy properties, it is still necessary to further investigate the transition from BC to RS from a morphological perspective. Also, recent works discuss the possibility of dense environments inducing morphological variations prior to SFR changes \citep{2009ApJ...707..250M,2014MNRAS.440..889S,2019MNRAS.486..868K}.

In Chapter~\ref{chap:sf_vs_morphology}, we address galaxy evolution from different perspectives. First, we use the SFMS diagram and the separation between BC, GV and RS to compare cluster and field galaxy properties. We focus on cluster galaxies and investigate their path from BC to RS as a function of stellar mass. We focus on understanding what defined different stages of galaxy evolution and address the question on whether SFR or morphology changes more quickly in cluster environments. Furthermore, we focus on the relation between location in the PPS and infall time to provide direct measurements of how galaxy properties vary with time. Lastly, we investigate how variations in morphology and star formation rate depend on stellar mass and environment.

We further extend the analysis from Chapter~\ref{chap:sf_vs_morphology} in Chapter~\ref{chap:timescales}. Combining the SFMS diagram with PPS, we provide insights into key questions regarding star formation quenching and morphological transitions for galaxies infalling into clusters. We quantify the temporal aspects of these processes and analyze their relation to the ``delayed-then-rapid'' quenching model. We present a comparative analysis of the timescales for star formation suppression and morphological transitions, aiming to provide a new perspective on the seemingly conflicting results in the literature \citep{2009ApJ...707..250M, 2019MNRAS.486..868K, 2022MNRAS.509..567S, 2024MNRAS.529.3651O}. Finally, we employ the BC and RS definitions to offer a dynamic estimation of the time galaxies spend in the GV after cluster infall and compare it to previous estimates \citep[e.g.,][]{2021MNRAS.508..157M}.

\subsection{The special case of central galaxies: fueling AGN activity through interactions}

A particularly interesting class of objects are the so-called `central galaxies'. These are systems located at that the centre of dark matter potential wells, and therefore at the focus of any possible gas accretion. Because they are the dominant galaxy in their dark-matter halo, it is reasonable to expect that the influence of external objects may be smaller than for satellite galaxies. If this is the case, internal processes may be relatively more dominant than external (environmental) ones, and therefore easier to study.

One of the main \textit{internal} drivers of galaxy evolution is the presence of a central SMBH. According to \cite{2013MNRAS.433.3297D} and \cite{2022PhR...973....1D}, the SMBH mass is one of the parameters that most effectively predicts whether a galaxy is star-forming or quenched. It has been suggested \citep[e.g.,][]{2014MNRAS.441.3055B} that the relation between SMBH mass and star-formation suppression is due to the effect an AGN has on the interstellar medium of the galaxy. The effect an AGN has will depend both on the mass of the SMBH at its centre and the amount and rate of `fuel' feeding it. 

In Chapter~\ref{chapter: central_galaxies} we investigate the demographics of optically-identified AGN in both isolated and group/cluster central galaxies. We study the interplay between internal and environmental properties driving AGN and fueling. To this end, we first separate galaxies according to their environment and use the BPT diagram to investigate the fraction of strong AGN (Seyfert galaxies) in different environments. Additional factors such as galaxy mass, morphology, central velocity dispersion, and host halo mass, are considered and linked, together with the environmental information, to the AGN and star-formation activity of the galaxies. Our aim is to identify the physical processes fueling AGN and driving star-formation activity, and causing both star-formation quenching and morphological transition. 

\subsection{The other side: the impact of galaxy accretion in clusters dynamical stage?}

Star formation quenching is conditional on clusters properties. For instance, RPS depends on the ICM density and temperature. In order to characterize different environments, substructure analyses in the optical (e.g. \citealt{1988AJ.....95..985D,1997ApJ...482...41G}) and X-ray (e.g. \citealt{2001A&A...378..408S,2009ApJ...699.1178Z}) shows that many clusters are not fully virialized. The degree of relaxeness is also related to galaxy's orbits. Infalling galaxies have highly radial orbits in the outskirts, while virialized objects show circular orbits within the virial radius.  

The use of N-body simulations supports that, when limited to velocities projected along the line of sight, the distribution better describing dynamical equilibrium after an accretion event is a Gaussian distribution \citep{merrall2003relaxation,2005NewA...10..379H}. Previous studies investigate the difference between clusters with projected velocity distributions well fit by a Gaussian (G) and those with a non-Gaussian (NG) velocity profile and find that: 1) NG clusters have an excess of star forming galaxies \citep{2010MNRAS.409L.124R}; 2) the stellar population parameters infalling and virialized galaxies in NG clusters are not well separated as in G clusters \citep{2013MNRAS.434..784R}; 3) there is evidence of a higher infall rate of pre-processed galaxies in NG clusters \citep{2017MNRAS.467.3268R,2017AJ....154...96D}; 4) the velocity dispersion profiles of members galaxy is significant different between G and NG systems \citep{2018MNRAS.473L..31C}; and 5) simulations show that NG systems suffered their last major merger more recently than G systems \citep{2019MNRAS.490..773R}. These results points towards a higher infall rate in NG clusters in comparison to G clusters. 

As an option to the velocity distribution, the full projected phase space can encompass additional information. In the PPS, the aforementioned G or NG velocity distribution is simply a projection along the y-axis. It is also important to highlight that environmental effects, such as RPS, are conditional on the velocity of the member galaxy and thus the PPS provide a more suitable to study environmental effects taking into account both position and velocity. 

In Chapter~\ref{chapter: G_vs_NG} we use the PPS approach to investigate further differences between G and NG clusters. We divide the PPS in a grid of pixels and build distributions of members galaxy density and median properties for each galactic environment. Specifically for stellar mass, we use the observed distribution to derive a rough estimate of the infall rate in NG clusters. We perform a statistical study about different regions and evaluate the global properties of the PPS of G and NG clusters. We also build a relation between galaxy properties and infall time using PPS regions presented in the right panel of Fig.~\ref{fig:infall_regions}.

%% file: Chapters/Chapter3.tex
\chapter{Data, sample and methods}
\label{chap:data}
In this project, we select galaxies from the Sloan Digital Sky Survey \citep{2009ApJS..182..543A}. The Sloan Digital Sky Survey (SDSS) is a comprehensive astronomical survey that has mapped over 14,000 square degrees of the sky, providing detailed photometric and spectroscopic measurements of over 500 million celestial objects, including stars, galaxies, and quasars. 

\begin{figure}[ht]
    \centering
    \includegraphics[width=0.8\linewidth]{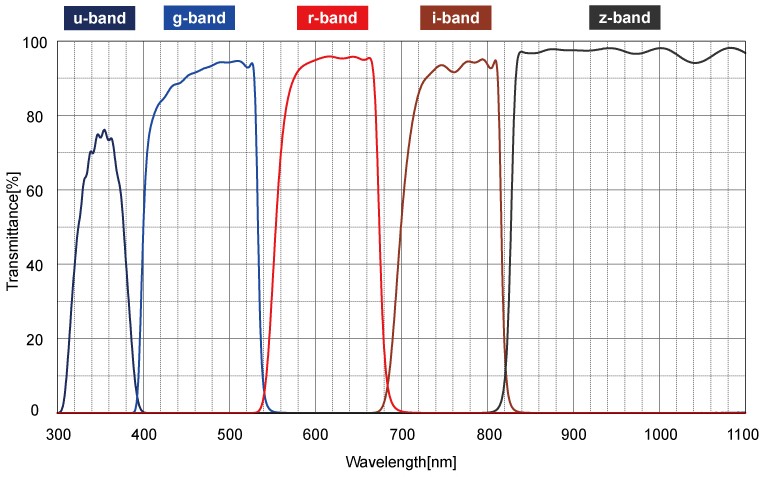}
    \caption{The 5 different optical broad bands used in the SDSS.}
    \label{fig:ugriz_bands}
\end{figure}

The SDSS photometric observations are made with a dedicated 2.5-meter telescope located at the Apache Point Observatory in New Mexico. The SDSS camera is a mosaic of 30 CCDs detectors, providing a total of 120 million pixels. The system captures images in five broad optical bands, u (ultraviolet), g (green), r (red), i (near-infrared), and z (infrared), the \textit{ugriz} system, whose wavelength coverage is shown in Fig.~\ref{fig:ugriz_bands}.

Regarding spectroscopic observations, the SDSS has used different spectrographs over its various phases. The original SDSS spectrograph coverages from 3,800 to 9,200 \AA~in wavelength with a resolution of $R = \lambda/\Delta \lambda \sim 1,800$. In a second phase, spectroscopic observations used the Baryon Oscillation Spectroscopic Survey (BOSS) spectrograph, which coverages from 3,600 to 10,400 \AA~with a $R \sim 2,000$ in the blue (3,600 to 6,300\AA) and $R \sim 2,500$ in the red (6,300 to 10,400 \AA). Still, both adopted fixed 3-arcsec fiber in spectroscopic observations, which can be relevant to the analysis of extended objects as galaxies, since only the light that falls within the 3-arcsecond aperture is captured. In other words, the fixed aperture size can introduce a sampling bias, as it might only capture the central region of a galaxy. For galaxies with significant radial gradients in properties like metallicity, star formation rate, or stellar population, this central region may not be representative of the galaxy as a whole.

We select only galaxies with both photometric and spectroscopic information from the SDSS database\footnote{Available at https://skyserver.sdss.org/casjobs/.}. We limit our sample to systems within the redshift range $\rm 0.03 \leq z \leq 0.1$ and with petrosian apparent magnitude in the r-band ($m_{\rm r}$) less than 17.78, which correspond to the spectroscopic limit of the survey at $z = 0.01$. The minimum redshift is applied to mitigate biasing the stellar population parameters due to the fixed 3-arcsec fiber used in the SDSS. Finally, by selecting only galaxies with zWarning=0zWarning=0, ensuring no anomalies in their spectra, we obtain a sample of 570,685 SDSS galaxies.

\section{Retrieving properties for SDSS galaxies}

A great advantage of the SDSS is that it provides a well established and organized database to characterize galaxy evolution in the local universe. The database contains an enormous amount of galaxy properties, which in combination with the large number of observations, provides a relevant framework that has been extensively explored in the last decades. 

\subsection{The Max Planck Institut f\"ur Astrophysik -- John Hopkins University value added catalog}
\label{sec:mpa_catalog}
The Max Planck Institut f\"ur Astrophysik -- John Hopkins University (MPA-JHU) value added catalog is a collection of spectrum derived galaxy properties based on data from the SDSS. We retrieve from The MPA-JHU catalog the integrated flux and equivalent width of several key emission lines, such as $\rm H{\alpha}$, $\rm H{\beta}$, $[{\rm OIII} ] \lambda 5007$, $[{\rm NII}]\lambda6584$, and $[{\rm SII}]\lambda6717,6731$. These lines are important to define the ionization state of galaxies, and are particularly relevant to the analysis presented in \autoref{chapter: central_galaxies}. It is noteworthy that the emission line properties are estimated after subtracting the continuum and absorption lines of the galaxy stellar component. The continuum and absorption lines are estimated using a spectrum fitting technique using \cite{2003MNRAS.344.1000B} models as SSPs.

Using emission-line data, the MPA–JHU catalog classifies galaxies based on their dominant ionizing mechanism using the BPT diagram, employing the lines described in subsection~\ref{subsec:diagnostic_diagrams}.  Galaxies are thus divided into six categories: 1) star forming, 2) low S/N star forming, 3) composite, 4) Seyfert, 5) LINER, and 6) unclassified. The last class comprises galaxies with no reliable classification in the BPT diagram due to at least one of the four relevant emission lines having S/N smaller than 3. Furthermore, in order to increase the robustness of our results, we only include the BPT classification for galaxies with $\rm H\alpha$ equivalent width ($EW(\text{H}\alpha$)) greater than 3\AA, which is the threshold used in the WHAN diagram to separate LINERs and Retired galaxies \citep{2010MNRAS.403.1036C}. The flux of weaker lines would be highly uncertain in the presence of underlying stellar absorption lines even after correcting for them. This way we avoid contaminating our sample with weak emission line galaxies that can be explained by the presence of old stellar population. 

It is important to point out that, while the emission lines of Seyfert galaxies are well explained by the presence of a strong AGN, there is considerable uncertainty -- and some controversy -- on the origin of the emission lines observed in LINERs. However, it is now generally accepted that, in addition to low-activity AGN, LINER emission may also be due to the presence of an old stellar population \citep{2010MNRAS.403.1036C} containing very hot low-mass stellar relics such as post-AGB stars. Thus, when needed, we will only consider galaxies in the Seyfert region of the BPT diagram as unequivocal optically-selected AGN.

From this catalog, we also retrieve stellar mass ($M_{\star}$), star formation rate (SFR), specific SFR ($\text{sSFR} = \text{SFR}/M_{\star}$) and central velocity dispersion ($\sigma_\text{galaxy}$). An important feature is the way they estimate the galaxies' SFR: it is first derived using the $\rm H\alpha$ emission-line flux for galaxies classified as `star-forming' in the BPT diagram, and then extrapolated to other galaxies using a calibration between SFR and the 4000\AA~break. This is particularly important in order to prevent the SFR estimates from being contaminated by possible AGN contribution to the $\rm H\alpha$ flux. We find reliable estimates for $\sim 98\%$ of the 570,685 galaxies.

\begin{table}[ht]
\caption{Quantiles, mean and standard deviation for the uncertainty distribution for $M_{\star}$, SFR and sSFR. We present uncertainties for the three stellar mass bins used.}
\label{Table:MPA-JHU_uncertainties}
\resizebox{\columnwidth}{!}{%
\begin{tabular}{c|c|c|c|c|c}
\hline
Uncertainty  & $X = \log(M_{\star}/{\rm M}_{\odot})$ & $ Q_{16\%}$ & $ Q_{50\%}$ & $ Q_{84\%}$ & $\mu_{\delta} \pm \sigma_{\delta}$ \\ \hline
\multirow{3}{*}{\begin{tabular}[c]{@{}c@{}}$\log(M_{\star}/{\rm M}_\odot)$\end{tabular}} & $9 \leq X < 10$ & 0.079 & 0.092 & 0.099 & $0.098 \pm 0.058$ \\ \cdashline{2-6}
 & $ 10 \leq X < 11$ & 0.088 & 0.092 & 0.098 & $0.099 \pm 0.051$ \\ \cdashline{2-6}
 & $ 11 \leq X < 12$ & 0.085 & 0.089 & 0.096 & $0.091 \pm 0.021$ \\ \hline
\multirow{3}{*}{\begin{tabular}[c]{@{}c@{}}$ \log(\text{SFR} / {\rm M}_{\odot} \, {\rm yr}^{-1})$\end{tabular}} & $ 9 \leq X < 10$ & 0.222 & 0.326 & 1.009 & $0.507 \pm 0.335$ \\ \cdashline{2-6}
 & $ 10 \leq X < 11$ & 0.354 & 1.028 & 1.071 & $0.844 \pm 0.119$ \\ \cdashline{2-6}
 & $ 11 \leq X < 12$ & 0.967 & 1.023 & 1.065 & $0.971 \pm 0.178$ \\ \hline
\multirow{3}{*}{\begin{tabular}[c]{@{}c@{}}$\log(\text{sSFR}/{\rm yr}^{-1})$\end{tabular}} & $ 9 \leq X < 10$ & 0.242 & 0.342 & 1.013 & $0.523 \pm 0.329$ \\ \cdashline{2-6}
 & $ 10 \leq X < 11$ & 0.372 & 1.029 & 1.078 & $0.854 \pm 0.312$ \\ \cdashline{2-6}
 & $ 11 \leq X < 12$ & 0.966 & 1.025 & 1.072 & $0.983 \pm 0.174$ \\ \hline
\end{tabular}%
}
\end{table}

The MPA-JHU catalog provide 5 percentiles (2.5, 16, 50, 84 and  97.5\%) for each estimate. Hereafter we use the 50\% percentile (median) as the desired estimate. Regarding the associated uncertainties, we estimate for MPA-JHU quantities from the 16 and 84\% quantiles as $0.5 \times (Q_{84\%} - Q_{16\%})$. In Table \ref{Table:MPA-JHU_uncertainties}, we present the quantiles, mean and standard deviation for the error distributions. We divide galaxies into three different bins of stellar mass: 1) $ 9 \leq \log(M_{\star}/{\rm M}_{\odot}) < 10$; 2) $10 \leq \log(M_{\star}/{\rm M}_{\odot}) < 11$; and 3) $11 \leq \log(M_{\star}/{\rm M}_{\odot}) < 12$, following works relating between galaxy properties and stellar mass.

\subsection{Properties obtained through spectrum fitting}

We select $Age$, Metallicity ($[Z/H]$) and stellar mass ($M_{\star}$) from \citealt{2017AJ....154...96D} (dC17, hereafter) galaxy catalog to characterize galaxy stellar population. The estimates are obtained via spectral fitting using the STARLIGHT code, which fits the input galaxy spectra using a combination of pre-defined SSPs. Galaxy stellar population $Age$ and Metallicity are estimated from the weighted sum of SSPs parameters used to fit the spectra. The estimates are derived only for galaxies with spectra that do not have any anomalies in their spectra. The adopted stellar models are based on the Medium resolution INT Library of Empirical Spectras (MILES - \citealt{2006MNRAS.371..703S}), which has an almost constant resolution of $\sim 2.5$\,\AA. The SSP basis grid has a constant $\log(Age/{\rm Gyr})$ steps of 0.2 from 0.07 to 14.2 Gyrs and includes SSPs with $[Z/H]$ = -1.71, -0.71, -0.38, 0.00, +0.20. 

Spectral fitting codes allow an arbitrary weight and there are two commonly adopted methods in literature: 1) luminosity-weighted parameters trace mainly younger stellar population properties; while 2) mass-weighted parameters are more closely related to the cumulative galaxy evolution \cite{2020MNRAS.491.5406T}. In the following we use luminosity-weighted parameters, such that $Age$ refers to the last episode of star formation.

\subsection{T--Type: a continuous way to trace morphology}

When the star-formation properties of galaxies change, their morphologies may also change. Here we use the T-Type parameter, first introduced by \cite{1963ApJS....8...31D}, to describe a galaxy's morphology. A T-Type value $\leq0$ indicates an early-type galaxy, while T-Type values $>0$ correspond to late-type morphologies. T-Type $=0$ means lenticular systems. We adopt the T-Type values from the work of \cite{2018MNRAS.476.3661D}, which was based on the application of a deep-learning convolutional neural network algorithm to~670,722 galaxies from the SDSS-DR7 database. To aid our analysis of galaxy evolution it is convenient to use this revised T-Type parameter due to its continuous variation from -3 to 6 instead of the original discrete values. The continuity is achieved by employing equation (7) from \cite{2015MNRAS.446.3943M}:
\begin{equation}
\text{T--Type} = -4.6P(\text{Ell}) -2.4P(\text{S0}) + 2.5P(\text{Sab}) + 6.1P(\text{Scd}),
\end{equation}
where $P({\rm X})$ denotes the probability of a galaxy being classified as a given morphology, with X representing Elliptical (Ell), lenticular (S0), A-B spiral (Sab), and C-D spiral (Scd)

\begin{figure}[ht]
    \centering
    \includegraphics[width = \textwidth]{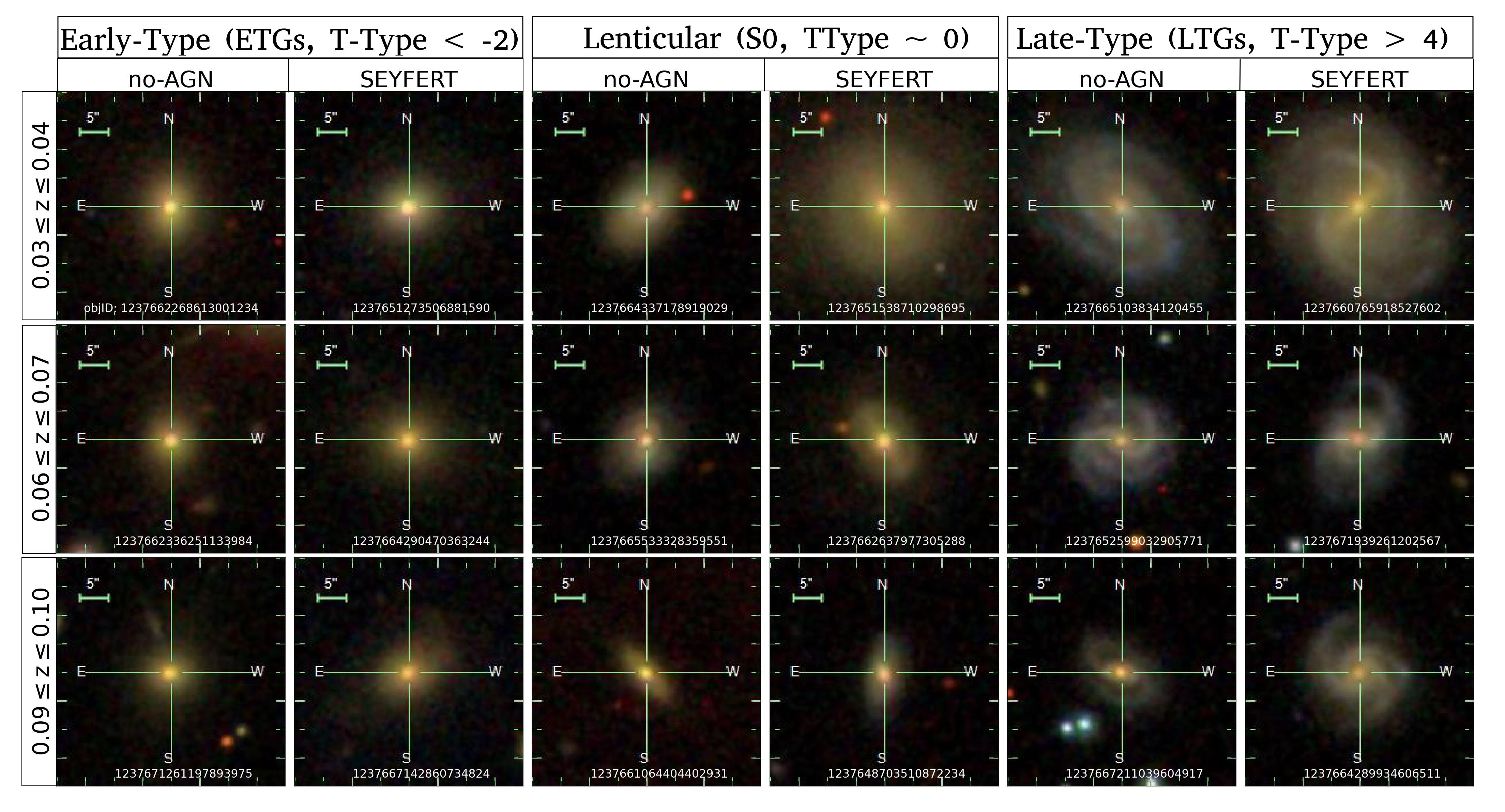}
   \caption{Comparison of the SDSS images of pairs of galaxies with different morphologies (T-Types) in three redshift bins (rows) for no-AGN (left image in each pair) and AGN (Seyfert, right image) galaxies classified using the BPT diagram.  It is apparent that the presence of an AGN doesn't alter the morphology of the host galaxy significantly. }
    \label{fig:TType_Mosaic}
\end{figure}

Furthermore, it is conceivable that the presence of a strong AGN may influence the morphological classification of the host galaxy. To ensure that such effect is not significant enough to alter the conclusions of this work, we examine the images of a subset of randomly selected galaxies classified as Seyfert and compare them with the images of non-Seyfert galaxies with similar morphologies. Fig.~\ref{fig:TType_Mosaic} shows this comparison for a sample of galaxies, Seyfert and non-Seyfert, for different morphologies and redshifts. Visual inspection indicates that that the presence of an AGN does not alter the morphology significantly, at least at optical wavelengths in these nearby galaxies. 

Finally, in order to check that the results of our analysis do not depend critically on the specific choice of morphological data, we repeated our analysis using independent visual morphologies from the the Nair and Abraham catalog \citep{2010ApJS..186..427N}. These morphologies do not rely on automated algorithms, being based on human inspection of the galaxy images. Notwithstanding the different sample sizes of both morphological catalogs, the results obtained are completely compatible. 

\subsubsection{Additional information from the Korea Institute for Advanced Study Value-Added Galaxy Catalog}

We added color gradient estimates to our catalog from the Korea Institute for Advanced Study Value-Added Galaxy Catalog (KIAS-VAGC - \citealt{2010JKAS...43..191C}). The KIAS-VAGC catalog provides galaxy spectroscopic and morphological information for $593,514$ galaxies from the SDSS-DR7 main galaxy catalog and $10,497$ from other galaxy catalogs (see \citealt{2005ApJ...635L..29P}). The color gradient is computed using the color indexes in the g and i band. The (g-i) color gradient is expressed as 
\begin{equation}
\label{eq:grad_gi}
\nabla ({\rm g}-{\rm i}) = ({\rm g}-{\rm i})_{0.5R_{\rm {p}}<r<R_{\rm {p}}} - ({\rm g}-{\rm i})_{r<0.5R_{\rm {p}}}, 
\end{equation}
where $\rm (g-i)_{x}$ denotes the g-i index where the condition x is satisfied, and $R_{\rm p}$ is the Petrosian Radius at the i-band. Equation \ref{eq:grad_gi} means that more negative values correspond to bluer colors in the galaxy outskirts.

We use a 1.5" threshold in a positional cross-match to select the appropriate estimates. The upper limit is defined empirically. We find reliable estimates for $98\%$ of our sample. 

\section{The Yang group catalog}

We adopt the Yang Catalog \citep{2007ApJ...671..153Y} to classify galaxies' environment. The catalog is build by applying a halo mass finder algorithm to the New York University - Value Added Catalog \citep[NYU-VAGC]{2005AJ....129.2562B}. Groups/clusters are then defined as galaxies in the same dark matter halo. 

\subsection{Sample of central galaxies}

The Yang catalog also enables the classification of galaxies into centrals (here we adopt the most massive cluster galaxy as central) and satellites, which is further extended to: 1) isolated centrals -- galaxies living in halos occupied by a single galaxy ($N_\text{members} = 1$); 2) binary system centrals -- the most massive galaxy of a pair hosted in the same halo ($N_\text{members}=2$); 3) group centrals -- central galaxies in halos occupied by at least 3 galaxies, but no more than 10 ($3 \leq N_\text{members} < 10$); and 4) cluster centrals -- the central system in halos populated by more than 10 member galaxies ($N_\text{members}>10$). For halos hosting more than one galaxy, we define the central system as the most massive one, with stellar mass calculated the following equation \cite{2007ApJ...671..153Y},
\begin{equation}
    \log \left ( \frac{M_{\star}}{h^{-2}{\rm M}_{\odot}} \right ) =  - 0.306 + 1.097 \, ^{0.0}({\rm g} - {\rm r}) - 0.1 - 0.4(^{0.0}M_{\rm r} - 5\log(h) -4.64),
\end{equation}
where $h$ is a parametrization of the Hubble constant, $H_{0} = 100 h \, {\rm km} \, {\rm s}^{-1} \, {\rm Mpc}^{-1}$, $^{0.0}({\rm g} - {\rm r})$ and $^{0.0}M_{\rm r} - 5\log(h)$ are the ($\rm g-r$) and r-band magnitude $K+E$ corrected to $z=0.0$, 4.64 is the r-band magnitude of the Sun in the AB system and the $-0.10$ term is a IMF dependents offset. We hereafter refer to this sample of central galaxies as ``central sample''. We tested how this sample would change if we selected the brightest cluster galaxy (BCG) instead; we find no significant differences, since 98.2\% of the most massive galaxies are also BCGs. 

\begin{figure}[ht]
    \centering
    \includegraphics[width = 0.8\textwidth]{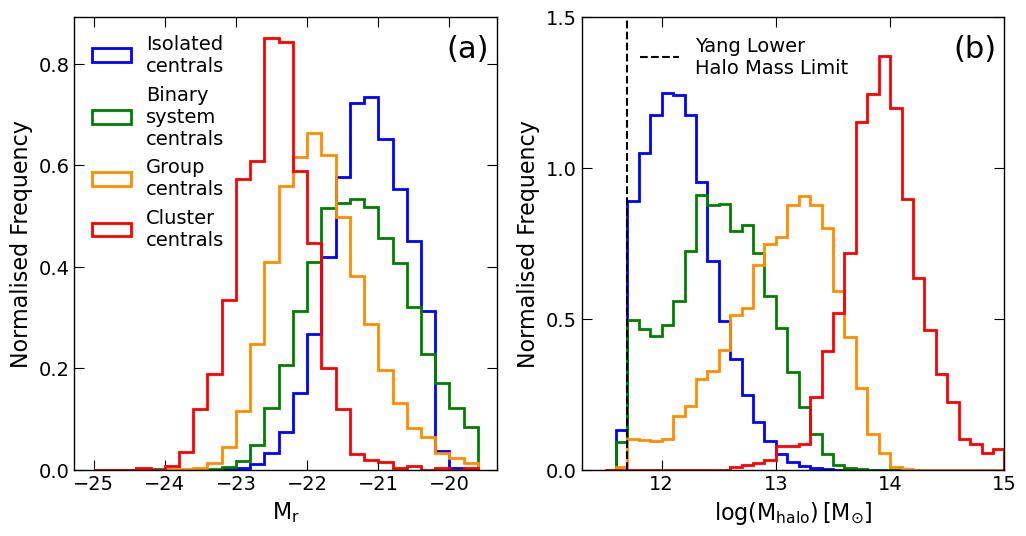}
   \caption{Distribution of the $r$-band absolute magnitude $M_\text{r}$ (left) and host-halo mass 
   $M_\text{halo}$ (right) for the four sub-samples of central galaxies: isolated centrals (blue line), centrals in binary system (green), in groups (orange), and in clusters (red). The vertical dashed line in the right panel show the lower halo-mass completeness limit from the Yang catalog.}
    \label{fig:Yang_Sample_Properties}
\end{figure}

We further guarantee completeness regarding the halo mass estimate by imposing the conservative threshold presented in \cite{2009ApJ...695..900Y}, 
\begin{equation}
     \log(M_\text{halo}/{\rm M}_{\odot}) \geq 12 + \frac{z_\text{c} - 0.085}{0.069},
    \label{eq:yang_mass_completeness}
\end{equation}
where $z_\text{c}$ denotes the median redshift of a given system (halo), and $M_\text{halo}$ is the halos mass estimated by \cite{2007ApJ...671..153Y}. In our analysis presented mainly in \autoref{chapter: central_galaxies} we take into account that, as a result of separating central galaxies according to the number of satellites in their host halo, each sub-sample has different magnitude (and therefore stellar mass), and host halo-mass distributions, as shown in Fig.~\ref{fig:Yang_Sample_Properties}.

The four sub-samples of central galaxies we have defined contain 59,605 isolated centrals, 11,639 in binary systems, 7,586 in groups, and 1,291 in clusters. 

\subsection{Galaxies in isolation -- a sample of field galaxies}

In addition to our central sample, we define a secondary sample of field galaxies to trace how galaxies evolve when strictly in isolation. First, we queried the SDSS-DR7 database for all galaxies with reliable spectroscopic redshift measurements. We then select field galaxies as follows: 1) we identify all groups from the Yang Catalog with halo masses greater than $\rm 10^{13} M_{\odot}$ -- a characteristic halo mass of poor groups ($ N_{\rm members} \leq 10$); 2) we use the scaling relation 
\begin{equation}
    R_{\rm vir} \sim 1.61 \, {\rm Mpc} \left ( \frac{M_{\rm halo}}{10^{14}{\rm M}_{\odot}} \right )^{1/3} \, (1+z_{\rm group})^{-1}
\end{equation}
to estimate the virial radius for every group; 3) we select galaxies beyond 5 $R_{200}$ of every structure with halo mass greater than $\rm 10^{13} M_{\odot}$ listed in the Yang Catalog \citep{2017MNRAS.471L..47T}; 4) we consider galaxies only in the main part of the SDSS footprint ($\rm 120 \leq RA \leq 250$, $\rm 0 \leq Dec \leq 60$) to minimize edge effects. This results in a set of 12,398 galaxies, which is dominated ($\rm \sim 93 \%$) by galaxies classified as Isolated Centrals in the Yang Catalog. To guarantee the reliability of our field sample, we discard all non isolated central galaxies ($\sim 7\%$). Yet, a word of caution is needed regarding fossil groups. These are groups that, after a series of merging events, end up as a single luminous central galaxy. \cite{2009AJ....137.3942L} shows that fossil groups are characterized by a bright galaxy with $ M_{\rm r} \leq -22$, where $M_{\rm r}$ is the absolute magnitude in the r-band. Therefore, we adopt this value as the absolute magnitude lower limit for our field sample in order to minimize the effects of fossil groups. This removes $\sim 5\%$ of the our sample and results in a final field sample of 11,674 galaxies, although our results do not change if we keep these galaxies in our sample. This follows from the fact that fossil groups can be considered rare events, which are characterized by a comoving number density of $ 2.83 \times 10^{-6} \, h^{3} \, {\rm Mpc}^{-3}$ \citep{2009AJ....137.3942L}.

\section{Galaxy Clusters from an updated version of the Yang Catalog}

The main objective of this work is to understand how different environment affect galaxy evolution. The third sample that we adopt in this work is a ``cluster sample''. However, assigning galaxy membership can be done via different methods. In particular, the method employed in the Yang Catalog is considerably restrictive, as almost there is no member galaxies beyond $R_{vir}$, which can be significantly impactful when tracing galaxy evolution within clusters. Therefore, we use an updated version of the Yang Catalog presented in dC17 to define our cluster sample, which is built using data from the SDSS-DR7 with the same redshift range and $ m_{\rm r}$ adopted in this work.

\subsection{Membership assignment through the shiftgapper technique}

Unlike in the original Yang Catalog, membership in dC17 is defined via a ``Shiftgapper'' technique, which is applied in clusters from the Yang Catalog with at least 20 members. It is also noteworthy that the shiftgapper technique avoids prior assumptions about the cluster's dynamical stage. 

The first step of the shiftgapper technique is to select galaxies at a maximum distance of 2.5$h^{-1}$ Mpc (3.47 Mpc for $h = 0.72$) and with a line-of-sight velocity in the range $\rm \pm \, 4000 km\, s^{-1}$ with respect to the clustercentric coordinates (RA, DEC and $z$) presented in the Yang Catalog\footnote{The clustercentric coordinates are the only information from the Yang Catalog used to define the clusters.}. \cite{2017AJ....154...96D} then apply a Gap Technique by placing galaxies in radial bins with a minimum size of 0.42$h^{-1}$ Mpc, which guarantees at least 15 galaxies per bin, and removing galaxies with a velocity gap greater than $\rm 1000 \, km \, s^{-1}$ with respect to the cluster mean velocity. A new center is then defined as the median RA, DEC and redshift of the remaining galaxies and the process is reiterated until no more galaxies are removed. This process results in a final list containing only member galaxies. 

Using the final list of member galaxies, dynamical quantities like virial radius ($R_{200}$), virial mass ($M_{200}$) and velocity dispersion along the line-of-sight ($\sigma_{\rm LOS}$) are estimated by dC17 through virial analysis (see Appendix \ref{Appendix_dynamics} for more details) for each cluster. A comparison between the Yang catalog and the shiftgapper technique shows differences in $M_{200}$ of less than 0.1, which are smaller than the related uncertainties in the Yang $M_{200}$ estimates ($\sim 0.15 \, {\rm dex}$, \citealt{2007ApJ...671..153Y}). By imposing a minimum number of 20 galaxies within the virial radius,\footnote{We define the clustercentric coordinates as the median $z$ and the luminosity weighted average of RA and DEC.} we define a sample of 319 massive clusters. By using the relation between $M_{200}$ and $N_{200}$, where the last term is the number of galaxies within $R_{200}$, we define a halo mass threshold, namely $M_{200}^{\rm thresh} = 10^{14} \, {\rm M}_{\odot}$. By considering clusters with $M_{200} \geq M_{200}^{\rm thresh}$, our sample decreases to 254 clusters. The halo mass completeness limit means we are probing clusters at the extreme tail of the halo mass function, namely the 5\% most massive systems with halo masses in the range $13 \leq log(M_{\rm halo}/{\rm M}_{\odot}) \leq 15.5$\footnote{We used the Halo Mass Function Calculator \citep{2013A&C.....3...23M} with Planck 15 cosmology \citep{2016A&A...594A..13P} and $z = 0.075$, which is the median redshift for our sample.}. Additionaly, we make a distinction between satellites and centrals according to the class assigned in the Yang Catalog. Here on we focus only on satellite galaxies, which consists of 20,191 galaxies.

\subsection{Further characterization of cluster environment}

In particular in Chapter \ref{chapter: G_vs_NG}, we investigate how galaxy properties can vary according to the dynamical state of clusters. Still, observationally characterizing the dynamical state of clusters is a long-standing problem in astrophysics. Based on the results from N-body simulations \citep{merrall2003relaxation,2005NewA...10..379H}, we further divide clusters according to the gaussianity of the members galaxy projected velocity distribution. However, comparing two distributions is not straightforward. We use the classification presented in dC17, which is done using the Hellinger Distance (HD) method. The HD measures the distance between two discrete distributions, $P_{1}$ and $P_{2}$, and is expressed as
\begin{equation}
\label{eq:HD_equation}
    HD^{2}(P_1, P_2)  = 2 \sum_{x} \left[\sqrt{p_{1}(x)} - \sqrt{p_{2}(x)}  \right]^{2},
\end{equation}
where $p_{1}$ and $\rm p_{2}$ are the two probability density functions (PDFs) and $x$ is a random variable. 

\begin{figure}[ht]
    \centering
    \includegraphics[width=\linewidth]{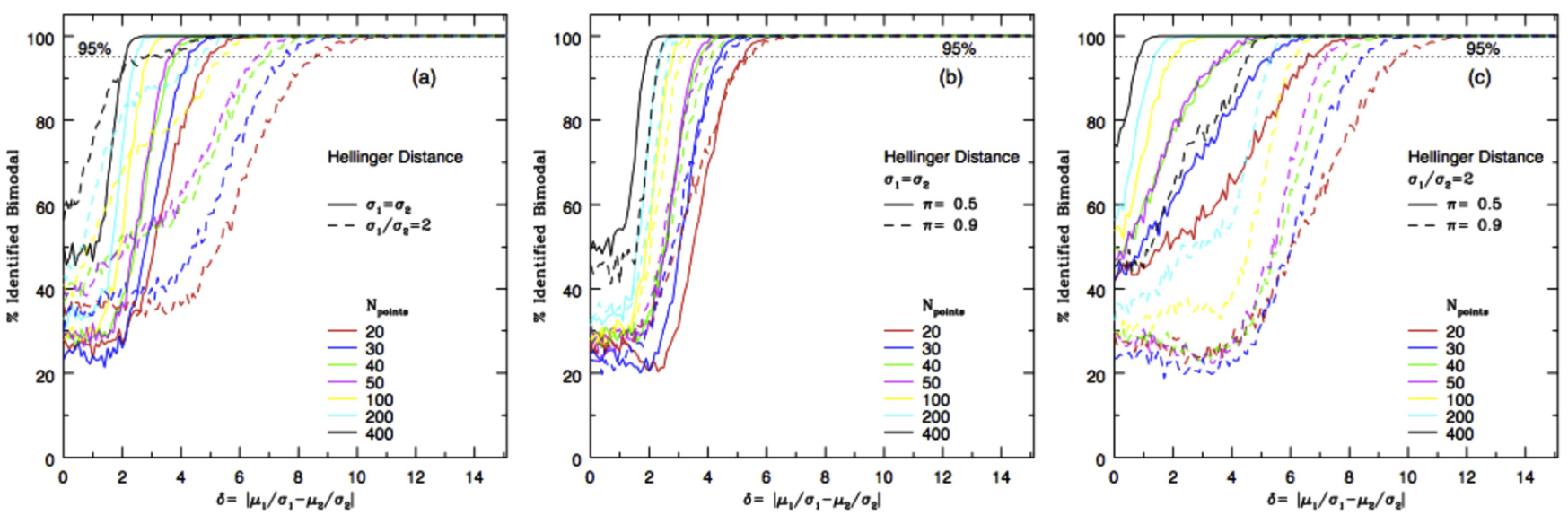}
    \caption{Performance of HD in a simulated bimodal data set, and its dependence on different sample size, $N_{\rm points}$; the proportion in one group, $\pi_{\rm i}$, which varies from 0.5 to 0.9, in steps of 0.1; and the Full Width at Half Maximum (FWHM or $\sigma_{\rm i}$) of the Gaussian. In all three panels, we display the percentage of identified bimodal distributions as a function of $\delta_{\rm modes}$. A dotted line is displayed in each panel, identifying the 95\% recovery. Different colors for dashed and solid lines indicate different $N_{\rm points}$ and conditions for $\sigma_{\rm i}$ and $\pi_{\rm i}$. Adapted from \citealt{2017AJ....154...96D}.}
    \label{fig:HD_performance}
\end{figure}

A detailed characterization of the HD performance in identifying the mixture of two normal distributions, characterized by amplitude, mean and variance given by [$\pi_{1}, \mu_{1}, \sigma_{1}$] and [$\pi_{2}, \mu_{2}, \sigma_{2}$]. We ``quantify'' the distance between the two distributions using 
\begin{equation}
    \delta_{\rm modes} = \left | \frac{\mu_{1}}{\sigma_{1}} - \frac{\mu_2}{\sigma_{2}} \right |.
\end{equation}
The performance of the HD method in a simulated bimodal data set and its dependence on different sample size, $N_{\rm points}$, as a function of $\delta_{\rm modes}$ is shown in Fig.~\ref{fig:HD_performance}

The method is shown robust to systems with at least 20 members within $R_{200}$, which translates to a cutoff in mass. The relation between $M_{200}$ and $N_{200}$ (where $N_{200}$ is the number of B galaxies inside $R_{200}$) yields a lower mass threshold of $ 10^{14} {\rm M}_{\odot}$ for our sample, similar to the threshold applied in the previous subsection. 

\begin{figure}[ht]
    \centering
    \includegraphics[width=0.7\linewidth]{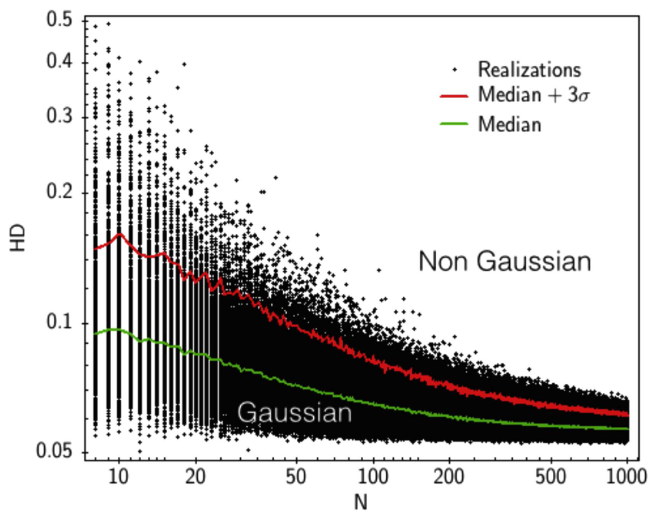}
    \caption{Calibration of the relation between HD and the number of points, $N_{\rm points}$, sampling the distribution. Green solid line indicates the median of HD for a given number of points. Red solid line displays the median of HD$ + 3\sigma$. This is the line used to separate G from NG systems. Adapted from \citealt{2017AJ....154...96D}.}
    \label{fig:HD_threshold}
\end{figure}

Furthermore, it is important to define HD threshold separating G from NG, and its dependence on the number of points representing the distribution. In this regard, the threshold is defined using 1000 realizations of a Gaussian distribution with $\mu_{1} = 0$ and $\sigma_{1} = 1$, a given number of points. Fig.~\ref{fig:HD_threshold} shows how
HD varies with the number of points defining the Gaussian
distribution. As can be seen, the median HD, computed from
the 1000 realizations, decreases with the number of points (green line). The final rule to establish whether a distribution, with a given number of points, is G or NG is median$+3\sigma$ (red line). 

We also restrict our dynamical state analysis to systems with at least $70\%$ reliability, measured with a bootstrap technique, on the gaussianity classification. This reduces the 254 clusters to 177 clusters (143 G and 34 NG). The ratio of G and NG clusters in our sample ($N_{\rm G}/N_{\rm Total} \sim 80 \%$; $N_{\rm NG}/N_{\rm Total} \sim 20 \%$) is in agreement with previous works (e.g. \citealt{2013MNRAS.434..784R}). We herafter define this sample of 177 clusters as ``GNG sample''.

\section{Assessing sample completeness}

When using large samples to draw conclusions from statistical inference, it is fundamental to assess the sample completeness. Completeness refers to the extent to which a survey/sample includes all relevant objects within a certain volume or parameter space, such as brightness or redshift. Incomplete samples can introduce biases, as certain types of galaxies (e.g., faint, distant, or low surface brightness galaxies) may be underrepresented. This can bias statistical analyses and lead to incorrect conclusions about the properties and distribution of galaxies. For instance, if a survey is incomplete at faint magnitudes, it may underestimate the number of low-luminosity galaxies, affecting studies on the luminosity function and galaxy formation models.

\subsection{Mass completeness}

\begin{figure}[ht]
    \centering
    \includegraphics[width = 0.5\textwidth]{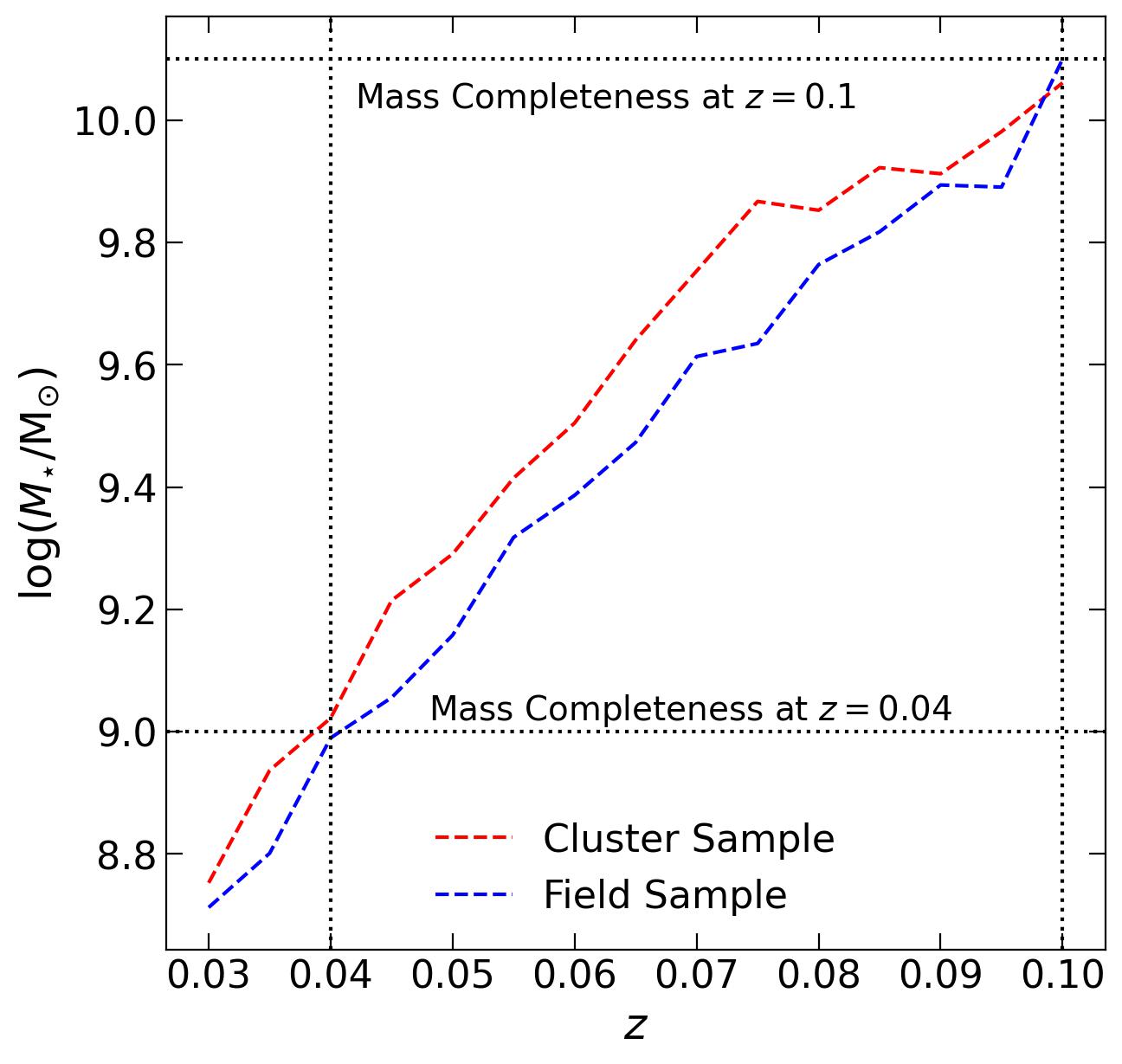}
    \caption{Stellar mass completeness as a function of redshift for the cluster (red curve) and field (blue) samples. The black dashed lines denote the mass completeness at $z=0.1$ (top line) and $z = 0.04$ (bottom line). In the following, when considering galaxies below $10^{10}M_{\odot}$, we restrict our analysis to $z \leq 0.04$.}
    \label{fig:mass_completeness}
\end{figure}

To guarantee that our sample is representative of the physical processes affecting galaxies' evolution, it is pivotal to assess the mass completeness as a function of redshift. This follows from galaxy evolution being strongly connected to their stellar mass. Therefore, we divide galaxies into redshift bins with a width of 0.005 and compute the 95\% quantile in the stellar mass distribution for each bin. Given that the cluster and field samples have different selection functions, this procedure is performed separately for both. In Fig.~\ref{fig:mass_completeness} we show our mass completeness as a function of redshift for both cluster (red) and field (blue) samples. Notably, there is an average offset of approximately 0.1 dex between the cluster and field samples.

In Figure~\ref{fig:mass_completeness}, the upper black dashed line indicates that our dataset is complete across the entire redshift range for galaxies with stellar masses greater than $10^{10.1} \rm{M_{\odot}}$, regardless of environmental factors. To ensure completeness for galaxies with stellar masses down to $10^{9} \rm{M_{\odot}}$, we establish a secondary redshift limit. The lower black dashed line sets this threshold at $z=0.04$. Consequently, we analyze the full redshift range for galaxies with stellar masses exceeding $10^{10}  \rm{M_{\odot}}$, while our analysis for galaxies with stellar masses between $10^{9} \rm{M_{\odot}}$ and $10^{10} \rm{M_{\odot}}$ is restricted to the redshift range $0.03 \leq z \leq 0.04$. Although different redshift ranges are applied to different stellar mass bins, we anticipate minimal evolution in the properties of galaxies within the studied range ($0.03 \leq z \leq 0.1$).

\subsection{Completeness at small radii}
\label{sec:compl_radii}

\begin{figure}
    \centering
    \includegraphics[width = \textwidth]{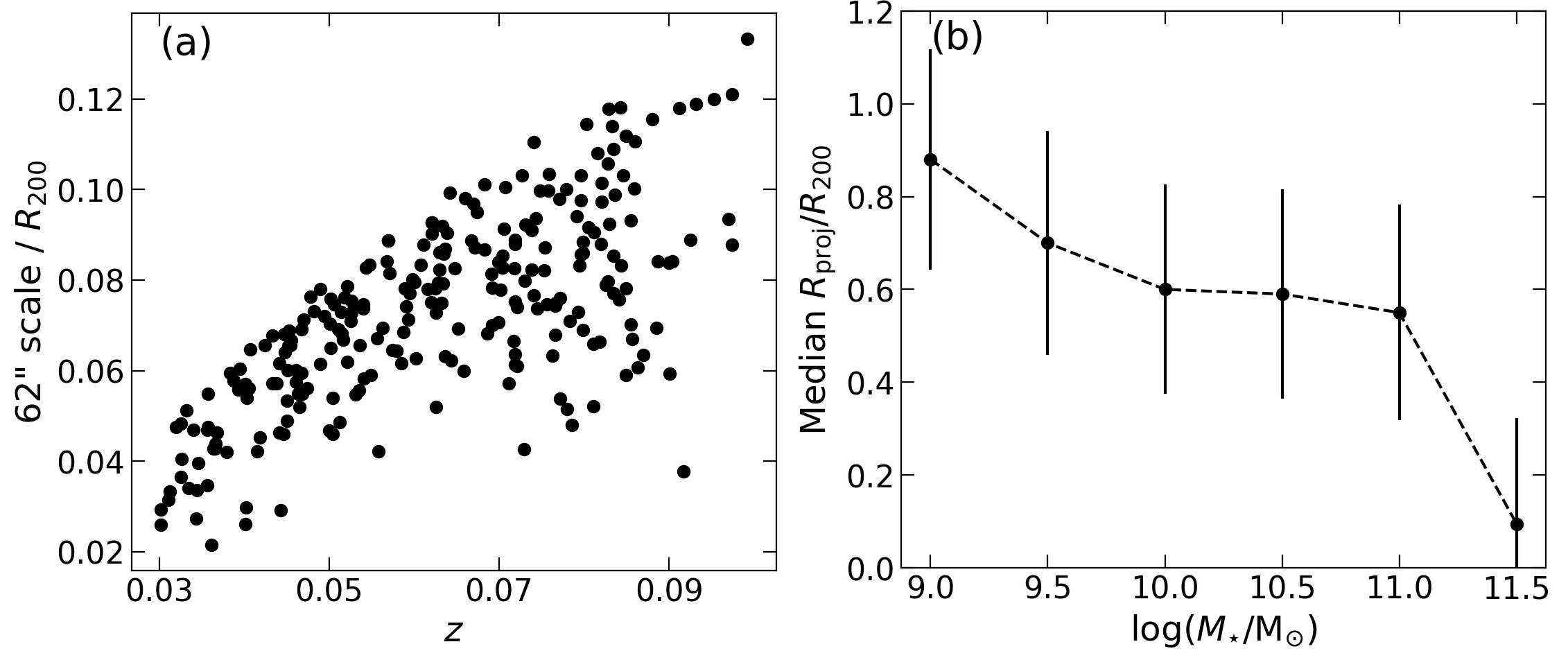}
    \caption{Top: ratio between the fiber collision limiting 62'' scale and $R_{200}$ for clusters at different redshifts. Bottom: median $R_{\rm proj}/R_{200}$ as a function of stellar mass. We limit our analysis to the $R_{\rm proj}/R_{200} \geq 0.1$ region in clusters (see text).}
    \label{fig:radii_completeness}
\end{figure}

In particular, for our cluster sample, the SDSS Legacy survey suffers from fiber collision effects, such that no two fibers can be placed closer than 62 arc-seconds. This limitation means that galaxies closer than a certain threshold, depending on redshift, cannot both be observed. To quantify this effect, we consider the ratio between the 62'' distance converted to kpc at each redshift and the $R_{200}$ for each cluster. Panel (a) of Figure~\ref{fig:radii_completeness} shows the distribution of this ratio as a function of redshift for the clusters in our sample. On average, the 62'' scale represents $7 \pm 2 \%$ of $R_{200}$.

In addition to the instrument limitations, the intracluster light in the inner region of clusters is dominated by the BCG, which can hinder the detection of faint, low-mass objects in the vicinity of the BCG. To address this issue, we investigate the median $R_{\rm proj}/R_{200}$ as a function of stellar mass. The results are shown in panel (b) of Figure~\ref{fig:radii_completeness} and suggest that the most massive galaxies primarily dominate the region with $R_{\rm proj}/R_{200} \leq 0.1$. This observation aligns with the principle of mass segregation \citep{1980ApJ...241..521C}, where massive galaxies are predominantly located in the cores of clusters, while less massive systems are found in the outskirts.

To minimize both instrumental and contamination from massive galaxies' light in our conclusions, we hereon limit our analysis to the $R_{\rm proj}/R_{200} \geq 0.1$ region of clusters.

\subsection{The special case of the GNG sample}
\label{subsec:Bright_Faint}

Our GNG sample is built to understand how different dynamical states affect clusters. However, due to observation limitations, we are not able to track every cluster with the same depth regarding the luminosity function. This follows from Fig.~\ref{fig:mass_completeness}, which shows that, while we have a complete sampling of the down to $10^{9} {\rm M}_{\odot}$ up to $z = 0.04$, after this threshold we gradually loose faint galaxies. The whole redshift range is complete only down to $10^{10} {\rm M}_{\odot}$. 

Thus, to guarantee the same sampling of the luminosity function for all clusters, we separate member galaxies into two different luminosity regimes 
\begin{itemize}
    \item Bright (B) means $0.03 \leq z \leq 0.1$ and $M_{\rm r} \leq -20.55 \sim M^{*} + 1$, where $M_{\rm r}$ is the limiting absolute magnitude in the r-band. This regime ensure completeness in galaxy mass down to $\sim 10^{10} {\rm M}_{\odot}$;

    \item 2) Faint (F) means $0.03 \leq z \leq 0.04$ and $ -20.55 < M_{\rm r} \leq -18.40 \sim M^{*} + 3$. This regime ensure completeness in galaxy mass between $10^{9} \leq M_{\star} \leq 10^{10} {\rm M}_{\odot}$, and \textit{it does not overlap the bright regime.} 
\end{itemize}
This ensures that we are sampling all the cluster to the same depth in the luminosity function and further provide robustness for the comparisons presented in \autoref{chapter: G_vs_NG}. Our GNG sample comprises 6578 B galaxies (4817 in G clusters and 1661 in NG) and 2205 F galaxies (907 in G and 1298 in NG).
 
\section{Chapter summary}

In this chapter we present the samples that will be used in the following analyses. We start by selecting galaxies with both reliable photometric and spectroscopic observations from the SDSS database. We impose completeness criteria, following the SDSS observational limitations. We then select galaxy properties from a variety sources, which includes: emission line properties, BPT classification and spectrum derived quantities ($M_{\star}$, SFR, sSFR and $\sigma_{\rm galaxy}$) from the MPA-JHU catalog; Morphological characterization (through the T--Type parameter) from the \citealt{2018MNRAS.476.3661D} catalog; and spectrum fitting derived $Age$ and stellar metallicity from \citealt{2017AJ....154...96D}. 

\begin{landscape}

    \begin{figure}[hpt]
        \centering
        \includegraphics[height = 0.8\textwidth]{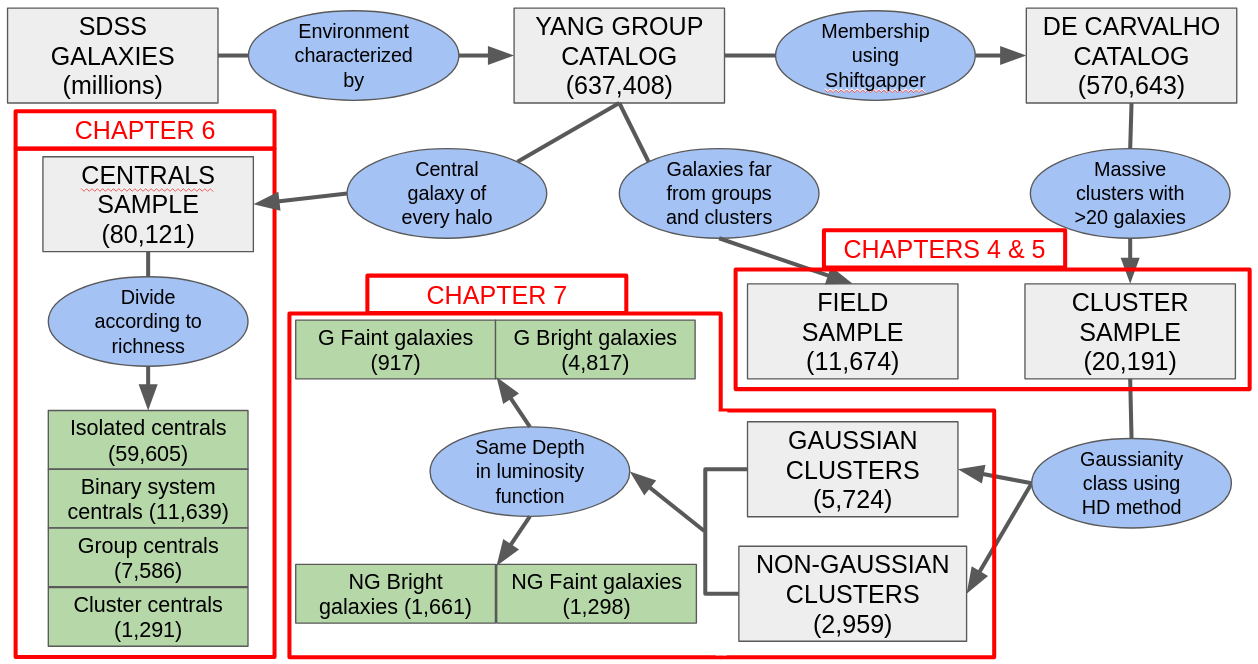}
        \caption{Diagram representing the selection of the three samples used in this project. The blue ellipses show the selection method applied between samples/subsamples. The red rectangles highlight in which chapter the sample will be used. Sub-samples from a larger sample are presented in green coloured rectangles.}
        \label{fig:samples_resume}
    \end{figure}
    
\end{landscape}

Three main samples that are investigated in this project: central sample, field sample and cluster sample. Each sample is defined with a pre-defined purpose, such that no following chapter will analyse all the three samples at once. In particular, the central and cluster sample are further divided into sub-samples. The former is classified into ``isolated centrals'', ``binary system centrals'', ``group centrals'' and ``cluster centrals'', based on the number of galaxy members occupying the same dark matter halo, following the Yang Catalog classification. The cluster sample is further divided according to the gaussianity of the member galaxy velocity distribution projected along the line-of-sight into ``Gaussian'' (G) and ``non-gaussian'' (NG). Also when analysing G and NG clusters, we separate member galaxies into two luminosity regimes (bright and faint), to guarantee the same sampling of the luminosity function for all clusters. We schematically summarize our samples definition in Fig.~\ref{fig:samples_resume}.

%% file: Chapters/Chapter4.tex
\chapter{Evolution of satellite galaxies in massive clusters}
\label{chap:sf_vs_morphology}

The text from this section is originally presented in 
\noindent \textit{Sampaio, V. M., De Carvalho, R. R., Ferreras, I., Aragón-Salamanca, A., \& Parker, L. C. (2022). From blue cloud to red sequence: evidence of morphological transition prior to star formation quenching. Monthly Notices of the Royal Astronomical Society, 509(1), 567-585.}

\thispagestyle{empty}

\noindent

\section{Overview}

We present a study of a sample of 254 clusters from the SDSS-DR7 Yang Catalog and an auxiliary sample of field galaxies to perform a detailed investigation on how galaxy star formation quenching depends on both environment and galaxy stellar mass. Our samples are restricted to 0.03$\leq$z$\leq$0.1 and we only consider clusters with $\log(M_{\rm halo}/{\rm M}_{\odot}) \geq 14$. Comparing properties of field and cluster galaxies in the Blue Cloud, Green Valley and Red Sequence, we find evidence that field galaxies in the red sequence hosted star formation events $\rm 2.1 \pm 0.7$ Gyr ago, on average, more recently than galaxies in cluster environments. Dissecting the star formation rate vs stellar mass diagram we show how morphology rapidly changes after reaching the green valley region, while the star formation rate keeps decreasing. In addition, we use the relation between location in the projected phase space and infall time to explore the time delay between morphological and specific Star Formation Rate variations. We estimate that the transition from late to early-type morphology happens in $\Delta t_{\rm inf} \sim$1 Gyr, whereas the quenching of star formation takes $\sim$3 Gyr. The time-scale we estimate for morphological transitions is similar to the expected for the delayed-then-rapid quenching model. Therefore, we suggest that the delay phase is characterized mostly by morphological transition, which then contributes morphological quenching as an additional ingredient in galaxy evolution.

\section{Environmental Impact on the Observed Bimodality}

\begin{figure}[ht]
    \centering
    \includegraphics[width = \textwidth]{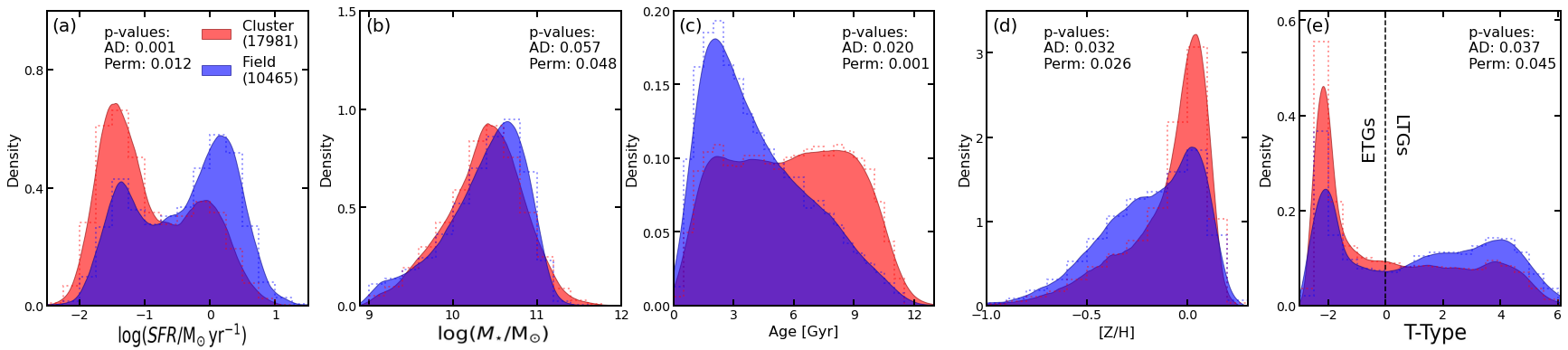}
    \caption{Distribution of SFR, $M_{\star}$, $Age$, $[Z/H]$ and T--Type. We separate galaxies according to their environment into clusters (red) and field (blue). The dashed lines represent the distributions histogram, while the filled area represent an epanechnikov kernel density estimate with bandwidth set equal to half of the dashed line histogram bins. The dark dashed vertical line in panel (e) denote the separation between early and late type morphology. We also add in each panel the resulting p-value of Anderson-Darling (AD) and Permutation (Perm) statistical tests. P-values smaller than 0.05 means the two distributions are statistically different.}
    \label{fig:Full_Properties}
\end{figure}

The observed bimodality in certain galaxy properties has been shown to depend on stellar mass. In particular, the bimodality is stronger at lower stellar masses. In this section we focus on exploring how the bimodality in different galaxy properties depend not only on stellar mass, but on environment as well. In Fig.~\ref{fig:Full_Properties} we show the distributions for SFR, $M_{\star}$, $Age$, $[Z/H]$ and morphology (T--Type) for galaxies in clusters (red) and in the field\footnote{Low interacting galaxies} (blue). The distributions are built using a Epanechnikov kernel density estimation with bandwidth set to 1.5 times the bin of the dotted histograms in Fig.~\ref{fig:Full_Properties}. We apply the kernel density estimate techniques directly on the data and only use the histograms for the choice of bandwidth. We statistically compare the distributions using 2-sample hypothesis tests. In each panel we display the resulting p-value of k-sample Anderson-Darling and Permutation tests. We decide to use two different statistical tests so that our results are free of any underlying hypothesis of such tests. We select Anderson-Darling and permutation tests due to the different approach they use to measure similarities between the distributions, which enhance the reliability in the derived p-values and distribution comparison. We adopt $\alpha = 0.05$ as the significance level.

Our results reinforce previously found trends. Properties of cluster and field galaxy are significantly different, which is statistically confirmed by the p-values shown in each panel. In panel (a), we find an excess of quenched galaxies ($\log(\text{SFR} / {\rm M}_{\odot} \, {\rm yr}^{-1}) < -1$ ) in clusters, whereas field galaxies have more active star formation, which can be seen by the excess of field galaxies with $\log(\text{SFR} / {\rm M}_{\odot} \, {\rm yr}^{-1}) > -0.5$. Note that the stellar mass distributions are marginally similar (see panel b), which show that the differences in star formation history and environment result in different stellar population properties, at fixed stellar mass. These differences are also seen in $Age$ (panel c), in which field galaxies have had a more recent star formation episode\footnote{$Age$ here is closely related to the last star formation episode.} (blue peak at lower ages) in comparison to clusters. These trends indicate how environment is directly affecting the observed bimodality and is further reinforced exploring $[Z/H]$ (panel d). Older galaxies are expected to have higher metallicities, for which we see an excess of more metal-poor galaxies in the field in comparison to clusters. Ultimately, environment affects morphology and we find a percentage excess of LTGs in the field in comparison to clusters, in agreement with the Morphology-Density relation.

In this section we investigate overall comparisons between galaxy properties inhabiting distinct environments. The reported results indicate how environment plays a major role in galaxy evolution even when considering global distributions such as Fig.~\ref{fig:Full_Properties}. Our results suggest that galaxies in the field hosted more recent star formation episodes in comparison to those in clusters. We call attention to the presence of a significant fraction of quenched galaxies in the field. This shows how quenching mechanisms unrelated to environment are sufficient to quench field galaxies, whilst environmental effects act like a catalyst for the quenching process, which happens faster in clusters. Namely, in dense environments: 1) galaxy formation can happen earlier due to larger density fluctuations; and 2) clusters add new channels for star formation quenching through interactions, which are not seen for galaxies evolving in isolation. The next step is to consider not only global distributions, but divide galaxies into BC, GV and RS subsamples.

\section{Tracing A Galaxy Path Towards the Red Sequence}

Galaxy evolution from the BC to the RS is directly related to the quenching of star formation. However, galaxy evolution is seen in other properties as well. Quiescent galaxies are mostly early-type \footnote{Note that 30\% of RS galaxies are LTGs and 10\% of BC galaxies are ETGs.}, older and more metal-rich than star-forming ones. It is hence important to understand whether popular definitions of the green valley cover all the properties expected for a transitioning galaxy. A common GV definition is based on the $\log(\text{SFR} / {\rm M}_{\odot} \, {\rm yr}^{-1}) \, vs \, \log(M_{\star}/{\rm M}_{\odot})$ plane, which traces mostly the current star formation, whereas the $\rm D4000$ break definition traces the current stellar population. We adopt the definition presented in \citet{2020MNRAS.491.5406T}, which use the relations
\begin{equation}
     \log(\text{SFR} / {\rm M}_{\odot} \, {\rm yr}^{-1}) = 0.7 \log(M_{\star}/{\rm M}_{\odot}) - 7.52
\end{equation}
to divide BC and GV galaxies and 
\begin{equation}
     \log(\text{SFR} / {\rm M}_{\odot} \, {\rm yr}^{-1}) = 0.7 \log(M_{\star}/{\rm M}_{\odot}) - 8.02
\end{equation}
to separate GV and RS. Further, galaxy evolution is known to be mass-dependent. We hence divide galaxies into three logarithmic stellar mass bins ($\log(M_{\star}/{\rm M}_{\odot})$ of 9 to 10, 10 to 11 and 11 to 12), which are selected to trace different star formation histories \citep{2012ApJ...752L..27T}. In Table \ref{tab:galaxies_in_each} we display the number of Cluster and Field galaxies in each stellar mass regime and location in the $ \log(\text{SFR} / {\rm M}_{\odot} \, {\rm yr}^{-1}) \, vs. \, \log(M_{\star}/{\rm M}_{\odot})$ plane. In a few words, the RS regions becomes more populated for increasing stellar mass and environment density. 
\begin{table}
\centering
\caption{The number (percentage) of cluster or field galaxies in a given stellar mass range and location in the $\log(\text{SFR} / {\rm M}_{\odot} \, {\rm yr}^{-1})$ vs $\log(M_{\star}/{\rm M}_{\odot})$ plane. BC, GV and RS are defined following \cite{2020MNRAS.491.5406T}.}
\label{tab:galaxies_in_each}
\resizebox{0.85\columnwidth}{!}{%
\begin{tabular}{cc|c|c|c}
\cline{3-5}
 &  & \multicolumn{3}{c}{Region in Plane} \\ \hline
\multicolumn{1}{c|}{Sample} & $X = \log(M_{\star}/{M}_{\odot})$ & BC (\%) & GV (\%) & RS (\%) \\ \hline
\multicolumn{1}{c|}{\multirow{3}{*}{Cluster}} & $9 \leq X < 10$ & 1727 (61\%) & 330 (11\%) & 816 (28\%) \\
\multicolumn{1}{c|}{} & $10 \leq X < 11$ & 1991 (21\%) & 970 (10\%) & 6519 (69\%) \\
\multicolumn{1}{c|}{} & $11 \leq X < 12$ & 55 (6\%) & 57 (6\%) & 855 (88\%) \\ \hline
\multicolumn{1}{c|}{\multirow{3}{*}{Field}} & $9 \leq X < 10$ & 2074 (87\%) & 122 (5\%) & 185 (8\%) \\
\multicolumn{1}{c|}{} & $10 \leq X < 11$ & 4064 (47\%) & 916 (11\%) & 3667 (42\%) \\
\multicolumn{1}{c|}{} & $11 \leq X < 12$ & 93 (13\%) & 93 (13\%) & 460 (74\%) \\ \hline
\end{tabular}
}
\end{table}

\begin{figure}[ht]
    \centering
    \includegraphics[width = 0.5\columnwidth]{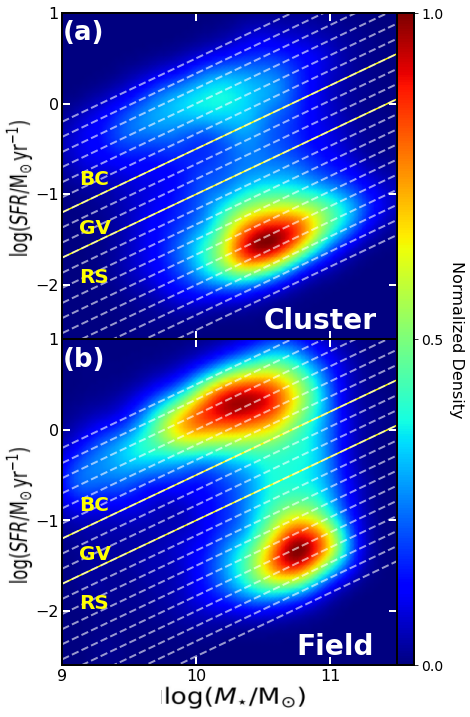}
   \caption{Kernel smoothed normalized density in the SFR vs $M_{\star}$ plane for cluster (panel a) and field (panel b) galaxies. Yellow lines denote the limits of BC, GV, and RS according to \citet{2020MNRAS.491.5406T}. The white dashed lines have slope equal to equation (1), but with varying intercept from 6.3 to 9.4 in steps of 0.17 (see Section \ref{sec:TTRD} for an explanation).}
    \label{fig:SFR_Mstellar_density}
\end{figure}

In Fig.~\ref{fig:SFR_Mstellar_density} we present an adaptive kernel density estimate for the cluster (panel [a]) and field (panel b) galaxy SFR vs $M_{\star}$ distributions. The limiting lines between BC/GV and GV/RS are shown in yellow. In panel (a), we find a single high density peak at the RS, which comprises $\sim$ 62\% of the cluster sample. This indicates that more than half of the cluster galaxies reached a passive state, where most of evolution is driven by stellar evolution. However, it is important to stress that galaxies can also experience an ``inverse evolution'', in which galaxies go from the RS to the BC due to rejuvenation processes \citep[e.g.][]{2016MNRAS.460.3925T}. Cluster Galaxies in the RS are characterized by median values of $10.73 \pm  0.23$ and $-1.64 \pm  0.19$ in $\rm log(M_{stellar})$ and $\log(\text{SFR} / {\rm M}_{\odot} \, {\rm yr}^{-1})$, respectively. On the other hand, the bimodality is more striking in the field sample. Panel (b) shows how most of the star forming galaxies are found in the field, namely $\sim 53\%$ and $\sim 38\%$ of field galaxies are in the BC and RS, respectively. It is clear that most of the active star formation is happening in galaxies with lower stellar mass, which is in agreement with the downsizing scenario, in which most of the star formation in the current universe is happening in low mass galaxies \citep{2006MNRAS.372..933N}.

\begin{figure}[ht]
    \centering
    \includegraphics[width = \textwidth]{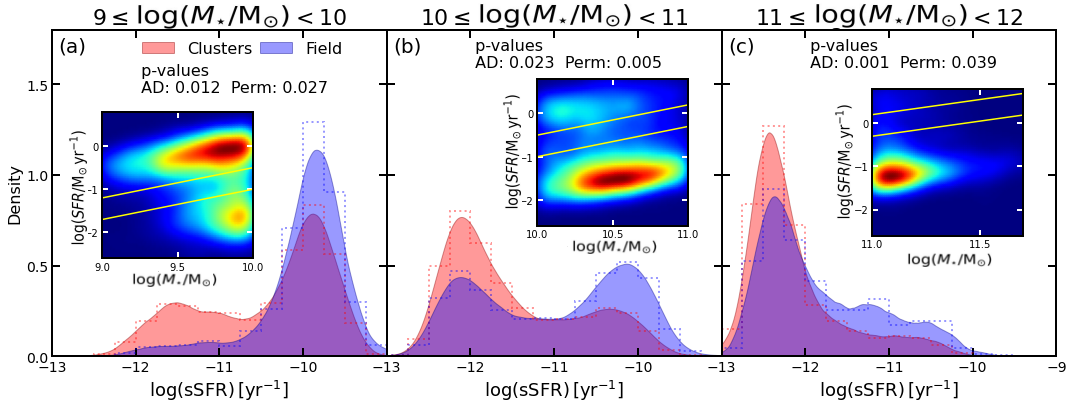}
   \caption{Distributions of sSFR for cluster (red) and field (blue) galaxies for the low (a), intermediate (b) and high (c) stellar mass regimes. We add the mass range in the top of each panel. The filled area is an epanechnikov kernel density estimate using bandwidth equal to 2 times the bin of the displayed histogram in dashed lines. In each panel we present the resulting p-values of a permutation (Perm) and Anderson-Darling (AD) statistical test. For comparison, we include an inset of the relevant part of Fig's.~\ref{fig:SFR_Mstellar_density} panel (a), representing cluster galaxies, for each panel.}
    \label{fig:sSFR_hist}
\end{figure}

A fundamental parameter in astrophysics is the sSFR. This quantity is usually reported in units of $\rm yr^{-1}$, hence its inverse gives an estimate of a time-scale the galaxy would take to form its current stellar component with its current SFR.  In Fig.~\ref{fig:sSFR_hist} we present the distribution of $\log(\text{sSFR}/{\rm yr}^{-1})$ for cluster (red) and field (blue) galaxies for each stellar mass bin. The shaded areas are built using an epanechnikov kernel density estimator with bin width equal to 0.25 in both cases. For comparison, we add a miniature version of the related SFR vs $M_{\star}$ plane for cluster galaxies in each panel. We compare the distributions using the permutation (perm) and Anderson-Darling (AD) two sample statistical tests and report the p-values in each panel. Exploring the low mass regime (panel a), we find $\sim$84\% galaxies with $\log(\text{sSFR}/{\rm yr}^{-1}) > -10.5$, in comparison to $\sim$62\% for cluster galaxies. We find an excess of low sSFR galaxies in clusters in comparison to the field (26\% and 9\%, respectively, have $\log(\text{sSFR}/{\rm yr}^{-1})<-11$). It is important to stress that, even in clusters, low mass galaxies are those predominantly in the BC as can be seen by the inset in panel a. In panel b and c, we see an increase of quenched fraction with increasing stellar mass. Panel b displays the distribution peaks at different $\log(\text{sSFR}/{\rm yr}^{-1})$, for which we find cluster galaxies to have dominantly low $\log(\text{sSFR}/{\rm yr}^{-1})$ in comparison to the peak we find for the field. Finally, in panel c, massive galaxies are predominantly quenched, independent of environment. This suggests that massive galaxies can rely mostly on internal mechanisms to halt their star formation and do not strongly depend on environmental effects as is the case for lower mass galaxies. Yet we do find statistically different distributions for cluster and field massive galaxies, which may indicate the major effect of environmental is to simply ``accelerate the proccess''.

\subsection{Cluster vs. Field Galaxy Properties}

\begin{figure}
    \centering
    \includegraphics[width = \textwidth]{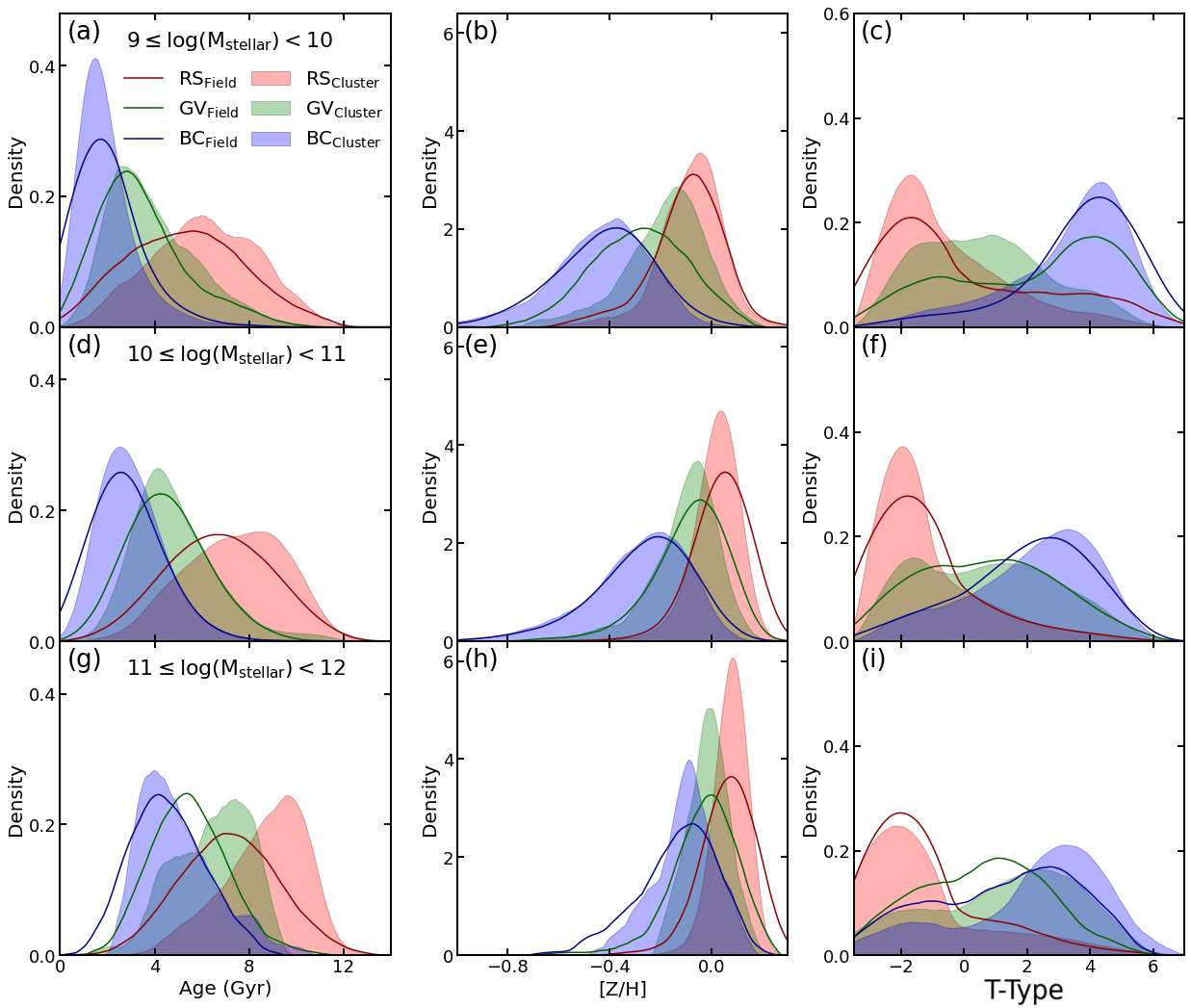}
    \caption{Each panel is an epanechnikov kernel density estimate for the distributions of of $Age$, $[Z/H]$ and T--Type for galaxies in RS (red), GV (green) and BC (blue). Cluster galaxy distributions are shown as filled areas, whereas solid lines  represent field distributions. Each row corresponds to a stellar mass bin, which is presented in the first column.}
    \label{fig:Cluster_vs_Field_properties}
\end{figure}

Galaxies in different environments are affected by different quenching mechanisms, which leave an imprint on galaxy properties. In Fig.~\ref{fig:Cluster_vs_Field_properties} we present the distributions of $Age$, $[Z/H]$ and T--Type of cluster and field galaxies. We separate them according to stellar mass regime and location in the SFR vs. $M_{\star}$ diagram (RS - red, GV - green and BC - blue). The filled areas represent the distribution for clusters, whereas solid lines denote its field counterpart. In Tab.~\ref{tab:p_values_field_cluster} we show the resulting p-values for AD and Permutation tests comparing cluster and field galaxy $Age$, $[Z/H]$ and T--Type distributions. The highlighted cells correspond to distributions statistically different according to AD and permutation tests.

\begin{table}
\caption{The resulting p-values of AD and Permutation tests comparing the distributions of cluster and field galaxy $Age$, $[Z/H]$ and T--Type. We separate galaxies according to luminosity and location in the SFR vs. $M_{\star}$ plane. We highlight in red the statistically different distributions.}
\label{tab:p_values_field_cluster}
\resizebox{\columnwidth}{!}{%
\begin{tabular}{ccccccccccc}
\cline{3-11}
 &  & \multicolumn{9}{c}{$\rm X = \log(M_{\star}/{\rm M}_{\odot})$} \\ \cline{3-11} 
 &  & \multicolumn{3}{c|}{$\rm 9 \leq X < 10$} & \multicolumn{3}{c|}{$\rm 10 \leq X < 11$} & \multicolumn{3}{c|}{$\rm 11 \leq X < 12$} \\ \cline{3-11} 
 &  & \multicolumn{1}{c|}{$Age$} & \multicolumn{1}{c|}{${[}Z/H{]}$} & \multicolumn{1}{c|}{T--Type} & \multicolumn{1}{c|}{$Age$} & \multicolumn{1}{c|}{${[}Z/H{]}$} & \multicolumn{1}{c|}{T--Type} & \multicolumn{1}{c|}{$Age$} & \multicolumn{1}{c|}{${[}Z/H{]}$} & T--Type \\ \hline
\multicolumn{1}{c|}{} & \multicolumn{1}{c|}{Perm} & \multicolumn{1}{c|}{0.204} & \multicolumn{1}{c|}{0.069} & \multicolumn{1}{c|}{0.073} & \multicolumn{1}{c|}{0.234} & \multicolumn{1}{c|}{0.403} & \multicolumn{1}{c|}{\cellcolor[HTML]{FFCCC9}0.012} & \multicolumn{1}{c|}{\cellcolor[HTML]{FFCCC9}0.003} & \multicolumn{1}{c|}{0.792} & 0.106 \\ \cdashline{2-11}
\multicolumn{1}{c|}{\multirow{-2}{*}{\begin{tabular}[c]{@{}c@{}}Blue\\ Cloud\end{tabular}}} & \multicolumn{1}{c|}{AD} & \multicolumn{1}{c|}{0.362} & \multicolumn{1}{c|}{0.088} & \multicolumn{1}{c|}{0.053} & \multicolumn{1}{c|}{0.435} & \multicolumn{1}{c|}{0.672} & \multicolumn{1}{c|}{\cellcolor[HTML]{FFCCC9}0.028} & \multicolumn{1}{c|}{\cellcolor[HTML]{FFCCC9}0.009} & \multicolumn{1}{c|}{0.603} & 0.163 \\ \hline
\multicolumn{1}{c|}{} & \multicolumn{1}{c|}{Perm} & \multicolumn{1}{c|}{\cellcolor[HTML]{FFCCC9}0.021} & \multicolumn{1}{c|}{\cellcolor[HTML]{FFCCC9}0.001} & \multicolumn{1}{c|}{\cellcolor[HTML]{FFCCC9}0.001} & \multicolumn{1}{c|}{0.304} & \multicolumn{1}{c|}{\cellcolor[HTML]{FFCCC9}0.015} & \multicolumn{1}{c|}{\cellcolor[HTML]{FFCCC9}0.013} & \multicolumn{1}{c|}{\cellcolor[HTML]{FFCCC9}0.006} & \multicolumn{1}{c|}{0.842} & \cellcolor[HTML]{FFCCC9}0.039 \\ \cdashline{2-11}
\multicolumn{1}{c|}{\multirow{-2}{*}{\begin{tabular}[c]{@{}c@{}}Green\\ Valley\end{tabular}}} & \multicolumn{1}{c|}{AD} & \multicolumn{1}{c|}{\cellcolor[HTML]{FFCCC9}0.018} & \multicolumn{1}{c|}{\cellcolor[HTML]{FFCCC9}0.002} & \multicolumn{1}{c|}{\cellcolor[HTML]{FFCCC9}0.001} & \multicolumn{1}{c|}{0.607} & \multicolumn{1}{c|}{\cellcolor[HTML]{FFCCC9}0.004} & \multicolumn{1}{c|}{\cellcolor[HTML]{FFCCC9}0.019} & \multicolumn{1}{c|}{\cellcolor[HTML]{FFCCC9}0.002} & \multicolumn{1}{c|}{0.352} & \cellcolor[HTML]{FFCCC9}0.018 \\ \hline
\multicolumn{1}{c|}{} & \multicolumn{1}{c|}{Perm} & \multicolumn{1}{c|}{\cellcolor[HTML]{FFCCC9}0.002} & \multicolumn{1}{c|}{0.386} & \multicolumn{1}{c|}{\cellcolor[HTML]{FFCCC9}0.009} & \multicolumn{1}{c|}{\cellcolor[HTML]{FFCCC9}0.001} & \multicolumn{1}{c|}{\cellcolor[HTML]{FFCCC9}0.003} & \multicolumn{1}{c|}{0.447} & \multicolumn{1}{c|}{\cellcolor[HTML]{FFCCC9}0.001} & \multicolumn{1}{c|}{0.053} & \cellcolor[HTML]{FFCCC9}0.001 \\ \cdashline{2-11}
\multicolumn{1}{c|}{\multirow{-2}{*}{\begin{tabular}[c]{@{}c@{}}Red\\ Sequence\end{tabular}}} & \multicolumn{1}{c|}{AD} & \multicolumn{1}{c|}{\cellcolor[HTML]{FFCCC9}0.003} & \multicolumn{1}{c|}{0.319} & \multicolumn{1}{c|}{\cellcolor[HTML]{FFCCC9}0.002} & \multicolumn{1}{c|}{\cellcolor[HTML]{FFCCC9}0.001} & \multicolumn{1}{c|}{\cellcolor[HTML]{FFCCC9}0.001} & \multicolumn{1}{c|}{0.274} & \multicolumn{1}{c|}{\cellcolor[HTML]{FFCCC9}0.001} & \multicolumn{1}{c|}{0.067} & \cellcolor[HTML]{FFCCC9}0.001 \\ \hline
\end{tabular}%
}
\end{table}

Exploring Fig.~\ref{fig:Cluster_vs_Field_properties}, we find increasing differences between cluster and field distributions with increasing stellar mass. Regarding age (first column panels), we find the most striking differences in the GV and RS distributions, in which field galaxies have an excess of young galaxies in comparison to clusters. For instance, 50.58\% of the most massive cluster galaxies in the RS have $Age > 8$ Gyr, while this percentage decreases to 28.06\% for field galaxies. In other words, this excess of younger GV/RS galaxies in the field suggests more recent star formation episodes in these systems in comparison to their counterpart in clusters. On the other hand, the differences we find in $[Z/H]$ (central column panels) are nuanced in comparison to those in $Age$. Qualitatively, field GV/RS $[Z/H]$ distributions have a slight excess of more metal-rich systems in comparison to clusters. However, the $[Z/H]$ distribution for the most massive GV galaxies in the field extends to lower metallicities in comparison to clusters. This is in agreement with previous works showing differences in spectral lines and structure of elliptical galaxies in clusters in comparison to those in the field \citep{1999ApJ...527...54B,2017A&A...597A.122S}. Regarding morphology, panel (c) indicates an excess of GV galaxies with lower values of T--Type in clusters, when compared to the field, which is further evidence of the environment acting on low-mass members.

\subsection{Towards the Red Sequence}
\label{sec:TTRD}
In the last section we detail, for a given galaxy environment and location in the SFR vs. $M_{\star}$ plane, significant differences in the $Age$, $[Z/H]$ and T--Type distributions between field and cluster galaxies. In Fig.~\ref{fig:Cluster_vs_Field_properties}, the intersections between the galaxy property distributions in the BC, GV and RS indicate that the definition from the SFR vs $M_{\star}$ perspective alone is insufficient to categorize galaxies. An intrinsic problem is that galaxy evolution is a continuous process, while the usual approach is to discretize galaxy populations to just three regions. In order to further understand how galaxies evolve from the BC towards the RS, we create 17 different regions in the SFR vs $M_{\star}$ plane for the cluster and field sample. The limits of each region are defined as:
\begin{equation}
    \log(\text{SFR} / {\rm M}_{\odot} \, {\rm yr}^{-1}) = 0.7 \log(M_{\star}/{\rm M}_{\odot}) - i
\end{equation}
with i varying from 6.5 to 9.5 in steps of 0.17, which guarantees a minimum of $\sim 50$ galaxies per slice. We present these regions as white dashed lines in Fig.~\ref{fig:SFR_Mstellar_density}. We then select galaxies within each slice and create normalized kernel density estimates for $Age$, $[Z/H]$ and T--Type. An important feature is the different number of galaxies in each slice. We then define a number-dependent bandwidth for the kernel density estimate, given as:
\begin{equation}
    \label{eq:BW}
    BW = 1.5 \times \frac{2 \times IQR}{N_{\rm slice}^{1/3}},
\end{equation}
where $BW$ is the bandwidth, $IQR$ is the interquartile range\footnote{Defined as the distance between the 75\% and 25\% quartiles.} and $N_{\rm slice}$ is the number of points in the slice. Equation \ref{eq:BW} is 1.5 times the optimal bin width of a histogram characterized by a given $IQR$ and $N_{\rm slice}$ \citep{scott1979optimal}. The factor 1.5 is empirically defined to slightly reduce the noise while preserving global trends. For each distribution, we trace the kernel peak density. However, as we adopt the normalized kernel, the peak density is directly related to the ``width'' of the observed distribution. Namely, high peaks denote narrower distributions, whereas low density values are related to broader distributions. We use a bootstrap technique with $N_{\rm bootstrap}=1000$ repetitions to assess errorbars. We define the SFMS as the slice containing the best linear fit of the BC galaxies. This procedure (in gif format\footnote{That can be viewed by opening the PDF in Adobe Reader 9.}) and results are presented in Figs.~\ref{fig:gif_evolution} and \ref{fig:gif_evolution_faint} for cluster and field galaxies, respectively. For better visualization, we further add an inset with the variation of the FWHM in each panel to quantify the mixture of galaxy population properties in each slice. Each curve covers roughly the same variation in the SFR vs. $ M_{\star}$ plane, namely from the beginning of the BC to the end of the RS. Nevertheless, since the comparison is done in the SFR vs. $ M_{\star}$ plane, we are not directly addressing the mechanism navigating galaxies from one slice to another. Rather, we are comparing galaxy properties with a similar SFR and $ M_{\star}$, but in different environments.

\begin{figure}
    \centering
    \animategraphics[width =  0.8\textwidth,controls,loop]{3}{Figures/Gif_Cluster/Hist_Evolution_Smooth_Cluster_No_Mass_Separation_v2_}{0}{17}
     \caption{
     This image is originally a gif. The image seen is the last frame/slice of the following procedure for cluster galaxies: 1) select galaxies within the solid black line slice highlighted in panel (a); 2) build a probability density histogram for each property ($Age$, $[Z/H]$ and T--Type). In this case, we use the Scott criteria \citep{scott1979optimal} to define the bin size in order to account for different number of points in each slice; 3) for each case we create a normalized epanechnikov kernel density estimate with bandwidth set to 1.5 times the histogram bin; 4) we then estimate the peak density, $\rm FWHM$ and related errorbars using a bootstrap technique (with $N_{\rm bootstrap} = 1000$ repetitions) for each slice. The curves in panels (b), (c) and (d) are the evolution of the peak density (large panel) and $\rm FWHM$ (miniature panel) as we progress from the top most slice to the bottom one. We color points and histograms according to the region in which the slice is (BC - blue, GV - green, RS - red). For completeness, as we evolve from top to bottom slice, we maintain the distributions from previous slices, which is shown as faded gray in each plot. The purple lines in panel (a) denote the slice containing the SFMS, which we then use the $\Delta {\rm SFMS}$ to report our results. In the same panel, we also add an arrow indicating increasing $\Delta {\rm SFMS}$. Finally, in panel (d) we stress through a black arrow the significant T--Type transition experienced by galaxies during the GV. Due to the use of a normalized kernel, there is a relation between peak density and FWHM, namely increasing peak density means decreasing FWHM.}
    \label{fig:gif_evolution}
\end{figure}

We next provide an overall description of the observed trends for galaxies in clusters. We quantify the trends using the slice's offset from the SFMS (shown in purple in panel (a)), namely $ \Delta {\rm SFMS}$, which is defined as the perpendicular distance with respect to the slices' slope (see black arrow in panel a). At the beginning of the BC ($\Delta {\rm SFMS} = 0.5)$, the age distribution (panel b) is highly peaked at $\sim 1.5$ Gyr, due to the active star formation in these systems. For the $[Z/H]$ (panel c), we find the opposite trend, namely the most star-forming BC galaxies have a broad $[Z/H]$ distribution peaking at low stellar metallicity ($\sim -0.5$). LTGs dominate galaxy morphology distribution at $\Delta {\rm SFMS} = 0.5$. The evolution towards $ \Delta {\rm SFMS} = 0.33$ (the last slice before GV) is then characterized by: 1) age distributions becoming broader and peaking at older values; 2) $[Z/H]$ distribution continues to be broad and evolve almost in a coeval way towards higher stellar metallicity; 3) there is a decrease in the peak value for T--Type distribution, but still characteristic of LTGs (T--Type > 0). The evolution in the GV ($-0.5 < \Delta {\rm SFMS} < -0.83$) indicate several processes happening simultaneously in galaxies. We find that age distribution have an almost constant peak age, but with increasing $\rm FWHM$, which suggest a mixed population with different recent star formation history. Regarding $[Z/H]$, the GV is roughly the region where distributions start to become narrower and with an increasing peak. However, we find the more striking result in morphology, for which the peak T--Type evolves from greater to smaller than zero in a single slice ($\Delta {\rm SFMS}$ from -0.67 to -0.83), despite a large $\rm FWHM$ and evidently bimodal T--Type distribution. From the GV on, age distributions have higher $\rm FWHM$ and an increasing peak age until the end of the RS ($\rm \Delta {\rm SFMS} \leq -2$), in which there is a decrease in the $\rm FWHM$. In panel (c), galaxies evolve towards an asymptotic value as they progresses in the RS, which is characterized by increasingly narrower distributions peaking at $[Z/H] \sim 0.1$. We find similar trends for morphology in the RS too. There is an increasingly excess of ETGs as we progresses towards the RS tail, as expected. In particular, at $\Delta {\rm SFMS} = -1.33$ most of the galaxies are ETGs, despite not yet being completely quenched. This suggests that galaxies may reach their final morphological stage before the full star formation quenching.

In Fig.~\ref{fig:gif_evolution_faint} we show the result for field galaxies. We find similar evolution for the cluster and field galaxies evolution across the SFR vs $M_{\star}$ plane, which indicates that their difference is more related to the acting quenching mechanisms instead of the galaxy property itself. The GV also denotes the region in which we find the transition towards broader and narrower age and $[Z/H]$ distributions, respectively, and a significant T--Type peak variation from LTG to ETG morphology. However, we stress that the difference we find for age distributions at the end of the RS ($\Delta{\rm SFMS} = -2.33$) is broader than what we find for clusters. This possibly highlights the clusters influence to quickly evolve galaxies towards higher ages (which means lack of star formation), whereas field galaxies may host more recent star formation episodes, resulting in broader distributions for field galaxies age at the end of the RS in comparison to clusters.

\begin{figure*}
    \centering
    \animategraphics[width =  0.8\textwidth,controls,loop]{3}{Figures/Gif_Field/Hist_Evolution_Smooth_Field_No_Mass_Separation_v2_}{0}{17}
     \caption{The same as Fig.~\ref{fig:gif_evolution}, but for field galaxies.}
    \label{fig:gif_evolution_faint}
\end{figure*}

Using the SFR vs $M_{\star}$ diagram, we dissect how galaxy properties evolve as they go from the BC to the RS. An important feature is the plateau followed by an increase we find in age for galaxies in clusters. It is characterized by a low peak density, which means that each region composing the plateau has a high $\rm FWHM$, thus indicating a mixture of galaxies with several ages in a given slice. The more striking result the broader distributions, but with increasing peak ages as we approach the RS tail. In other words, moving from one slice to the following one just moves the entire distribution towards higher ages. This is a first suggestion on how the cluster environment feeds the GV region in a continuous form due to its infall rate. Galaxies seems then to evolve in a coeval manner from the GV to the RS, which results in the plateau of equal peak density and increasing age. Moreover, ``the end of the line'' is the tail of the RS, for which the cluster environment accumulates quenched galaxies. With respect to $[Z/H]$, our results suggests an upper achievable limit due to the lack of further star formation events when reaching the RS. At last, galaxy morphology changes towards early-type shapes in just a few ``slices'' for both field and cluster environment, which suggests that most of the morphological transformation (if it happens) takes place in the green valley region. Finally, the plateau in age occurs exactly when we find peak densities towards lower values of T--Type.

\subsection{Dependence on Galaxy Stellar Mass}

\begin{figure}
    \centering
    \includegraphics[width = \textwidth]{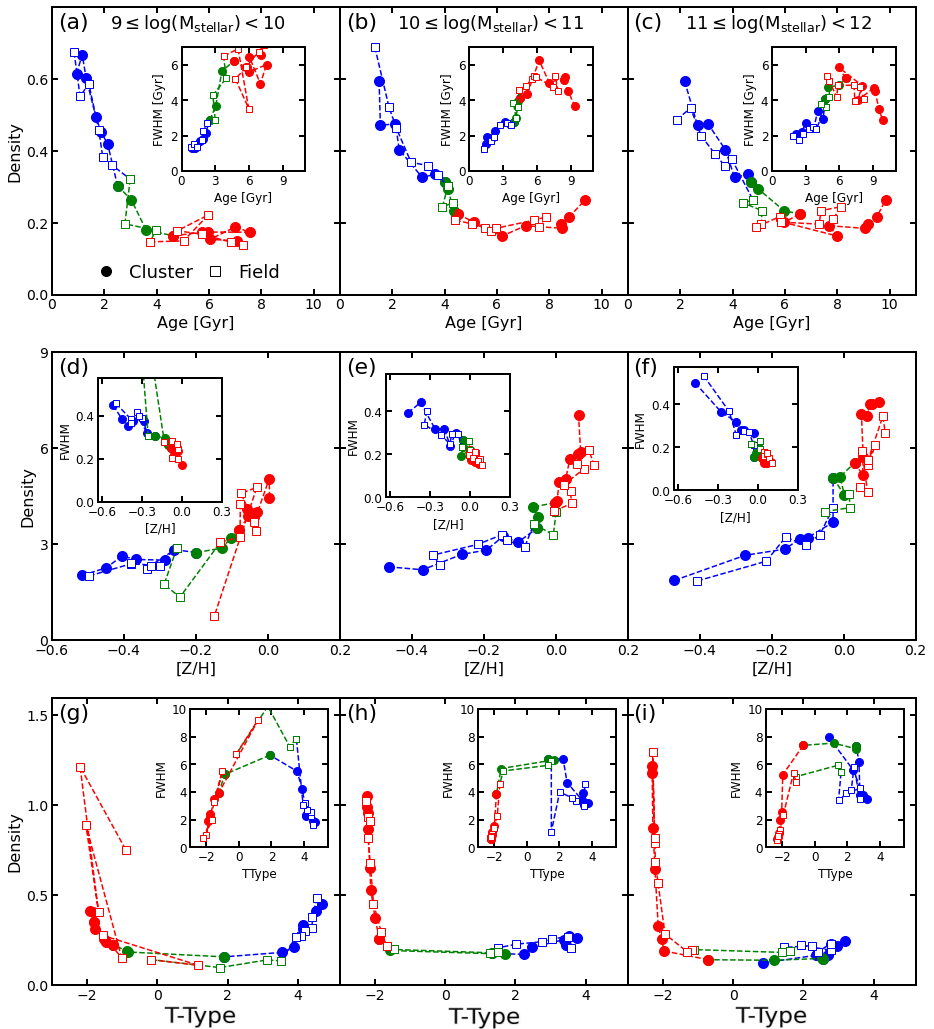}
    \caption{The resulting curves of the same procedure shown in Figs.~\ref{fig:gif_evolution} and \ref{fig:gif_evolution_faint} for cluster (filled circle) and field (white-faced squares) galaxies according to their stellar mass (each column). Colors denote if galaxies belong to BC (blue), GV (green) or RS (red). While the large panels display the peak of the normalized epanechnikov kernel density distribution as galaxies progresses from BC to RS, the inset in each panel shows the FWHM variation for each case.}
    \label{fig:hists_results}
\end{figure}

Environmental mechanisms such as RPS and TML depend on both environment and galaxy halo mass. More massive galaxies are expected to be less prone to environmental effects of cluster environments. Here we use galaxy stellar mass as a proxy for full galaxy mass. In this section, in addition to the field-cluster separation, we separate galaxies according to stellar mass bins, similar to  Fig.~\ref{fig:Cluster_vs_Field_properties}. We then follow the same procedure as Figs.~\ref{fig:gif_evolution} and \ref{fig:gif_evolution_faint}. The results are presented in Fig.~\ref{fig:hists_results}. The stellar mass bins are chosen to guarantee robust statistics in every slice, while probing a sufficient variation to observe possible galaxy stellar mass related effects. 

Our results suggest that galaxies in different stellar mass bins experience different galaxy property variations during the same transition in the SFR vs. $M_{\star}$ plane. In panels (a), (b) and (c), the BC galaxy age distribution peaks are increasing with increasing stellar mass. Namely, for low mass galaxies (panel a), the top BC (first ``slice'') age distribution peaks at $\sim 1.5$ Gyr, while it peaks at $\sim 2$ and $\sim 3$ Gyr for intermediate (panel b) and high (panel c) mass galaxies. We find similar trends for $[Z/H]$. Panels (d), (e) and (f) indicate an increasing peak density of $[Z/H]$ with stellar mass, which indicates an excess of galaxies with higher metallicity ($[Z/H] > 0.0$) at the end of the RS (``last slice''). With respect to morphology (panels g, h and i), the low mass bin is the only case where the peak density in the BC is greater than those at the end of the RS. The peak density at the end of the RS increases towards greater masses. 

After describing differences with respect to stellar mass, we now focus in the comparison between cluster and field systems. Although similar for low mass galaxies (panel a), age distribution peaks for intermediate (panel b) and high (panel c) mass galaxies in clusters reach older ages at the end of the RS in comparison to their field counterparts, despite starting roughly at the same peak density in the beginning of the BC. On the other hand, $[Z/H]$ differences in peak densities happens mostly in the beginning of the BC (at $\Delta {\rm SFMS} > 0$). Clusters galaxies have lower $[Z/H]$ peak densities in comparison to the field in the first $\sim 1-2$ slices. However, they converge roughly to the same peak density as we approach the RS tail. Finally, we do not find significant differences regarding morphology (T--Type). This suggests that the differences between field and cluster galaxy properties are mostly relevant in massive systems, which probably are more efficient in using stellar feedback to form new stars when in low density environment, whereas the cluster environment removes the galaxy's gas and prevents further star formation. This results in the plateau (more mixed distribution) in age observed for the field at the end of the RS in comparison to cluster environment.

Again, we stress that, in this case, we are not tracing the environmental/internal quenching mechanism itself, but the variations on galaxy properties with respect to location in the SFR vs. $M_{\star}$ diagram. The trends we find for different stellar masses are in agreement with works indicating that more massive galaxies are less affected by their environment in comparison to low mass systems, which is expected since more massive systems form and assemble their stellar population at earlier times. Interestingly, in panel (a) of Fig.~\ref{fig:hists_results} we do not find significant differences between cluster and field galaxy ages. This may indicate: 1) in the field, AGN and stellar feedback driven gas outflows are sufficient to quench low mass galaxies, which is in agreement with \cite{2020MNRAS.491.5406T}; 2) the similar start and end points we observe indicate the cluster environment causing a more rapid quenching, while low mass galaxies have a similar ``fate'' indistinctly of the operating mechanism. The more striking results concerning cluster vs. field systems are those for intermediate and high mass galaxies in the RS, for which we find more broader $Age$ distributions for the tail-end RS galaxies in the field in comparison to clusters. This may follow from more massive systems in the field being able to produce a small amount (small enough to be RS galaxy), that otherwise is shutdown when in clusters due to the presence of environmental quenching mechanisms. Additionally, an important result comes from the T--Type panels (g), (h) and (i), namely galaxies suffer a rapid morphological transition in a single ``step'' of the GV. This provides further insight on whether SFR or Morphology changes ``faster''. Thus, a pivotal tool to understand galaxy is evolution especially in clusters is the focus of the next section: the infall time.

\section{Environment at Work: Cumulative Quenching in Galaxies}

Environmental quenching includes multiple mechanisms which are inherently non-linear and complex. A fundamental parameter to understand how long a galaxy has been experiencing the cluster environment, traced by the time since infall. An important difference between galaxies experiencing the cluster for the first time and long-time members are their orbits. This fact translates to a well defined trajectory in the Phase-Space (see Fig.~1 of \citealt{2017ApJ...843..128R}), which enables an estimate of the infall time from galaxy position in the Phase Space. However, observationally we are limited to the projected phase space (PPS), but combined with N-body and cosmological simulations we can relate the galaxy's location in the PPS and time since infall.

We build PPS diagrams following the prescription of \cite{2017ApJ...843..128R}. The x-axis correspond to the projected radial distance and y-axis the velocity along the line of sight, both with respect to the clustercentric coordinates. We normalize the x and y-axis by the cluster's virial radius and velocity dispersion, respectively, in order to enable comparisons between clusters with different properties and create a stacked version of the PPS. With respect to the relation between location in the PPS and infall time, we adopt the ``New Zones'' (PNZs, hereafter) presented in \cite{2019MNRAS.484.1702P}. In this case, time  since infall is defined as the time since galaxy first crossed the virial radius. They used the YZiCS simulation to define quadratic functions fitting the observed time since infall distribution in the PPS. However, their region 7 is quite narrow and thus we decided to join regions 6 and 7 into a single region characterized by the mean infall time of both regions (see their Fig.~4 and Table 1). The PNZs are defined within one virial radius, thus the following analysis is limited to this radial limit. Still, a word of caution is needed regarding the way the PPS is discretized. Backsplash galaxies -- galaxies that have already completed their first pericentric passage and were thrown back to the cluster outskirts -- may influence our results when we examine the PPS since this population may suffer partial quenching in comparison to those first infalling. Even with cosmological simulations, the location in the PPS dominated by backsplash galaxies is not well defined. For instance, backsplash galaxies are mainly found in the $[1 < R_{\rm proj}/R_{200} < 1.5] \times [0 < |V_{\rm LOS}|/\sigma_{\rm LOS} < 1]$ region, but even in this region they only account for 30\% of the whole galaxy population \citep{2011MNRAS.416.2882M}. An important feature of our analysis is the limitation to $R_{\rm proj}/R_{200} = 1$, which we expect to minimize the effect of backsplash galaxies. \cite{2021MNRAS.503.3065S} conclude that backsplash galaxy property distributions do not strongly affect the PPS based results.

\subsection{Galaxy Properties as a Function of Infall Time}
\label{sec:Galaxy_Properties_Infall_Time}
\begin{figure}[ht]
    \centering
    \includegraphics[width = 0.7\textwidth]{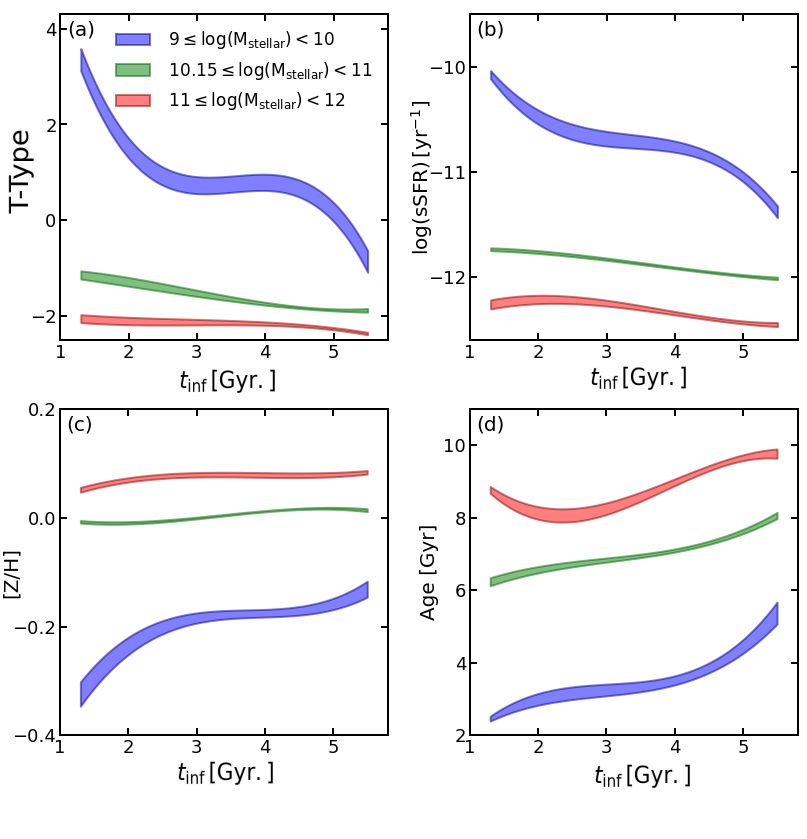}
    \caption{Relation between galaxy properties and time since infall. We present the relations for T--Type (left), sSFR (center-left), [Z/H] (center-right) and age (right). We divide galaxies according to their stellar mass (cyan, purple and orange colors). See panel [a] for the stellar mass intervals.}
    \label{fig:infall_time_relations}
\end{figure}

We use the mean infall time of each PNZ to establish how galaxy properties evolve when infalling into clusters. We separate galaxies according to the same stellar mass bins of Fig.~\ref{fig:Cluster_vs_Field_properties}, but here the analysis is limited to galaxies within one virial radius. We compute the galaxy property median value in each of the 7 PNZs. To estimate the associated variance, we use a bootstrap technique: 1) we select $N_{\rm data}$ values of the observed distribution in a random way and with replacement; 2) We calculate the variance using the interquartile range as $Q_{\sigma} \sim 0.7407 \times IQR$; 3) we repeat this procedure $N_{\rm bootstrap} = 1,000$ times; 4) we then define the variance as the median of the $ Q_{\sigma}$ distribution. We present the results for T--Type, sSFR, $[Z/H]$ and $Age$ in Fig.~\ref{fig:infall_time_relations}. For comparison, we add the median values of the field sample as colored stars. As a first approximation, we quantify variations and differences via a linear fit. However, this is only a first approximation and does not take into account the full behavior of the relations. We present the results for the linear fits in Table \ref{tab:linear_fit}. First, we notice how field galaxies are younger, more star-forming, more metal-poor and higher T--Type (LTGs) in comparison to those in clusters, irrespective of stellar mass. Moreover, differences are larger with increasing stellar mass. 

Next we detail the observed trends in Fig.~\ref{fig:infall_time_relations}. With respect to T--Type (panel (a), the more massive galaxies (purple and orange) have early-type morphologies, regardless of the time since infall. This trend is even more evident for galaxies with higher stellar mass, which at $t_{\rm inf} \sim 1.42$ Gyr have ${\text{T--Type}}<-2$. On the other hand, less massive galaxies (cyan) show a quick transition from late-type to early-type morphology. Namely, we find a $\rm \Delta {\text{T--Type}} \sim 3$ in $\Delta t_{\rm inf} \sim 1 {\rm Gyr}$. In panel (b), we find similar trends for sSFR, which is almost constant for more massive galaxies and indicate more significant variations for the low mass regime. Furthermore, regarding low mass systems, the variations in T--Type and sSFR as a function of the time since infall, have a steeper slope (see Tab.~\ref{tab:linear_fit}) for the former one. This is in agreement with galaxies reaching early-type morphologies before being full quenched. We focus exclusively on this topic in Section \ref{sec:ttype_ssfr}. With respect to $[Z/H]$ (panel c), more massive galaxies have little variation with time since infall, suggesting these galaxies already reach their upper metal enhancement limit and lack gas for further star formation events. On the other hand, an important result is the trend we find for low mass galaxies. Namely, low mass galaxies show an increase in metal richness at times of of $\sim ~1.5-2.5$ Gyr since infall and then level off to a constant value. This may indicate that galaxies quickly lose their gas component due to environmental effects, which prevents further enhancement in metallicity and results in the observed plateau at lower $[Z/H]$. Yet, the results in panel (d) are a consequence of the age-metallicity relation, in which the increasing $[Z/H]$ means increasing age.

\begin{table}[ht]
\caption{Resulting slope and intercept of a linear fit for each curve in Fig.~\ref{fig:infall_time_relations}.}
\label{tab:linear_fit}
\resizebox{\columnwidth}{!}{%
\begin{tabular}{c|c|c|c|c|c}
\hline
$X = \log(M_{\star}/{\rm M}_{\odot})$ & $y = ax + b$ & T--Type & \begin{tabular}[c]{@{}c@{}}$\log(\text{sSFR} / {\rm yr}^{-1})$\end{tabular} & ${[}Z/H{]}$ & \begin{tabular}[c]{@{}c@{}}$Age / {\rm Gyr}$\end{tabular} \\ \hline
\multirow{2}{*}{$9 \leq X < 10$} & $a$ & $-0.60 \pm 0.12$ & $-0.26 \pm 0.03$ & $0.032 \pm 0.011$ & $0.45 \pm 0.08$ \\ \cdashline{2-6}
 & $b$ & $2.98 \pm 0.47$ & $-9.83 \pm 0.11$ & $-0.281 \pm 0.036$ & $1.96 \pm 0.30$ \\ \hline
\multirow{2}{*}{$10 \leq X < 11$} & $a$ & $-0.17 \pm 0.03$ & $-0.07 \pm 0.01$ & $0.007 \pm 0.001$ & $0.40 \pm 0.05$ \\ \cdashline{2-6}
 & $b$ & $-1.01 \pm 0.14$ & $-11.62 \pm 0.02$ & $-0.02 \pm 0.004$ & $5.57 \pm 0.21$ \\ \hline
\multirow{2}{*}{$11 \leq X < 12$} & $a$ & $-0.09 \pm 0.03$ & $-0.07 \pm 0.02$ & $0.002 \pm 0.002$ & $044 \pm 0.08$ \\ \cdashline{2-6}
 & $b$ & $-1.82 \pm 0.13$ & $-12.07 \pm 0.07$ & $0.072 \pm 0.008$ & $7.15 \pm 0.33$ \\ \hline
\end{tabular}%
}
\end{table}

The evolution of galaxy properties with time since infall depends strongly on galaxy stellar mass. Less massive systems are more affected by environmental effects, namely we find, for the same infall time interval, increasing variations with decreasing stellar mass. We investigate via a linear fit (as first approximation) that galaxies with different stellar mass reach the cluster environment with significantly different properties. This can be seen by the differences in the intercepts shown in Table \ref{tab:linear_fit}, in which all the intercepts are, at least, more than $1-\sigma_{\rm scatter}$ different between galaxy population with different stellar mass. We find evidence of galaxies reaching an asymptotic value in $[Z/H]$, which decreases with decreasing stellar mass. Additionally, an important feature is the time it takes for the galaxy properties to reach the asymptotic value. We find that more massive galaxies reach more rapidly a constant value than low mass systems. This is directly related to how environment affect galaxy star formation. A quick halt of star formation disables further metal enrichment for the stellar component, resulting in an asymptotic value. The differences between the asymptotic value for different stellar masses then may be related to the amount of gas available for removal in these systems. Therefore, our results show how morphological and sSFR transitions happens at a different pace. Exploring the slope of the relations, we find that changes in T--Type are always steeper in comparison to sSFR. However, the difference depends on the galaxy's stellar mass too. This is an important piece of evidence on whether morphology or sSFR changes ``faster'' when galaxies infall in cluster environments, as discussed in the next section.

\begin{figure}[ht]
    \centering
    \includegraphics[width = 0.7\columnwidth]{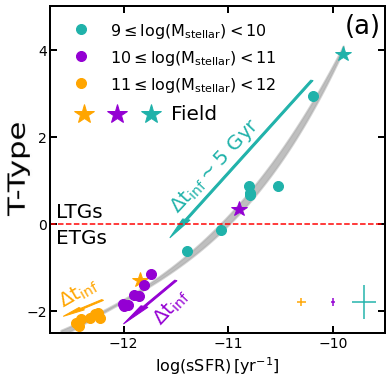}
    \caption{Relation between T--Type and sSFR for a fixed interval of time since infall ($\sim ~5$ Gyr). Each point represents the median T--Type and sSFR for a single PNZ (and hence infall time) in the PPS. In other words, we compress the results regarding T--Type and sSFR shown in Fig.~\ref{fig:infall_time_relations} in a single space, since they roughly cover the same time since infall range. We divide galaxies according to stellar mass (cyan, purple and orange). We present the mean errors in each case in the lower left corner. The colored arrows illustrate increasing infall time for each stellar mass bin. We add a spline fit and $1-\sigma_{\rm scatter}$ as a gray dashed area. The median values of the field counterpart is displayed as the colored stars and the separation between LTGs and ETGs is denoted by the red dashed horizontal line.}
    \label{fig:ssfr_vs_ttype}
\end{figure}

\subsection{Addressing Variations in Morphology and Star Formation Rate}
\label{sec:ttype_ssfr}
In this section we focus strictly on clusters to address the question of whether infalling galaxies reach their ``final stage'' in morphology or sSFR first? To do so we use the relations of T--Type and sSFR as a function of infall time. This way we can trace variations in T--Type and sSFR for a fixed time interval. We divide galaxies just as in Section \ref{sec:Galaxy_Properties_Infall_Time} to probe both stellar mass driven and environmental effects. We present the results in Fig.~\ref{fig:ssfr_vs_ttype}. Each point of each color corresponds to a single PNZ and, consequently, to a given time since infall. Colors represent the same stellar mass bins as Fig.~\ref{fig:infall_time_relations} (see legend in Fig.~\ref{fig:ssfr_vs_ttype}). At the bottom right we show the associated mean errors. We also include an illustrative arrow indicating the increase of infall time according to our results. The red dashed line represents the separation between LTGs and ETGS. Lastly, the grey dashed area and its width are the result of a spline fit and the $1-\sigma_{\rm scatter}$ error, respectively.

As an overall trend, more massive galaxies first infalling in the cluster environment have T--Type smaller than zero and low sSFR. We find decreasing overall variations in both sSFR and T--Type with increasing stellar mass. A linear fit in the T--Type vs. sSFR relations results in slopes of  $2.74 \pm 0.36$ and $1.51 \pm 0.43$ for the high (orange) and intermediate (purple) stellar mass galaxies. In the former case, galaxies already enter the cluster as ETGs and then environmental effects (excluding mergers) do not cause significant variations in either T--Type or sSFR. Fig.~\ref{fig:ssfr_vs_ttype} results show unequivocally a higher variation in both T--Type and sSFR for less massive galaxies, which are more affected by the environment. Galaxies appear to have an especially rapid change in their morphology. However, comparing morphology and sSFR is not straightforward. We then make use of the fixed $\Delta t_{\rm inf}$ to detail these trends. Quantitatively, in a $\Delta t_{\rm inf} \sim 1$ Gyr, low mass (cyan) galaxies experience a variation in T--Type (sSFR) corresponding to $\sim 60\%$ ($\sim 28\%$) of the variation observed during the entire $\Delta t_{\rm inf} \sim 5$ Gyr. Finally, it is noticeable that low mass galaxies reach the T--Type = 0 before being completely quenched. Our results hence provide strong evidence of, with decreasing stellar mass, morphology rapidly changing due to the environment, rather than sSFR.

\section{Chapter Summary}

In this chapter we investigate the dependence of galaxy evolution on its host environment and provide a detailed view of the galaxy transition from the Blue Cloud to the Red Sequence. We use a sample of cluster member galaxies from an updated version of the Yang Catalog. In addition, we identify a field sample from the SDSS-DR7 database by finding all galaxies located at least 5 $R_{200}$ away from any structure with $\log(M_{\rm halo}/{\rm M}_{\odot}) \geq 13$ listed in the Yang catalog. First, we recover a series of already known results regarding cluster vs. field galaxies and explore the mass dependence of such results. 

We then compare galaxy property distributions for galaxies in the Blue Cloud, Green Valley and Red Sequence according to their stellar mass and host environment. Red Sequence and Green Valley field galaxies hosted more recent episodes of star formation in comparison to their counterparts in clusters, for which we find an increasing relative difference with increasing stellar mass. These differences are then confirmed by dissecting the diagram in 17 different slices of $\rm \Delta {\rm SFMS}$. This may indicate that massive systems in the green valley and red sequence are able to keep a small amount of star formation, otherwise halted in dense environments. We suggest that field galaxies are more prone to reuse their gas content, which would be removed in clusters, which also prevents further metal enhancement. Additionally, the greater difference is found in high mass galaxies, which appears to be more efficient in reusing stellar feedback. Our analysis provides additional observational evidence of a considerable fraction of galaxies undergoing a transition from early to late-type morphologies in the Green Valley. Yet, it is important to stress that galaxy evolution is not linear and galaxies can reform a disk after a merger \citep{2017MNRAS.465..547P}.

We directly assess how infalling galaxy properties vary as a function of infall time through the projected phase space. Galaxies with different stellar masses enter the cluster environment with different properties, for which we note an increasing environmental influence with decreasing stellar mass. We highlight the following results:
\begin{itemize}
    \item Low mass galaxies infalling in clusters in the local universe suffer a quick ($\sim$ 1 Gyr) morphological transition (LTG to ETG) just after crossing the virial radius for its very first time, while specific star formation keeps decreasing for a longer period of time ($\sim 3$ Gyr);
    \item Different stellar mass galaxies reach distinct plateau values in stellar metallicity, which are increasing for increasing stellar mass. 
\end{itemize}
Finally, in the context of the ``delayed-then-rapid'' quenching model, we suggest that the delay phase is mostly characterized by morphological transition, while star formation rate decreases at a slower pace. After, a combination of environmental effects (Ram Pressure Stripping, for example) further quench low mass galaxies by removing most of their gas content in a short time-scale ($\sim$ 1 - 3 Gyr). In particular, we interpret our results regarding stellar metallicity to follow directly from environmental quenching mechanisms removing all the gas of less massive, more metal-poor galaxies and disabling further metal enhancement via star formation in these systems.

%% file: Chapters/Chapter5.tex
\chapter{Quantifying time-scales for galaxy evolution in clusters}
\label{chap:timescales}
The text from this section is originally presented in 
\noindent \textit{Sampaio, V. M., De Carvalho, R. R., Aragón-Salamanca, A., Merrifield, M. R., Ferreras, I., \& Cornwell, D. J. (2024). Exploring galaxy evolution time-scales in clusters: Insights from the projected phase space. Monthly Notices of the Royal Astronomical Society, stae1533.}

\thispagestyle{empty}

\noindent

\section{Overview}

Galaxies infalling into clusters undergo both star-formation quenching and morphological transformation due to environmental effects. We investigate these processes and their time-scales using a local sample of 20,191 cluster and 11,674 field galaxies from SDSS. By analysing morphology as a function of distance from the star-formation main sequence, we show that environmental influence is especially pronounced for low-mass galaxies, which emerge from the green valley with early-type morphologies before their star formation is fully suppressed. Using the galaxies' positions in the clusters' Projected Phase Space, we examine the evolution of blue cloud, green valley, and red sequence fractions as a function of time since infall. Interestingly, the green valley fraction remains constant with time since infall, suggesting a balanced flow of galaxies in and out of this class. We estimate that galaxies less massive than $10^{10}\rm M_{\odot}$ spend approximately 0.4 Gyr in the green valley. By comparing quenched and early-type populations, we provide further evidence for the ``slow-then-rapid'' quenching model and suggest that it can also be applied to morphological transitions. Our results indicate that morphological transformation occurs at larger radii than complete star-formation quenching. About 75\% of galaxies undergoing morphological transition in clusters are spirals evolving into S0s, suggesting that infalling galaxies retain their disks, while massive ellipticals are relics of early merger events. Finally, we show it takes approximately 2.5 and 1.2 Gyr after the delay-time ($\sim 3.8 {\rm Gyr}$) for the population of low mass galaxies in clusters to reach a 50\% threshold in quenched and early-type fraction, respectively. These findings suggest morphological transition precedes full star formation quenching, with both processes possibly being causally linked.

\section{Parameter space of galaxy evolution}

\begin{figure}
    \centering
    \includegraphics[width = 0.5\textwidth]{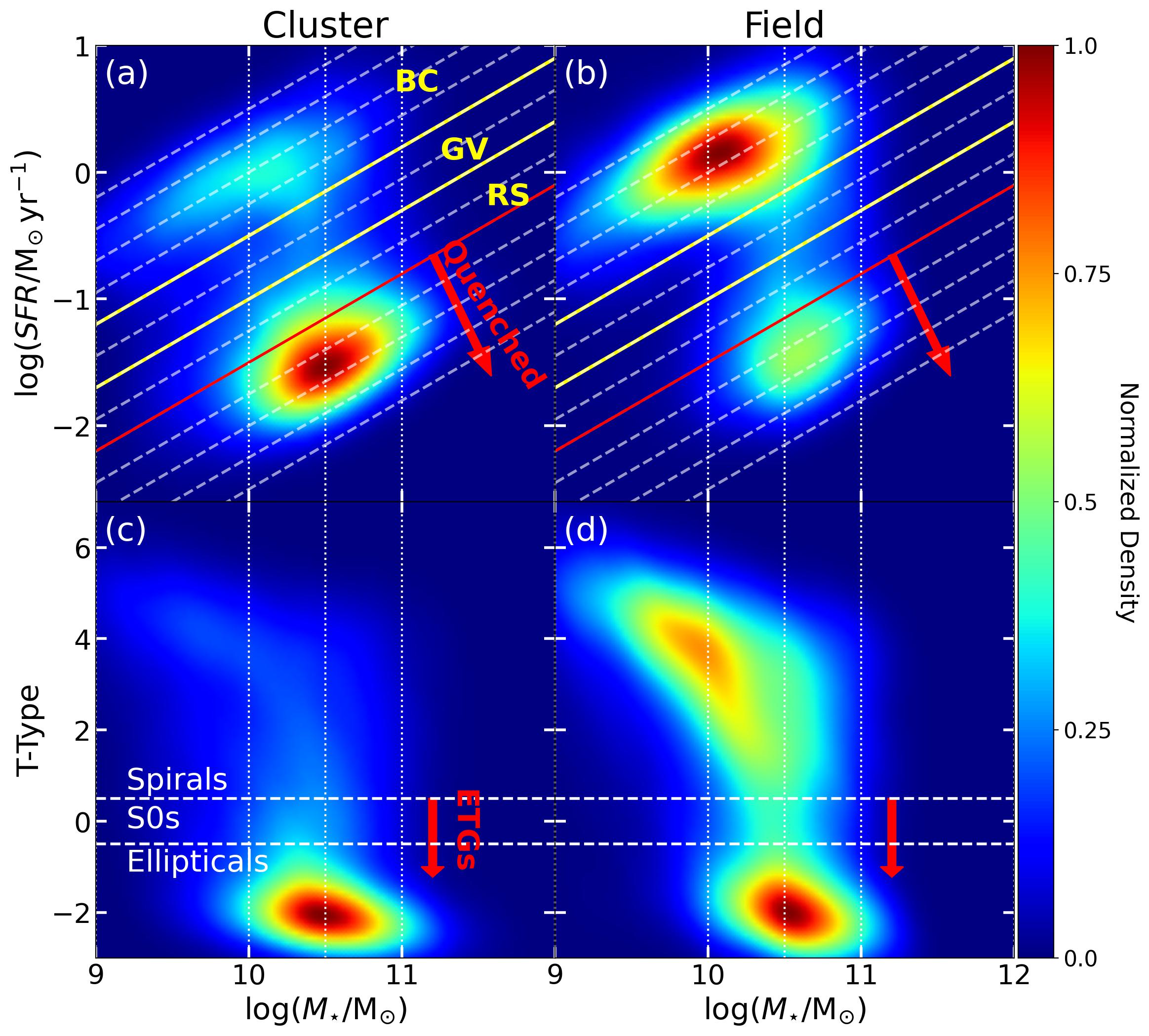}
    \caption{Smoothed distributions in the SFMS (top row) and T--Type diagram (bottom row) for cluster (left column) and field (right) samples. The smoothing is done using a kernel density estimator with width equal to 1\% of the X- and Y-axis ranges. In the SFMS diagram we also show the limiting lines between BC, GV and RS in yellow, slices with the same slope but varying intercept as the SFMS as white dashed lines, and a red line and arrow denoting what we select as ``quenched galaxies". In the T--Type diagram we include the threshold lines between Spiral, S0 and Elliptical morphologies, alongside a red arrow denoting what we call ``early-type'' galaxies.}
    \label{fig:SFR_TType_Distribution}
\end{figure}

Galaxies can be characterized within a parameter space defined by observable properties, aiding in the understanding of their evolutionary stages. The parameters considered in this study include stellar mass, star formation rate (SFR), and morphology (T--Type). In Fig.~\ref{fig:SFR_TType_Distribution}, we present the distribution of cluster (left column) and field (right column) samples in the SFMS diagram and T--Type versus Stellar Mass (bottom row). These distributions were smoothed using a Gaussian kernel density estimator with a width corresponding to 1\% of the ranges of the X and Y axes. In the SFMS panels, the yellow dividing lines BC/GV and GV/RS are shown, determined respectively by the equations \citep{2020MNRAS.491.5406T, 2022MNRAS.509..567S}:
\begin{equation}
    \log(\text{SFR}/{\rm M}_{\odot} {\rm yr}^{-1}) = 0.7 \log(M_{\star}/{\rm M_{\odot}}) - 7.5
\end{equation}
\begin{equation}
    \log(\text{SFR}/{\rm M}_{\odot} {\rm yr}^{-1}) = 0.7 \log(M_{\star}/{\rm M_{\odot}}) - 8.
\end{equation}
For the T--Type diagram, we indicate the separation spiral/S0 and S0/elliptical as white dotted lines in the bottom panels of Fig.~\ref{fig:SFR_TType_Distribution}. In all panels, vertical lines denote the boundaries of the four adopted stellar mass bins. Henceforth, we define ``quenched'' and ''early-type'' galaxies as those below the thresholds indicated by the red arrows or lines in the relevant panels.

Examining the SFMS diagram, we observe that cluster galaxies predominantly occupy a single high-density region within the RS, whereas field galaxies exhibit two distinct regions: a primary one in the BC and a secondary, less pronounced one in the RS. This disparity between field and cluster galaxies suggests environmental effects are driving star-forming galaxies towards a quiescent state. Notably, this effect is primarily observed in galaxies with stellar masses $\log(M_{\star}/\rm M_{\odot}) < 10$. Above this mass threshold, the differences between cluster and field galaxies diminish, indicating that more massive galaxies are less susceptible to environmental influences and rely predominantly on internal mechanisms to suppress star formation. Consequently, we define stellar mass bins to distinguish populations whose evolution is governed by different mechanisms.

When examining morphology, we find that the comparison between cluster and field galaxies yields results consistent with those observed in the SFMS analysis. Specifically, there is a notable similarity in the stellar mass range where peak densities are found for elliptical galaxies and RS galaxies, pointing to an overabundance of massive, quenched, elliptical galaxies in clusters. In contrast, field galaxies exhibit a greater diversity of star-forming systems, reflected in a distribution extending towards higher T--Types. This implies that star formation in field environments occurs across a broader range of spiral morphologies.

\begin{figure}
    \centering
    \includegraphics[width = 0.5\textwidth]{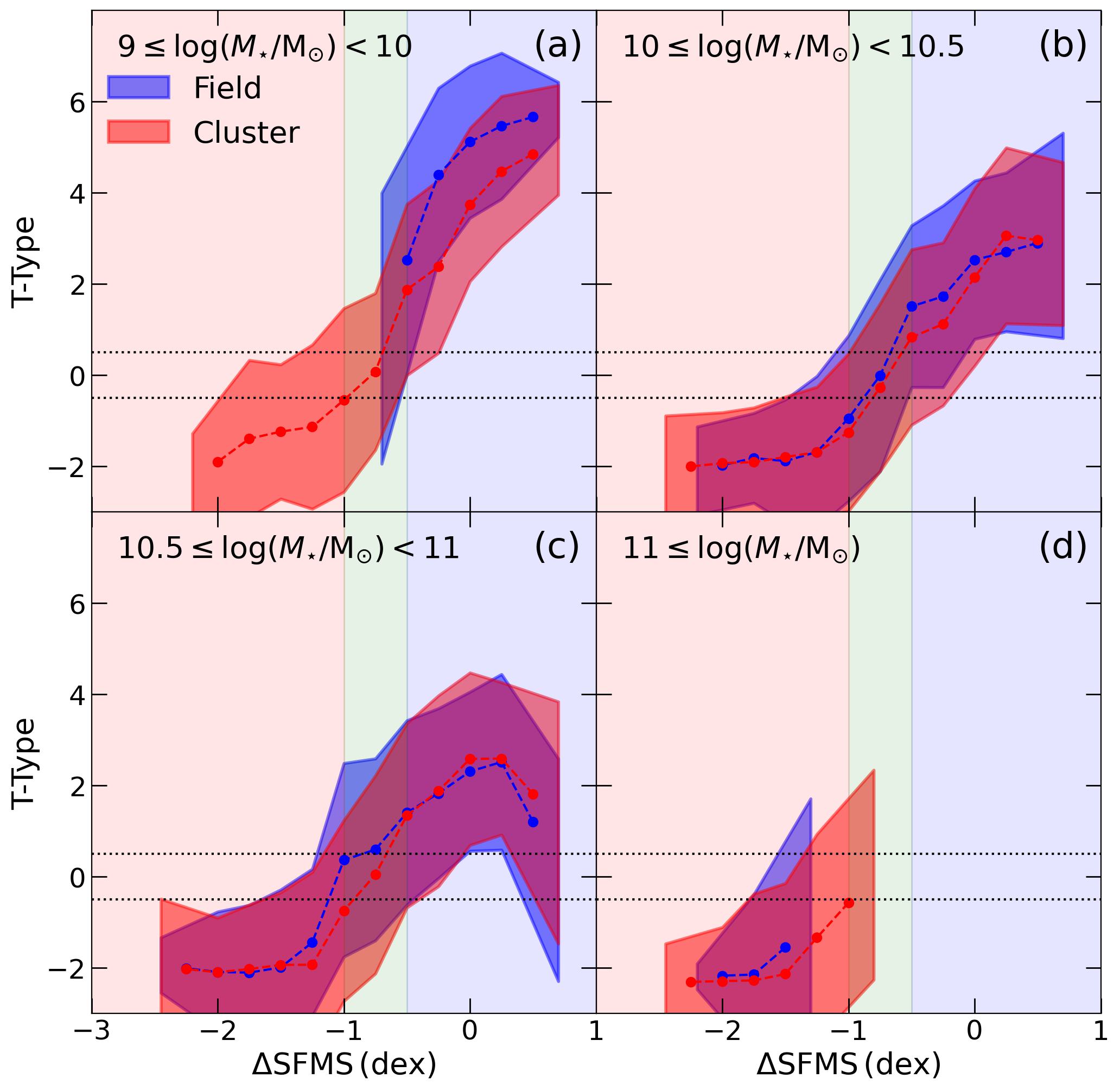}
    \caption{T--Type as a function of the distance from the SFMS for cluster (red) and field (blue) samples, separating galaxies according to their stellar mass. The circles joined by solid lines represent the median of the distributions, while the shaded areas correspond to the observed 1-$\sigma$ scatter. The background colors denote the BC, GV and RS regions and the horizontal lines are the morphological thresholds presented in Fig.~\ref{fig:SFR_TType_Distribution}.}
    \label{fig:TType_deltaSFMS_evolution}
\end{figure}

The broad range of morphologies exhibited by star-forming systems, contrasted with the relatively narrow range observed in quenched systems, underscores the non-linear relationship between morphology and star formation rate. While the SFMS diagram clearly delineates star-forming from quenching or quenched galaxies across different stellar masses, the correlation between morphology and stellar mass in the T--Type diagram prevents such distinct separation. Critically, the dependence of morphology on stellar mass suggests that ``morphological transition'' may entail different variations in T--Type across various stellar mass bins. For instance, it is posited that high-mass elliptical galaxies are remnants of major merger events in the early universe. Additionally, it has been proposed that spirals, even those undergoing suppression of star formation, may retain their stellar disks and evolve into S0s \citep{1996hst..prop.6480D, 1997hsth.conf..185D}.

Hence, it is crucial to establish a method for comparing star formation rate and morphology effectively. We utilize the large number of galaxies in our sample to segment the SFMS diagram into more than three discrete regions, facilitating the tracking of T--Type variations from the BC to the RS. The regions we analyze are delineated by white dashed lines in panels (a) and (b) of Fig.~\ref{fig:SFR_TType_Distribution} and are described by the equation
\begin{equation}
\log({\mathrm{SFR} / \mathrm{M}_{\odot} \mathrm{yr}^{-1}}) = 0.7\log(M_{\star}/{\rm M_{\odot}}) - i,    
\end{equation}
with $i$ varying from 6.5 to 9.5 in steps of 0.25. We further define the SFMS as the best linear fit to all galaxies (cluster and field) within the BC. This unified SFMS ensures that subsequent analyses reference the same baseline for both cluster and field galaxies, particularly when describing the distance from the SFMS. The SFMS zero-point is determined to be
\begin{equation}
    \log({\mathrm{SFR} / \mathrm{M}_{\odot} \mathrm{yr}^{-1}}) = (0.7 \pm 0.05) \log(M_{\star}/{\rm M_{\odot}}) - (7 \pm 0.2).
\end{equation}

In Fig.~\ref{fig:TType_deltaSFMS_evolution}, we display the median (dashed lines) and 1-sigma scatter (shaded regions) of the T--Type distribution as a function of the perpendicular distance from the SFMS ($\Delta \mathrm{SFMS}$), segregated by environment (red for cluster galaxies and blue for field galaxies) and divided into four stellar mass bins (each panel). It is noteworthy that testing different numbers of bins revealed no significant changes in the observed trends. The background colors represent the BC, GV, and RS regions as defined in Fig.~\ref{fig:SFR_TType_Distribution}. Horizontal lines indicate the T--Type thresholds for Spiral, S0, and Elliptical galaxies.

Comparison between cluster and field galaxies reveals decreasing differences in T--Type with increasing stellar mass. The average differences are $\Delta \mathrm{T{-}Type}(\mathrm{Cluster{-}Field}) = -0.83 \pm 0.05$, $-0.16 \pm 0.03$, $0.09 \pm 0.03$, $0.02 \pm 0.03$\footnote{Errors denote the statistical error in the mean}, respectively. Low-mass systems are more susceptible to losing components through RPS and/or tidal mass loss. However, in the highest mass bin, the minimal differences between cluster and field galaxies may suggest that these systems have experienced similar evolutionary paths. Nonetheless, it is important to consider that we are analyzing a low-redshift sample, implying that most of the massive systems had undergone significant evolution at higher redshifts, as predicted by the downsizing scenario. Consequently, drawing definitive conclusions about the physical mechanisms influencing these galaxies is challenging. For instance, due to the high-density environment, massive galaxies in the cores of clusters may have experienced a higher merger rate in early epochs compared to their field counterparts.

More importantly, it is noticeable that galaxies emerge from the GV with $\text{T--Type} \leq 0$, irrespective of stellar mass and environment. This consistent trend suggests that star formation suppression is intertwined with the fading and/or removal of prominent star-forming disks in spiral galaxies. This observation aligns with the reported excess of S0 galaxies in the GV \citep{2018MNRAS.477.4116K, 2022MNRAS.509..567S}. Additionally, it indicates that galaxies reach early-type morphologies while there is still room for a gradual decrease in the star formation rate towards the bottom of the RS. Although we adopt $\text{T--Type} \sim 0$ to define galaxies that have undergone morphological transition, our conclusion remains valid even if we use a more restrictive $\text{T--Type} = -2$ as the morphological transition boundary.

These results suggest that the fading and/or suppression of spiral arms occurs prior to a galaxy reaching its final quenched state, with this removal being more likely for low-mass galaxies in clusters, at least in the local universe. However, when investigating the SFMS diagram, time evolution is not directly measurable, as we are limited to observing the present state of a given galaxy. Therefore, to derive time-scales related to the transition between different regions, we invoke the dynamical evolution expected for galaxies within clusters, viewed through the perspective of Projected Phase Space.

\begin{figure}
    \centering
    \includegraphics[width = 0.5\textwidth]{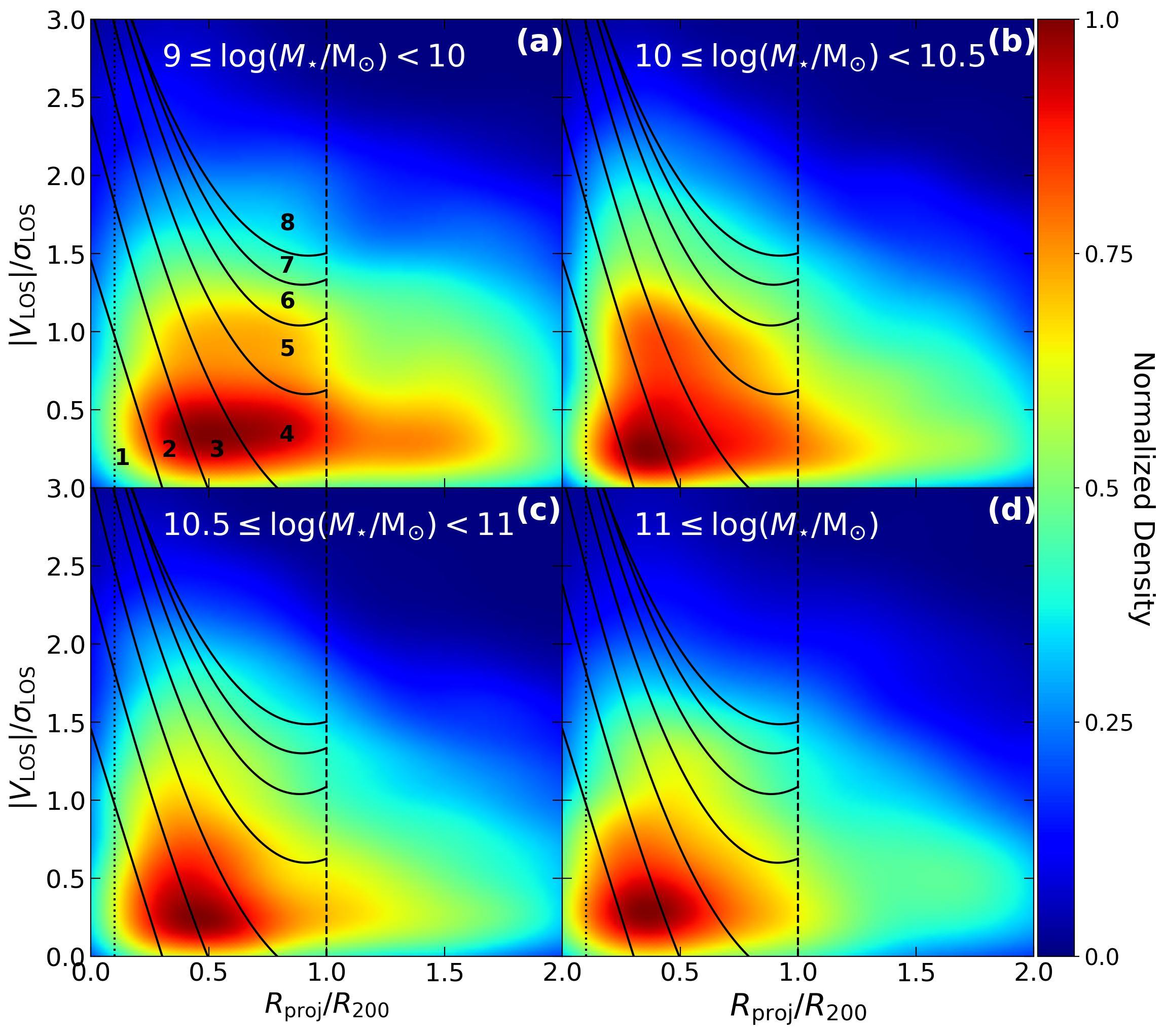}
    \caption{Normalized density distribution of galaxies in the PPS. We separate galaxies into four different stellar mass bins. The solid black lines show the different regions we analyse and relate to a given infall time \citep{2019MNRAS.484.1702P}. These regions are limited to within $R_{200}$, which is shown as a vertical dotted line.}
    \label{fig:PPS_distribution}
\end{figure}

\section{The projected phase space}

\begin{figure}
    \centering
    \includegraphics[width = 0.5\textwidth]{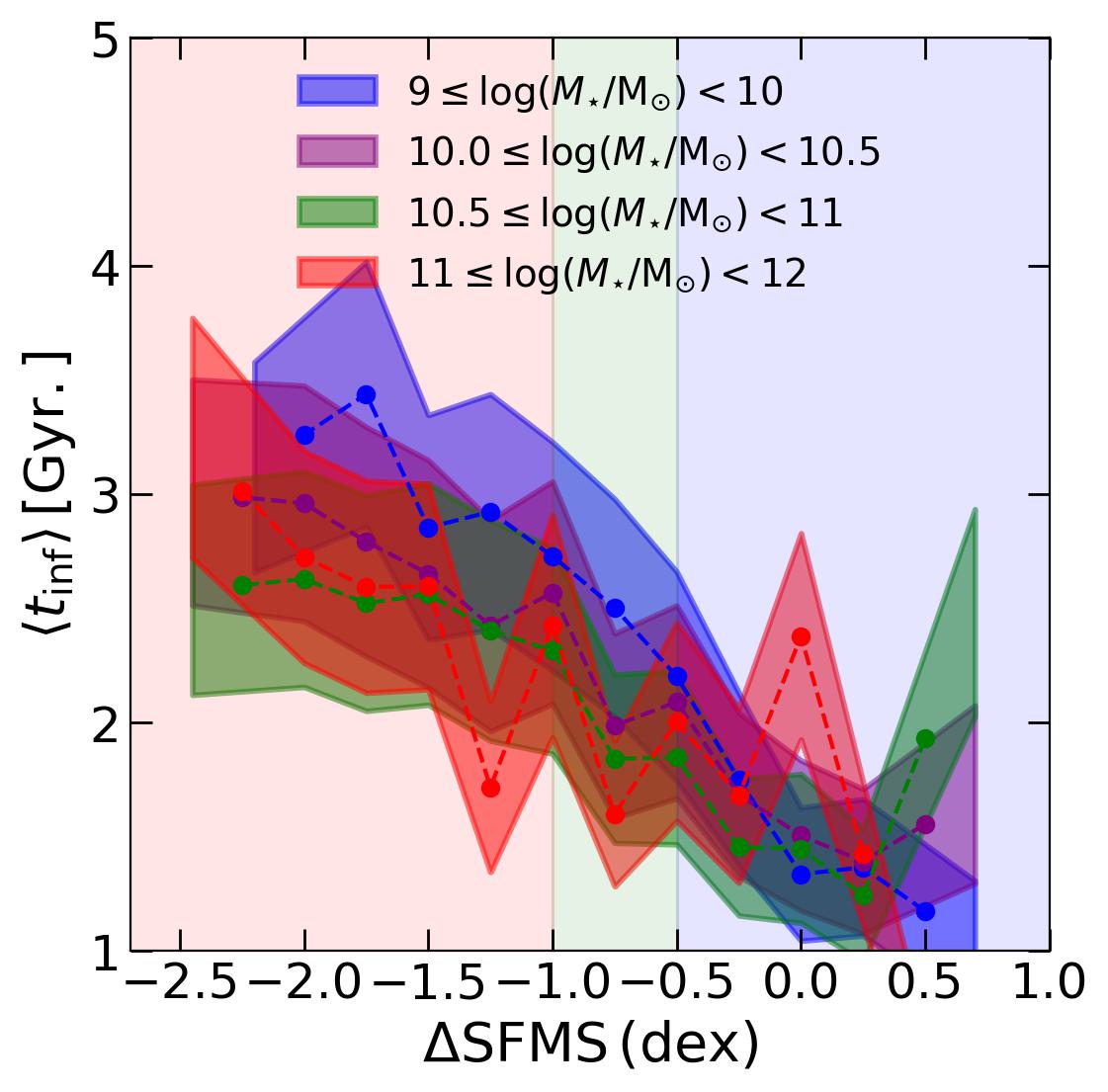}
    \caption{Average infall time as a function of $\Delta \rm SFMS$ for four different stellar mass bins. The circles joined by solid lines represent the median of the distributions, while the shaded areas correspond to the observed 1-$\sigma$ scatter. The background colors represent the BC, GV and RS regions.}
    \label{fig:tinf_vs_deltaSFMS}
\end{figure}

The evolution of galaxies infalling into clusters can be understood through their trajectories in Phase Space. This allows us to establish a relationship between the time since infall ($ t_{\rm inf}$), defined as the time since a galaxy first crossed the $R_{200}$ radius, and its location in Phase Space. Despite limitations and additional uncertainties, simulations also enable average infall times to be associated with different regions of the Projected Phase Space (PPS), where only line-of-sight velocities and projected sky positions are available. Here, we use the slicing method presented in \citet{2019MNRAS.484.1702P}. Utilizing the YZiCS zoom-in simulation \citep{2017ApJ...837...68C}, \citet{2019MNRAS.484.1702P} investigated the time-since-infall distribution in the projected phase space and defined regions that constrain galaxies to a narrow time-since-infall range. These regions are constructed for $R_{\rm proj} \leq R_{200}$, but for completeness, we also qualitatively examine galaxy properties beyond this threshold. Additionally, we tested our results with two alternative methods of dividing the PPS: (1) \citet{2017ApJ...843..128R}, which assigns a probability of a galaxy having a given time-since-infall within each region; and (2) \citet{2020ApJS..247...45R}, which employs a square grid in the PPS. We find no significant differences in the results across these methods. In Fig.~\ref{fig:PPS_distribution}, we present the distribution of galaxies in the PPS. We categorize galaxies into different stellar mass bins and smooth the distribution using a Gaussian kernel density estimator with a width equal to 1\% of the X- and Y-axis ranges. The solid black lines delineate regions containing galaxies with different $t_{\rm inf}$ values: 1.42, 2.24, 2.77, 3.36, 3.89, 4.50, 5.18, and 5.42 Gyr, respectively, for regions 8 down to 1 \citep{2019MNRAS.484.1702P}. The dotted line indicates the lower radial threshold adopted to avoid contamination by very massive (bright) galaxies in the cluster core (see Section \ref{sec:compl_radii}).

Comparison between the different stellar mass bins shows that the high-density envelope (red/orange colors) of low-mass galaxies (panel a) extends beyond $R_{200}$, whereas most of the massive systems are primarily constrained within $R_{200}$. This spatial distribution is consistent with the concept of mass segregation \citep{1980ApJ...241..521C} and is in agreement a dynamical structure divided into two components: a virialized core, dominated by the most massive, red, and dead galaxies; and an outer semi-equilibrium region where most of the infalling and/or backsplash galaxies are found \citep{1997ApJ...476L...7C, 2001ApJ...547..609E}.

\begin{figure*}
    \centering
    \includegraphics[width = \textwidth]{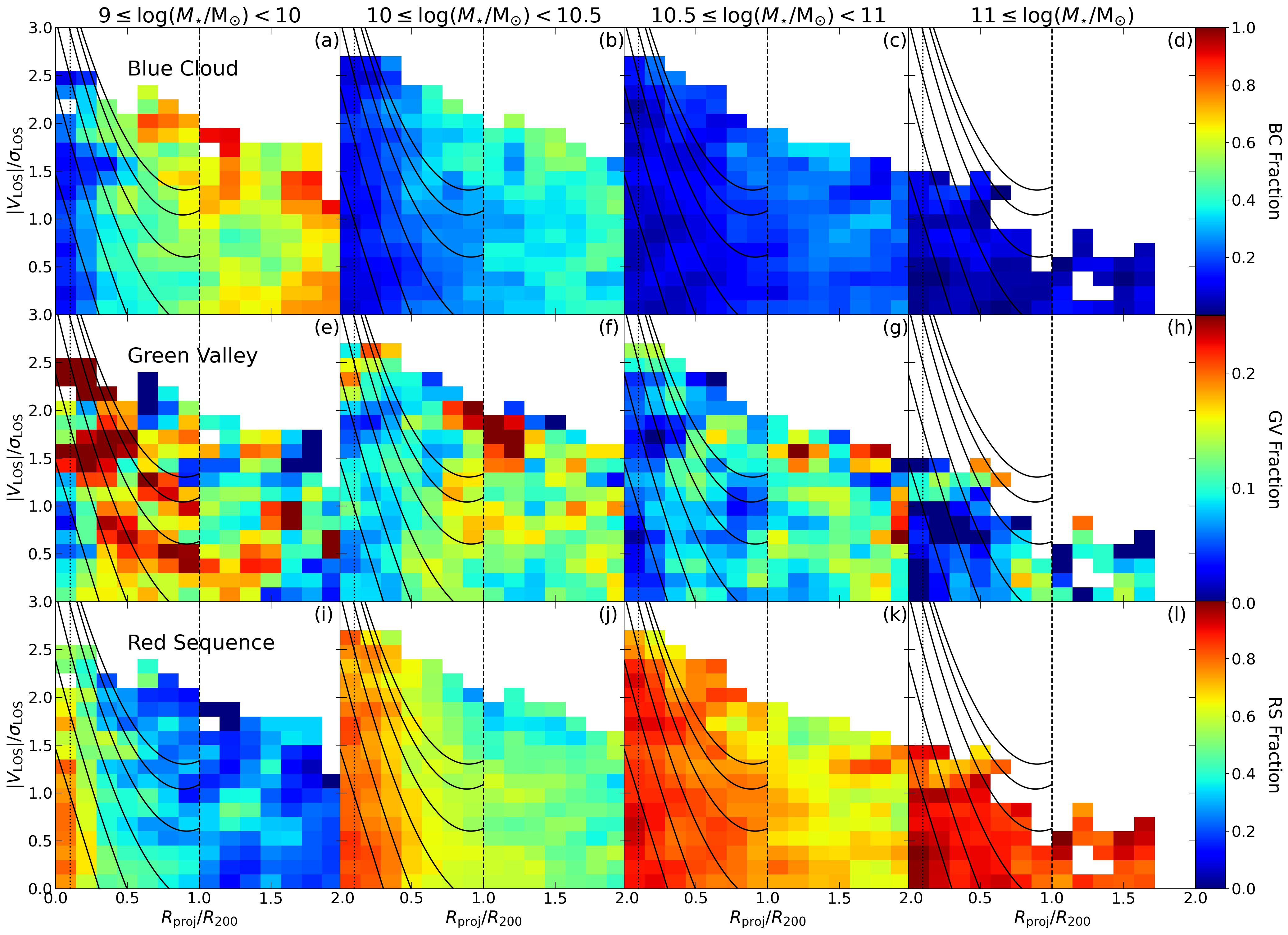}
    \caption{Distribution of (from top to bottom) BC, GV and RS galaxies in the Projected Phase Space. Galaxies are divided according to stellar mass. The solid black lines define values of constant infall time. The vertical dotted line denotes the $R_{200}$ threshold, which is the upper limit in radii in which the defined regions are valid. The color-bar range is selected to highlight relevant trends. We highlight that the fraction of galaxies in the green valley is considerably smaller to the observed fraction range observed in the BC and the RS.}
    \label{fig:pps_bc_gv_rs}
\end{figure*}

Our analysis relies on the connection between time since infall and the evolutionary time sequence of galaxies. However, galaxies in region 8 are not necessarily the progenitors of those observed in region 1. Nevertheless, our upper time since infall is 5.42 Gyr, which corresponds to $z \sim 0.5$. Galaxies formed most of their stellar mass ($\sim 75\%$) within $0.5 \leq z \leq 2.5$ \citep{2014ARA&A..52..415M}, indicating that we do not expect significant variations in the mass assembly history of galaxies within the analyzed infall time range. Additionally, the variation in the SFMS from $z \sim 0.5$ to $z \sim 0.1$ is $\leq 0.3$ dex \citep[e.g.][]{2014ApJ...795..104W, 2016ApJ...817..118T, 2022MNRAS.509.5382C}, suggesting small variations in the properties of galaxies being accreted onto clusters over the analyzed time-since-infall range. Therefore, as a first approximation, we can consider the time-since-infall as an evolutionary sequence for infalling galaxies.

To demonstrate the correspondence between the evolution in the SFMS diagram and the PPS, we present Fig.~\ref{fig:tinf_vs_deltaSFMS}, which illustrates the relationship between $t_{\rm inf}$ and $\Delta \rm SFMS$ across four distinct stellar mass bins. Similar to Fig.~\ref{fig:TType_deltaSFMS_evolution}, the background colors indicate the BC, GV, and RS regions. It is noticeable that $t_{\rm inf}$ increases as $\Delta \rm SFMS$ decreases, indicating progression from the BC to the RS. The minor variations across different stellar mass bins are consistent with findings from \citet{2019MNRAS.484.1702P}, suggesting that projection effects blend galaxies with varying infall times. Notably, this relationship offers a rough estimate of the time galaxies spend in the GV, which appears to be approximately 0.8 Gyr. However, the inherent uncertainties in $t_{\rm inf}$ limit our analysis to a qualitative assessment. Thus, we infer that the GV represents a relatively brief phase in a galaxy's evolutionary timeline, compared to the BC and RS.

\subsection{Evolution from blue cloud to red sequence}
\begin{figure*}
    \centering
    \includegraphics[width = \textwidth]{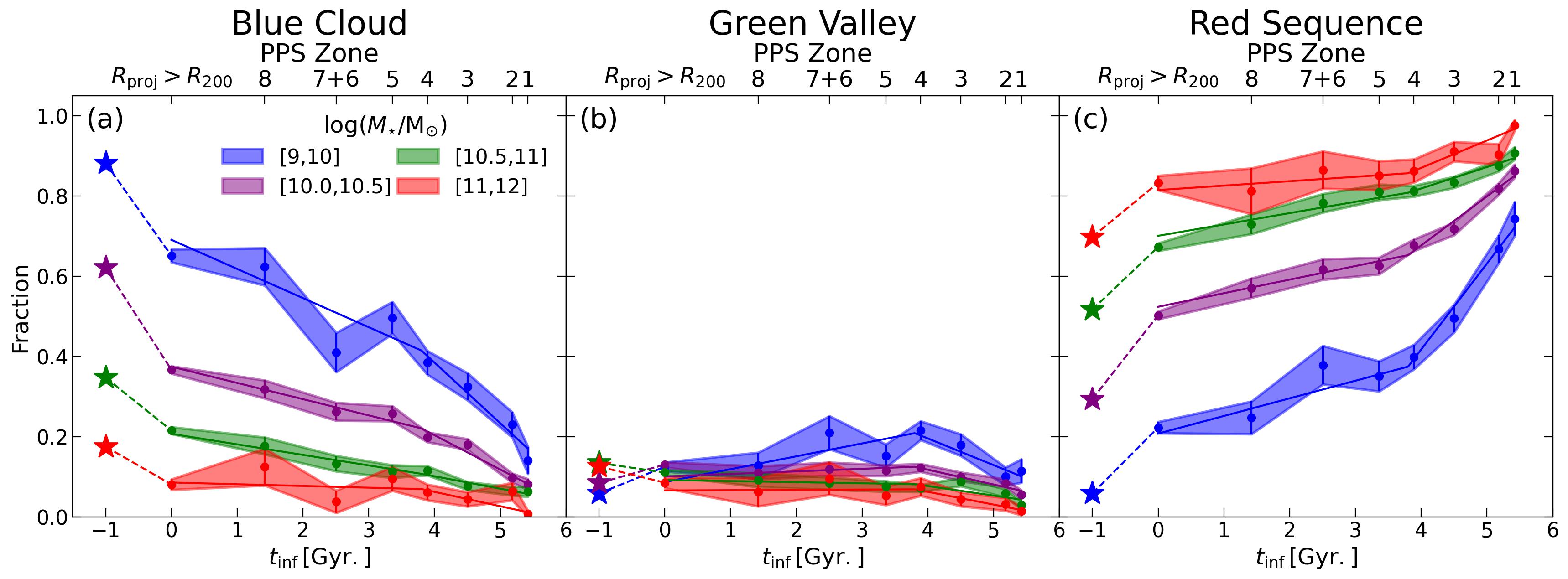}
    \caption{Relation between fraction of galaxies in the BC (left), GV (center) and RS (right) as a function of infall time. For each stellar mass bin, the error-bars and shaded areas the median and 1-sigma scatter. For completeness, we also include two extra points in the relations: 1) the field counterpart value, symbolized with a star of the respective colour (at $ t_{\rm inf} = -1$ Gyr.); and 2) the fraction of BC, GV and RS galaxies beyond $R_{200}$ (at $ t_{\rm inf} = 0$ Gyr.). The double linear model fitted in each curve is shown as solid lines.}
    \label{fig:tinf_bc_gv_rs}
\end{figure*}
Our initial investigation examines the distribution of BC, GV, and RS galaxies within the Projected Phase Space (PPS). In Fig.~\ref{fig:pps_bc_gv_rs}, we present these distributions, segmented by stellar mass. We only include pixels containing at least 20 galaxies to guarantee robustness, and adjust the color-bar range to emphasize relevant trends. It is noteworthy that the fraction of green valley galaxies is consistently smaller than that of the BC and RS galaxies, regardless of stellar mass. However, distinct trends in the green valley distribution fractions are observed across different mass bins. In panel (e), a relatively high object density is found between regions 3 to 7 (as labeled in Fig.~\ref{fig:PPS_distribution}), corresponding to intermediate infall times and suggesting environmental influences on these galaxies. This peak, indicative of intermediate times since infall, is statistically significant and is observed exclusively in the lower stellar mass bin. Additionally, the absence of the most massive galaxies at higher radii and velocities is consistent with mass segregation.

In our analysis of galaxy fractions within the Projected Phase Space (PPS), we discern an inverse symmetry between the distributions of BC and RS galaxies across all stellar masses. This symmetry is primarily due to the relatively constant fraction of green valley (GV) galaxies, with only slight variations observed throughout the PPS. A particularly notable trend is the marked increase in the fraction of RS galaxies beyond the $R_{200}$ threshold during the transition from BC to RS. These results suggest two potential scenarios: first, the influence of the cluster environment on galaxy evolution may extend to larger radii, likely driven by harassment or tidal effects \citep{2017MNRAS.467.3268R}; second, this observation may reflect the peak in the fraction of backsplash galaxies within the 1-2 $R_{200}$ range \citep{2011MNRAS.416.2882M, 2020MNRAS.492.6074H, 2023MNRAS.519.4884F}. As galaxies move beyond $R_{200}$, the decline in BC galaxy fractions appears to align with regions characterized by varying times since infall, thereby suggesting a significant link between the infall process and subsequent galaxy evolution.

Another relevant aspect of our findings is the pronounced reduction in the fraction of BC galaxies, which is concurrently accompanied by an increase in the fraction of RS galaxies. This transition becomes particularly noticeable within the inner regions of the cluster, specifically at approximately $\sim 0.3R_{200}$. At this radial distance, the fraction of BC galaxies drops below 20\%, while the fraction of RS galaxies rises above 50\%. This observation aligns with the expectation that environmental mechanisms, such as ram pressure stripping, become increasingly effective at removing galactic components as galaxies move closer to the cluster core. The correlation between this radial threshold and the steep increase in environmental density further supports this scenario \citep{2019ApJ...873...42R}.

\subsection{Addressing the Slow-then-Rapid Quenching model}

In this section, we focus in analyzing the fraction of BC, GV, and RS galaxies as a function of time-since infall, in order to assess galaxy evolution over time. Figure \ref{fig:tinf_bc_gv_rs} illustrates the variation in the fractions of BC (panel a), GV (panel b), and RS (panel c) galaxies as a function of infall time for different stellar mass bins, each represented by distinct colors. To provide a comprehensive overview, we introduce two supplementary points in each panel: 1) the fraction in the field sample, denoted by the same color star symbol at $t_{\rm inf} = -1$; and 2) the fraction at larger radii ($R_{\rm proj} > R_{200}$), marked at $t_{\rm inf} = 0$\footnote{This infall time is manually set, acknowledging that the outer regions of clusters may host a significant fraction of back-splash galaxies \citep{2013MNRAS.431.2307O, 2023MNRAS.519.4884F}.}. The x-axis coordinates for these two points are chosen for illustrative purposes and will not be employed in any quantitative analyses.

Utilizing Fig.~\ref{fig:tinf_bc_gv_rs}, we provide a quantitative view of the trends observed in Fig.~\ref{fig:pps_bc_gv_rs}. We highlight the symmetry between BC and RS fractions, coupled with a GV fraction that remains relatively constant over time. The distinctions between the fractions of BC and RS galaxies in both the field and at small infall times ($t_{\rm inf} < 2$ Gyr) tend to amplify with decreasing stellar mass, conforming to expectations where BC galaxies are more prevalent in the field, while RS galaxies dominate in clusters. Notably, the GV fraction remains approximately constant across all infall times, albeit with some mass dependence: for low-mass galaxies, there is a discernible increase in the fraction of GV galaxies with intermediate $t_{\rm inf}$, suggesting that for these galaxies, the environment plays a role in quenching their star formation.

Significantly, the behavior of BC and RS fractions relative to infall time reveals a consistent trend across all stellar mass bins. We observe two distinct phases: initially, before a critical time $t_{\rm delay}$, the BC and RS fractions exhibit minimal variation. However, post $t_{\rm delay}$, there is a sharp decline in the BC fraction, mirrored by an increase in the RS fraction. This transition is strongly dependent on stellar mass, with the most pronounced changes occurring in the lower mass bins.

Our extended analysis of the PPS beyond $R_{200}$ supports a two-step quenching process. Initially, as galaxies cross $R_{200}$, they encounter minimal environmental influences, with tidal mass loss emerging as the likely mechanism during this stage \citep{2017ApJ...843..128R, 2019MNRAS.490..773R}. This tidal interaction may reduce the stellar mass of galaxies, potentially causing a shift to lower mass bins compared to their progenitors. This phase is characterized by slow transformation. Subsequently, a more rapid transformation phase occurs around $\sim 0.6R_{200}$ (approximately $\sim R_{500}$), where the denser ICM facilitates strong RPS, effectively removing the majority of the galaxy's gas component. This two-step process aligns with the slow-then-rapid quenching model \citep[e.g.][]{2012MNRAS.424..232W}, and we can pinpoint the infall time at which the transition from the slow to the rapid quenching phase occurs.

\subsection{The in-between: green valley time-scale}

The results shown in Fig.~\ref{fig:tinf_bc_gv_rs} allow us to estimate the timescale galaxies spend in the GV. The relatively constant GV fraction indicates a balance between the inflow and outflow of galaxies within this region. It is important to note that, although galaxies can undergo rejuvenation \citep[e.g.][]{2010ApJ...714L.171T, 2015Galax...3..192M, 2019MNRAS.489.1265M}, most of their mass is already assembled in the adopted redshift range. Furthermore, the cluster environment is expected to predominantly drive the quenching of star formation, making episodes of rejuvenation rare. Thus, galaxy evolution can be statistically approximated as a unidirectional transition from the BC to the RS. Consequently, we can utilize the BC and RS fractions to estimate the duration of environmental effects required to populate the GV with the observed fractions.

We model the curves as a function of time since infall using a double line profile:
\begin{equation}
\centering
\label{eq:double_line}
F(t_{\rm inf}) = 
     \begin{cases}
       a_{1}t_{\rm inf} + b, & \text{if} \, \, t_{\rm inf} < t_{\rm delay} \\
       a_{2}t_{\rm inf} + (a_{1} - a_{2})t_{\rm delay} + b, & \text{if} \,\, t_{\rm inf} \geq t_{\rm delay},
     \end{cases}
\end{equation}
where the second and third terms in the $t_{\rm inf} \geq t_{\rm delay}$ line ensure continuity at $t_{\rm inf} = t_{\rm delay}$. It is important to note that our dataset contains a limited number of data points, and allowing all variables to be free parameters may result in overfitting. To address this, we reduce the degrees of freedom by fixing $t_{\rm delay} = 3.8$ Gyr. This value is chosen based on fitting procedures conducted for both the BC and RS fraction relations, with the average of the resulting $t_{\rm delay}$ distribution used. The model results are presented in Table \ref{tab:bc_gv_rs_model} and are presented as solid lines in Fig.~\ref{fig:tinf_bc_gv_rs}.

\begin{table}[ht]
\caption{Double line model fitting results for the different curves shown in Fig.~\ref{fig:tinf_bc_gv_rs}. $a_{1}$ and b denote the slope and intercept, respectively, of the line fitting the $ t_{\rm inf} < t_{\rm delay}$ region. $a_{2}$ denotes the slope of the line fitting the $ t_{\rm inf} > t_{\rm delay}$. Both $a_{1}$ and $a_{2}$ are given in the units of $\rm Gyr^{-1}$. We divide the sample according to the SFMS region and into different stellar mass bins.}
\label{tab:bc_gv_rs_model}
\resizebox{\textwidth}{!}{%
\begin{tabular}{c|ccc|ccc|ccc}
\hline
\multirow{2}{*}{$\log(M_*/{\rm M}_\odot)$} & \multicolumn{3}{c|}{Blue cloud}                       & \multicolumn{3}{c|}{Green valley}                    & \multicolumn{3}{c}{Red sequence}                    \\ \cline{2-10} 
                                           & $a_1$            & $b$             & $a_2$            & $a_1$           & $b$             & $a_2$            & $a_1$           & $b$             & $a_2$           \\ \hline
{[}9,10{]}                                 & $-0.07 \pm 0.01$ & $0.70 \pm 0.03$ & $-0.15 \pm 0.03$ & $0.04 \pm 0.03$ & $0.07 \pm 0.03$ & $-0.06 \pm 0.02$ & $0.04 \pm 0.03$ & $0.21 \pm 0.01$ & $0.21 \pm 0.04$ \\
{[}10,10.5{]}                              & $-0.04 \pm 0.01$ & $0.38 \pm 0.04$ & $-0.09 \pm 0.02$ & $0.01 \pm 0.02$ & $0.10 \pm 0.04$ & $-0.03 \pm 0.03$ & $0.03 \pm 0.02$ & $0.52 \pm 0.03$ & $0.12 \pm 0.02$ \\
{[}10.5,11{]}                              & $-0.02 \pm 0.02$ & $0.20 \pm 0.04$ & $-0.03 \pm 0.02$ & $0.00 \pm 0.01$ & $0.09 \pm 0.04$ & $-0.02 \pm 0.02$ & $0.03 \pm 0.02$ & $0.70 \pm 0.02$ & $0.05 \pm 0.03$ \\
{[}11,12{]}                                & $0.00 \pm 0.02$  & $0.08 \pm 0.05$ & $-0.03 \pm 0.02$ & $0.00 \pm 0.02$ & $0.06 \pm 0.02$ & $-0.01 \pm 0.01$ & $0.01 \pm 0.01$ & $0.81 \pm 0.01$ & $0.07 \pm 0.02$ \\ \hline
\end{tabular}
}
\end{table}

The slope for the $t_{\rm inf} < t_{\rm delay}$ region is approximately zero, irrespective of galaxy mass and the region considered (BC, GV, or RS). For the intercept $b$, we observe a decreasing trend with increasing stellar mass for BC galaxies, and an increasing trend for RS galaxies, as anticipated. In the GV, the intercept value remains nearly constant across different masses, barely exceeding the 10\% mark. For $t_{\rm inf} \geq t_{\rm delay}$, we observe an increase in the steepness of the BC and RS relations regardless of stellar mass, whilst the slope modelling the GV remains virtually zero. This supports the hypothesis of a balance between the inflow and outflow of galaxies in this region.

The observed trends allow us to estimate the time galaxies spend in the GV ($\Delta t_{\rm inf}^{\rm GV}$) by combining the $a_{2}$ slopes of the BC and RS relations:
\begin{equation}
    F^{\rm GV} = \Delta t_{\rm inf}^{\rm GV} (a_{2}^{\rm BC} - a_{2}^{\rm RS}),
\end{equation}
where $F^{\rm GV}$ denotes the average fraction of galaxies in the GV. Our analysis focuses on separating environmental effects from the internal evolution of galaxies, highlighting that the derived time-scale mainly reflects environmentally-driven quenching processes. Concentrating on the two lower stellar mass bins, which are most affected by the environment, we find that galaxies would cross the GV due to environmental effects in approximately $0.5$ and $1.1$ Gyr, respectively, for the two lower stellar mass bins. This indicates a shorter time-scale for galaxies with lower stellar mass. Our estimates are consistent (within 1-sigma uncertainty) with time-scales derived for clusters at $1 \leq z \leq 1.4$ \citep{2021MNRAS.508..157M}. However, the trend we observe with mass in significantly different from their work. While we find a decreasing time-scale with stellar mass, their estimate remain roughly constant. This may follow from several possibilities: 1) large uncertainty in their estimate at higher masses; 2) adopted different stellar mass bins; and 3) at higher redshifts, more massive galaxies were actively forming stars, such that the variation caused by the environment is more significant in comparison to the local universe, where it seems to affect only low stellar mass systems.

\section{Comparing the time-scales for star-formation suppression and morphological transformation}

We now compare the time-scales for star formation suppression and morphological transition. These processes are closely linked, and defining the boundaries for quenched galaxies and those that have undergone morphological transition is crucial. This analysis is intended to also consider the underlying morphology-density relation in galaxy clusters \citep{Dressler}.

\subsection{Insights from the distribution of quenched and early-type Galaxies}

\begin{figure}[ht]
    \centering
    \includegraphics[width = \textwidth]{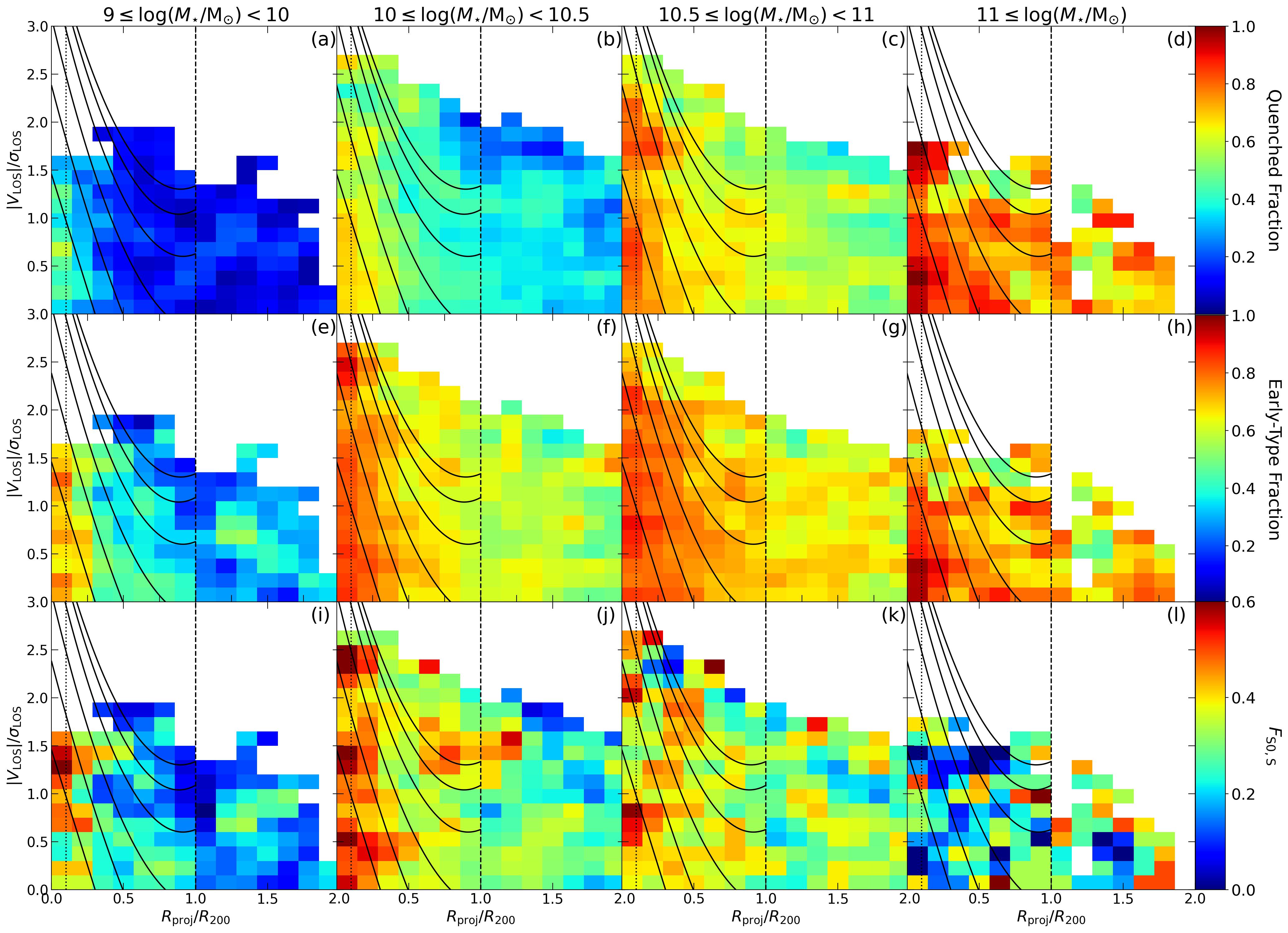}
    \caption{The distribution of quenched (top row) and early-type (bottom row) fraction in the PPS. Each column represent a different stellar mass bin. The thresholds for calling a galaxy quenched and/or early-type are shown in Fig.~\ref{fig:SFR_TType_Distribution}. The solid black lines show the different regions comprising galaxies with a narrow range of time since infall. For calculating the fractions in a given pixel we impose a minimum number of 20 galaxies to guarantee robustness.}
    \label{fig:Quenched_Fraction_PPS}
\end{figure}

In Fig.~\ref{fig:Quenched_Fraction_PPS}, we present PPS diagrams showing the fractions of quenched galaxies (first row) and early-type galaxies (second row) across different stellar mass bins (each column). Additionally, we display the ratio $F_{\rm S0,S} = N_{\rm S0}/(N_{\rm S0}+ N_{\rm S})$ in the last row, where $N_{\rm S0}$ represents the number of S0 galaxies ($-0.5 \leq\, \text{T--Type} \,\leq 0.5$) and $N_{\rm S}$ denotes the number of spiral galaxies ($0.5 \leq\, \text{T--Type}$). To ensure statistical robustness, each pixel includes at least 20 galaxies. Notably, for the highest stellar mass bin, galaxies primarily occupy the $\sim 0.5 R_{200}$ region, as shown in panels (d), (h), and (l). These trends support models suggesting that clusters have a virialized inner core predominantly composed of massive elliptical quenched galaxies.

Panels (a), (b), (c), and (d) reveal an increasing quenched fraction with stellar mass, consistent with the downsizing scenario. We observe that the quenched fraction reaches its maximum within the $R_{\rm proj} < 0.4 R_{200}$ range, regardless of stellar mass. Quantitatively, the average differences in quenched fractions between the $R_{\rm proj} < 0.4 R_{200}$ and $R_{\rm proj} > R_{200}$ regions are $19 \pm 3\%$, $23 \pm 5\%$, $16 \pm 6\%$, and $14 \pm 8\%$, respectively, for increasing stellar mass.

The early-type fraction, shown in panels (e), (f), (g), and (h), increases both with stellar mass and with decreasing $R_{\rm proj}$. These trends mirror those observed for quenched fractions, suggesting a link between star formation quenching and morphological transition. Notably, in the lowest stellar mass bin (panel e), there is a marked increase in the fraction of early-type galaxies after crossing $R_{200}$. Specifically, the average fraction beyond $R_{200}$ is $27 \pm 8\%$, compared to $40 \pm 7\%$ within $R_{\rm proj} < R_{200}$, indicating that $R_{200}$ may act as an initial threshold for morphological transition. However, this trend needs further investigation, as backsplash galaxies, which are primarily found near $R_{200}$, could influence this pattern.
\begin{figure*}
    \centering
    \includegraphics[width = \textwidth]{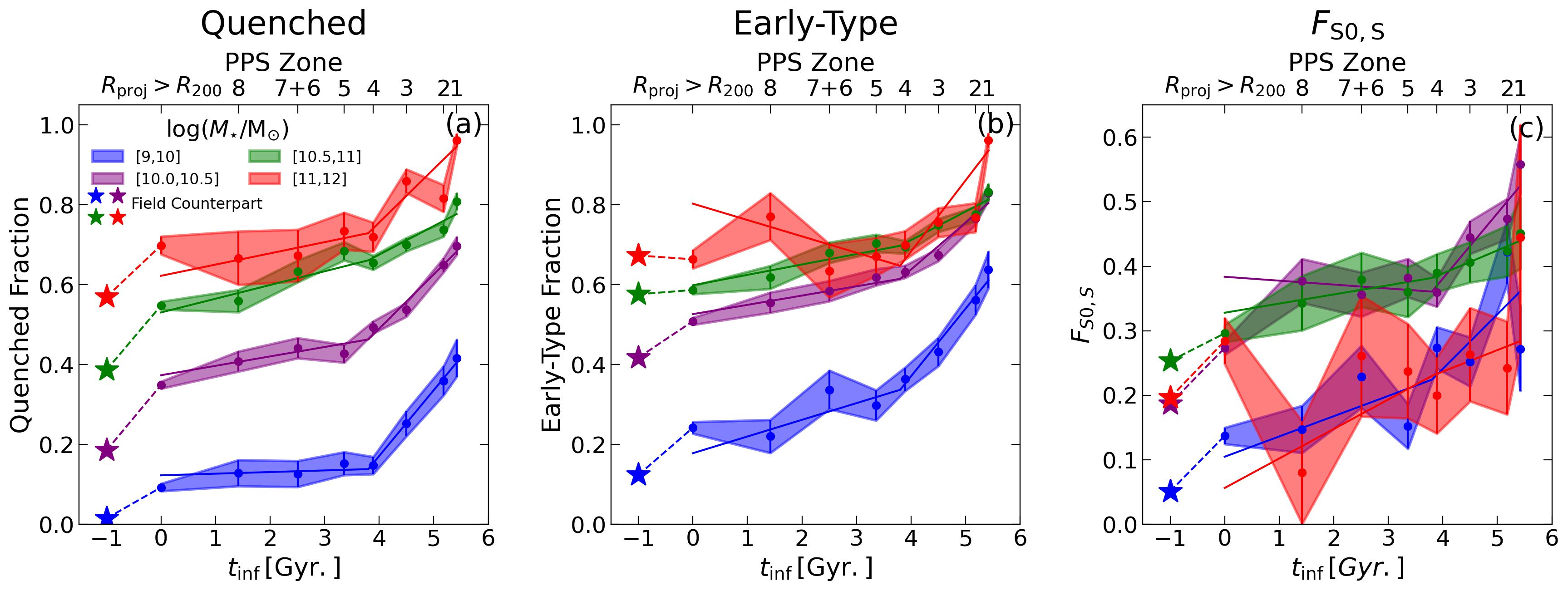}
    \caption{Similar to Fig.~\ref{fig:tinf_bc_gv_rs}, but showing the quenched (left) and early-type (center) fractions, and the $F_{\rm S0,S}$ ratio (right).}
    \label{fig:tinf_q_e_s0}
\end{figure*}

Panels (i) to (l) offer valuable insights into whether galaxies infalling into clusters evolve towards elliptical morphologies or retain a remnant disk, thus becoming S0s. By comparing the early-type fraction (panels e to h) with the $F_{\rm S0,S}$ ratio (panels i to l), we observe mass-dependent trends. For the two lower stellar mass bins, the trends in the early-type fraction align with those in the $F_{\rm S0,S}$ ratio, suggesting that the increase in early-type fraction is primarily due to a rise in the number of S0s. Conversely, in the most massive bins, a significant disparity between $F_{\rm S0,S}$ and the early-type fraction suggests that these massive early-type galaxies are predominantly ellipticals.

Our findings indicate a possible coupling between star formation quenching and morphological transition for galaxies of varying stellar mass across different PPS regions. From a qualitative analysis, we suggest that morphological transition may be initiated at larger radii compared to star formation quenching and is more pronounced with decreasing stellar mass. The similarities between early-type fraction and $F_{\rm S0,S}$ for low-mass galaxies suggest that these galaxies are primarily evolving into S0s. Conversely, for the most massive systems, the trends observed in the early-type fraction distributions do not align with those in the $F_{\rm S0,S}$ distribution. This disparity suggests that the morphological evolution leading to massive ellipticals differs significantly from the process experienced by spiral galaxies infalling into clusters. In the next section, we will quantitatively explore these differences and derive the associated time-scales.

\subsection{Quantifying environmental effects}

Fig.~\ref{fig:tinf_q_e_s0} illustrates the variation of quenched (panel a) and early-type (panel b) fractions, along with the $F_{\rm S0,S}$ ratio (panel c) as a function of infall time. The error bars and filled areas represent the median and 1-sigma scatter. We divide our sample according to stellar mass, and include $R_{\rm proj} > R_{200}$ and field estimates, similar to Fig.~\ref{fig:tinf_bc_gv_rs}. As in Fig.~\ref{fig:tinf_bc_gv_rs}, we begin our analysis by comparing field and cluster galaxies. The quenched fraction (panel a) comparison between cluster and field environments mirrors the trends observed in panel (c) of Fig.~\ref{fig:tinf_bc_gv_rs}. However, we observe a mass dependence in the comparison of early-type fractions between cluster and field environments. For the two lower stellar mass bins, there is a significant excess ($>10\%$) of early-type galaxies in clusters compared to their field counterparts. Conversely, this excess is only seen in the two most massive bins in the $t_{\rm inf} \geq 4$ Gyr region, whereas in the $t_{\rm inf} < 4$ Gyr region the early-type fraction in the field is similar to that observed in clusters. This enhanced fraction of early-type galaxies in the $t_{\rm inf} > 4$ Gyr region may result from massive galaxies being affected by the cluster environment only in inner regions, or it may reflect a pre-existing cluster structure where the most massive ellipticals are found in the core. This mass dependence highlights the role of the cluster environment in driving the morphological transition of low-mass galaxies. In contrast, massive systems either rely primarily on internal mechanisms to alter their morphology, or they enter the cluster environment after experiencing pre-processing effects in medium-density environments, such as small groups or filaments.

Exploring the relations with infall time, we find trends consistent with the slow-then-rapid quenching model. More importantly, the similar trends observed in both the quenched and early-type fractions suggest two major points: 1) the slow-then-rapid quenching model can be extended to include morphological variation; and 2) there may exist a causal relationship between quenching and morphological transition, where the removal of outer components of galaxies in clusters plays a crucial role in star formation suppression.

To quantify the observed trends, we adopt a similar double linear relation to the one presented in equation~\ref{eq:double_line}. The parameter fit results are presented in Table~\ref{tab:q_e_s0_model}. The fitted lines are also presented as solid lines in Fig.~\ref{fig:tinf_q_e_s0}.

\begin{table*}
\caption{Similar to Tab.~\ref{tab:bc_gv_rs_model}, but for curves shown in Fig.~\ref{fig:tinf_q_e_s0}. While $b$ is dimensionless, $a_{1}$ and $a_{2}$ are given in $\rm Gyr^{-1}$.}
\label{tab:q_e_s0_model}
\resizebox{\textwidth}{!}{%
\begin{tabular}{c|ccc|ccc|ccc}
\hline
\multirow{2}{*}{$\log(M_*/{\rm M}_\odot)$} & \multicolumn{3}{c|}{Quenched fraction}              & \multicolumn{3}{c|}{Early-type fraction}             & \multicolumn{3}{c}{$F_{\rm S0,S}$}                  \\ \cline{2-10} 
                                           & $a_1$           & $b$             & $a_2$           & $a_1$            & $b$             & $a_2$           & $a_1$           & $b$             & $a_2$           \\ \hline
{[}9,10{]}                                 & $0.00 \pm 0.01$ & $0.12 \pm 0.03$ & $0.16 \pm 0.03$ & $0.04 \pm 0.02$  & $0.18 \pm 0.01$ & $0.17 \pm 0.02$ & $0.03 \pm 0.02$ & $0.11 \pm 0.04$ & $0.11 \pm 0.03$ \\
{[}10,10.5{]}                              & $0.02 \pm 0.01$ & $0.38 \pm 0.04$ & $0.13 \pm 0.02$ & $0.02 \pm 0.02$  & $0.53 \pm 0.04$ & $0.11 \pm 0.03$ & $0.00 \pm 0.02$ & $0.38 \pm 0.05$ & $0.10 \pm 0.02$ \\
{[}10.5,11{]}                              & $0.03 \pm 0.01$ & $0.53 \pm 0.05$ & $0.08 \pm 0.02$ & $0.03 \pm 0.03$  & $0.59 \pm 0.03$ & $0.07 \pm 0.02$ & $0.01 \pm 0.01$ & $0.33 \pm 0.04$ & $0.04 \pm 0.03$ \\
{[}11,12{]}                                & $0.03 \pm 0.02$ & $0.62 \pm 0.07$ & $0.10 \pm 0.02$ & $-0.04 \pm 0.02$ & $0.81 \pm 0.08$ & $0.15 \pm 0.01$ & $0.05 \pm 0.03$ & $0.05 \pm 0.06$ & $0.03 \pm 0.03$ \\ \hline
\end{tabular}
}
\end{table*}

In general, we observe consistent patterns with those presented in Table \ref{tab:q_e_s0_model}. Notably, $a_{1}$ is approximately zero, independent of stellar mass, while $a_{2}$ declines with stellar mass. The $b$ parameters increase with stellar mass in both the quenched and early-type fractions. However, for the $F_{\rm S0,S}$ ratio, the $b$ parameter reaches a maximum value at intermediate stellar masses (purple and green curves in panel c). These trends indicate that massive systems have largely evolved before entering the cluster. Thus, we posit that most cluster-driven morphological transitions in the local universe occur in intermediate and low stellar mass galaxies.

By comparing the ratio of the rate of increase of $F_{\rm S0,S}$ to the early-type fraction (i.e., the ratio of the corresponding $a_{2}$ values), we gain insight into whether galaxies evolve towards elliptical morphologies or retain a remnant disk. Focusing on the two lower stellar mass bins, we find that $71 \pm 5 \%$ and $83 \pm 4 \%$ of galaxies undergoing morphological transition do so by transforming into S0s. While these fractions depend slightly on the exact T--Type boundaries used to classify morphological types, the results highlight a significant distinction in the origins of cluster S0s -- environmentally-driven transformation of relatively low-mass infalling spirals -- and the origins of massive elliptical galaxies, which were likely present in the cores of clusters from earlier times.

The time-scales for star formation suppression and morphological transition can be compared using the $a_{2}$ slopes corresponding to the quenched and early-type fractions. Although the $a_{2}$ slopes are very similar in both cases, the initial $b$ parameter differs significantly. To estimate the time-scale for 50\% of the galaxies to be quenched or transformed into early-type, we use the equation $\tau_{50\%} = (0.5 - b)/a_{2}$, where $b$ and $a_{2}$ represent the results for quenching or morphological transformation in a given stellar mass bin. For the lowest stellar mass bins, it takes approximately, after the delay time, $2.4$ Gyr for star formation suppression and $1.2$ Gyr for morphological transition. This suggests a potential causal connection between these processes, with the removal of spiral arms being a critical step in the suppression of star formation.

\section{Chapter Summary}

In this chapter, we have explored the impact of environmental factors on the suppression of star formation and the morphological transformation of galaxies infalling into clusters, with a particular emphasis on estimating the time-scales for these processes. Our primary focus has been on cluster galaxies and those entering these environments, while also including a comparison sample of field galaxies that evolve in isolation. Our results indicate that cluster galaxies, particularly those with lower stellar masses, are transitioning from the blue cloud to the green valley and red sequence due to a combination of environmental mechanisms. This environmental influence is also reflected in the morphology of these galaxies, as low-mass field galaxies exhibit higher T--Type values (later morphologies) compared to their cluster counterparts at the same position in the Star Formation Main Sequence diagram.

By examining the morphology of galaxies as a function of their deviation from the Star Formation Main Sequence (Fig.~\ref{fig:tinf_vs_deltaSFMS}), we reveal a strong correlation between T--Type and the perpendicular distance from the star formation main sequence ($\Delta \rm SFMS$). Our findings indicate that galaxies emerge from the green valley with early-type morphologies ($\text{T--Type} < 0$), regardless of their environment and stellar mass. This strongly suggests a close physical link between morphological transformation and the quenching of star formation.

To incorporate time into our analysis, we leverage the statistical relationship between a galaxy's position in the Projected Phase Space and the time since its infall, defined as the time elapsed since it first crossed $R_{200}$. Notably, the transition from the blue cloud to the red sequence corresponds to an increase in the average time since infall (Figs.~\ref{fig:PPS_distribution} and \ref{fig:tinf_vs_deltaSFMS}). Key findings from our time-scale analysis using the Projected Phase Space include:

\begin{itemize}

    \item By analyzing the distribution of blue cloud, green valley, and red sequence galaxies across the Projected Phase Space and as a function of infall time (Figs.~\ref{fig:pps_bc_gv_rs} and \ref{fig:tinf_bc_gv_rs}), we provide evidence supporting the slow-then-rapid quenching model \citep{2012MNRAS.424..232W, 2019ApJ...873...42R}. Notably, the variation in the fractions of galaxies at different evolutionary stages primarily occurs after a delay time of approximately 3.8 Gyr since infall. We claim that this delay corresponds to the time required for galaxies to reach an intracluster medium density sufficient for efficient ram pressure stripping of their gas \citep{2019ApJ...873...42R}. Similar results are obtained when considering morphological transformation, suggesting that the slow-then-rapid model also applies to morphological transition, reinforcing a possible connection between star formation suppression and morphological evolution;

    \item Our analysis shows an approximately constant fraction of green valley galaxies as a function of infall time, suggesting a balanced flow of galaxies into and out of this class as they fall into the cluster. Using a simple model to quantify the observed trends, we find that the most substantial variation occurs in low stellar mass galaxies, yielding a time-scale of approximately 0.5 Gyr for their transition through the green valley;

    \item By examining the fraction of quenched and early-type galaxies in the Projected Phase Space, our analysis reveals notable differences that become more pronounced with decreasing stellar mass (Figs.~\ref{fig:Quenched_Fraction_PPS}). In the lower stellar mass bin, we observe a higher early-type fraction at $0.3 < R_{\rm proj}/R_{200} < 1$ compared to the quenched fraction (35\% vs. 15\%), supporting the notion that morphological transition precedes full star formation quenching;

    \item Comparing the trends between the early-type fraction and the $F_{\rm S0,S} = N_{\rm S0}/(N_{\rm S0} + N_{\rm S})$ ratio across different masses and environments, we suggest that the increase in early-type galaxies in the two lower stellar mass bins is primarily due to a rise in the number of S0 galaxies relative to spirals. This highlights significant differences in morphological transitions between low and high mass galaxies. While massive ellipticals appear to be remnants of major merger events at higher redshifts, environmental effects such as ram pressure stripping on low-mass spiral galaxies transform them into S0s.

\end{itemize}

In summary, we demonstrate that the slow-then-rapid quenching model is applicable to both the suppression of star formation and the morphological transformation of galaxies infalling into clusters, possibly due to a shared causal connection between these processes. By quantifying time-scales using the relationship between location in the Projected Phase Space and time since infall, we provide a first, dynamically motivated estimate of the time galaxies spend in the green valley due to environmental effects. Our analysis also reveals that the fractions of quenched and early-type galaxies remain nearly unchanged until a delay time of $t_{\rm delay} \sim 3.8$ Gyr. We estimate that it takes approximately 2.4 and 1.2 Gyr, respectively, after the delay time for the fractions of quenched and early-type galaxies in clusters to surpass 50\% in low-mass galaxies. These time-scales reinforce the notion that morphological transition precedes complete star formation quenching and suggest that the physical mechanisms driving these changes may be interconnected.

%% file: Chapters/Chapter6.tex
\chapter{A special case of galaxy evolution: the interplay between central and satellite galaxies}
\label{chapter: central_galaxies}
The text from this section is originally presented in 
\noindent \textit{Sampaio, V. M., Aragón-Salamanca, A., Merrifield, M. R., De Carvalho, R. R., Zhou, S., \& Ferreras, I. (2023). The co-evolution of strong AGN and central galaxies in different environments. Monthly Notices of the Royal Astronomical Society, 524(4), 5327-5339.}

\thispagestyle{empty}

\noindent

\section{Overview}

We exploit a sample of 80,000 SDSS central galaxies to investigate the effect of AGN feedback on their evolution. We trace the demographics of optically-selected AGN (Seyferts) as a function of their internal properties and environment. We find that the preeminence of AGN as the dominant ionising mechanism increases with stellar mass, overtaking star-formation for galaxies with $M_{\star} \geq 10^{11}M_\odot$. The AGN fraction changes systematically with the galaxies' star-formation activity. Within the blue cloud, this fraction increases as star-formation activity declines, reaching a maximum near the green valley ($\sim 17 \pm 4\%$), followed by a decrease as the galaxies transition into the red sequence. This systematic trend provides evidence that AGN feedback plays a key role in regulating and suppressing star formation. In general, Seyfert central galaxies achieve an early-type morphology while they still host residual star formation. This suggests that, in all environments, the morphology of Seyfert galaxies evolves from late- to early-type before their star formation is fully quenched. Stellar mass plays an important role in this morphological transformation: while low mass systems tend to emerge from the green valley with an elliptical morphology ($\text{T--Type}\sim -2.5 \pm 0.7$), their high-mass counterparts maintain a spiral morphology deeper into the red sequence. In high-stellar-mass centrals, the fraction of Seyferts increases from early- to late-type galaxies, indicating that AGN feedback may be linked with the morphology and its transformation. Our analysis further suggests that AGN are fuelled by their own host halo gas reservoir, but when in group centrals can also increase their gas reservoir via interactions with satellite galaxies.

\section{Star formation and AGN activity as a function of stellar mass and environment}
\label{sec:SF_AGN_stellar_mass_environment}
\begin{figure}[ht]
    \centering
    \includegraphics[width = \textwidth]{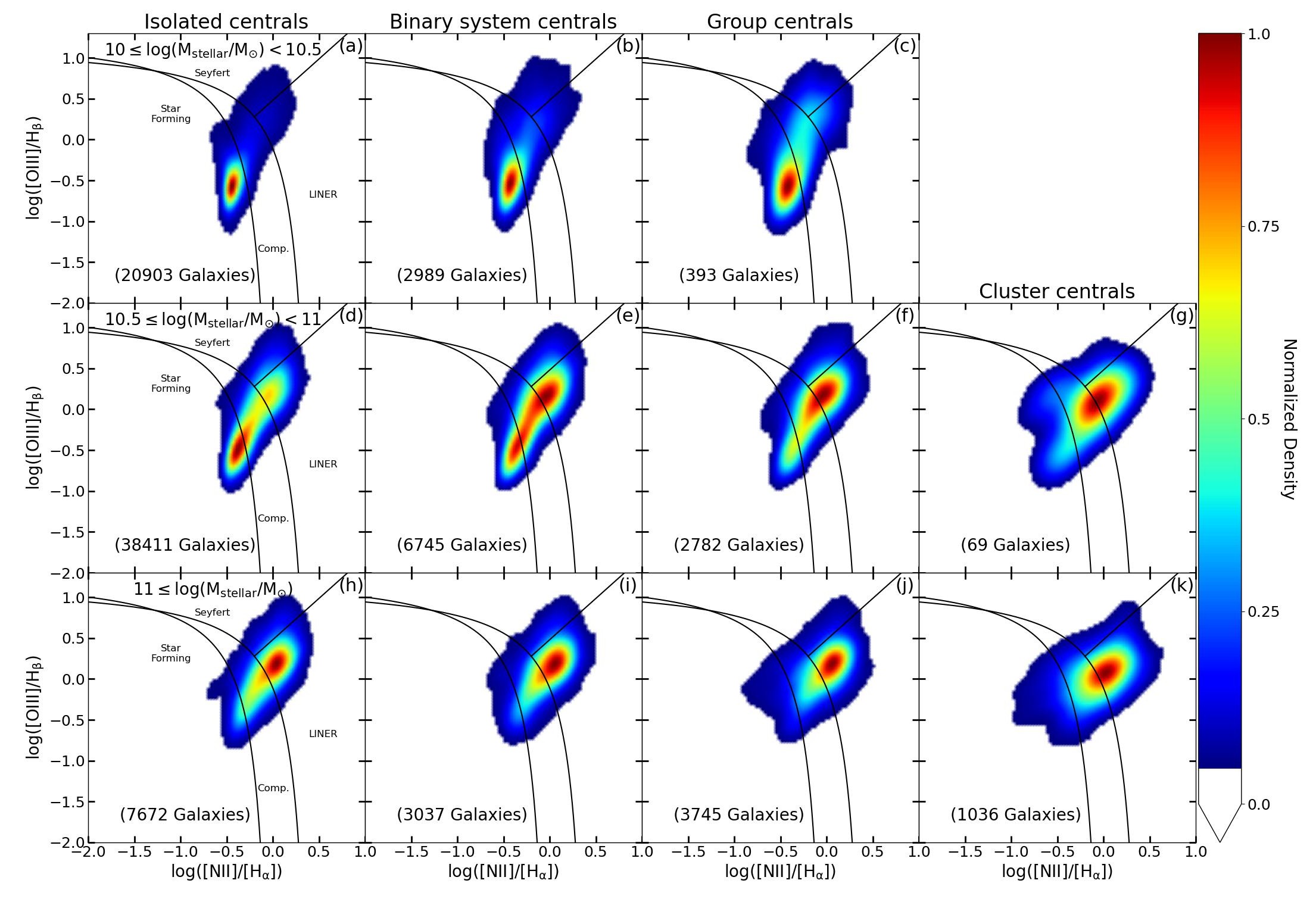}
    \caption{Normalized Gaussian-kernel-smoothed distribution of central galaxies in the BPT diagram for different environments (from left to right, and in order of `environmental richness', isolated centrals, centrals in binary system, in groups, and in clusters) and in three mass bins (from top to bottom, $10^{10} \leq M_{\star}/{\rm M}_{\odot} < 10^{10.5}$; $10^{10.5} \leq M_{\star}/{\rm M}_{\odot} < 10^{11}$; and $M_{\star}/{\rm M}_{\odot} \geq 10^{11}$). The lines separating different regions of the diagram are described in section~\ref{sec:mpa_catalog}. The number of galaxies in each sub-sample is indicated towards the bottom left of each panel. Note that our sample does not contain cluster centrals with log$(M_{\star}/{\rm M}_{\odot}) < 10.5$.} 
    \label{fig:BPT_Smooth}
\end{figure}

Fig.~\ref{fig:BPT_Smooth} shows the distribution of central galaxies in the BPT diagram for the different environments we are considering (isolated centrals, centrals in binary system, in groups, and in clusters), in three stellar-mass bins ($10^{10} \leq M_{\star}/M_{\odot} < 10^{10.5}$; $10^{10.5} \leq M_{\star}/M_{\odot} < 10^{11}$; and $M_{\star}/M_{\odot} \geq 10^{11}$).

We observe a decreasing fraction of star-forming galaxies for increasing stellar mass, irrespective of environment. This is in agreement with the so-called `downsizing' scenario, and suggests that, as star formation decreases, other ionizing mechanisms become more relevant and prominent at high stellar masses. In particular, the increasing fraction of massive galaxies where collisionally driven ionisation dominates suggests that AGN activity may be involved in quenching star formation \citep{2017FrASS...4...10C, 2020MNRAS.499..230B}. Within a given stellar mass bin, we find a decreasing fraction of star-forming galaxies as `environmental richness' increases. For instance, in the low stellar-mass bin the fraction of star-forming galaxies decreases from 31\% in isolated centrals to 17\% in cluster centrals.

\begin{figure}
    \centering
    \includegraphics[width = 0.5\textwidth]{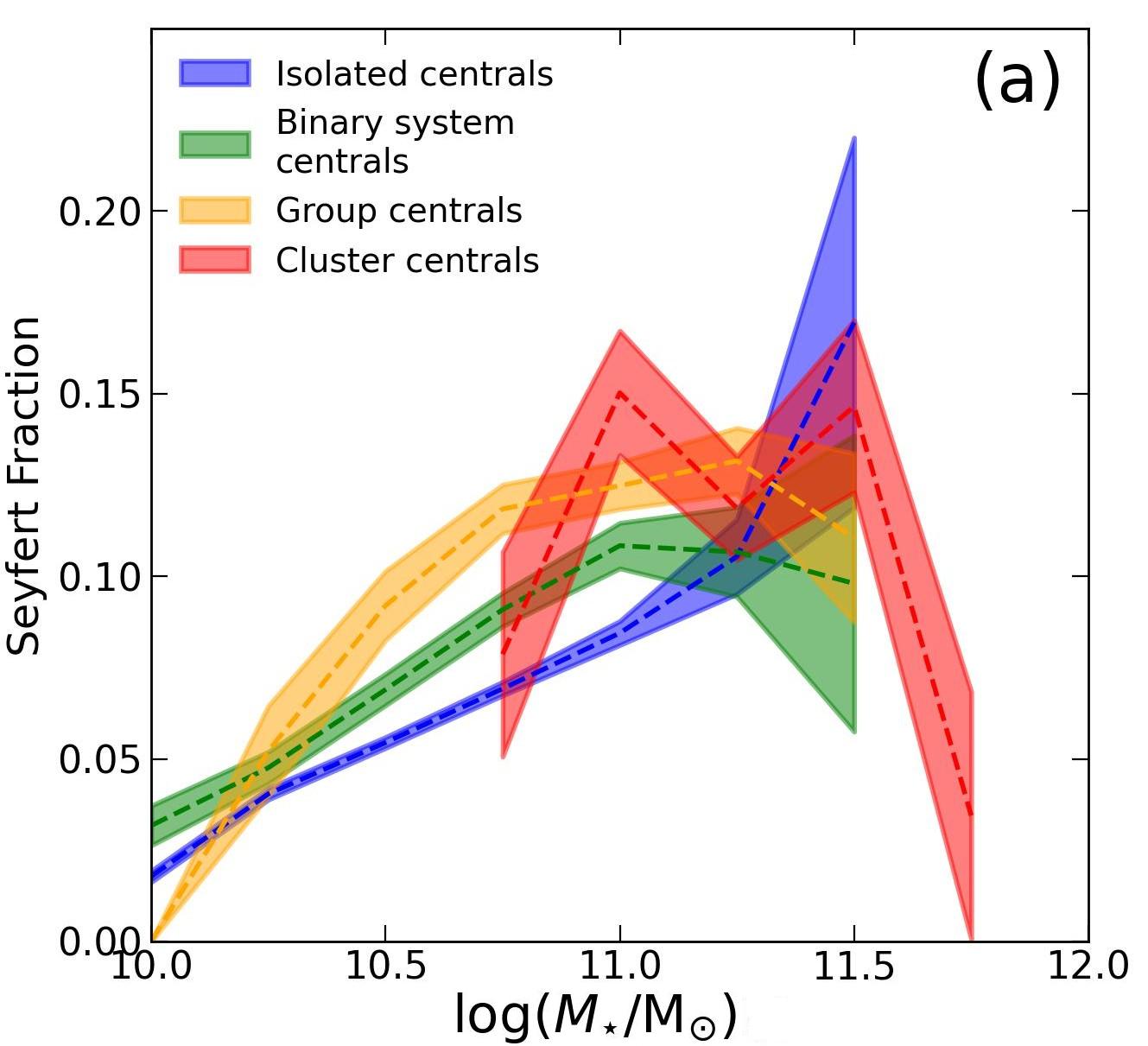}
    \caption{Seyfert fraction as a function of stellar mass, separating galaxies according to environment. Dashed lines and shaded areas represent medians and 1-sigma uncertainties, respectively.}
    \label{fig:seyfert_overall_fraction}
\end{figure}

Moreover, Fig.~\ref{fig:BPT_Smooth} indicates that AGN presence also depends on environment. For instance, for low stellar mass centrals there is a higher Seyfert fraction in group centrals (11\%), compared with 6\%, 8\% and 7\% for isolated, binary system and cluster centrals, respectively. Although AGN are always a small minority in each category, the change is sizeable. In more detail, Fig.~\ref{fig:seyfert_overall_fraction} shows the Seyfert fraction as a function of stellar mass, separating galaxies according to their environment. We calculate the fraction in bins of 0.25 dex, from $\log (M_{\star}/{\rm M}_\odot)=10$ to~12, limiting the calculation to bins with at least 20 galaxies. We interpolate linearly between neighbouring bins. Here and in subsequent figures we estimated the uncertainties using a standard bootstrap technique with 1000 re-samplings. We find that the Seyfert fraction increases with stellar mass for isolated, binary system and group centrals, whereas cluster centrals show an almost constant fraction ($\sim 12.5 \pm 4 \%$ from $\rm 10^{11}$ to $\rm 10^{11.5}M_{\odot}$ followed by a steep decrease towards $\rm 10^{11.75}M_{\odot}$ ($\rm 3 \pm 6 \%$ in the last bin).  
Comparison between different environments shows an increasing Seyfert fraction for increasing richness in the $\rm 10^{10.25}$ and $\rm 10^{11.25}M_{\odot}$ range. This suggests there may be a significant environmental effect on the fueling of AGN activity, which supports that in order to understand AGN feedback on central galaxies we need to analyse the complex interplay between internal and external factors in detail.

\section{The demographics of AGN as a function of the internal and environmental properties of central galaxies}
\label{sec:demographics}

AGN activity depends on the availability of gas that can be accreted onto the central SMBH. It is generally accepted that most galaxies in the mass range we are considering contain a SMBH at their centre.  The fact that only a small fraction of galaxies contain a strong AGN (e.g., Seyfert galaxies) can be attributed to the large amounts of gas required to fuel strong AGN  \citep{2014MNRAS.437.1199C, 2016MNRAS.458L..34E}. By tracing the fraction of Seyferts as a function of the galaxies' properties and environment we may be able to shed light on the combination of internal and environmental factors needed to drive AGN fueling in central galaxies, and consequently investigate the effect of AGN activity on galaxy evolution.

\subsection{The relation between AGN and the morphology of their host galaxies}
\label{sec:AGN_morphology}

\begin{figure*}
    \centering
    \includegraphics[width = \textwidth]{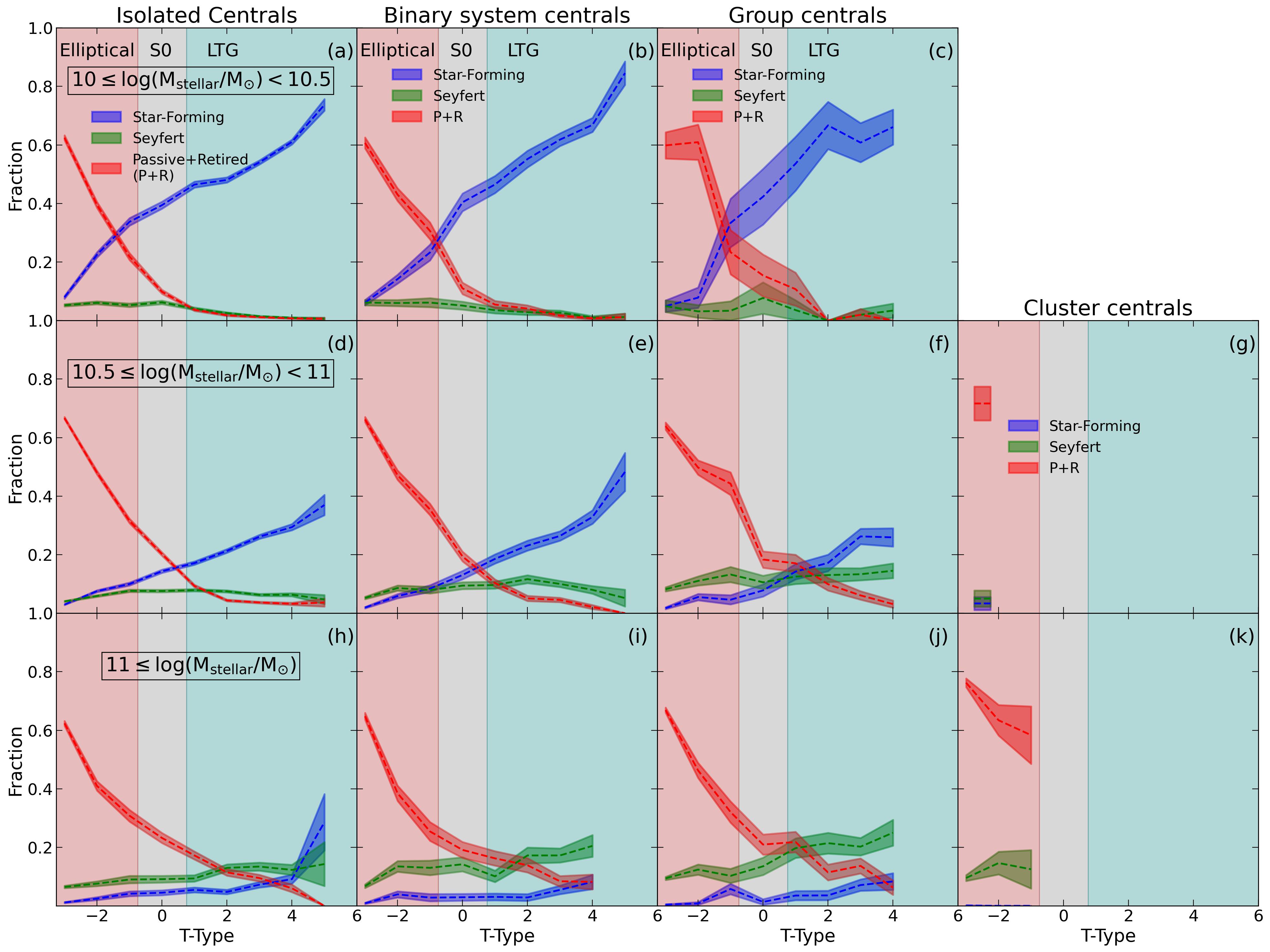}
    \caption{Fraction of star-forming (blue), Seyfert (green) and P+R (red) galaxies as a function of morphological T--Type. As in Fig.~\ref{fig:BPT_Smooth}, galaxies are divided according to their environment (columns) and stellar mass (rows). The red, gray and aqua colored backgrounds correspond to  the T--Type ranges of ellipticals, lenticulars, and spirals, respectively. Median error bars are shown on centre right of each panel. }
    \label{fig:Fract_vs_TType}
\end{figure*}

In Fig.~\ref{fig:Fract_vs_TType} we show the fraction of Seyferts as a function of the T--Type of their host galaxies. To provide context, we also show the fraction of star-forming and P+R galaxies. To guarantee reliable results, we also impose a minimum of 20 galaxies per bin to calculate the relevant fractions. For reference, we define three T--Type ranges: $-3 \leq \text{T} <  -0.75$ corresponds to elliptical galaxies; $-0.75 \leq \text{T} \leq 0.75$ to lenticular (S0) galaxies; and $0.75 < \text{T} \le 6$ to spiral galaxies. These limits were chosen to be consistent with the uncertainties in T--Type ($\Delta \text{T--Type} \sim 0.5$); changes in the bin boundaries of this size will not significantly change our results. Elliptical and lenticulars are often combined under the label ETGs, while spirals are LTGs. Overall, the decreasing fraction of star-forming systems for decreasing T--Type values (from LTGs to ETGs) simply reflects the fact that most LTGs host significant star formation, whereas the opposite trend observed for P+R galaxies indicates that most ETGs do not form many stars. Although this basic picture is true for all galaxy masses and environments, the actual fraction of galaxy type changes both with mass and environment. 

The fraction of Seyfert galaxies is always relatively small, increasing with the galaxies' stellar mass in all environments. In agreement with the results of previous works, we find that low-mass galaxies are less likely to host an AGN than massive ones \citep{2008MNRAS.391..785W}. This can be interpreted as an indication that the fueling of an AGN depends on the amount of gas retained by the galaxy's host halo, which correlates with the halo mass. For the two lowest stellar mass bins, we find an almost constant fraction of Seyferts for morphologies in all environments, although the small numbers make this invariance somewhat uncertain. This low fraction may indicate that, regardless of their morphology, these systems are rarely able to hold a large enough gas reservoir to both fuel star formation and AGN activity. In this case, as soon any AGN feedback is triggered, all the neutral hydrogen gas is removed, preventing any subsequent AGN activity \citep{2022ApJ...933L..12G}.

For high stellar masses, the Seyfert fraction is significantly higher than in low-mass galaxies. Quantitatively, the average over all environments for the lower stellar mass bin gives a Seyfert fraction of $\rm 5 \pm 3\%$, in comparison with $\rm 13 \pm 4\%$. In the most massive bin, Seyferts are more abundant than star-forming galaxies in all environments for virtually all morphologies. There are more than twice as many high-mass galaxies classified as Seyferts than as star-forming even for late-type spirals.  This could simply suggest that, in the absence of star formation as an ionising mechanism, even a relatively modest level of AGN activity becomes apparent. But it could also mean that AGN activity is linked to (and perhaps the cause of) star-formation suppression in massive galaxies. 

Furthermore, the decreasing fraction of Seyferts when we transition from spirals to ETGs provides some support to the idea that AGN feedback plays some role in the morphological transition process, and in generating the morphological diversity we observe in present-day galaxies \citep{2009MNRAS.399.2172R, 2016MNRAS.463.3948D, 2022arXiv220909270V}. 
Our analysis further shows that the slope of the relation between AGN activity and morphology depends on environment, increasing with `environmental richness': it is flatter for isolated centrals and for those in binary systems than for group and cluster centrals. The numerical values of the slopes of the relation between Seyfert fraction and T--Type in Fig.~\ref{fig:Fract_vs_TType} are $0.009 \pm 0.003$, $0.015 \pm 0.009$, $0.022 \pm 0.010$ and $0.042 \pm 0.009$ for isolated, binary, group, and cluster centrals respectively, showing that environment plays a role. 

\subsection{The relation between AGN, halo mass, and central velocity dispersion of their host galaxies}
\label{sec:halo_mass_dispersion}
\begin{figure}[ht]
    \centering
    \includegraphics[width = 0.7\columnwidth]{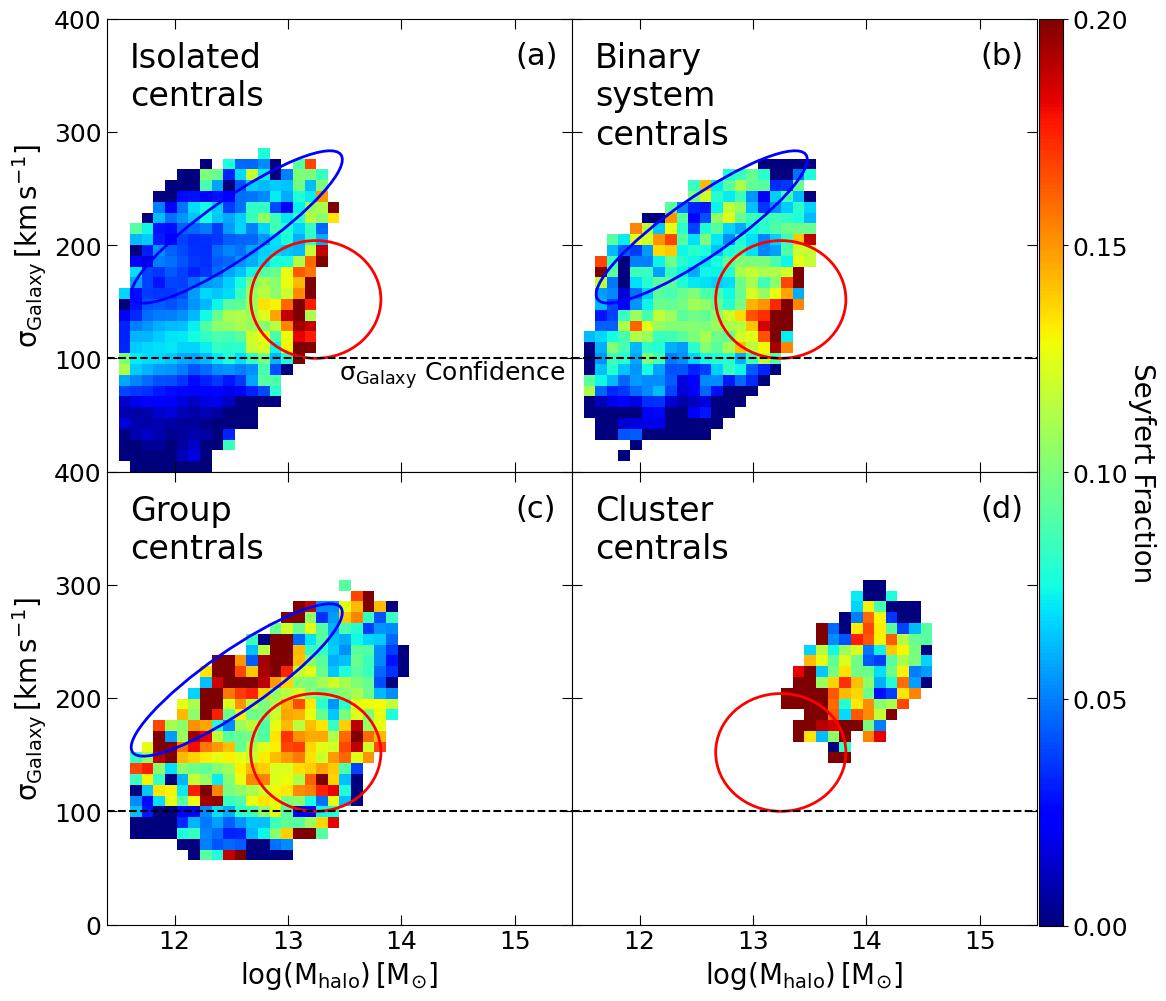}
    \caption{Seyfert fraction in the galaxies' central velocity dispersion ($\sigma_\text{galaxy}$) versus Halo Mass ($M_\text{halo}$) diagram. We separate central galaxies according to their environment into isolated centrals (top left), centrals in binary system (top right), in groups (bottom left) and in clusters (bottom right). The black dashed line shows the limit below which $\sigma_\text{galaxy}$ values from the MPA--JHU catalog become unreliable. Note that galaxy stellar mass correlates reasonably well with $M_\text{halo}$. Circles and ellipses indicate regions of special interest (see text for details). 
    }
    \label{fig:Fraction_Sigma_vs_Mhalo}
\end{figure}

The availability of gas and the mass of the central SMBH are both key factors in AGN activity. The total gas mass is related, albeit in a complex way, to the mass of the host galaxy halo, while the mass of the SMBH correlates with the central velocity dispersion of the galaxy ($\sigma_\text{galaxy}$; \citealt{2002ApJ...574..740T,2009ApJ...698..198G}). To bring all these parameters together, we show in Fig.~\ref{fig:Fraction_Sigma_vs_Mhalo} the median fraction of Seyferts in the $\sigma_\text{galaxy}$ vs.\ $M_\text{halo}$ plane for central galaxies in different environments. For each pixel, we derive the Seyfert fraction median and its uncertainty using a bootstrap technique, as before. We will limit our analysis to galaxies with $\sigma_\text{galaxy}>100\,$km/s, since below this value the galaxy velocity dispersions provided by the MPA--JHU catalog become unreliable due to the resolution of SDSS spectra. In order to confirm the robustness of this trend, we repeated this analysis using 50\% larger bins to reduce the statistical uncertainties. None of the trends described below changed. Furthermore, we applied Gaussian smoothing to the density maps to enhance the visual appearance of the observed overdensities. This also confirmed our conclusions below. We decided to present the raw binned density maps without smoothing since they provide a fairer representation of the data and their uncertainties. 

Panel (a) of Fig.~\ref{fig:Fraction_Sigma_vs_Mhalo} shows that the largest Seyfert fraction in isolated centrals is found in galaxies residing in the largest mass halos ($M_\text{halo} \geq 10^{13} M_{\odot}$) and at intermediate velocity dispersions ($100 \leq \sigma_\text{galaxy} \leq 200\,$km\,s$^{-1}$). A similar locus with a relatively high fraction of Seyfert galaxies (highlighted by the red circles) is also found in the other panels. Quantitatively, this region has median Seyfert fractions of $\rm 17 \pm 9 \%$, $\rm 15 \pm 7\%$, and $\rm 14 \pm 3\%$ in isolated, binary system, and group centrals respectively. A high $M_\text{halo}$ suggests a large reservoir of gas \citep{2014ApJ...795..144C,2019MNRAS.485.3783D}, which may be used to fuel the AGN. However, a direct relation between the size of the gas reservoir and AGN fueling still needs to be verified by observations. The intermediate velocity dispersion range occupied by these galaxies corresponds to the transition region between late- and early-type morphologies, providing additional circumstantial evidence of the effect strong AGN may have on the galaxies' morphological transition.

Panel (c) of Fig.~\ref{fig:Fraction_Sigma_vs_Mhalo} shows, for group centrals, a second region with high Seyfert fraction (indicated by the blue ellipse). These are galaxies with relatively high velocity dispersion for their halo masses. The fraction of Seyferts in the same location of the diagram decreases systematically when we move to centrals in binary systems and isolated ones, with median values of $2 \pm 2\%$, $ 9 \pm 4\%$ and $18 \pm 7\%$ for isolated, binary system and group centrals, respectively. Clearly, in this region of the diagram, the environment affects AGN fuelling. For group centrals, at a given halo mass Seyferts are mainly found in galaxies with high $\sigma_\text{galaxy}$, indicating that, even if the gas reservoir is the same, a larger SMBH mass increases the probability of AGN activity. These high velocity dispersion galaxies tend to be ETGs. That this happens mostly in the group environment may suggest that group centrals are more likely to have experienced a recent interaction with other galaxies (their satellites) than isolated centrals and centrals in binary systems. Such interactions can drive gas to the central system, boosting its fuel reservoir. In support of this, \cite{2014MNRAS.445.1977L} show that central galaxies in groups hosted a more recent star formation episode in comparison to their isolated counterparts, indicating a recent supply of gas. Additionally, \cite{2012A&A...548A..37K} show and increase in [OIII]5007\AA\ when moving closer to the central galaxy. Finally, \cite{2018PASJ...70S..31O} show an enhanced galaxy overdensity around quasar pairs, which are more common in more massive halos. Together with our results, these results provide evidence that the interaction between galaxies can enhance AGN activity. Nevertheless, a more direct test to support this scenario would be the observation of the gas component of a large sample of central galaxies covering a wide range of halo masses, with enough resolution to detect inflows and outflows.

Moving to cluster centrals, panel (d) of Fig.~\ref{fig:Fraction_Sigma_vs_Mhalo} shows the highest fraction of Seyferts at the low-halo-mass end of the distribution ($\rm 20 \pm 10\%$). The innermost regions of the most massive clusters are already ``fully developed'' \citep{1978ApJ...219...18B, 1980ApJ...241..521C, 1997ApJ...476L...7C, 2021MNRAS.503.3065S}, and the probability of satellite galaxies interacting with the cluster central -- and thus providing fuel -- is lowest for the most massive clusters due to the very high relative velocities of the satellite galaxies \citep{1943ApJ....97..255C}. 
\begin{figure}[ht]
    \centering
    \includegraphics[width = 0.7\columnwidth]{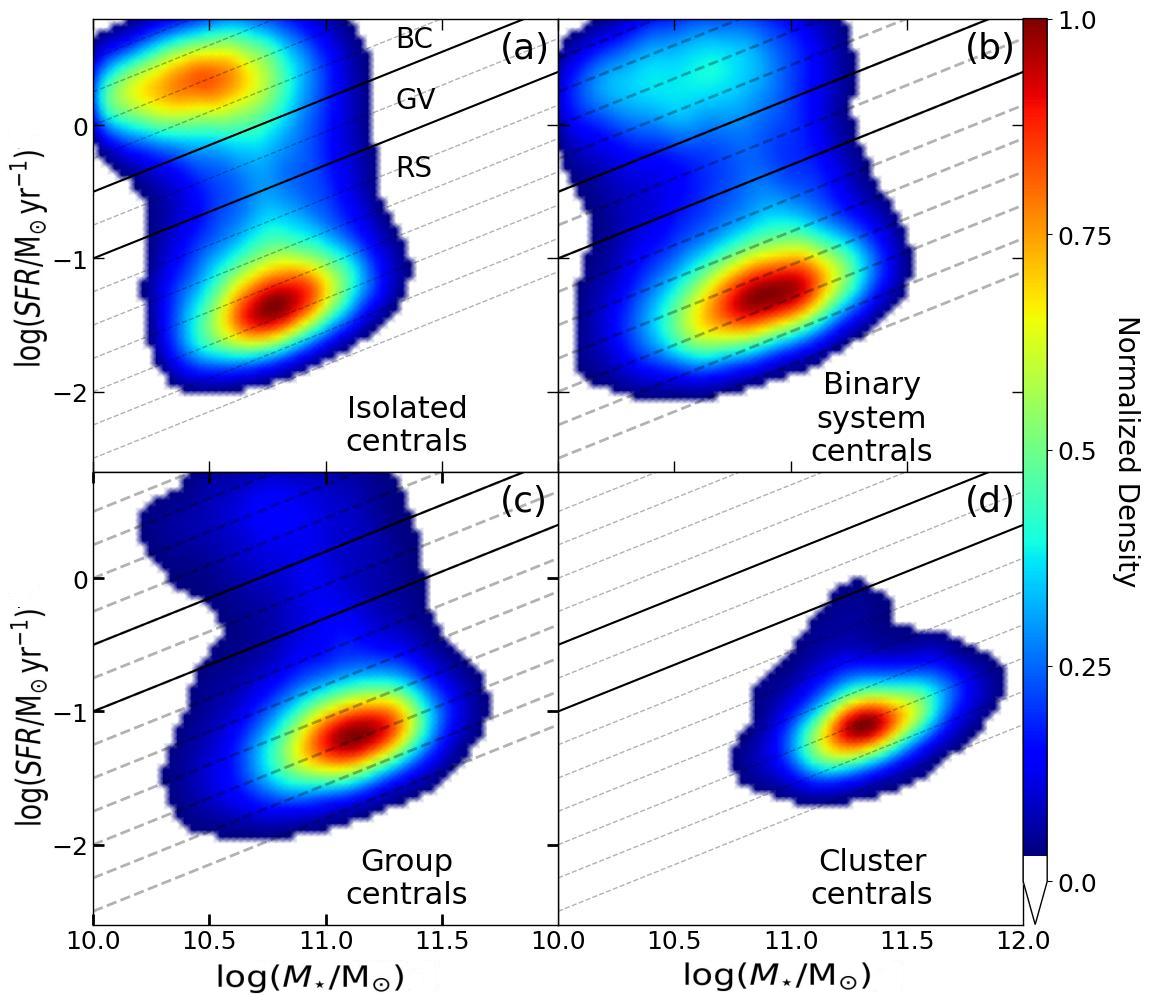}
   \caption{Gaussian kernel smoothed distribution of central galaxies in the star formation rate versus stellar mass plane. The separating lines between blue cloud, green valley, and red sequence regions are shown as yellow solid lines. The white dashed lines have the same slope as the yellow lines, but with varying intercept. Central galaxies are divided into isolated centrals (upper left), centrals  in binary systems (upper right), in groups (lower left), and in clusters (lower right).}
    \label{fig:Smooth_Density}
\end{figure}

From the AGN demographics presented in this section, we have found that, although stellar mass is the strongest statistical predictor of AGN activity (c.f.\ Sections~\ref{sec:SF_AGN_stellar_mass_environment} and~\ref{sec:AGN_morphology}), other factors such as morphology, environment, SMBH mass, and halos mass also play important roles. For massive galaxies, the relatively high AGN fraction in LTGs and the dependence of AGN frequency on T--Type suggests that feedback from the AGN is relevant in the morphological transformation of the galaxies. The details of the distribution of Seyfert fractions in the $M_\text{halo}$ versus $\sigma_\text{galaxy}$ plane provides further insight into how the environment may drive the fueling of AGN activity. For a given halo mass range, the high fraction of Seyferts in group centrals is found at high velocity dispersion, suggests that interactions with satellite group galaxies can enhance the gas reservoirs that fuel the AGN of group central galaxies, which are predominantly ellipticals. 

\begin{figure}[ht]
    \centering
    \includegraphics[width = \textwidth]{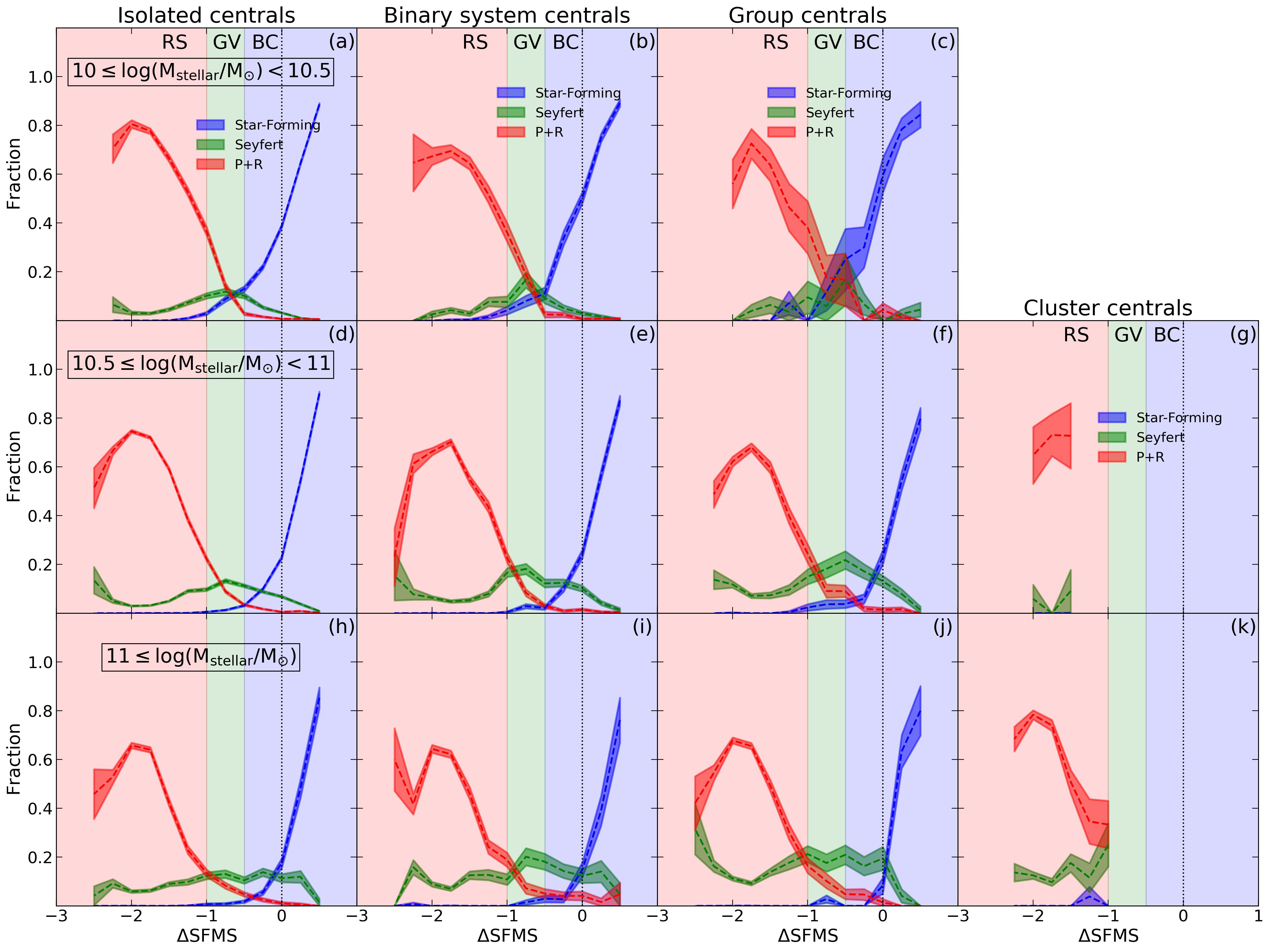}
   \caption{Fraction of central galaxies classified as star forming (blue), Seyferts (green) and P+R galaxies (red) as a function of the vertical distances $\rm \Delta SFMS$ from the SFMS for galaxies in different environments (columns) and stellar-mass ranges (rows). The blue, green, and red backgrounds represent the blue cloud, green valley, and red sequence regions, as defined in Fig.~\ref{fig:Smooth_Density}. 1-sigma uncertainties are presented as shaded regions. Positive $\rm \Delta SFMS$ values correspond to galaxies with higher star-formation activity than galaxies in the SFMS, while negative values indicate star-formation suppression relative to the SFMS.
   }
    \label{fig:Dissection_SFR_Mstellar_Fraction}
\end{figure}

\section{The role of AGN on the star-formation activity of central galaxies}
\label{sec:AGN_SFR}

In this section we focus on how the interplay between AGN activity, stellar mass, and environment drives the evolution of the star-formation activity of central galaxies. We quantify the galaxies' star-formation activity through the the SFR vs.\ $M_{\star}$ diagram, shown in Fig.~\ref{fig:Smooth_Density} for central galaxies in different environments. The BC, GV, and RS regions are defined by two dividing lines: $\log(\text{SFR} / {\rm M}_{\odot} \, {\rm yr}^{-1}) =  0.7 \log(M_{\star}/{\rm M}_{\odot}) - 7.5$ separates the BC from the GV, and  $\log(\text{SFR} / {\rm M}_{\odot} \, {\rm yr}^{-1}) =  0.7 \log(M_{\star}/{\rm M}_{\odot}) - 8.0$ separates the GV from the RS \citep{2020MNRAS.491.5406T, 2022MNRAS.509..567S}. To avoid confusion, we refer to galaxies located in the BC region of this diagram `Blue-Cloud galaxies', reserving the term `star-forming galaxies' for those classified as such in the BPT diagram. 

Consistent with the results of previous sections, we find that the fraction of galaxies in the BC, i.e. with significant star formation, decreases with stellar mass and environmental richness, while the RS shows the opposite trend.

In order to provide a more detailed continuous description of the transition from the BC to the RS, we divide the SFMS plane into 12 zones separated by the white dashed lines in Fig.~\ref{fig:Smooth_Density}. The dividing lines have the same slope as the lines separating the BC/GV/RS regions, but with intercepts varying from $-6.5$ to $-9.5$ in steps $0.25$ wide. It is important to highlight that by adopting a slicing procedure rather than using three discrete zones, we largely remove the dependence of our results on the specific boundaries chosen for the BC/GV/RS. Nevertheless, despite being a somewhat arbitrary choice, the adopted boundary definition has proven to have physical meaning, since it is shown to coincide with the regions where most of the morphological transition happens \citep{2022MNRAS.509..567S}. Each zone corresponding to different vertical distances $\rm \Delta SFMS$ from the SFMS \citep{2007ApJ...670..156D, 2007ApJ...660L..43N, 2011ApJ...730...61K}, defined by the line $\log(\text{SFR} / {\rm M}_{\odot} \, {\rm yr}^{-1}) =  0.7 \log(M_{\star}/{\rm M}_{\odot}) - 7.0$. The resulting variation in galaxy type fraction as a function of the vertical distance to the star-forming main sequence is shown in Fig.~\ref{fig:Dissection_SFR_Mstellar_Fraction} for different environments and stellar-mass ranges. As before, we limit our analysis to bins with at least 20 galaxies.
\subsection{Red but not dead: AGN in massive central red-sequence galaxies in groups}
\label{sec:AGN_in_ellipticals}

ETGs are mainly found in the red sequence and are usually assumed to be passively evolving, such that their star-formation has ceased and their stellar populations simple age with time. They are often called ``red and dead''. However, by tracing the Seyfert fraction as a function of $\rm \Delta SFMS$, we find that, depending on environment, not all central galaxies in the red sequence are quite as dead as previously thought. In panel (h) of Fig.~\ref{fig:Dissection_SFR_Mstellar_Fraction} we find that a relatively high fraction (about one third) of the most massive group galaxies with the lowest star-formation activity are AGN (panel h). This high AGN frequency is in agreement with our previous results, suggesting that massive central galaxies in groups (usually ellipticals) can be fed with gas via interactions with satellite galaxies, generating AGN activity.

\begin{figure}[ht]
    \centering
    \includegraphics[width = 0.7\columnwidth]{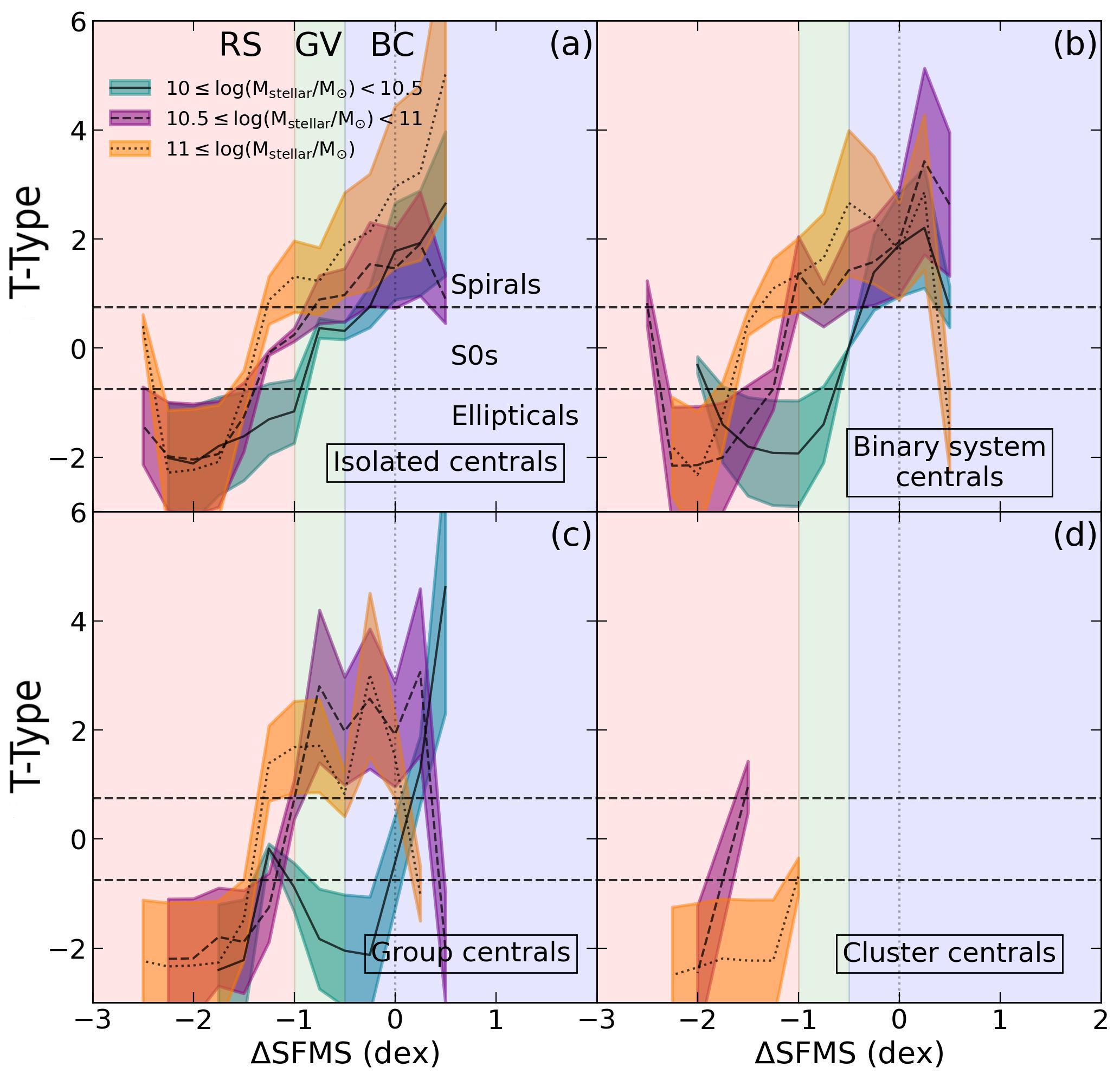}
   \caption{Median T--Type as a function of $\rm \Delta SFMS$ for central galaxies classified as Seyferts in the BPT diagram. We divide galaxies according to stellar mass (different colours and line styles) and environmental richness (different panels). Background colors correspond to the BC, GV and RS regions, as in Fig.~\ref{fig:Dissection_SFR_Mstellar_Fraction}. 1-sigma errors are shown as shaded areas.}
    \label{fig:Dissection_SFR_Mstellar_Morphology}
\end{figure}

\subsection{The AGN-driven transition from the blue cloud to the red sequence}
\label{sec:AGN_RS_BC}

The results shown in Fig.~\ref{fig:Dissection_SFR_Mstellar_Fraction} provide a new perspective on the possible effect of AGN activity on the evolution of central galaxies. The increasing Seyfert fraction across the BC as we move away from the SFMS coincides with a corresponding  decrease in the star-forming fraction, providing evidence that, despite representing only 10--20\% of the central galaxy population, AGN feedback could be extremely efficient at quenching star formation for all stellar masses. In this scenario, the AGN is responsible for decreasing the galaxies' star formation, but there is a time delay between the triggering of the AGN and its radiation becoming the main source of the ISM ionisation. During this process, the AGN activity becomes `unmasked' after the associated decrease in star formation reduces the ionisisng effect of the high-mass stars. High spatial resolution integral-field spectroscopic data from instruments such as MUSE may provide the necessary evidence to test this scenario.

These results are consistent with the idea that the transition from the BC to the RS is caused by AGN-driven outflows \citep{2020MNRAS.491.5406T}. Interestingly, the peak of the Seyfert fraction occurs in the vicinity of the GV, a region associated with both star-formation suppression and morphological transition \citep{2022MNRAS.509..567S}, suggesting, once again, that AGN activity is linked with galaxy transition from spirals to early-type galaxies.

The subsequent decrease in Seyfert fraction from the GV to the RS may be due to the combination of two processes, AGN feedback ejecting gas from central galaxies, which depletes the gas fuel reservoir needed to feed the SMBH, and the morphological transition from disks to spheroids, in which the gas is more stable and less likely to be accreted by the SMBH \citep{2009ApJ...707..250M}. In other words, the AGN itself is ultimately responsible for shutting down its own activity. 

\subsection{Morphological Transition vs. Star Formation Quenching}

By tracking the morphology of Seyfert galaxies as a function of their star-formation activity, we are also be able to address the question of whether AGN feedback changes the star formation activity of their host galaxies before or after their morphology changes. The median T--Type of Seyfert galaxies as a function $\rm \Delta SFMS$ is shown in Fig.~\ref{fig:Dissection_SFR_Mstellar_Morphology}, where the galaxies have been separated according to stellar mass and environmental richness. For group and cluster centrals, we show only the curves with a minimum number of 10 galaxies per bin to ensure the statistical uncertainties are not too large. 

This figure shows that, statistically, Seyfert galaxies reach elliptical morphologies (median T--Type $\rm \sim -2$) before their star-formation is completely suppressed (i.e., before they reach the core of the RS region, $\rm \Delta SFMS \sim -2$). There is also a mass dependence in the morphological transition of Seyfert galaxies: while low-mass galaxies typically emerge from the GV with elliptical morphologies (irrespective of their environmental richness), intermediate-mass systems emerge, on average, as S0s, and high mass systems tend to retain spiral morphologies beyond the GV. This provides evidence suggesting that the AGN activity is more relevant in the morphological transition of low-mass galaxies, whilst more massive systems hold their morphology ``longer''. 

In summary, our results show that morphological transition and star-formation suppression happen at different pace, in agreement with \cite{2022MNRAS.509..567S}. Although the detailed connection between AGN activity and the physical processes by which it affects morphological transition are still not wholly understood, it is noticeable that the sequencing we observe suggests that, irrespective of stellar mass and environment, the morphology of Seyfert galaxies evolves from late- to early-type before their star formation is fully quenched.

\section{Chapter Summary}

With the aim of shedding light understanding how AGN affect the evolution of their host galaxies, we have investigated the relationship between the properties of central galaxies, their environment, and the presence or absence of an AGN at their centres. We focus on central galaxies because, as the dominant object in their dark-matter halo, external influences may be weaker than for satellite galaxies, making them ideal candidates for studying the effect of internal processes such as AGN feedback.   

We have traced the effect of the environment by dividing the galaxies into isolated centrals and centrals in pairs, groups, and clusters. SDSS spectroscopy and the BPT diagram allow us to determine whether the dominant ionising mechanism of the galaxies' interstellar medium is star-formation or AGN activity. We analyse this information as a function of internal galaxy properties such as their stellar and dark-matter halo mass, morphology, and central velocity dispersion (correlated with their SMBH mass). In doing so we explore how the interplay between different combinations of internal and external factors may be linked to the fueling of the central SMBH and thus AGN activity, and how this activity may affect the properties and evolution of its host galaxies. The main results of this analysis are:

\begin{itemize}

    \item In all environments, the preeminence of AGN activity as the dominant ionising mechanism increases with stellar mass, overtaking star-formation for galaxies with $\rm M_{\star} \geq 10^{11}M_\odot$ (Figs.~\ref{fig:seyfert_overall_fraction} and \ref{fig:Fract_vs_TType}). It is noticeable that, at these high masses, the fraction of LTGs with Seyfert activity reaches $\rm 21 \pm 4\%$ in group centrals, while only $\rm 7 \pm 3\%$ of these galaxies have their interstellar medium ionised by star formation.
    
    \item In high stellar mass central galaxies, the fraction of Seyferts increases with T--Type from $\rm 5 \pm 1\%$ at $\text{T--Type}\sim-2.5$ (ellipticals) to $\rm 18 \pm 3\%$ at $\text{T--Type}\sim4$ (late spirals). This increase in Seyfert fraction with morphology suggest that AGN feedback is linked with the morphology of these galaxies and, perhaps, its transformation. The low fraction of Seyfert nuclei in low-mass galaxies, independent of morphology and environment, results from the inability of these systems to fuel star-formation and AGN simultaneously.

    \item The separate variation of the Seyfert fraction with the galaxies' velocity dispersion and halo mass suggests two different ways to fuel AGN activity. First, central galaxies can rely on their own host halo gas reservoir to fuel their AGN, which is increasingly efficient with increasing halo mass; and second, central galaxies can increase their gas reservoir via interactions with satellite systems. We find that the second mechanism is most important for the evolution of high-mass centrals in groups and low-mass clusters, where such interactions will be more common. 

    \item The Seyfert fraction changes systematically as a function of star-formation activity in their host galaxies (Fig.~\ref{fig:Dissection_SFR_Mstellar_Fraction}). Within the blue cloud this fraction increases as star-formation activity declines, reaching a maximum when the galaxies enter the green valley. The peak fraction depends on both stellar mass and environmental richness. For instance, high mass group centrals show a peak Seyfert fraction of $\rm 18 \pm 4\%$ near the green valley. Subsequently, the Seyfert fraction decreases as the galaxies transition into the red sequence. This sequence strongly suggest that AGN feedback plays a key role in regulating and suppressing star-formation.     

    \item Tracing the morphology of Seyfert central galaxies as a function of their star-formation activity, we find that they typically achieve an early-type morphology  while they still host some residual star formation. This suggests that, in all environments, the morphology of Seyfert galaxies evolves from late- to early-type before their star formation is fully quenched. We additionally show that stellar mass plays an important role in the morphological evolution of Seyfert galaxies: low mass systems tend to emerge from the green valley with an elliptical morphology, whereas their high-mass counterparts maintain a spiral morphology deeper into the red sequence.

\end{itemize}

In summary, we have found strong evidence that AGN activity in central galaxies is intimately linked with the morphology and star-formation of their hosts, and that this link depends on both the internal properties of these galaxies and their environment.

%% file: Chapters/Chapter7.tex
\chapter{On the other side: tracing imprints of accretion history in cluster properties}
\label{chapter: G_vs_NG}
The text from this section is originally presented in \noindent\textit{Sampaio, V. M., de Carvalho, R. R., Ferreras, I., Laganá, T. F., Ribeiro, A. L. B., \& Rembold, S. B. (2021). Investigating the projected phase space of Gaussian and non-Gaussian clusters. Monthly Notices of the Royal Astronomical Society, 503(2), 3065-3080.}




\section{Overview}

We investigate from the projected phase space perspective the relation between galaxy properties and cluster environment in a subsample of the Yang Catalog. We separate clusters according to the gaussianity of the members galaxy velocity distribution into gaussian (G) and non-gaussian (NG). Our sample is limited to massive clusters with $M_{200} \geq 10^{14} {\rm M}_{\odot}$ and $0.03\leq z \leq 0.1$. NG clusters are more massive, less concentrated and have an excess of faint galaxies compared to G clusters. NG clusters show mixed distributions of galaxy properties in the PPS compared to the G case. Using the relation between infall time and locus in the PPS we estimate the infall rate of NG clusters and find that they on average accreted $\sim 10^{12}\, {\rm M}_{\odot}$ more stellar mass in the last 5 Gyr than G clusters. Examining the relation between galaxy properties and infall time we show that the relation is significant different for galaxies in G and NG systems. The more mixed distribution in NG clusters PPS translates to more flattened relations between galaxy properties and infall time. Faint galaxies infalling in NG clusters are older and more metal rich in comparison to its counterpart in G systems and already started their morphological transition towards elliptical shapes. All these results suggest that NG clusters experience an accretion of pre-processed galaxies, assembling the galactic component of NG clusters in comparison to G systems. Finally, we find evidence that bright galaxies are less affected by their environment and are mostly quenched by internal processes.

\section{Structure and Composition of G and NG Clusters}

\begin{figure}[ht]
    \centering
    \includegraphics[width = 0.7\columnwidth]{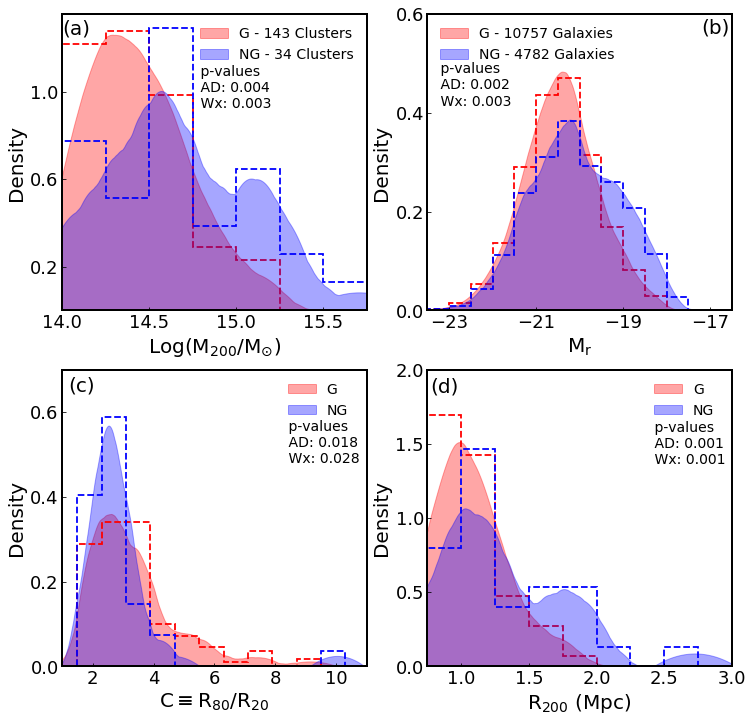}
    \caption{Comparison of the $M_{200}$ (top-left), $M_{\rm r}$ (top-right), $R_{80}/R_{20}$ (bottom-left) and $R_{200}$ (bottom-right) distribution according to the gaussianity classification (G or NG cluster). In each plot we show the resulting p-value of a Anderson-Darling (AD) and Wilcoxon (Wx) statistical test. NG clusters are more massive, less concentrated (see text for the concentration measurement here adopted), larger virial radius and have an excess of fainter galaxies with respect to G systems. }
    \label{fig:concentration}
\end{figure}

In this work, the differences between G and NG clusters plays a major role. In a first step we focus our attention to characterize the structure and galaxy member properties distribution for each cluster class, separately. In Fig.~\ref{fig:concentration} we show the distributions of: a) Logarithmic virial mass ($ \log(M_{200}/{\rm M}_{\odot})$); b) r-band absolute magnitude of members galaxy; c) a proxy for concentration, defined as $R_{80}/R_{20}$, where $R_{x}$ is the projected radius at which the sum of stellar mass within it is equal to x\% of the total stellar mass within $R_{200}$; and d) virial radius (in Mpc) of G and NG clusters. We compare the distributions using two  different statistical tests, Anderson-Darling (AD) and Wilcoxon Rank Test (Wlx, see \citealt{engmann2011comparing} and \citealt{gehan1965generalized} for a review on both)\footnote{The adopted significance level threshold is $\alpha = 0.05$}. The results are shown in each panel. We find that G and NG clusters have statistically different distributions for the studied parameters. In panel (a) we note that NG clusters tend to have higher values of $M_{200}$ in comparison to G clusters. Namely we find that 32.4\% (11/34) of NG clusters have$\log(M_{200}/{\rm M}_{\odot}) \geq 14.75$, while this percentage decreases to $6.3\% \, (9/143)$ in G clusters. The majority of G systems ($\sim 71.3\%$) have $\log(M_{200}/{\rm M}_{\odot}) < 14.5)$. Panel (b) shows an excess of fainter galaxies\footnote{Not to be confused with the faint regime defined in Section \ref{subsec:Bright_Faint}} in NG clusters in comparison to G systems. We find that $30.45\%$ of NG clusters members galaxy have $M_{\rm r} \geq -19.5$, while in G systems galaxies with $M_{\rm r} \geq -19.5$ corresponds to $14.08\%$. In panel (c), a comparison of G and NG clusters concentration shows that NG clusters are less concentrated than G clusters. Only one NG cluster have $C > 4.7$. Finally, in panel (d) we note significant differences also for the virial radius distribution of G and NG clusters. NG clusters tend to have larger virial radius than G clusters. Quantitatively, the distribution for NG (G) clusters have an mean value of 1.46 (1.06) Mpc. In other words, we find that NG clusters have larger virial mass and radius, have an excess of fainter galaxies and are less concentrated in comparison to G systems.

\begin{figure}[ht]
    \centering
    \includegraphics[width = 0.85\textwidth]{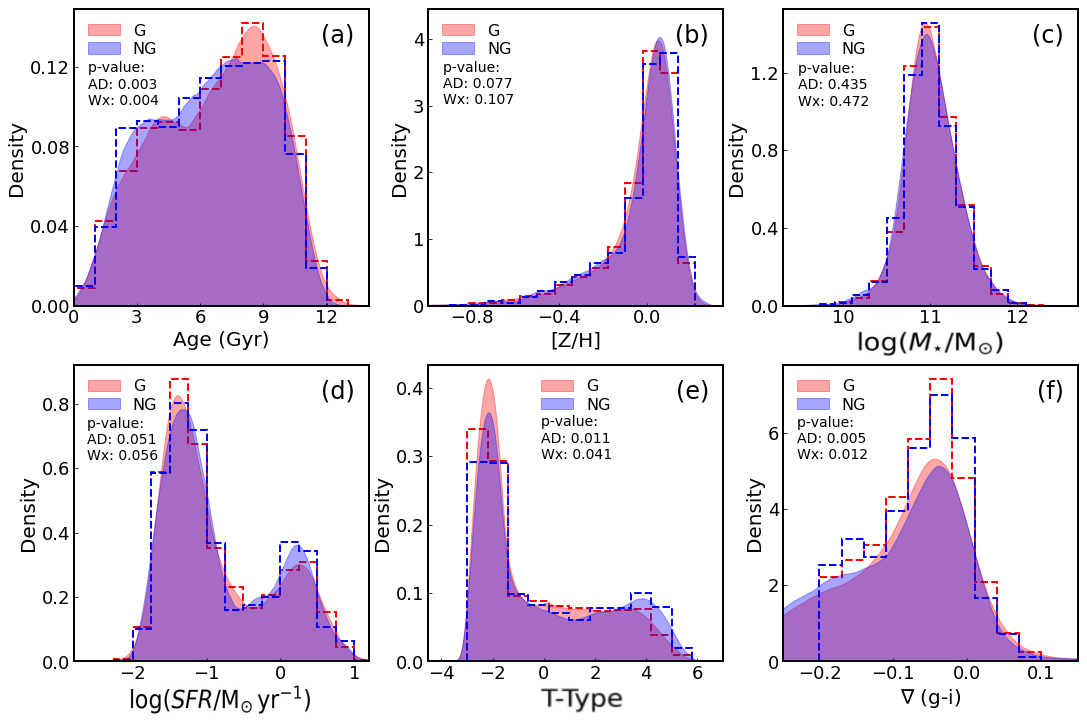}
    \caption{Distributions of $Age$ (a), $[Z/H]$ (b), $M_{\star}$) (c), SFR (d), T--Type (e) and $\nabla$(g-i) (f) for bright galaxies in G and NG clusters. In each panel we also show the resulting p-values of an AD and Wx statistical test. The distributions of $Age$ and T--Type of BG are statistically different between G and NG clusters.}
    \label{fig:bright_hist}
\end{figure}

\begin{figure}[ht]
    \centering
    \includegraphics[width = 0.85\textwidth]{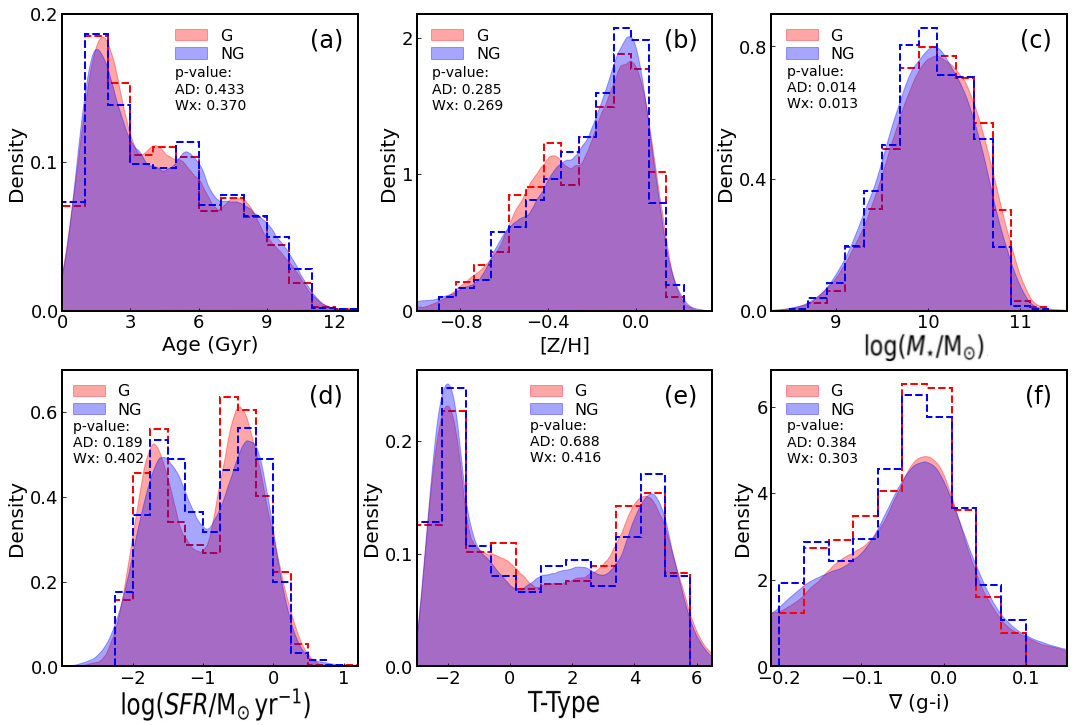}
    \caption{The same as Fig.~\ref{fig:bright_hist}, but for F galaxies. Differently from the bright regime case, only $ \log(M_{\star}/{\rm M}_{\odot})$ have statistically different distributions.}
    \label{fig:faint_hist}
\end{figure}

Regarding galaxy properties, we compare the distribution of $Age$, $[Z/H]$, $M_{\star}$, SFR, T--Type and Color Gradient ($\rm \nabla (g-i)$) of G and NG clusters members galaxy in Figs.~\ref{fig:bright_hist} (B galaxies) and \ref{fig:faint_hist} (F galaxies). In the bright case, we notice that G and NG clusters have statistically different distributions of $Age$, T--Type and $\rm \nabla (g-i)$. In panel (a) of Fig.~\ref{fig:bright_hist} we notice in G clusters an excess of B galaxies with $Age > 7.5$ Gyr in comparison to NG systems. Regarding galaxies with $Age < 7.5$ Gyr, the excess is seen in NG systems. In panel (b), (c) and (d) we note that G and NG clusters have a similar distribution of galaxy members Metallicity, $M_{\star}$ and SFR. In panel (e) we observe that NG clusters have an excess of B galaxies with ${\rm T--Type} > 3$, which translates to an excess of B galaxies with spiral morphology. An analysis of the distributions in panel (f) suggests that both distributions are statistically different (${\rm p-values} < 0.05$). However, the differences in the distribution of $\rm \nabla (g-i)$ G and NG clusters are not straightforward. 

Extending the same analysis to F galaxies, shown in Fig.~\ref{fig:faint_hist}, we find that $M_{\star}$ is the only parameter to show statistically different distributions. However, as in the case of $\rm \nabla (g-i)$ for B galaxies, the differences are not straightforward to characterize. In panels (a), (b), (d), (e) and (f) we notice that the p-value analysis is unable to reject the null hypothesis that both distributions represent two samples of the same underlying distribution.

\section{The Projected Phase Space}
\label{sec:thePPS}

In this Section we describe how we define the PPS for G and NG clusters and present an comparison between G and NG clusters. The PPS is constructed with the projected radial distance to the clustercentric coordinates\footnote{We define the cluster center as the mean RA and DEC weighted by galaxy flux.} ($R_{\rm proj}$) in the x-axis and the velocity projected along the line of sight in the y-axis. Here we adopt the $v_{\rm LOS}$ absolute value in the y-axis, following a commonly approach due to the existing symmetry along the line of sight (e.g. \citealt{2011MNRAS.416.2882M,2017ApJ...843..128R,2019MNRAS.484.1702P,2020ApJS..247...45R}). The radial axis is normalized by $R_{200}$ and the velocity axis is normalized by the cluster's velocity dispersion ($\sigma$) to enable comparisons between different systems. The normalization is based on a homology hypothesis, which assumes that systems are structurally equal, aside from a characteristic radius and velocity dispersion. 

A major concern in the PPS construction is that clusters in our sample have a large variety of richness, ranging from 25 to 900 galaxies. To avoid biases due to the different number of members galaxy we stack all the members galaxy of a given cluster class and a luminosity regime into a single PPS. Hence we end up with four stacked PPSs: 2 for G clusters (B and F) and 2 for NG clusters (B and F).

\subsection{Discretizing the Projected Phase Space}

The PPS approach enable a discrimination of members galaxy according to the infall time. Although in the 3D phase space galaxies has a well defined trajectory while entering in clusters (e.g. \citealt{2010MNRAS.408.2442W}), in the PPS this trajectory is degenerated due to effects of projection along the line of sight. The trace of galaxies with different time since infall cannot be done with observational data. Cosmological simulations are then used to define different regions in the PPS corresponding to galaxies with different time since infall. We here make use of two different ways of slicing the PPS: 1) following \cite{2017ApJ...843..128R}, that use the YZiCS simulation to study the numeric density of galaxies with different time since infall occupying specific regions in the PPS. We refer to these regions as ``Rhee Region'' hereafter. They provide a probabilistic approach to each region in the PPS; and 2) the separation presented in \citealt{2019MNRAS.484.1702P} (P19, hereafter), which also use the YZiCS simulation, but define analytic quadratic functions to fit the observed distribution of time since infall along the simulated stacked PPS. These P19's New Zones (PNZs, hereafter) constrain the time since infall of galaxies occupying a given locus in the PPS. PNZs have a decreasing characteristic infall time from PNZ 1 (most inner one, $t_{\rm inf} \sim 5.42$) to PNZ 8 ($t_{\rm inf} \sim 1.42$). Our results are typically independent on the way we slice the PPS (see Appendix A), hence in the following analysis we present results obtained using the PNZs.

\subsection{Galaxy Distribution in the PPS}
\label{sec:PPS_distribution}

In a first approximation, differences in cluster galaxy population may translate to a different distribution in the PPS. In this section we explore how members galaxy are distributed in the PPS of G and NG clusters. Table \ref{table:galaxies_distribution} show the number of galaxies in each PNZ for G and NG systems, for both B and F regimes. Column (1) lists the cluster class; column (2) lists  the luminosity regime; columns (3) to (10) shows the number of galaxies in PNZs 1 to 8, respectively; and column (11) lists how many galaxies are beyond $R_{200}$. 41\% of the B galaxies in NG clusters are beyond $R_{200}$, while in G clusters this percentage decreases to 28\%. F galaxies also show a similar trend: 28\% and 34\% of the F galaxies are beyond the $R_{200}$ of G and NG clusters, respectively. These numbers show unequivocally that NG clusters show an excess of galaxies beyond $R_{200}$. 

In Fig.~\ref{fig:PPS_Density} we show the normalized density of B and F galaxies in the PPS of G and NG clusters. We limit the PPS to $R_{200}$, where the PNZs are valid and enable a connection between infall time and locus in the PPS. We divide the PPS in bins of $0.15 |V_{\rm LOS}|/\sigma_{\rm LOS} \times 0.05 R_{\rm proj}/R_{200}$ to guarantee square pixels where the the PNZs are valid ($[0,3] \times [0,1]$). For each pixel we assign the number of galaxies within it. We then perform a convolution in the resulting matrix using a Gaussian Kernel with $\rm FWHM = 1.88$ ($\rm \sigma_{\rm std} = 1$)\footnote{The FWHM/standard deviation was empirically defined, but does not affect the trends here shown.}. Bins without observations are carefully treated. We apply an interpolation in ``empty'' bins that have at least 50\% of their vicinity occupied by valid observations. It is noticeable that by applying the interpolation we are assuming that transitions in the PPS occupation are sufficiently smooth.

\begin{table}

\caption{Number of galaxies in each PNZ.}
\label{table:galaxies_distribution}
\resizebox{\textwidth}{!}{\begin{minipage}{\textwidth}
\centering
\begin{tabular}{c|c|ccccccccc}
\hline
                    & PNZ & 1   & 2   & 3   & 4   & 5   & 6   & 7  & 8   & $R>R_{200}$ \\ \hline
\multirow{2}{*}{G}  & BG  & 392 & 611 & 723 & 824 & 391 & 176 & 93 & 255 & 1352            \\
                    & FG  & 58  & 125 & 249 & 150 & 66  & 33  & 19 & 41  & 289             \\ \hline
\multirow{2}{*}{NG} & BG  & 101 & 157 & 126 & 229 & 141 & 66  & 26 & 69  & 623             \\
                    & FG  & 47  & 128 & 182 & 249 & 119 & 52  & 22 & 58  & 441             \\ \hline
\end{tabular}

\end{minipage}}
\end{table}

We note in Fig.~\ref{fig:PPS_Density} that indistinctly for B or F galaxies and G or NG clusters there is an offset of approximately $\sim 0.35 R_{\rm proj}/R_{200}$ between the cluster center and the higher density envelope (redder colors). This offset may be due to projection effects or a limitation on defining the center of the systems using only galaxies, which represent a minor fraction of the cluster mass compared to the gas and dark matter components. Despite this offset, the galaxy distributions are in agreement with both observational and simulated data from \cite{2020ApJS..247...45R}, which guarantee that this offset do not represent a sample bias.

Comparison between galaxy distribution in the PPS of G and NG clusters shows that: 1) the higher density envelope (redder colors) in NG clusters extends to higher velocities than in the G case, for both B and F (see the dashed line in Fig.~\ref{fig:PPS_Density}); and 2) in the faint regime, we note an excess of galaxies in PNZ 8 of NG clusters compared to G systems (see red ellipses in Fig.~\ref{fig:PPS_Density}). In the bright regime we note an excess of B galaxies in the PNZ 8 of G clusters, however this may be due to the larger number of B galaxies in G clusters. An excess of infalling galaxies translates to a higher rate of galaxies with high velocities and the PNZ 8 is closely related to galaxies first entering the cluster. Thus these trends are in agreement with a higher infall rate in NG.

\section{Connecting Galaxy's Properties and Locus in the Projected Phase Space}
\label{sec:locus_properties}

\begin{figure}
    \centering
    \includegraphics[width = 0.5\textwidth]{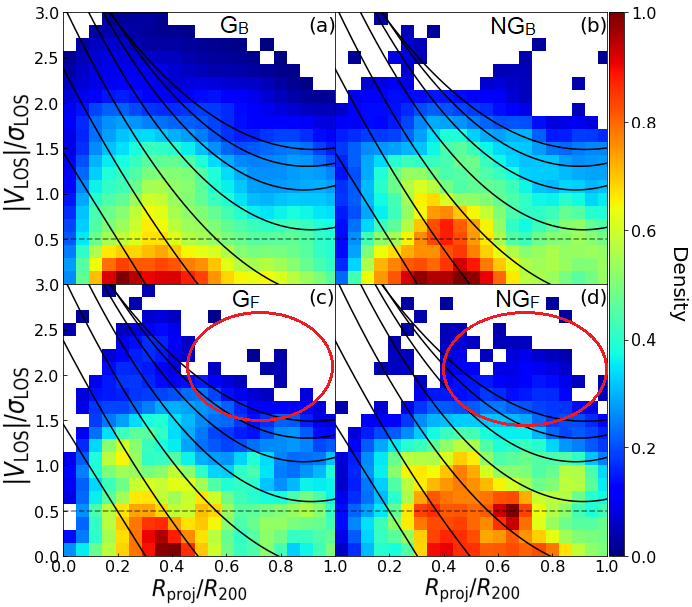}
    \caption{Normalized density distribution of galaxies in the PPS of G clusters (left) and NG clusters (right). We separate galaxies accordingly to the luminosity regimes: bright on top panels and faint in the bottom panels. We note that for NG cluster, higher density regions extend to higher velocities in comparison to G cluster. We highlight this trend by including a $|V_{\rm LOS}|/\sigma_{\rm LOS} = 0.5$ dashed line in the four panels. The red ellipses highlight the excess of F galaxies in the PNZ 8 of NG clusters.}
    \label{fig:PPS_Density}
\end{figure}

Environmental quenching drastically affect galaxy properties. Previous works show that the quenched fraction of galaxies within clusters is a function of the clustercentric radius. However, RPS and tidal mass loss are also conditional on the incoming velocity of the infalling galaxy (see R19 for an example). The PPS provide a powerful tool to understand how galaxy properties is affected by high density environments taking into account both velocity and position. In this section we study the distribution of median $Age$, $[Z/H]$, SFR, T--Type, $\rm \nabla (g-i)$ $M_{\star}$ in the PPS of G and NG clusters. We use a similar approach of Fig.~\ref{fig:PPS_Density} to study the distribution of galaxy properties. 

\begin{figure*}
    \centering
    \includegraphics[width = 0.9\textwidth]{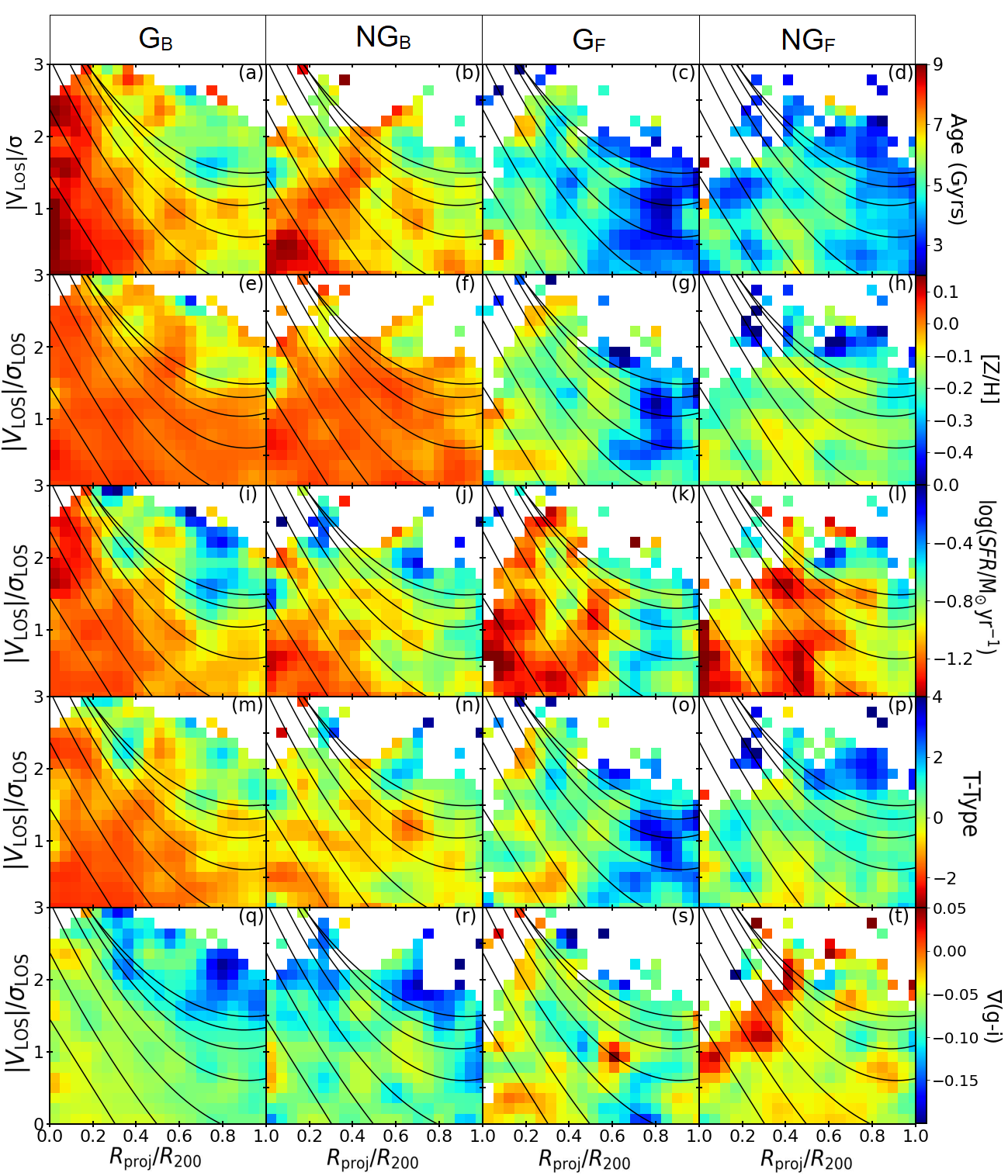}
    \caption{From top to bottom we present
    the distribution of $Age$, $[Z/H]$, SFR, T--Type and $\nabla$(g-i), respectively, over the PPS of G and NG clusters and luminosity regime (B or F).}
    \label{fig:full_grid}
\end{figure*}

Fig.~\ref{fig:full_grid} displays (from top to bottom) the distribution of median $Age$, $[Z/H]$, $\log(\text{SFR} / {\rm M}_{\odot} \, {\rm yr}^{-1})$, T--Type and $\rm \nabla(g-i)$ for B (two first columns) and F (two last columns) galaxies in the PPSs of G and NG clusters. PNZs are shown in solid lines and separate galaxies with different infall time. Comparison of panels (a) and (b) shows that the PNZs $\leq$ 4 of NG cluster have galaxies with $Age < 7$ Gyr, while in the PPS of G clusters these regions are occupied mainly by B galaxies with $Age > 7$ Gyr. Quantitatively, 70\% of the B galaxies in the PNZ $\leq$ 4 region of G cluster's PPS have $Age > 7$ Gyr, while in NG clusters this percentage is 53\%. In panels (c) and (d) we note a quite similar trend for F galaxies. However, the differences are highlighted according to the clustercentric distance. F galaxies within $0.5R_{200}$ in NG clusters are on average 0.31 Gyr younger than in G clusters. On the other hand, F galaxies beyond $0.5R_{200}$ in NG clusters are older by, on average, 0.84 Gyr than in G clusters.     

Panels (e) and (f) show the distribution of B galaxies median $[Z/H]$ in the PPS of G and NG clusters. We find that both G and NG clusters PPS are mainly occupied by $\rm $[Z/H]$>-0.1$ B galaxies. In G and NG cluster's PPS the mean $[Z/H]$ of B galaxies are -0.016 and -0.010, respectively. We highlight that the differences between G and NG mean $[Z/H]$ comes mainly from the PNZ 8, relative to recent infalling galaxies. B galaxies in the PNZ 8 of G clusters are, on average, 0.028 more metal-poor than its counterpart in NG clusters. Significant differences between G and NG clusters members galaxy $[Z/H]$ distribution are mainly found in the faint regime, shown in panels (g) and (h). Galaxies beyond 0.6$R_{200}$ in G clusters are more-metal-poor by on average 0.091 dex than in NG clusters. However, F galaxies within $0.6R_{200}$ do not show significant differences between G and NG clusters. Namely we find that F galaxies within 0.6$R_{200}$ of G clusters are on average 0.01 dex more metal rich than those in NG clusters, which is comparable to the uncertainty in the estimates derived through spectrum fitting. 

The action of quenching process translates to a decrease in star formation. Examining the distribution of SFR in the PPS of G and NG clusters, shown in panels (i) to (l), lead to further evidence on the role of environmental effects. Indistinctly for B and F galaxies, both G and NG clusters show a trend of decreasing SFR with increasing infall time (decreasing PNZ). This highlights the cumulative effect of environmental quenching with time since infall of galaxies in clusters. Despite this overall behavior, comparison of panels (i) and (l) show that, in G clusters, B galaxies (panel i) with $\log(\text{SFR} / {\rm M}_{\odot} \, {\rm yr}^{-1}) < -0.6$ occupy mainly PNZs 1 to 4. On the other hand, B galaxies with $\log(\text{SFR} / {\rm M}_{\odot} \, {\rm yr}^{-1}) < 0.6$ in NG clusters (panel j) occupy only PNZs 1 to 3. In the PNZ 4 of NG clusters we note an excess of galaxies with higher SFR in comparison to its counterpart in G systems. Regarding F galaxies, a trend of a core with quenched galaxies with a tail to higher velocities is visible from panels (k) and (l). In panel (k) we note that PNZs 1 and 2 of G clusters are occupied mainly by F galaxies with $\log(\text{SFR} / {\rm M}_{\odot} \, {\rm yr}^{-1}) < -1.2$. Additionally we note a tail of galaxies with $\log(\text{SFR} / {\rm M}_{\odot} \, {\rm yr}^{-1}) < -1.0$ that extends up to $\sim 1.7 |V_{\rm LOS}|/\sigma_{\rm LOS}$. In NG clusters this tail is more highlighted. NG clusters have a more mixed distribution. Namely, we note a column of quenched galaxy in $\rm 0.5R_{200} < R_{\rm proj} < 0.75R_{200}$ that extends to $\rm \sim 1.7 |V_{\rm LOS}|/\sigma_{\rm LOS}$ and that only PNZ 1 is fully occupied by galaxies with $\log(\text{SFR} / {\rm M}_{\odot} \, {\rm yr}^{-1}) < -1.2$.     

Comparison of the T--Type distribution of members galaxy in G and NG clusters PPS (panels m to p) shows similarities with the SFR and $[Z/H]$ distribution (panels i to l). This provides further evidence that the quenching of star formation is also related with a morphological transition. For example, the green/blue region in panel (k) and (g) corresponds to galaxies with TTYpe > 1 in panel (o). B galaxies in the PNZ $\leq$ 4 region of G clusters an average T--Type value of -0.81, which is lower than the average value found for the same region in NG systems (-0.36). For the PNZ > 4 the mean difference between G and NG decreases to $\rm \bar{\Delta} T--Type(G-NG) \sim 0.18$ ($\bar{T--Type_{PNZ > 4}} = $ -0.12 and 0.06 for G and NG clusters, respectively). Similarly to SFR and $[Z/H]$, differences in the faint regime are better described using a radial separation. F Galaxies within 0.6 $R_{200}$ of G have an slight lower average T--Type value, 0.39, in comparison to its counterpart in NG clusters, 0.55. On the other hand, galaxies beyond 0.6$R_{200}$ in G clusters have higher T--Types compared to the same subset in NG clusters. Namely we find an average T--Type value of 1.98 and 1.49 for G and NG clusters, respectively. 

The distribution of $\rm \nabla (g-i)$ in G and NG clusters are similar for B galaxies (panels q and f). Galaxies with $\rm \nabla (g-i) < -0.10$ are found mainly in PNZ 8 and at $\rm |V_{LOS}|/\sigma > 1.7$. Lower values of color gradient means that galaxy is bluer on the outside according to the definition presented in Equation (3). This trend hence suggests that B galaxies enter clusters with bluer colors on the outside and then this difference decreases after PNZ 6. In the faint regime (panels s and t) we note an overall higher value of $\rm \nabla (g-i)$ than in the bright regime, which translates to a lack of bluer colors. Furthermore, NG clusters have an excess of F galaxies with $\rm \nabla (g-i)$ > 0. This subset of galaxies are found in every PNZ, which may indicate that these F galaxies entered the NG cluster already with positive values of color gradients.

\begin{figure}[ht]
    \centering
    \includegraphics[width = 0.5\textwidth]{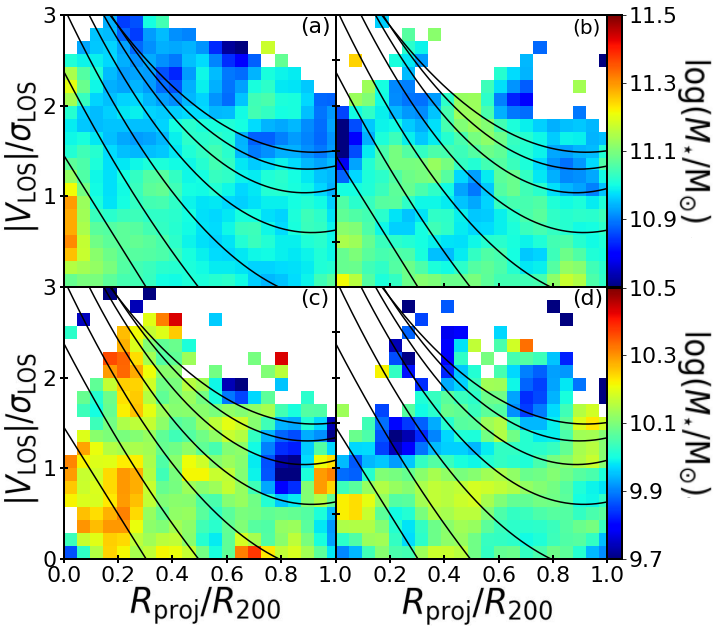}
    \caption{Panels (a) and (b): B galaxies $M_{\star}$ distribution over the PPS of G and NG clusters, respectively. Panels (c) and (d): the same, but for F galaxies.}
    \label{fig:grid_mstellar}
\end{figure}

In Fig.~\ref{fig:grid_mstellar} we present the median $\rm M_{\star}$ distribution in the PPS of G and NG clusters, with galaxies divided into B and F. We separate this specific parameter from the last 5 discussed, since in this case B and F translates to different ranges in stellar mass. Therefore, we select suitable ranges for each luminosity regime, while guaranteeing that both have the same relative difference in $\log(M_{\star}/{\rm M}_{\odot}$). Ranges are defined based on the distribution of $\log(M_{\star}/{\rm M}_{\odot})$ shown in Figs.~\ref{fig:bright_hist} and \ref{fig:faint_hist}. Comparing panels (a) and (b), we see that the distribution of $\log(M_{\star}/{\rm M}_{\odot})$ in G and NG clusters for B galaxies are quite similar. We observe that in both G and NG clusters, only $\sim 34.5$\% of B galaxies have $\log(M_{\star}/{\rm M}_{\odot}) > 11.1$. In the faint regime, we note that approximately 48\% of the F galaxies have $\log(M_{\star}/{\rm M}_{\odot}) > 10.1$, corresponding to the upper part of the defined range. In Section \ref{sec:Discussion} we discuss this difference between bright and faint galaxies related to the downsizing scenario \citep{2006MNRAS.372..933N}. In the faint regime we also note significant differences between G and NG clusters. Panel (c) shows that within $0.3R_{200}$ there is an excess of F galaxies with $\log(M_{\star}/{\rm M}_{\odot}) > 10.2$ in G clusters. Additionally, in panel (d) we note within $0.4R_{200}$ NG clusters an excess of F galaxies with $ \log(M_{\star}/{\rm M}_{\odot}) < 10.0$.

\section{The Relation Between Galaxy Properties and Infall Time in G and NG Clusters}

\begin{figure}[ht]
    \centering
    \includegraphics[width = \textwidth]{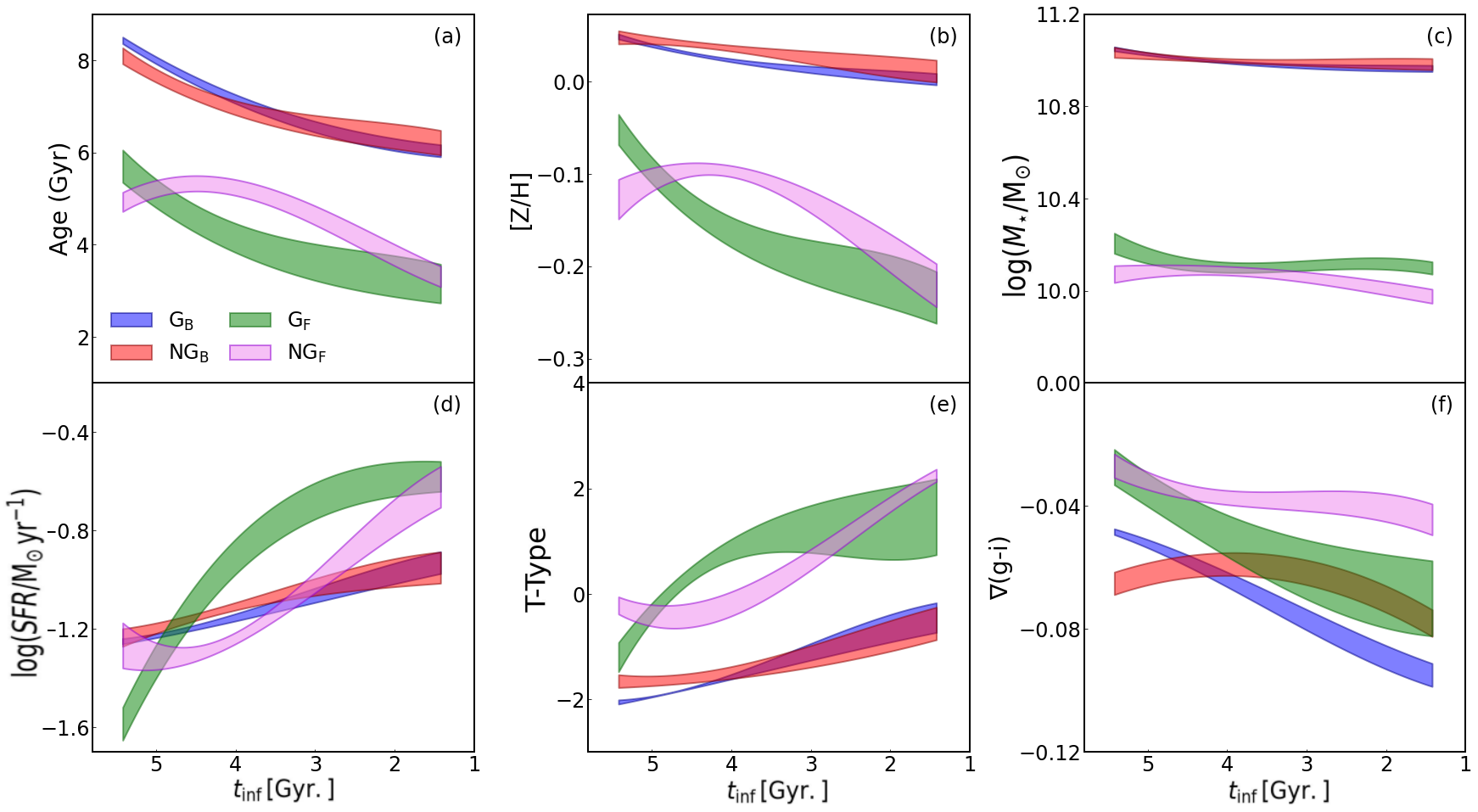}
    \caption{From panel (a) to (f): Relation between time since infall and $Age$, $[Z/H]$, $\log(M_{\star}/{\rm M}_{\odot})$, $\log(\text{SFR} / {\rm M}_{\odot}\,{\rm yr}^{-1})$, T--Type and $\nabla$(g-i) of galaxy members of G and NG clusters. Galaxy members are separated into bright (red) or faint (blue). The shaded regions correspond to the median variance calculated using the bootstrap technique.}
    \label{fig:tinfall_relation}
\end{figure}

In Section \ref{sec:locus_properties} we investigate differences in the way galaxy properties are distributed in the PPS of G and NG clusters. The delay then rapid model proposes that galaxies with different $t_{\rm inf}$ are quenched upon different mechanisms. In this section we probe environmental effects on galaxies by studying the relation between $t_{\rm inf}$ and galaxy properties in G and NG systems, separating galaxy population into B and F. We consider that all galaxies in a single PNZ are well represented by the mean $t_{\rm inf}$ presented in P19. We show in Fig.~\ref{fig:tinfall_relation}, in panels (a) to (f),  the median values of $Age$, $[Z/H]$, $\log(M_{\star}/{\rm M}_{\odot})$, $\log(\text{SFR}/{\rm M}_{\odot} \, {\rm yr}^{-1})$, T--Type and color gradients, respectively, for galaxies in each PNZ. We estimate the medians variance using a bootstrap technique: 1) for each PNZ we randomly select N values (where N is the number of galaxies in that locus), with replacement, from the observed distribution; 2) calculate the variance using the new distribution $Q_{\rm sigma}$\footnote{The variance from quartiles is calculated as $Q_{\rm sigma} = 0.74 \times (Q_{75\%} - Q_{25\%})$.}; 3) we repeat this procedure 1000 times; and 4) consider the variance as the median of the $Q_{\rm sigma}$ distribution. 

In a first approximation, we adopt a linear relation between $ t_{\rm inf}$ and galaxy properties. However, we highlight that this is a functional approach and the relations do not need necessarily to be indeed linear. This approach is mainly to quantify quantities such as variations with infall time. We fit the observed relations using the SciDavis statistical analysis tool \citep{benkert2014scidavis}, which takes into account errors in both x and y for the fit. The resulting linear and angular coefficients are shown in Table \ref{table:linear_coeficients}. In column (1) we list cluster class and luminosity regime; in column (2) we list the correspondent coefficient; columns (3) to (8) show the results and associated errors for $Age$, $[Z/H]$, $\log(M_{\star}/{\rm M}_{\odot})$, $\log(\text{SFR} / {\rm M}_{\odot}\,{\rm yr}^{-1})$, T--Type and $\rm \nabla(g-i)$, respectively. We compare the significance level of the differences between G and NG clusters as:
\begin{equation}
  \sigma = \frac{\sqrt{(\sigma^{\chi}_{\rm i,G})^{2} + (\sigma^{\chi}_{\rm i,NG})^{2}}}{2},
\end{equation}
where $\sigma^{\chi}_{i,G}$ and $\sigma^{\chi}_{i,NG}$ are the errors associated with the coefficient i (linear or angular) of the parameter $\chi$ ($Age$, for example). In Table~\ref{table:linear_coeficients} we highlight in red relevant differences in the angular coefficients of G and NG clusters. Dark red means differences greater than 2-sigma and light red differences greater than 1-sigma. We highlight differences in the linear coefficients in the same way, but using blue. 

With respect to global trends, we see from Fig.~\ref{fig:tinfall_relation} that the median stellar population parameters (panels a, b and c) differs significantly between B and F galaxies, both in G and NG systems. For instance, we find a mean difference of 2.18 Gyr, 0.28 and 0.99 dex between the intercept parameter of B and F galaxies for $Age$, $[Z/H]$ and $\log(M_{\star}/{\rm M}_{\odot})$, respectively. Additionally, the relation with respect to SFR of B galaxies have a slope closer to zero  than F galaxies. We find an average slope\footnote{Between G and NG clusters.} of -0.082 and -0.20 for the SFR of B and F galaxies, respectively. These results suggests unequivocally that B and F galaxies are distinctly affected by environmental quenching. However, we find that indistinctly in G and NG clusters, both B and F galaxies show slopes smaller than 0.1 dex in stellar mass, which is approximately a constant relation with infall time. This last result may suggest that galaxy quenching is mostly related to the removal of gas component instead of stellar mass itself.

\begin{table}
\centering
\caption{Angular and Linear Coefficients from Linear Fit.}
\label{table:linear_coeficients}
\centering
\resizebox{\textwidth}{!}{%
\begin{tabular}{c|c|cccccc}
\hline
                             & $y = ax + b$ & $Age$ (Gyr)                                 & $[Z/H]$                                      & $\rm Log(M_{stellar}/M_{\odot})$         & $\log(\text{SFR} / {\rm M}_{\odot} \, {\rm yr}^{-1})$ ($\rm M_{\odot}/yr$)                & T--Type                                      & $\rm \nabla (g-i)$                          \\ \hline
                             & a          & \cellcolor[HTML]{FFCCC9}$0.701 \pm 0.052$ & $0.014 \pm 0.002$                          & {\color[HTML]{333333} $0.025 \pm 0.006$} & \cellcolor[HTML]{FFCCC9}$-0.078 \pm 0.011$  & $-0.333 \pm 0.042$                         & \cellcolor[HTML]{FFCCC9}$0.012 \pm 0.001$  \\
\multirow{-2}{*}{$\rm G_B$}  & b          & $4.468 \pm 0.219$                         & $-0.031 \pm  0.008$                        & $10.896 \pm  0.023$                      & \cellcolor[HTML]{DAE8FC}$-0.832 \pm 0.051$  & \cellcolor[HTML]{DAE8FC}$-0.266 \pm 0.201$ & \cellcolor[HTML]{34CDF9}$-0.115 \pm 0.006$ \\ \hdashline
                             & a          & \cellcolor[HTML]{FFCCC9}$0.538 \pm 0.105$ & $0.013 \pm 0.004$                          & $0.019 \pm 0.010$                        & \cellcolor[HTML]{FFCCC9}$-0.100 \pm 0.020$  & $-0.455 \pm 0.168$                         & \cellcolor[HTML]{FFCCC9}$0.003 \pm 0.006$  \\
\multirow{-2}{*}{$\rm NG_B$} & b          & $4.826 \pm 0.448$                         & $-0.022 \pm 0.014$                         & $10.925 \pm 0.041$                       & \cellcolor[HTML]{DAE8FC}$-0.706 \pm 0.089$  & \cellcolor[HTML]{DAE8FC}$0.553 \pm 0.643$  & \cellcolor[HTML]{34CDF9}$-0.078 \pm 0.009$ \\ \hline
                             & a          & \cellcolor[HTML]{FFCCC9}$0.626 \pm 0.068$ & \cellcolor[HTML]{FD6864}$0.052 \pm 0.010$  & $0.037 \pm 0.012$                        & \cellcolor[HTML]{FD6864}$-0.249 \pm 0.027 $ & $-0.611 \pm 0.151$                         & $0.009 \pm 0.004$                          \\
\multirow{-2}{*}{$\rm G_F$}  & b          & \cellcolor[HTML]{DAE8FC}$2.043 \pm 0.705$ & \cellcolor[HTML]{34CDF9}$-0.360 \pm 0.040$ & $9.973 \pm 0.043$                        & \cellcolor[HTML]{34CDF9}$-0.139 \pm 0.110$  & $2.497 \pm 0.661$                          & \cellcolor[HTML]{DAE8FC}$-0.080 \pm 0.020$ \\ \hdashline
                             & a          & \cellcolor[HTML]{FFCCC9}$0.511 \pm 0.101$ & \cellcolor[HTML]{FD6864}$0.021 \pm 0.010$  & $0.035 \pm 0.012$                        & \cellcolor[HTML]{FD6864}$-0.141 \pm 0.038$  & $-0.755 \pm 0.079$                         & $0.005 \pm 0.002$                          \\
\multirow{-2}{*}{$\rm NG_F$} & b          & \cellcolor[HTML]{DAE8FC}$2.888 \pm 0.446$ & \cellcolor[HTML]{34CDF9}$-0.192 \pm 0.039$ & $9.928 \pm 0.046$                        & \cellcolor[HTML]{34CDF9}$-0.692 \pm 0.162$  & $3.257 \pm 0.275$                          & \cellcolor[HTML]{DAE8FC}$-0.057 \pm 0.010$ \\ \hline
\end{tabular}
}
\end{table}

In Fig.~\ref{fig:tinfall_relation} panel (a) we see two noticiable trends regarding galaxy properties evolution with $t_{\rm inf}$ in G and NG clusters. First, B galaxies with $t_{\rm inf} \geq 3.8$ Gyr in G clusters are on average 0.34 Gyr older than in NG clusters. Furthermore, examining F galaxies we do observe that $t_{\rm inf} < 4.8$ Gyr galaxies in NG clusters are on average 0.62 Gyr older than in G clusters. Table \ref{table:linear_coeficients} third column shows that in both bright and faint regime the slope of the relation between $Age$ and $t_{\rm inf}$ in NG clusters are more than 1-sigma lower than in G clusters. In panel (b) we see that $[Z/H]$ behave similarly. Despite no significant differences are seen for B galaxies, F galaxies with $t_{\rm inf} < 4.8$ Gyr in NG clusters are 0.047 more metal rich than in G clusters. We also note a difference greater than 2-sigma in the slope for the $[Z/H]$ relation in G and NG clusters. Namely, we find a more flat relation in NG clusters compared to G systems (see Table \ref{table:linear_coeficients} fourth column). In other words, the trends found in panels (a) and (b) suggest that galaxies infalling in NG clusters are older and more metal rich than in G clusters. On the other hand, the excess of B galaxies with lower $Age$ with high infall time in NG clusters may suggest that G clusters have a better defined virialized core, as we will discuss in Section \ref{sec:Discussion}. In panel (c) we do see that $M_{\star}$ is approximately constant with infall time, indistinctly of cluster class and luminosity regime. This translates to slopes lower than 0.1 dex in all cases. In panel (d) we exhibit the SFR, as a function of infall time. SFR behave similar to $[Z/H]$. In the bright regime, we note that galaxies do not show significant differences in SFR in G and NG clusters. However, we find that the SFR slope for G and NG cluster do differ more than 1-sigma. The differences between G and NG clusters members galaxy SFR evolution are highlighted in the faint regime. That is for $t_{\rm inf} < 4.8$ Gyr F galaxies in NG clusters are less star forming by, on average, 0.207 dex than in G clusters. Also we do observe that the slopes between G and NG clusters SFR relation differ by more than 2-sigma. As $Age$ and $[Z/H]$, we find that NG clusters show closer to 0 slopes, which translates to a flatter relation between SFR and infall time. In panel (e) we do observe similarities between the trends of T--Type (morphology) and SFR, which further suggests the relation between star formation quenching and morphological transition. F galaxies with $t_{\rm inf} < 4.8$ Gyr in NG clusters have lower values of T--Type, 0.271 on average, in comparison to G clusters. Finally, in panel (f) we show the relation between $\rm \nabla (g-i)$ and infall time. We do observe significant differences for both B and F galaxies. In the bright regime, we see that galaxies in G clusters show a constantly increasing color gradient with infall time, while in NG clusters we note a kind of plateau after $t_{\rm inf} \sim 2.5$ Gyr. The slopes of these two relations differs more than 1-sigma, as can be seen in Table \ref{table:linear_coeficients}. In the faint regime we note that F galaxies in NG clusters have unequivocally greater color gradients than in G clusters. Namely we find a mean difference of 0.026, favoring NG clusters.

\section{An Estimate of the Infalling Rate in NG Clusters}
\label{sec:infall_rate_estimate}

Several works relate the Non-Gaussianity of the velocity distribution with a higher infall rate in NG cluster compared to G systems. In this section we present further evidence of a higher infall rate in NG systems by detailing the distribution of stellar mass in the PPS. Galaxies in the PNZ have an average infall time of 1.42 Gyr and are mainly first infallers. However, the discretization of the PPS presented by P19 is limited to $R_{200}$ and it is expected that galaxies first infalling in clusters are also found beyond this threshold. Thus we decide to also use in our analysis the Rhee Region A in order to account for galaxies beyond $R_{200}$. In the former work the Rhee Region A is occupied mostly by interlopers. However, it is expected that the Shiftgapper Technique returns a catalog with members only. Hence, we consider that galaxies in the Rhee Region A corresponds to the second most probable population, namely first infallers.

Differently from the previous analysis, here we consider the PPS for each cluster separately. We calculate the sum of stellar mass in the PNZ 8 (or Rhee Region A) for each cluster and then take an average value for a given cluster class (G or NG) and luminosity regime (B or F). In the bright regime, we find that both in PNZ 8 and Rhee Region A NG clusters have an excess of stellar mass of respectively $\rm 0.51 \times 10^{11}M_{\odot}$ and $\rm \times 0.84 \times 10^{12} M_{\odot}$ with respect to G clusters. In the faint regime we also note an excess of $\rm 0.33 \times 10^{11}M_{\odot}$ and $\rm \times 0.31 \times 10^{12} M_{\odot}$ in the PNZ 8 and Rhee Region A of NG clusters in comparison to G clusters. Summing the contributions of B and F galaxies, we find that NG clusters have an excess of $\sim 10^{11} M_{\odot}$ in the PNZ 8 and $\sim 10^{12} M_{\odot}$ in the Rhee Region A. This unequivocally shows that there are more galaxies infalling in NG clusters in comparison to G clusters. 

\begin{figure}[ht]
    \centering
    \includegraphics[width = 0.8\textwidth]{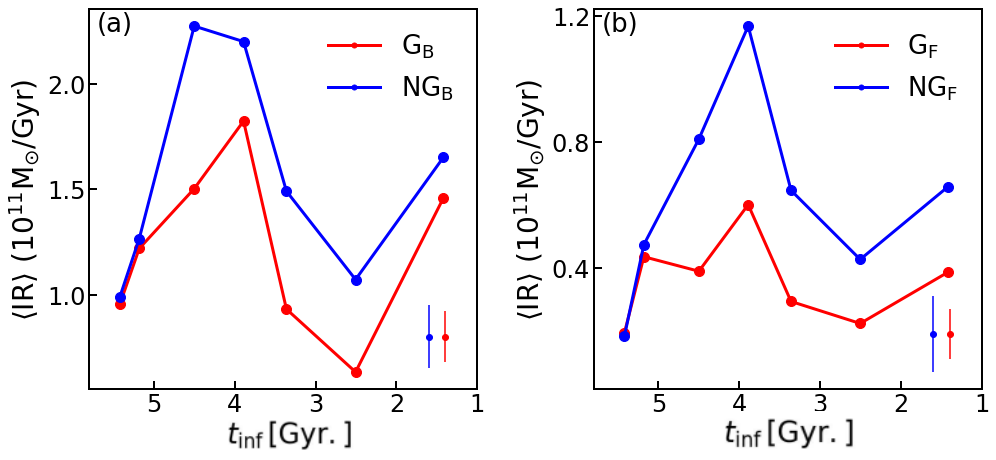}
    \caption{Estimate of mean infall rate of B (left) and F (right) luminosity regime as a function of $t_{\rm inf}$ for G and NG clusters. Non-Gaussian systems show higher values of infall rate in both in comparison to G systems.}
    \label{fig:infall_rate_estimate}
\end{figure}

Using the relation between locus in the PPS and infall time we can derive a rough estimate of the infall rate in NG clusters. We estimate the mean infall rate ($\rm \langle IR \rangle$) as it follows: 1) For each cluster we sum the stellar mass in a single PNZ and divide it by the characteristic infall time of that PNZ; 2) we take the average value for each PNZ for a given cluster class and luminosity regime and 3) then it provides an evolution of the estimate of the infall rate with respect to the infall, namely an infall history for G and NG clusters. The results are shown in Fig.~\ref{fig:infall_rate_estimate}, where panels (a) and (b) correspond to B and F galaxies, respectively. We note that NG clusters unequivocally have a larger infall rate across the infall time. We find a mean difference of $\rm \langle \langle IR \rangle_{NG} - \langle IR \rangle_{G} \rangle = 0.35 \times 10^{11}M_{\odot}$ and $0.26 \times 10^{11}M_{\odot}$ in the B and F regimes, respectively. An integration of the relation between $\rm \langle IR \rangle$ and $\rm t_{inf}$ suggest that NG clusters accreted $\rm (1.49 \pm 0.82) \times 10^{12} M_{\odot}$ more stellar mass in the last 5.42 Gyrs than G systems. For comparison, this mass corresponds roughly to the stellar mass of the local group, $\rm \sim 10^{12} M_{\odot}$.

A possible problem comes from the different virial mass distribution from G and NG clusters, as shown in panel (a) of Fig.~\ref{fig:concentration}. In order to guarantee that NG clusters have a higher infall rate indistinctly of their mass, we separate our sample in bins of $\log(M_{200}/{\rm M}_{\odot})$ from 14 to 14.75 in steps of 0.25. This range is chosen due to a limitation for G clusters faint component. Namely, we do not find G clusters with z<0.04 and $\log(M_{200}/{\rm M}_{\odot} > 14.75$. Thus this range guarantees that we have G and NG clusters in every bin for both luminosity regimes. This analysis show that in the three bins NG clusters have accreted more mass in the last 5.42 in comparison to G clusters. Taking the average over the three virial mass bins we find that NG clusters accreted an excess of $(1.36 \pm 0.78) \times 10^{11} M_{\odot}$ and $(0.91 \pm 0.51) \times 10^{11} M_{\odot}$ in the B and F regimes, respectively. These trends show that most of the difference are found indeed in the tail end of the halo distribution, but do confirm that the NG classification is directly related to a higher infall rate independent on the virial mass.

\section{Chapter Summary}
\label{sec:Discussion}

In the last decades several contributions indicate that environment plays a major role in galaxy evolution. The morphology-density relation \citep{Dressler} marks the starting point to the study environmental effects in galaxy evolution. Galaxy transition from low to high density environments is usually characterized by a quenching in star formation and can be studied in systems covering different orders of magnitude in galactic density, such as galaxy clusters. Cluster's outskirts provide densities comparable to low density fields, while higher densities are found towards the cluster core. 

However, clusters are not unique and provide a variety of situations to study galaxy evolution. A detailed characterization of the cluster environment hence is crucial to understand how it affects galaxy members evolution. Density profiles, the presence of a cool-core and, more recently, the dynamical state of the cluster are examples of methods employed to characterize such systems. The later follows from several works that suggest that galaxy clusters can be modeled by two components: 1) a virialized dominated by old galaxies; plus 2) a second one in quasi-equilibrium, containing mainly young galaxies (e.g. \citealt{1997ApJ...476L...7C,2001ApJ...547..609E}). The second is of particular interest to study galaxy evolution, as it likely results from galaxy accretion through filaments and show distinct features from those virialized. Furthermore, cosmological N-body simulations suggest that a Maxwell-Boltzmann velocity distribution translates to a Gaussian profile when considered the members galaxy projected velocity distribution. Differences of clusters with a Gaussian projected velocity profile (G Clusters) and clusters with Non-Gaussian velocity profiles (NG Clusters) has been the focus of several studies \citep{2010MNRAS.409L.124R,2013MNRAS.434..784R,2014AAS...22335806C,2017AJ....154...96D,2019MNRAS.490..773R,2019MNRAS.487L..86D}. In this work we study a sample of  143 G and 34 NG Clusters defined in dC17 using a more informative space than the velocity distribution alone, that is the projected phase space. We discretize the PPS in order to separate galaxies according to their $t_{\rm inf}$ to investigate further differences in G and NG Clusters. 

\subsection{G and NG Clusters: Different Scenarios to Study Galaxy Evolution}

Comparison of G and NG clusters indicate that these two groups have significant differences in both structure and galaxy properties distribution. We can summarize the differences we find in the following way:

\begin{itemize}
    \item NG Clusters are more massive and sparse, less concentrated and show an excess of fainter galaxies in comparison to G clusters (Fig.~\ref{fig:concentration});

    \item Fig.~\ref{fig:PPS_Density} highlights that higher density regions in the PPS of NG Clusters extends to higher velocities in comparison to G Clusters, a possible signature of an excess of high velocity infalling galaxies;

    \item Galaxy properties distribution is systematically more mixed in the PPS of NG clusters than in G clusters (Figs.~\ref{fig:full_grid} and \ref{fig:grid_mstellar}).  We highlight that this is true for both luminosity regimes, but more noticeable for F galaxies;

    \item Fig.~\ref{fig:infall_rate_estimate} indicate that G and NG clusters has a different accretion history in the last $\sim 5$ Gyr. Our findings suggest that NG clusters have accreted on average $\rm \sim 10^{12} M_{\odot}$ more stellar mass than G systems. The trend of higher accretion rate in NG clusters is true also when comparing systems with similar $ M_{200}$;
    
\end{itemize}

These results suggest that G and NG clusters provide two different scenarios to study galaxy evolution and environmental quenching. N-body simulations show that more massive clusters show a higher fusion rate \citep{2009ApJ...701.2002G}. In other words, massive clusters are usually formed by the interaction of two groups and/or clusters. Our sample is built upon massive clusters and hence it is expected that a large fraction of them had experienced such fusion events in the past. It is expected a higher fraction of fusion events in NG clusters due to its excess of massive systems in comparison to G groups. Fusion events are perturbations in the dynamical equilibrium of a cluster. From the PPS point of view this translates to a more mixed distribution of galaxy properties in comparison to the unperturbed state. Our results points to NG clusters having more mixed distributions in the PPS. Finally, \cite{2019MNRAS.490..773R} show that NG cluster suffered their last major merger more recently than G clusters. These trends show that NG clusters are statistically in a more unrelaxed state in comparison to G clusters and thus provide a different scenario to study galaxy evolution.

Previous works indicate that NG clusters also show a higher infall rate. Here we quantify this infall rate by tracing the stellar mass in different regions of the PPS and estimate that in the last 5 Gyr NG clusters accreted roughly a massive group in comparison to G systems. This infall rate is mostly due the accretion of low luminosity galaxies (FG), consistent with works that assign the non-gaussianity in the projected velocity distribution as a result of an accretion of low luminosity galaxies (e.g. \citealt{2017AJ....154...96D,2019MNRAS.490..773R}). Examining Table \ref{table:linear_coeficients} we note that the relation between galaxy properties and $\rm t_{inf}$ in NG clusters show lower gradients (i.e. slopes closer to zero) in comparison to G clusters. These gradients reflect a more mixed galactic population in NG systems. However, we highlight that these results are found by considering large samples of clusters to guarantee that the effects of projection along the line of sight influence both cluster classes in the same way. \cite{2018MNRAS.473L..31C} study the galaxy velocity dispersion and indicates that B and F galaxies are more separated in G clusters. It follows from the mass-segregation phenomenon in clusters \citep{1980ApJ...241..521C} that relaxed systems show a clear separation between high and low mass galaxy properties, hence a more mixed population of bright and faint galaxies provide further support for NG clusters being in a more unrelaxed state in comparison to G systems. Here we also provide further evidence that the non-gaussianity in the projected velocity distribution is connected to a higher infall rate of faint galaxies, in agreement with \cite{2019MNRAS.490..773R}. However, a more detailed characterization of one-by-one cluster dynamical state should be given by combining further dynamic probes such as caustic curves analysis \citep{gifford2013systematic,2012ApJ...747L..42D}, X-Ray characterization \citep{2001A&A...378..408S}, velocity dispersion profiles \citep{2019MNRAS.482.5138B} and galaxy spatial distribution \citep{2006A&A...450....9F}.

\subsection{Low \emph{vs.} High Mass Galaxies Quenching in Dense Environments}

One of the primary results of this work is the unequivocal relations between  
G and NG clusters and galaxy members properties, shown in Fig.~\ref{fig:tinfall_relation}. In this regard, we highlight:

    \begin{itemize}
        \item B galaxies $Age$, $[Z/H]$, SFR and T--Type show slopes closer to zero in comparison to F galaxies, indistinctly in G or NG clusters. These trends evidence how B and F galaxies are differently affected by their environment. Namely, B galaxies are more massive and hence less affected by effects such as RPS;
        
        \item $Age$ and $[Z/H]$ have very similar relations with infall time, which reinforces the closeness between these two parameters. $Age$ here means the time since the last star formation episode and thus higher $Age$ translates to more time to stars evolve and increase the metallicity of the ISM;
        
        \item The similarity between the relations of SFR and T--Type strongly suggest that quenching of star formation and morphological transformations may be causally connected;

        \item The slope for the relation regarding FG is consistently lower in NG clusters in comparison to the case in G systems, for most of the six-parameter set here adopted to characterize member galaxy properties;
    
        \item F Galaxies with time since infall lower than 4.5 Gyr in NG  clusters are younger, more metal poor and show higher star formation than it counterpart in G systems;

    \end{itemize}

These relation are of particular interest, since the quenching time scale is pivotal to understand environmental effects in galaxy evolution \citep{2016ApJ...816L..25P}. Galaxies moving within and/or towards the cluster are quenched after they reach a threshold density, which enables ram pressure stripping to remove the gas component of low mass galaxies in a short time scale ($\sim$ 3 Gyr) and quickly halt star formation (R19). However, the scenario for high mass galaxies seems to be quite different. The flatter relations between $[Z/H]$, SFR or T--Type and $t_{\rm inf}$ for B galaxies in comparison to F ones indicate that massive (and hence more luminous) galaxies are less affected by the environment, indistinctly in G or NG clusters. This result is in agreement with massive galaxies being quenched mostly due to internal mechanisms such as AGN and stellar feedback \citep{1974MNRAS.169..229L}. Galaxy star formation quenching seems to be related also to a morphological transformation \citep{2019MNRAS.486..868K}. The trends found in Fig.~\ref{fig:tinfall_relation} thus support the close SFR - morphology relation and extends it also to $Age$ and $[Z/H]$, providing further insight in how stellar population parameters reflect galaxy evolution. An interesting result comes from the almost constant relation between infall time and $M_{\star}$, indicating that the conversion efficiency of gas into stars do not depend on the galaxy environment, but mainly on the availability of gas content \citep{2013ApJ...776...71C}.

Additionally, in the faint regime we also note the more striking differences between G and NG clusters, again suggesting that bright galaxies are less affected by their environment. We find that galaxies with $t_{\rm inf} < 4.5$ Gyr in NG systems are older and more metal rich than its counterpart in G systems. Recent infalling galaxies with higher $Age$ and $[Z/H]$ than the low density field is an evidence of pre-processed galaxies (P19). Our result hence points to an excess of such systems in NG clusters. This is in agreement with recent works that suggest that the excess of faint galaxies in NG systems is caused by a time dependent accretion of pre-processed galaxies. This is reassured by the results found for SFR, T--Type and color gradient, namely we find galaxies with lower SFR and T--Type in NG systems. Also, faint galaxies in NG clusters show higher color gradients (e.g. redder on the outskirts) than faint galaxies in comparison to the same population in G systems. This later possibly indicates that galaxies entering in NG clusters already started a morphological transformation towards elliptical shapes. Complementary to the results here presented, \cite{2019MNRAS.487L..86D} show that faint spiral galaxies are the ones that suffer major environmental effects.

%% file: Chapters/Chapter8.tex
\chapter{Future perspectives: aimming at higher redshifts}

With the recent advent of the James Webb Space Telescope (JWST), the universe at high redshift became the focus of astrophysical research. Alongside current available high redshift observations, new facilities will be built to provide an even more detailed view of the early-universe. Among the different future instruments, the Multi-Object Spectrograph for Astrophysics, Intergalactic medium (IGM), and Cosmology (MOSAIC) will potentially be the one coupled to a Extremely Large Telescope (ELT) that will provide the deepest and most complete insight regarding the physics of the first galaxies. In terms of science products, the MOSAIC is designed to provide high quality spectra for galaxies spamming a wide range of redshift. Moreover, the instrument will also provide integral-field-unit observations (IFU) mode, which will be optimized to investigate the galaxies during the reionization epoch and serve as the most efficient follow-up of the JWST. Still, MOSAIC will only be ready to start collecting data ~10 years from now. Therefore, we propose this project as a series of analyses that will act as a relevant framework to the scientific goals of the MOSAIC project.

\thispagestyle{empty}

\noindent

\section{Morphological classification towards higher redshift (>3) with JWST}
\label{sec:JWST_morphology}
\begin{figure}[ht]
    \centering
    \includegraphics[width = \textwidth]{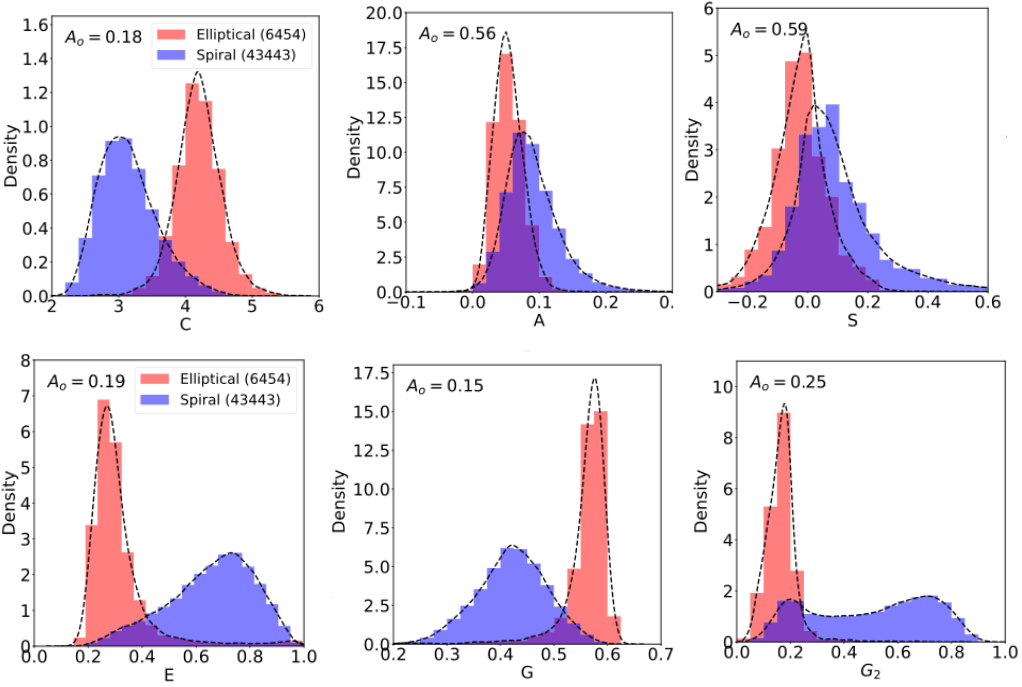}
    \caption{Distribution of the CAS (top row) and EGG (bottom row) indexes for a sample of approximately 50,000 galaxies with morphology defined from the Galaxy Zoo 1. Blue and red histograms show the distribution for elliptical and spiral galaxies, respectively. The $A_0$ parameter on the top left of each panel shows the overlap between the two different classes. Comparison shows that the EGG non-parametric system is better in discretizing between disk and bulge dominated systems. Adapted from \cite{2024MNRAS.528...82K}.}
    \label{fig:cas_vs_egg}
\end{figure}

In previous works, we investigate different methods to distinguish between disk and spheroid galaxies. Starting with galaxies in the local universe,  we compared the Concentration + Asymmetry + Smoothness (CAS) system, introduced by \cite{2003ApJS..147....1C}, with our proposed Entropy + Gini Index + Gradient Field Asymmetry (EGG) system, as described by \cite{2024MNRAS.528...82K}. Fig.~\ref{fig:cas_vs_egg} shows the comparison using bona fide samples of elliptical and spiral galaxies from Galaxy Zoo 1 \cite{2011MNRAS.410..166L}. The EGG system proved more accurate in distinguishing between bulge- and disk-dominated galaxies.

Using the EGG metrics as input, we thus defined a hybrid method combining supervised and unsupervised deep learning. In a few words, the metrics are used Using as input for a Self-Organizing Map \cite[SOM,][]{kohonen1991self} clustering algorithm, we defined training samples with high purity. This sample was then used with an unsupervised CNN algorithm to classify disk and spheroidal galaxies, achieving a 95\% accuracy rate.

\begin{figure}[ht]
    \centering
    \includegraphics[width = 0.8\textwidth]{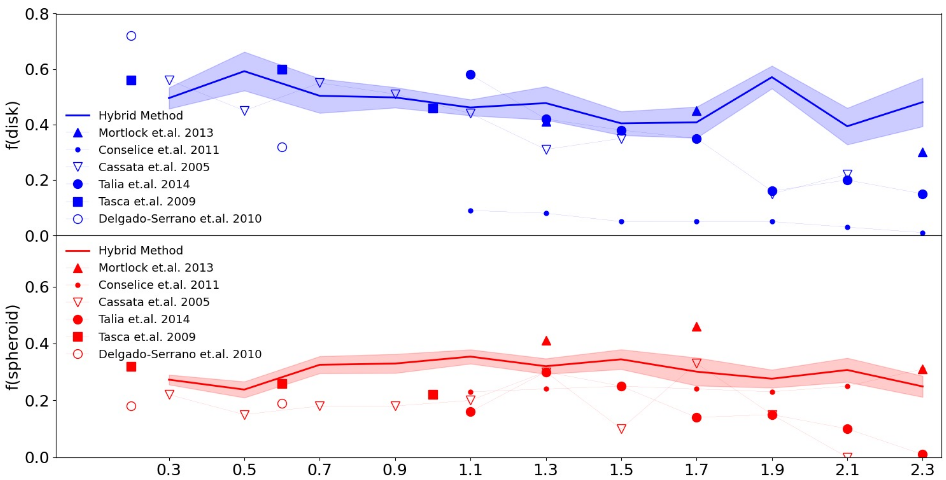}
    \caption{Fraction of disk (top panel) and spheroid (bottom panel) galaxies as a function of redshift. The solid line and shaded area denote our estimate, whereas the different points show the estimates from different works (shown in the Legend). Adapted from Kolenikov et al.(submitted).}
    \label{fig:disk_spheroid_fraction}
\end{figure}

We then extended our hybrid method to higher redshifts, applying it to galaxies in The Cosmic Assembly Near-infrared Deep Extragalactic Legacy Survey (CANDELS). This survey covers five fields: COSMOS, UDS, EGS, and GOODS-N/S, observed in multi-wavelength modes using the Hubble Space Telescope's WFC3 and ACS for optical and infrared wavelengths. We analyzed the morphology of CANDELS galaxies in the redshift range $0.2 \leq z \leq 2.4$, with H-band apparent magnitudes brighter than 24 and stellar masses greater than $10^{9} \mathrm{M}_{\odot}$, using stellar mass data from \citep{2015ApJ...801...97S}. Unlike other methods, such as \citep{2024ApJ...962..164T}, we fine-tuned the CNN algorithm across different redshift bins, enhancing the reliability and accuracy of our disk/spheroid classification. Fig.~\ref{fig:disk_spheroid_fraction} shows our results for the fraction of disk and spheroid galaxies as a function of redshift, with the solid line and shaded area representing the median and 1-sigma scatter of our estimates, and different symbols showing estimates from previous studies.

One primary goals is to extend the morphological classification of galaxies to higher redshifts. Building on the success of the hybrid method in the CANDELS fields, we plan to refine and apply this approach using data from the Cosmic Evolution Early Release Science (CEERS) survey. CEERS will cover 100 square-arcminutes of the EGS field with JWST's Near Infrared Camera (NIRCAm). Although CEERS has a smaller coverage area compared to CANDELS, its higher depth (around 29 apparent magnitude for 5-sigma detections) will ensure a sufficiently large sample for reliable statistical analysis. These observations will allow an extension of the analysis to redshifts up to $z \sim 8$, approaching the era of the first galaxies.

To adapt our EGG system for these higher redshifts, we will incorporate additional parameters that account for the redshift-induced shifting of spectral features into the infrared and the general dimming of galaxies at these distances, such as morphological k-correction. This will refine our machine learning models, particularly the unsupervised CNN, making them sensitive to the subtler and more varied galaxy features prevalent in the early universe.

\section{The bimodality in spheroid galaxies specific star formation rate}

\begin{figure}[ht]
    \centering
    \includegraphics[width = \textwidth]{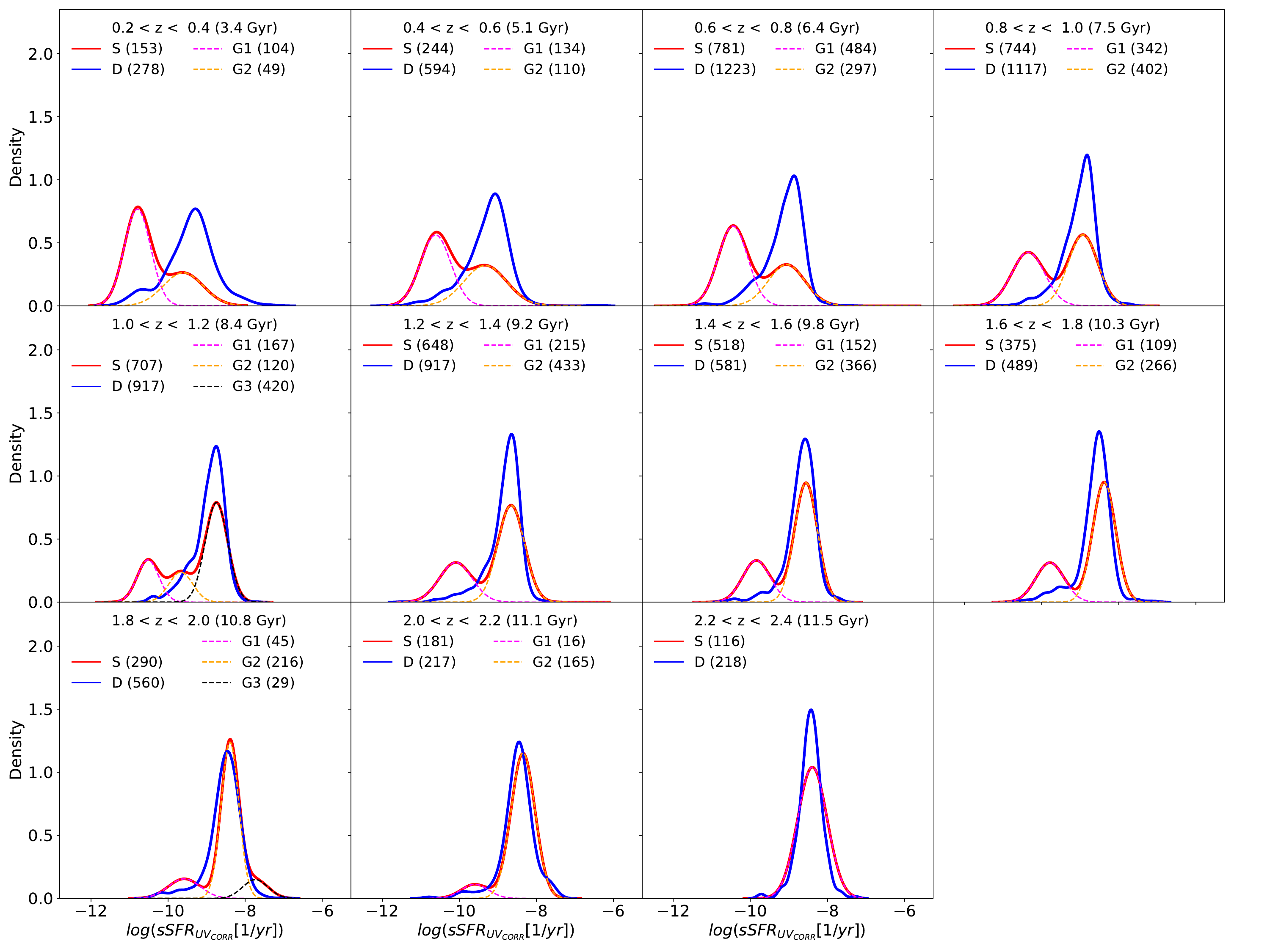}
    \caption{sSFR distribution for disk (blue) and spheroid (red) galaxies, separated into different redshift bins. We show as pink and orange dashed lines the two components of the spheroid bimodality. Made by the author.}
    \label{fig:ssfr_bimodality}
\end{figure}

A secondary preliminary result of our morphological analysis of CANDELS galaxies is the notable bimodality in the distribution of specific star formation rate ($\rm sSFR = SFR/M_{\star}$) for spheroid galaxies. Fig.~\ref{fig:ssfr_bimodality} illustrates the distribution of sSFR for CANDELS galaxies classified as disk (in blue) and spheroid (in red) across different redshift bins. The SFR and $\rm M_{\star}$ are estimated using methods described in \citep{2019ApJS..243...22B} and \citep{2015ApJ...801...97S}, respectively, based on ultraviolet (UV) observations. Notably, there is a redshift-dependent bimodality in the distribution for spheroids. At high redshift ($z \sim 2.2$), there is no significant difference between the specific star formation distribution for disks and spheroids. As redshift decreases, we observe an increasingly prominent second component of spheroids with low sSFR. In a preliminary approach, we applied a Gaussian mixture model to the spheroid galaxy distribution, which reinforces this bimodal characteristic.

\begin{figure}[ht]
    \centering
    \includegraphics[width = \textwidth]{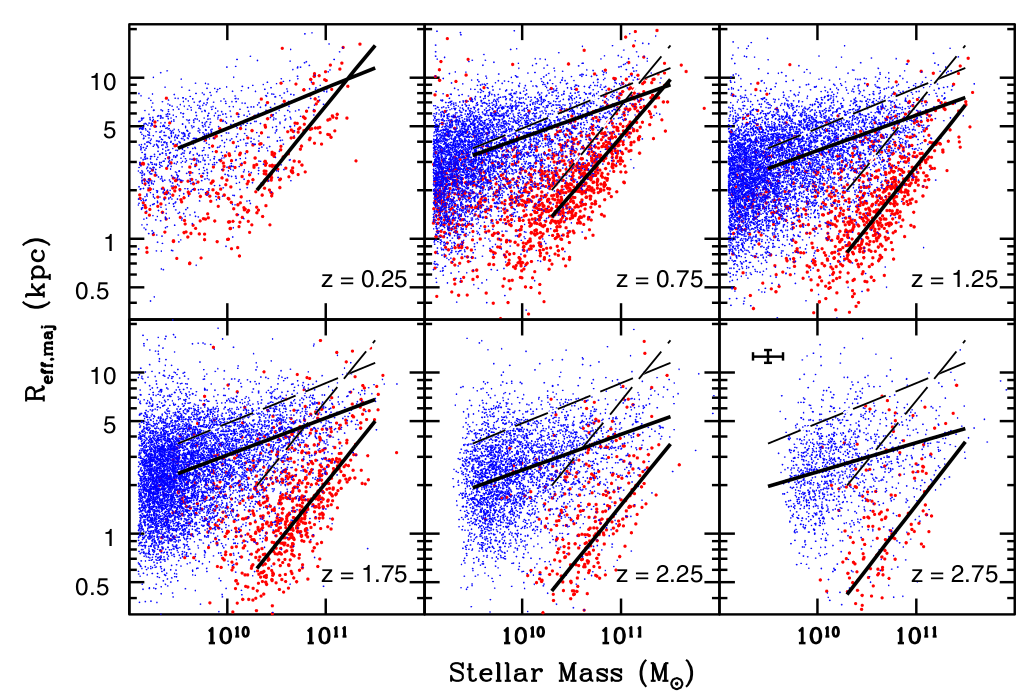}
    \caption{Size–stellar mass distribution of late-, in blue, and early-, in red, type galaxies, separated into different redshift bins. The solid lines shows the best linear fit for each redshift, whereas the dashed blas lines show the best linear fit for the z<0.25 universe.}
    \label{fig:size_mass_relation}
\end{figure}

We thus plan to investigate both spheroidal components and contrast them with the expectations from the inside-out galaxy growth model. As illustrated in Fig.~\ref{fig:size_mass_relation}, early-type (in red) and late-type (in blue) galaxies exhibit distinct mass-size relations. Additionally, there is strong evolution in the intercept of the size-mass relation for early-type galaxies, while moderate evolution is observed for late-type galaxies.

\section{How to trace ex-situ stellar component?}

Also related to ex-situ components, simulations offer predictions about the connection between the fraction of ex-situ stellar components and various galaxy properties. One key difference between in-situ and ex-situ stars is that ex-situ stars typically have distinct chemical enrichment profiles. Moreover, ex-situ stars often display different orbital characteristics compared to their in-situ counterparts.

\begin{figure}[ht]
    \centering
    \includegraphics[width = \textwidth]{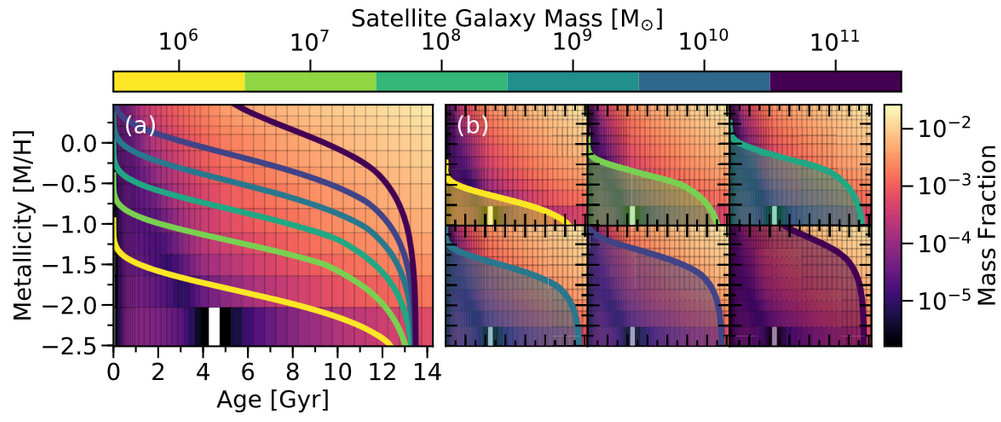}
    \caption{The relation between location in the age-metallicity grid and (possible) accreted satellite galaxy mass. Adapted from \cite{2020MNRAS.491..823B}.}
    \label{fig:exsitu_grid}
\end{figure}

In our initial approach, we will characterize the fraction of ex-situ components using spectrum fitting. This is based on the expectation that galaxies evolving in isolation should exhibit a tight correlation between stellar mass and stellar metallicity. Deviations from this expected relation can indicate the presence of ex-situ stellar components. \ref{fig:exsitu_grid} illustrates the relationship between position in the age-metallicity grid and the potential mass of accreted satellite galaxies, derived using the EAGLE simulation as presented in \citep{2020MNRAS.491..823B}. We will characterize the ex-situ fraction of galaxies from the Large Early Galaxy Astrophysics Census (LEGA-C, \cite{2021ApJS..256...44V}). Our criteria for selection are as follows: 1) galaxies must have $\rm S/N \geq 20$ to ensure robust spectrum fitting; 2) we restrict our analysis to the redshift range $0.6 \leq z \leq 0.8$ to ensure that all three emission lines ([OIII]$\lambda 5007$, [OII]$\lambda 3727$, and H$\beta$), necessary for estimating gas-phase metallicity following the \cite{2004ApJ...617..240K} method, fall within the LEGA-C coverage; and 3) we require reliable measurements of emission line properties, ensuring a line $\rm S/N > 5$ and an equivalent width of $\rm H \beta$ greater than 2\AA. This results in a sample of 152 galaxies at a redshift of approximately 0.7.

A more detailed analysis involves tracing the spatial distribution of stellar population parameters using integral field spectroscopy (IFU). We plan to submit observation proposals to the Large Binocular Telescope (LBT, \cite{2010ApOpt..49D.115H}) and the Multi-Espectrógrafo en GTC de Alta Resolución para Astronomía (MEGARA, \cite{2014SPIE.9147E..0OG}) to obtain IFU observations of galaxies at $z \sim 0.7$. The focus of these observations will be on characterizing the spatial distribution of stellar metallicity. A significant fraction of ex-situ stellar components can imprint noticeable gradients in the stellar metallicity of galaxies. This approach is particularly relevant for the upcoming MOSAIC survey, which will allow extending such analysis to even higher redshifts ($z \sim 2.5 - 3$).

%% file: Chapters/Chapter9.tex
\chapter{Post script - experiences during the PhD}

This chapter is, to some extent, an extra -- and informal -- part, in which I describe my personal/professional experience during the PhD. Completing a PhD is a significant journey, and reaching the point of submission comes with a sense of relief. However, the end of one phase inevitably marks the start of another. Transitioning from the focused work of writing a thesis to navigating job applications, planning publications, and preparing for conferences comes always of some uncertainty in the future, but constantly working towards the next goal. The PhD years have shaped me significantly, not just as a scientist but also in a broader sense, fostering greater maturity and resilience in all aspects of life. 

First, at the end of my Master, the decision of continuing my research with Prof Dr. Reinaldo R. de Carvalho, in detriment of doing my PhD in a more renowned institution, is probably one of the best decisions I've made so far, especially due to the amount of knowledge, critical thinking and insights I've gathered along the way. In the end, my work speaks for itself. The first paper I've co-authored, \cite{2020MNRAS.497.3251W}, came out right before the start of my PhD, and it showed me how is the publication process and the specific way to write in scientific papers. Scientific writing is particularly hard for Portuguese speakers, as we follow ``why write in a sentence, when you could use a whole paragraph'', since the language is overwhelmingly verbose. This was a problem when writing my first first-author paper, \cite{2021MNRAS.503.3065S}, which ironically is the last one presented in this thesis. In addition, it was the first time I presented a finished project in the form of a paper to other researchers/professors. In particular, Prof. Dr. I. Ferreras was present at that time, and he was through all my journey, always giving me good advices and comments on how I could improve both in specific questions and in overall as a researcher. I'm thankful for that. In the first PhD year (mid 2020 to mid 2021), I've also completed all the necessary classes and applied for a FAPESP grant, which I was awarded, thus becoming one of the 7 students in São Paulo with such a scholarship to do research in astrophysics.

After my first year in the PhD, I had no more classes and could focus mainly on my research. I've co-authored two papers, \cite{2022MNRAS.509.3889D} and \cite{2022MNRAS.509.3470M}, which was a sign to me that people were interest in my work and potential. Still, my highlight from that year was my second publication as first author, \cite{2022MNRAS.509..567S}, which started as a simple question: ``when infalling into clusters, what changes faster in galaxies? Morphology or (specific) star formation rate?'', and in three months I had everything done and ready to submit -- I really delved into it. By far, this is the paper I've written the fastest, so far. I believe the second first-author paper was a turning point, as we decided to invite two renowned international researchers (besides Prof. Dr. I. Ferreras), Prof. Dr. L. C. Parker and Prof. Dr. A. Aragón-Salamanca, to collaborate, and both agreed. This was an unique chance to show my work to a broader international audience. 

The collaboration with Prof. Dr. A. Aragón-Salamanca turned into a 1-year stay at the University of Nottingham with him and Prof. Dr. M. R. Merrifield as my supervisors, since I awarded another scholarship from FAPESP to do a doctoral stay. During this period, I've experience a different academic environment, and a lot happened. I joined the WEAVE collaboration during my stay, I gave talks on important universities such as University of Nottingham, Oxford, Harvard, etc., and learned a lot from interacting with both my supervisors. Not only with my supervisors, but I also interacted a lot with Dr. S. Zhou, a postdoc there at that time. In terms of scientific production, I took part in three papers, \cite{2023MNRAS.519.4884F}, \cite{2023MNRAS.524.5327S}, and \cite{2024MNRAS.527.1935Z}. Not only scientifically, but the experience of having to live in a different country on my own for a whole year, made me gather important skills, in particular when dealing with people. Of course, I also enjoyed my stay to travel a bit. 

Since my return to Brazil, I'm more self-confident and this reflected on the way I behave in seminars, conferences, or even at the institute. Facing a different reality here in Brazil, I tried to implement two different activities in the university: 1) a journal club; and 2) a coding club. The former is usual in European universities, but it seems that every time it is implemented here in Brazil, irrespective of the institution, it slowly dies. The second follows from a noticeable difficulty people have in dealing with programming. Unfortunately, just as in the past, both slowly died (but people still ask me for help in programming). After my return, I was happy to know that my previous work motivated a paper, \cite{2024MNRAS.529.3651O}, in which I'm also a co-author. Following a similar idea, I extended my work comparing star formation quenching and morphological transition to a more quantitative perspective, leading to the publication of \cite{2024MNRAS.532..982S}. 

In the last bit of my PhD, we changed our focus. Despite most of my work have been done using local universe data, with the advent of JWST the focus became predominantly at higher redshifts. This impacted a lot my postdoc applications, in which 80-90\% was to deal with JWST and, even applying for 30-40 places, I wasn't selected for a single one. It shook me a little (lie, it shook me a lot), but I kept trying to do my best. Although may sound weird, even being a PhD student, I became the monitor of the ``introduction to cosmology'' discipline, in which I solved exercises and questions from other PhD students. I also started co-supervising a undergraduate research student. All this put me back on feet, and restored my self confidence. In the end, we published two (in my opinion) amazing works exploring the morphology of galaxies in the local and high redshift universe, \cite{kolesnikov2023unveiling} and \cite{2024arXiv241203778K}, and here I should also highlight the wonderful computational work done by Dr. I. Kolesnikov\footnote{Fun fact: once I. Kolesnikov directly said to me ``I know it works, but I hate the way you code''.}. The latter is still waiting for a reviewer, but it is already submitted to arxiv. I'm also finishing a related paper in which, by exploring the properties of disk and spheroidal galaxies as function of cosmic time, we show the existence of different pathways for the evolution of spheroidal galaxies (hopefully in arxiv soon!). 

Looking back, every interaction I had to other person, and not only researchers but also family, friends, colleagues, even unknowns, shaped my way of thinking. I'm proud of who I've become, and, more than that, I thankful to all that helped me through this journey. I guarantee that I will strictly follow the message from Ludwig Edward Boltzmann, shown at the beginning of this thesis: Bring forward what is true, write it so that it is clear, defend it to your last breath\footnote{Let's face it, no need to be that extreme, it's okay -- and even helpful -- to be wrong sometimes.}!

%% file: referencias.tex
\newpage
\bigskip

\renewcommand{\refname}{References}
\makeatletter
\renewcommand{\bibsection}{%
   \section*{\refname%
            \@mkboth{\MakeUppercase{\refname}}{\MakeUppercase{\refname}}%
   }
}
\makeatother

\bibliography{references}{}

\newpage

%% file: Chapters/Appendix.tex
\chapter{A brief description of the cluster dynamical property estimates}
\label{Appendix_dynamics}
In this Appendix we briefly describe how virial mass, virial radius and velocity dispersion are estimated for each cluster. The velocity dispersion is estimated using the shiftgapper output final members list. Depending on the number of member galaxies, the cluster velocity dispersion estimate is then derived using gapper ($N_{\rm galaxies} < 15$) or biweight (otherwise, \citealt{1990AJ....100...32B}) estimators, which is then corrected for velocity errors following \citep{1980A&A....82..322D} and resulting in a first estimate of the ``projected virial radius'' ($R_{\rm PV}$). This radius is then used to derive a first estimate of virial mass, \citep{1998ApJ...505...74G}:
\begin{equation}
    M_{\rm V} \sim \frac{3 \pi \sigma_{\rm LOS}^{2} R_{\rm PV}}{2G},
\end{equation}
where $3\pi / 2 $ is the deprojection factor and $G$ is gravitational constant.
After, we apply a correction in the mass estimate due to the surface pressure term, for which we assume a Navarro-Frenk-White dark matter profile \citep[NFW]{1997ApJ...490..493N} with concentration given by
\begin{equation}
    c_{\rm NFW} = 4 \times \left(\frac{M}{M_{\rm KBM}}\right)^{-0.102},
\end{equation}
where the slope and normalization ($M_{\rm KBM}$) are taken from \cite{2004A&A...416..853D} and \cite{2004ApJ...600..657K}, respectively. $R_{200}$ is then through the equation \citep{1997ApJ...478..462C}
\begin{equation}
    R_{200} = \frac{\sqrt{3}\sigma_{\rm LOS}}{10 H(z)}.
\end{equation}
Next, it is necessary to apply C-Correction \citep{1998ApJ...505...74G} in order debias the virial mass estimate due to the concentration of extended objects, such as clusters, which is given by
\begin{equation}
     C_{\rm correction} = M_{\rm V} 4\pi R_{\rm PV}^{3} \frac{\rho (R_{\rm PV})}{\int_{0}^{R_{\rm PV}}} 4\pi r^{2} \rho dr \left[ \frac{\sigma_{\rm LOS}(R_{\rm PV})}{\sigma_{\rm LOS}(r < R_{\rm PV})} \right]^{2}
\end{equation}. 
Therefore, if $M_{\rm V}$ is the virial mass after correcting for the surface pressure term in a volume of radius $R_{\rm A}$, $R_{200}$ can be written as
\begin{equation}
    R_{200} = R_{\rm A} \left [  \frac{\rho_{\rm V}}{200 \rho_{\rm cd}(z)} \right]^{1/2.4},
\end{equation}
where $\rm \rho_{\rm V} = 3M_{\rm V}/(4\pi R_{\rm A}^{3})$ and $\rho_{\rm cd}(z)$ is the critical density at a given redshift z. Finally, $M_{200}$ is calculated by interpolating the the virial mass from $R_{\rm A}$ to $R_{200}$. 

\chapter{Comparing the activity of central galaxies using alternative diagnostic diagrams}

\begin{figure}[ht]
    \centering
    \includegraphics[width = \textwidth]{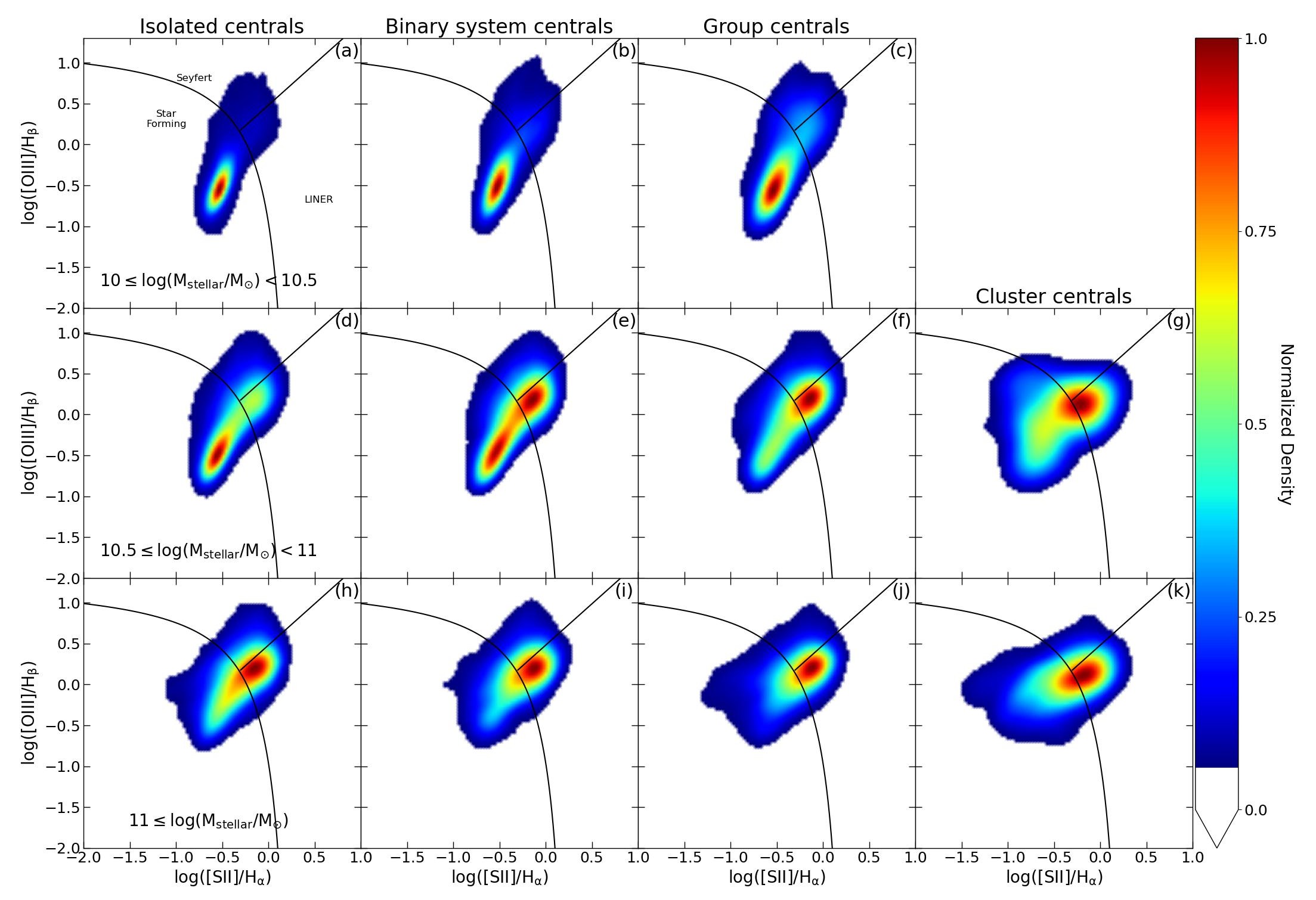}
    \caption{Gaussian kernel smoothed distribution of galaxies in the NII-BPT diagram. We divide systems according to stellar mass and environmental richness.}
    \label{fig:BPT_SII}
\end{figure}

The main underlying hypothesis of \ref{chapter: central_galaxies} is that the BPT diagram we have chosen is able to reliably identify the main mechanism ionizing the insterstellar gas in central galaxies. However, the literature provides alternative diagnostic diagrams to achieve the same goal using a variety of emission-line ratios. Since the question ``which diagnostic diagram is the best for this purpose?'' doesn't have a simple answer (see \citealt{2010MNRAS.403.1036C} for a comprehensive comparative study of several diagnostic diagrams), we have tested whether our findings depend on the specific diagnostic diagram we choose. As an example, we show in Fig.~\ref{fig:BPT_SII} the distribution of galaxies in an alternative BPT diagram that uses the [SII]/H$\alpha$ ratio instead of [NII]/H$\alpha$, as we did in Fig.~\ref{fig:BPT_Smooth}. Comparing both figures we find that the galaxy distribution is similar using the alternative diagnostic. Therefore, we are confident that our results do not depend significantly on the specific diagnostic diagram adopted.

\chapter{An alternative way to slice the PPS of G and NG clusters}
\begin{figure}[ht]
    \centering
    \includegraphics[width = \textwidth]{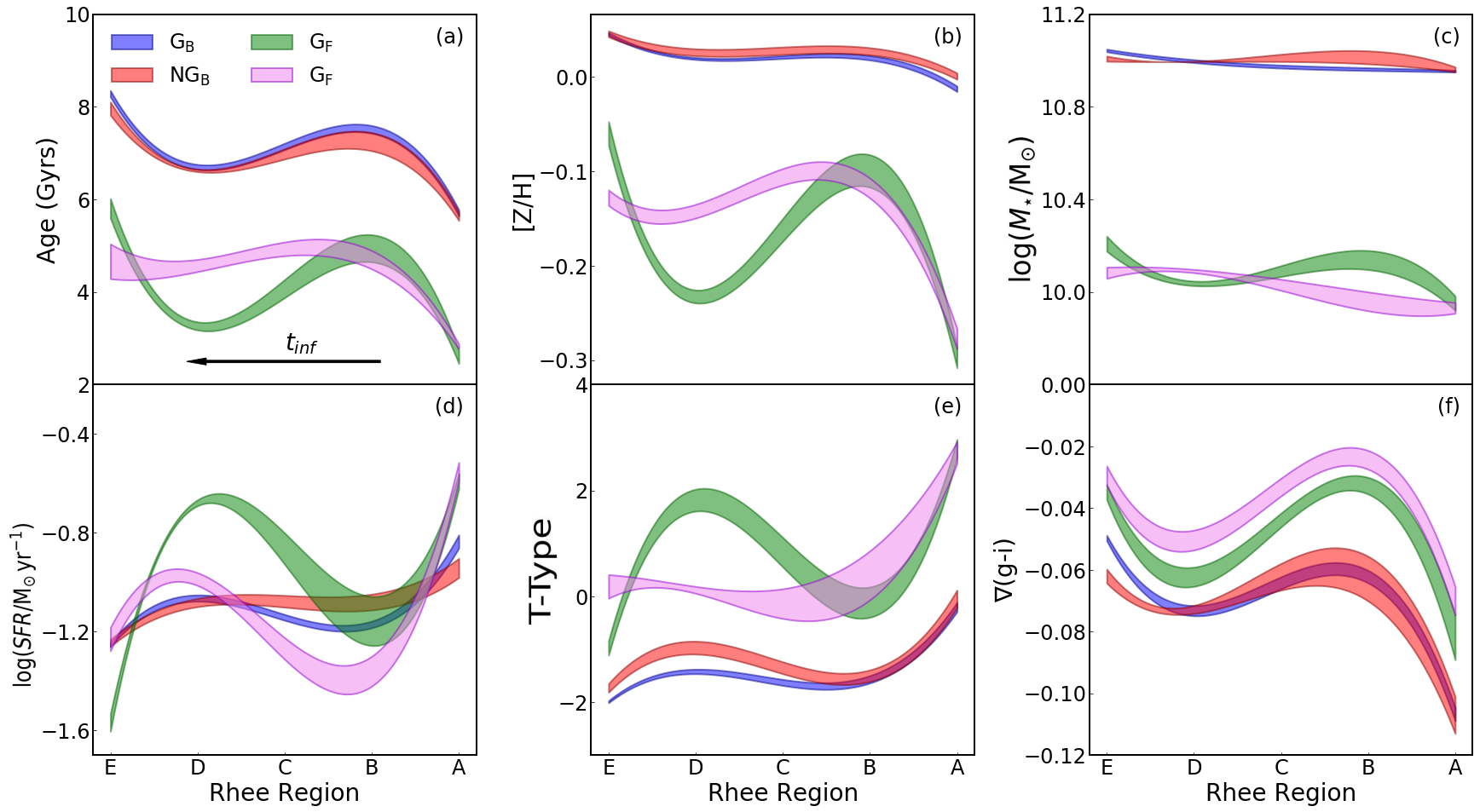}
    \caption{Same as Figure \ref{fig:tinfall_relation}, but using Rhee Regions in order to connect $t_{inf}$ and locus in the PPS.}
    \label{fig:rhee_relation}
\end{figure}

There are different ways to separate galaxies in the projected phase space. The results described in \ref{chapter: G_vs_NG} lay on slicing the PPS in PNZs, which divide galaxies according to their infall time. This choice, however, is not unique. PNZs are build to be valid until $R_{200}$. We then adopted the Rhee Regions as a second way of dividing the PPS tracing galaxies infall time. This second way of discretizing the PPS is based on a probabilistic approach and hence here we consider that the PNZs provide a more straightforward relation between infall time and galaxy properties. In Figure \ref{fig:rhee_relation} we show the relation between time since infall and galaxy properties when using Rhee Regions to discretize the PPS. We find that the trends observed using Rhee Regions are quite similar to those we get with the PNZs (see Figure \ref{fig:tinfall_relation}). We then performed all the suitable analysis using the PNZs and, when the $\rm R_{200}$ limitation play a major role (Section \ref{sec:infall_rate_estimate}, for instance), we explore the appropriate Rhee Regions.